\documentclass[longauth]{aa}  

\usepackage{mathptmx}
\usepackage{amsmath,amsfonts,amssymb}
\usepackage{mathrsfs}
\usepackage{makeidx}
\usepackage{graphicx}
\usepackage{epstopdf}
\usepackage[colorlinks=True]{hyperref}
\usepackage{multirow}
\usepackage{rotating}
\usepackage{color}
\usepackage{tablefootnote}
\usepackage{lscape}
\usepackage{tablefootnote}
\usepackage[T1]{fontenc}
\usepackage{aecompl}
\usepackage[normalem]{ulem}
\usepackage{siunitx}

\usepackage{CJKutf8}

\newcommand{\LCDM}{\ensuremath{\Lambda\textrm{CDM}}}
\newcommand{\OmegaM}{\ensuremath{\Omega_{\mathrm{M}}}}

\newcommand{\Omegab}{\ensuremath{\Omega_{\mathrm{b}}}}
\newcommand{\Hnow}{\ensuremath{H_{0}}}
\newcommand{\sigmaeight}{\ensuremath{\sigma_{8}}}
\newcommand{\Msun}{\ensuremath{\mathrm{M}_{\odot}}}
\newcommand{\Msunh}{\ensuremath{h^{-1}\mathrm{M}_{\odot}}}

\newcommand{\Mpch}{\ensuremath{h^{-1}\mathrm{Mpc}}}

\newcommand{\Rfiveoo}{\ensuremath{R_{500}}}
\newcommand{\Mfiveoo}{\ensuremath{M_{500}}}

\newcommand{\redshift}{\ensuremath{z}}
\newcommand{\dif}{\ensuremath{\mathrm{d}}}

\newcommand{\rhocrit}{\ensuremath{\rho_{\mathrm{c}}}}

\newcommand{\CHANDRA}{\emph{Chandra}}
\newcommand{\XMMNEWTON}{XMM-\emph{Newton}}

\newcommand{\eROSITA}{\emph{eROSITA}}
\newcommand{\ROSAT}{\emph{ROSAT}}
\newcommand{\XXL}{\emph{XXL}}

\newcommand{\fcont}{\ensuremath{f_{\mathrm{cont}}}}
\newcommand{\rich}{\ensuremath{\lambda}}

\newcommand{\zcl}{\ensuremath{z_{\mathrm{cl}}}}
\newcommand{\zs}{\ensuremath{z_{\mathrm{s}}}}

\newcommand{\MPIV}{\ensuremath{M_{\mathrm{piv}}}}
\newcommand{\ZPIV}{\ensuremath{z_{\mathrm{piv}}}}
\newcommand{\Ez}{\ensuremath{E(z)}}
\newcommand{\Ezpiv}{\ensuremath{E(z_{\mathrm{piv}})}}

\newcommand{\rshear}{\ensuremath{\gamma_{+}}}
\newcommand{\gshear}{\ensuremath{g_{+}}}
\newcommand{\eplus}{\ensuremath{e_{+}}}

\newcommand{\gcros}{\ensuremath{g_{\times}}}
\newcommand{\ecros}{\ensuremath{e_{\times}}}
\newcommand{\lensingw}{\ensuremath{w}}
\newcommand{\sigmashape}{\ensuremath{\sigma_{\mathrm{shape}}}}
\newcommand{\deltaSigma}{\ensuremath{\Delta{\Sigma}_{\mathrm{m}}}}
\newcommand{\Sigmam}{\ensuremath{{\Sigma}_{\mathrm{m}}}}

\newcommand{\bz}{\ensuremath{b_{z}}}
\newcommand{\bwl}{\ensuremath{b_{\mathrm{WL}}}}
\newcommand{\Mwl}{\ensuremath{M_{\mathrm{WL}}}}
\newcommand{\sigmacrit}{\ensuremath{\Sigma_{\mathrm{crit}}}}
\newcommand{\lensingeff}{\ensuremath{\beta}}

\newcommand{\Xlabel}{\ensuremath{\mathcal{X}}}

\newcommand{\fmis}{\ensuremath{f_{\mathrm{mis}}}}
\newcommand{\Pmis}{\ensuremath{P_{\mathrm{mis}}}}
\newcommand{\Pcen}{\ensuremath{P_{\mathrm{cen}}}}
\newcommand{\sigmacen}{\ensuremath{{\sigma}_{\mathrm{cen}}}}
\newcommand{\sigmamis}{\ensuremath{{\sigma}_{\mathrm{mis}}}}

\newcommand{\fcl}{\ensuremath{f_{\mathrm{cl}}}}
\newcommand{\Ncl}{\ensuremath{N_{\mathrm{cl}}}}

\newcommand{\Lx}{\ensuremath{L_{\mathrm{X}}}}
\newcommand{\Lb}{\ensuremath{L_{\mathrm{b}}}}
\newcommand{\Tx}{\ensuremath{T_{\mathrm{X}}}}
\newcommand{\Yx}{\ensuremath{Y_{\mathrm{X}}}}
\newcommand{\Mg}{\ensuremath{M_{\mathrm{g}}}}

\newcommand{\rate}{\ensuremath{\eta}}
\newcommand{\truerate}{\ensuremath{\hat\eta}}
\newcommand{\brate}{\ensuremath{b_{\eta}}}

\newcommand{\Aeta}{\ensuremath{A_{\eta}}}
\newcommand{\Beta}{\ensuremath{B_{\eta}}}
\newcommand{\deltaeta}{\ensuremath{\delta_{\eta}}}
\newcommand{\gammaeta}{\ensuremath{\gamma_{\eta}}}
\newcommand{\sigmaeta}{\ensuremath{\sigma_{\eta}}}

\newcommand{\AF}{\ensuremath{A_{\mathrm{f}}}}
\newcommand{\BF}{\ensuremath{B_{\mathrm{f}}}}
\newcommand{\deltaF}{\ensuremath{\delta_{\mathrm{f}}}}
\newcommand{\gammaF}{\ensuremath{\gamma_{\mathrm{f}}}}
\newcommand{\sigmaF}{\ensuremath{\sigma_{\mathrm{f}}}}

\newcommand{\Awl}{\ensuremath{A_{\mathrm{WL}}}}
\newcommand{\Bwl}{\ensuremath{B_{\mathrm{WL}}}}

\newcommand{\gammawl}{\ensuremath{\gamma_{\mathrm{WL}}}}
\newcommand{\sigmawl}{\ensuremath{\sigma_{\mathrm{WL}}}}

\newcommand{\Blx}{\ensuremath{B_{L_{\mathrm{X}}}}}
\newcommand{\deltalx}{\ensuremath{\delta_{L_{\mathrm{X}}}}}
\newcommand{\gammalx}{\ensuremath{\gamma_{L_{\mathrm{X}}}}}
\newcommand{\sigmalx}{\ensuremath{\sigma_{L_{\mathrm{X}}}}}

\newcommand{\Blb}{\ensuremath{B_{L_{\mathrm{b}}}}}
\newcommand{\deltalb}{\ensuremath{\delta_{L_{\mathrm{b}}}}}
\newcommand{\gammalb}{\ensuremath{\gamma_{L_{\mathrm{b}}}}}
\newcommand{\sigmalb}{\ensuremath{\sigma_{L_{\mathrm{b}}}}}

\newcommand{\Bmg}{\ensuremath{B_{M_{\mathrm{ICM}}}}}
\newcommand{\deltamg}{\ensuremath{\delta_{M_{\mathrm{ICM}}}}}
\newcommand{\gammamg}{\ensuremath{\gamma_{M_{\mathrm{ICM}}}}}
\newcommand{\sigmamg}{\ensuremath{\sigma_{M_{\mathrm{ICM}}}}}

\newcommand{\At}{\ensuremath{A_{T_{\mathrm{X}}}}}
\newcommand{\Bt}{\ensuremath{B_{T_{\mathrm{X}}}}}
\newcommand{\deltat}{\ensuremath{\delta_{T_{\mathrm{X}}}}}
\newcommand{\gammat}{\ensuremath{\gamma_{T_{\mathrm{X}}}}}
\newcommand{\sigmat}{\ensuremath{\sigma_{T_{\mathrm{X}}}}}

\newcommand{\By}{\ensuremath{B_{Y_{\mathrm{X}}}}}
\newcommand{\deltay}{\ensuremath{\delta_{Y_{\mathrm{X}}}}}
\newcommand{\gammay}{\ensuremath{\gamma_{Y_{\mathrm{X}}}}}
\newcommand{\sigmay}{\ensuremath{\sigma_{Y_{\mathrm{X}}}}}

\newcommand{\ansAeta}{\ensuremath{ 0.124^{+0.022}_{-0.019} }}
\newcommand{\ansBeta}{\ensuremath{ 1.58^{+0.17}_{-0.14} }}
\newcommand{\ansdeltaeta}{\ensuremath{ 1.0^{+1.0}_{-1.4} }}
\newcommand{\ansgammaeta}{\ensuremath{ -0.44^{+0.81}_{-0.85} }}
\newcommand{\anssigmaeta}{\ensuremath{ 0.301^{+0.089}_{-0.078} }}

\newcommand{\ansAlx}{\ensuremath{ 3.36^{+0.53}_{-0.49} }}
\newcommand{\ansBlx}{\ensuremath{ 1.44^{+0.14}_{-0.13} }}
\newcommand{\ansdeltalx}{\ensuremath{ -0.07^{+1.26}_{-0.79} }}
\newcommand{\ansgammalx}{\ensuremath{ -0.51^{+0.93}_{-0.75} }}
\newcommand{\anssigmalx}{\ensuremath{ 0.120^{+0.138}_{-0.060} }}
\newcommand{\ansrhowllx}{\ensuremath{ 0.24^{+0.38}_{-0.67} }}

\newcommand{\ansrholxeta}{\ensuremath{ 0.35^{+0.29}_{-0.45}  }}

\newcommand{\ansAlb}{\ensuremath{ 9.2^{+1.6}_{-1.3} }}
\newcommand{\ansBlb}{\ensuremath{ 1.59\pm 0.14 }}
\newcommand{\ansdeltalb}{\ensuremath{ 0.2^{+1.3}_{-1.1} }}
\newcommand{\ansgammalb}{\ensuremath{ -0.45^{+1.00}_{-0.86} }}
\newcommand{\anssigmalb}{\ensuremath{ 0.102^{+0.143}_{-0.043} }}
\newcommand{\ansrhowllb}{\ensuremath{ 0.61^{+0.26}_{-0.58} }}

\newcommand{\ansrholbeta}{\ensuremath{ 0.38^{+0.28}_{-0.67}  }}

\newcommand{\ansAmg}{\ensuremath{ 1.08\pm 0.13 }}
\newcommand{\ansBmg}{\ensuremath{ 1.190^{+0.099}_{-0.118} }}
\newcommand{\ansdeltamg}{\ensuremath{ 0.40^{+0.78}_{-0.70} }}
\newcommand{\ansgammamg}{\ensuremath{ 0.32^{+0.59}_{-0.61} }}
\newcommand{\anssigmamg}{\ensuremath{ 0.074^{+0.063}_{-0.019} }}
\newcommand{\ansrhowlmg}{\ensuremath{ 0.08^{+0.60}_{-0.43} }}
\newcommand{\ansrhowletamg}{\ensuremath{ -0.30^{+0.32}_{-0.38} }}
\newcommand{\ansrhomgeta}{\ensuremath{ 0.63^{+0.23}_{-0.49}  }}

\newcommand{\ansAt}{\ensuremath{ 3.27^{+0.26}_{-0.31} }}
\newcommand{\ansBt}{\ensuremath{ 0.65\pm 0.11 }}
\newcommand{\ansdeltat}{\ensuremath{ -0.02^{+0.66}_{-0.70} }}
\newcommand{\ansgammat}{\ensuremath{ -1.03^{+0.54}_{-0.75} }}
\newcommand{\anssigmat}{\ensuremath{ 0.069^{+0.061}_{-0.014} }}
\newcommand{\ansrhowlt}{\ensuremath{ 0.47^{+0.36}_{-0.59} }}
\newcommand{\ansrhowletat}{\ensuremath{ 0.00^{+0.27}_{-0.57} }}
\newcommand{\ansrhoteta}{\ensuremath{ 0.65^{+0.22}_{-0.59}  }}

\newcommand{\ansAy}{\ensuremath{ 6.9^{+1.5}_{-1.2} }}
\newcommand{\ansBy}{\ensuremath{ 1.78^{+0.24}_{-0.20} }}
\newcommand{\ansdeltay}{\ensuremath{ -1.1^{+1.9}_{-1.2} }}
\newcommand{\ansgammay}{\ensuremath{ -1.43^{+1.20}_{-0.93} }}
\newcommand{\anssigmay}{\ensuremath{ 0.106^{+0.171}_{-0.047} }}
\newcommand{\ansrhowly}{\ensuremath{ 0.12^{+0.53}_{-0.50} }}

\newcommand{\ansrhoyeta}{\ensuremath{ 0.794^{+0.079}_{-0.344}  }}

\newcommand{\ansArch}{\ensuremath{ 36.2^{+3.7}_{-3.5} }}
\newcommand{\ansBrch}{\ensuremath{ 0.881^{+0.077}_{-0.088} }}
\newcommand{\ansdeltarch}{\ensuremath{ -0.56^{+0.91}_{-0.40} }}
\newcommand{\ansgammarch}{\ensuremath{ -0.46^{+0.54}_{-0.51} }}
\newcommand{\anssigmarch}{\ensuremath{ 0.274^{+0.078}_{-0.055} }}

\newcommand{\percent}{\ensuremath{\%}}

\newcommand{\appropto}{\mathrel{\vcenter{
  \offinterlineskip\halign{\hfil$##$\cr
    \propto\cr\noalign{\kern2pt}\sim\cr\noalign{\kern-2pt}}}}}

\makeatletter
\newcommand*\bigcdot{\mathpalette\bigcdot@{.6}}
\newcommand*\bigcdot@[2]{\mathbin{\vcenter{\hbox{\scalebox{#2}{$\m@th#1\bullet$}}}}}
\makeatother

\hypersetup{linkcolor=blue,filecolor=magenta,citecolor=cyan}
\begin{document} 
\begin{CJK*}{UTF8}{bsmi}

%
%

\title{
The eROSITA Final Equatorial-Depth Survey (eFEDS):
}

\subtitle{
X-ray observable-to-mass-and-redshift relations of galaxy clusters and groups with weak-lensing mass calibration from \\
the Hyper Suprime-Cam Subaru Strategic Program survey
}

%
%

\author{
I-Non~Chiu （邱奕儂）\inst{1}\fnmsep\inst{2}\fnmsep\inst{3}\fnmsep\thanks{e-mail: inchiu@sjtu.edu.cn}
\and
Vittorio~Ghirardini\inst{4}
\and
Ang~Liu\inst{4}
\and
Sebastian~Grandis\inst{5}
\and
Esra~Bulbul\inst{4}
\and
Y.~Emre~Bahar\inst{4}
\and\\
Johan~Comparat\inst{4}
\and
Sebastian~Bocquet\inst{5}
\and
Nicolas~Clerc\inst{6}
\and
Matthias~Klein\inst{5}
\and
Teng~Liu\inst{4}
\and
Xiangchong~Li\inst{7}\fnmsep\inst{8}
\and
Hironao~Miyatake\inst{9}\fnmsep\inst{10}\fnmsep\inst{11}\fnmsep\inst{8}
\and
Joseph~Mohr\inst{5}\fnmsep\inst{4}
\and
Surhud~More\inst{12}\fnmsep\inst{8}
\and
Masamune~Oguri\inst{13}\fnmsep\inst{8}\fnmsep\inst{7}
\and
Nobuhiro~Okabe\inst{14}\fnmsep\inst{15}\fnmsep\inst{16}
\and
Florian~Pacaud\inst{17}
\and
Miriam~E.~Ramos-Ceja\inst{4}
\and
Thomas~H.~Reiprich\inst{17}
\and
Tim~Schrabback\inst{17}
\and
Keiichi~Umetsu\inst{3}
}
\institute{
Tsung-Dao Lee Institute, and Key Laboratory for Particle Physics, Astrophysics and Cosmology, Ministry of Education, Shanghai Jiao Tong University, Shanghai 200240, China
\and
Department of Astronomy, School of Physics and Astronomy, and Shanghai Key Laboratory for Particle Physics and Cosmology, Shanghai Jiao Tong University, Shanghai 200240, China
\and
Academia Sinica Institute of Astronomy and Astrophysics (ASIAA), 11F of AS/NTU Astronomy-Mathematics Building, No.1, Sec. 4, Roosevelt Rd, Taipei 10617, Taiwan
\and
Max Planck Institute for Extraterrestrial Physics, Giessenbachstrasse 1, 85748 Garching, Germany
\and
Faculty of Physics, Ludwig-Maximilians-Universit\"{a}t, Scheinerstr. 1, 81679, Munich, Germany
\and
IRAP, Universite de Toulouse, CNRS, UPS, CNES, Toulouse, France
\and
Department of Physics, University of Tokyo, 7-3-1 Hongo, Bunkyo-ku, Tokyo 113-0033 Japan
\and
Kavli Institute for the Physics and Mathematics of the Universe (WPI), The University of Tokyo Institutes for Advanced Study (UTIAS), The University of Tokyo, 5-1-5 Kashiwanoha, Kashiwa-shi, Chiba, 277-8583, Japan
\and
Kobayashi-Maskawa Institute for the Origin of Particles and the Universe (KMI), Nagoya University, Nagoya, 464-8602, Japan
\and
Institute for Advanced Research, Nagoya University, Nagoya, 464-8601, Japan
\and
Division of Physics and Astrophysical Science, Graduate School of Science, Nagoya University, Nagoya 464-8602, Japan
\and
The Inter University Centre for Astronomy and Astrophysics, Ganeshkhind, Pune 411007, India
\and
Research Center for the Early Universe, University of Tokyo, Tokyo 113-0033, Japan
\and
Core Research for Energetic Universe, Hiroshima University, 1-3-1, Kagamiyama, Higashi-Hiroshima, Hiroshima 739-8526, Japan
\and
Physics Program, Graduate School of Advanced Science and Engineering, Hiroshima University, 1-3-1 Kagamiyama, HigashiHiroshima, Hiroshima 739-8526, Japan
\and
Hiroshima Astrophysical Science Center, Hiroshima University, 1-3-1 Kagamiyama, Higashi-Hiroshima, Hiroshima 739-8526, Japan
\and
Argelander-Institut f\"{u}r Astronomie (AIfA), Universit\"{a}t Bonn, Auf dem H\"{u}gel 71, 53121 Bonn, Germany
}

\date{}

%
%

\abstract{
We present the first weak-lensing mass calibration and X-ray scaling relations of galaxy clusters and groups selected in the \eROSITA\ Final Equatorial Depth Survey (eFEDS) observed by Spectrum Roentgen Gamma/\eROSITA\ over a contiguous footprint with an area of $\approx140$~deg$^2$, using the three-year (S19A) weak-lensing data from the Hyper Suprime-Cam (HSC) Subaru Strategic Program survey.
In this work, we study a sample of $434$ optically confirmed galaxy clusters (and groups) at redshift $0.01\lesssim\redshift\lesssim1.3$ with a median of $0.35$, of which $313$ systems are uniformly covered by the HSC survey to enable the extraction of the weak-lensing shear observable.
In a Bayesian population modeling, we perform a blind analysis for the weak-lensing mass calibration by simultaneously modeling the observed count rate \rate\ and the shear profile \gshear\ of individual clusters through the count-rate-to-mass-and-redshift (\rate--\Mfiveoo--\redshift) relation and the weak-lensing-mass-to-mass-and-redshift (\Mwl--\Mfiveoo--\redshift) relation, respectively, while accounting for the bias in these observables using simulation-based calibrations. 
As a result, the count-rate-inferred and lensing-calibrated cluster mass is obtained from the joint modeling of the scaling relations, as the ensemble mass spanning a range of $10^{13}\Msunh\lesssim\Mfiveoo\lesssim10^{15}\Msunh$ with a median of $\approx10^{14}\Msunh$ for the eFEDS sample.
With the mass calibration, we further model the X-ray observable-to-mass-and-redshift relations, including the rest-frame soft-band and bolometric luminosity (\Lx\ and \Lb), the emission-weighted temperature \Tx, the mass of intra-cluster medium \Mg, and the mass proxy \Yx, which is the product of \Tx\ and \Mg.
Except for \Lx\ with a steeper dependence on the cluster mass at a statistically significant level, we find that the other X-ray scaling relations all show a mass trend that is statistically consistent with the self-similar prediction at a level of $\lesssim1.7\sigma$.
Meanwhile, all these scaling relations show no significant deviation from the self-similarity in their redshift scaling. 
Moreover, no significant redshift-dependent mass trend is present.
This work demonstrates the synergy between the \eROSITA\ and HSC surveys in preparation for the forthcoming first-year \eROSITA\ cluster cosmology.
}

%
%


%
%

\titlerunning{HSC weak-lensing mass calibration and X-ray scaling relations}
\authorrunning{Chiu et al.}

\maketitle

%
%

\section{Introduction}
\label{sec:introduction}

The probe based on the abundance of galaxy clusters plays a crucial role in constraining cosmology by utilizing the sample of clusters selected in the optical \citep{costanzi18,costanzi21,to21}, X-rays \citep{mantz15,schellenberger17,pacaud18}, and millimeter-wavelength bands \citep{PlanckCollaboration2015b,bocquet15,deHaan16,bocquet19} through the Sunyaev-Zel'dovich \citep[SZ;][]{sunyaev72} effect.
The number density of galaxy clusters as a function of mass, the  so-called halo mass function, is sensitive to \OmegaM, the density parameter of matter in the Universe, and \sigmaeight, the dispersion of linear density fluctuations on a comoving scale of $8\Mpch$.
Therefore, measurements of halo mass functions over a wide range of mass and redshift  essentially constrain the history of the cosmic expansion as well as the growth of large-scale structures, enabling a cosmological tool that is independent of and is as competitive as other methods.

A necessary ingredient in cluster cosmology is a large sample of galaxy clusters with a well-understood selection function.
With recent wide and deep surveys in the optical, such as the Dark Energy Survey \citep{des05}, a large sample of galaxy clusters and groups has been constructed out to redshift $\redshift\approx0.6$ and over a footprint with an area of more than a thousand square degrees.
A sample of optically selected clusters is now even extended to a much higher redshift at $\redshift\gtrsim1$ \citep[e.g.,][]{oguri18}.
Although a sizable sample of optically selected clusters has largely improved cosmological constraints, modeling the selection function of the clusters is challenging, mainly because of the projection effect \citep{zu17,costanzi19,sunayama20}, and could result in a systematic bias in cosmological parameters \citep{desclustercosmology20}.

On the other hand, searching for clusters based on the intra-cluster medium (ICM), either in X-rays or through the SZ effect, provides a clean sample for cosmological studies.
The large SZ survey carried out by the South Pole Telescope  \citep{bleem15}  has enabled the construction of a highly pure and complete sample of clusters out to redshift $\redshift\approx1.8$, based on which unbiased cosmological constraints have been obtained \citep{bocquet19}.
An even larger sample of SZ-selected clusters has recently been released by the Atacama Cosmology Telescope survey \citep{hilton21}.
However, current SZ surveys are limited to massive clusters because of the sensitivity of detectors, or only scan a small fraction of the sky, resulting in a sample of relatively small size.
To extend the success of the SZ-based cosmology to the low-mass regime, it is therefore necessary to combine the SZ sample with those from other surveys \citep[e.g.,][]{costanzi21}.
In terms of medium-size X-ray surveys with advanced telescopes, such as the \XXL\ survey \citep{peirre16}, a large number of clusters down to a low-mass regime has be obtained \citep{adami18}, although mainly at low redshift because of the cosmological dimming of bremsstrahlung emissions from clusters.
Meanwhile, the first all-sky survey in X-rays carried out by the \ROSAT\ mission \citep{rosat} had delivered the largest X-ray cluster catalog ever across the whole sky \citep{bohringer01,bohringer04,piffaretti11,boller16}, but the poor resolution and shallow depth have limited cosmological studies in practice.
However, this situation is no longer true with the ongoing all-sky X-ray survey carried out by the state-of-the-art telescope, \eROSITA.

The \eROSITA\ \citep[\textit{e}xtended \textit{RO}entgen \textit{S}urvey with an \textit{I}maging \textit{T}elescope \textit{A}rray;][]{merloni12,predehl21} is an X-ray space telescope on board the Russian-German ``Spectrum-Roentgen-Gamma'' (SRG) satellite \citep{sunyaev21}, which was successfully launched on July 13, 2019.
The main goal of the \eROSITA\ mission is to unveil the nature of dark energy by carrying out the \eROSITA\ All-Sky Survey (eRASS), which is to map the whole sky in X-rays for four years.
By doing so, the eRASS will deliver the largest sample of ICM-selected galaxy clusters to date and discover more than 100 000 clusters at the end of the four-year all-sky survey \citep{borm14}.
With this revolutionary sample, cluster-based cosmological analyses carried out as part of the \eROSITA\ mission will provide unprecedented power with which to constrain cosmology \citep{pillepich12,pillepich18}.

The key to the success of \eROSITA, and cluster cosmology in general, is an accurate mass calibration on an observable-to-mass-and-redshift scaling relation, which links the observed mass proxy to the  true cluster mass at the cluster redshift in order to construct unbiased halo mass functions \citep{pratt19}.
The technique of weak gravitational lensing (hereafter weak lensing) has been considered as the most optimal way to calibrate cluster mass, with huge success in practice \citep{umetsu14,vonderlinden14b,vonderlinden14a,hoekstra15,okabe16,
schrabback18,dietrich19,okabe19,mcclintock19}.
A forecast from  \cite{grandis19} shows that the inclusion of weak-lensing mass calibrations using advanced optical surveys could significantly tighten the constraints obtained from the \eROSITA\ sample, with uncertainties on \OmegaM, \sigmaeight, and the dark energy equation of state $w$ at levels of a few percent.

\begin{figure*}
\centering
\resizebox{\textwidth}{!}{
\includegraphics[scale=1]{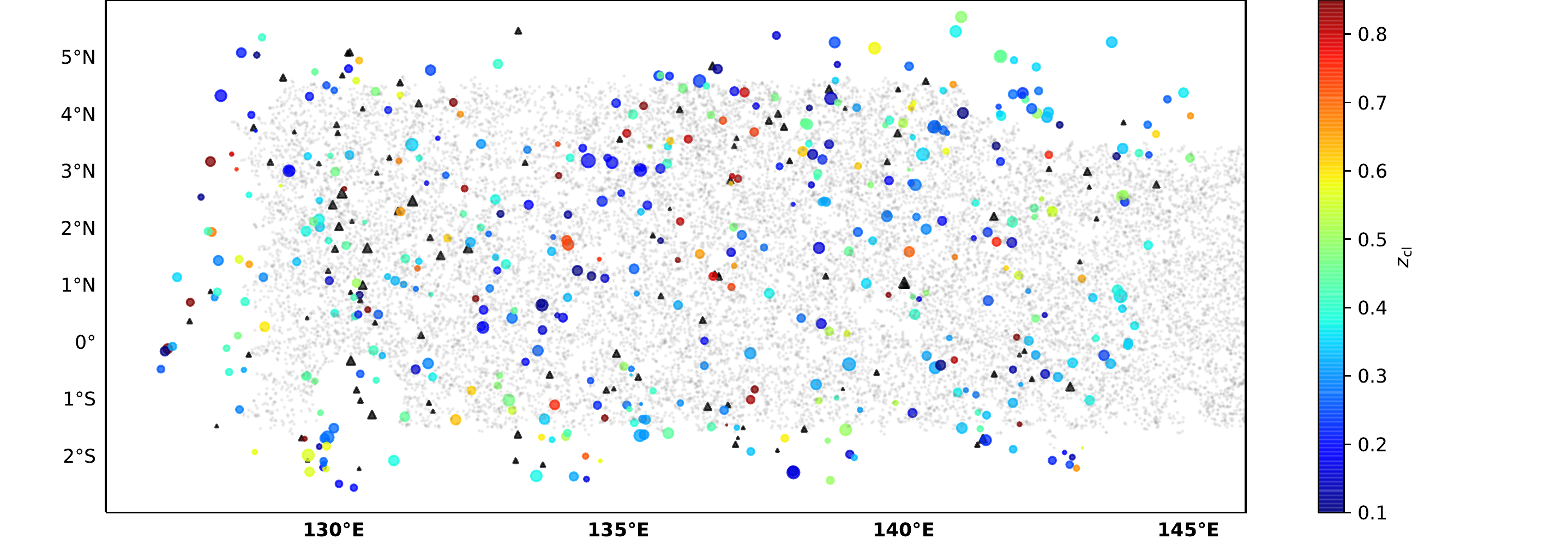}
}
\caption{
Angular distributions of eFEDS clusters and the available weak-lensing data set from the HSC survey.
Out of 542 eFEDS clusters, there are $434$ clusters as a secure sample, i.e., $\fcont<0.2$ (see Sect.~\ref{sec:clustersample}).
These clusters are represented by the circles, color-coded by their observed redshifts \zcl, with sizes proportional to their observed count rate \rate.
The other clusters with $\fcont\geq0.2$ are represented by the triangles. 
The underlying gray points represent a subset of sources that are randomly drawn from the HSC three-year (S19A; see Sect.~\ref{sec:hsc}) weak-lensing catalog.
}
\label{fig:footprint}
\end{figure*}

However,  a weak-lensing analysis of high-redshift clusters is challenging, requiring substantially deep imaging with good resolution to resolve background sources at even higher redshift.
It is made feasible by using pointing observations from space \citep{jee11,schrabback18,schrabback21} or deep high-resolution near-infrared imaging \citep{schrabback18b} for small samples, but not for a large number of clusters.
It is also important to stress that a uniform weak-lensing mass calibration  across the whole redshift range is needed; otherwise systematic errors due to inhomogeneity would be introduced \citep{chiu16a}.
To achieve a high-quality weak-lensing mass calibration for high-redshift clusters ($\redshift\gtrsim0.7$) that is as consistent as the calibration available for low redshift, the ongoing Hyper Suprime-Cam (HSC) Subaru Strategic Program survey \citep{aihara18a} is currently the only available resource for a large sample over the \eROSITA\ footprint.
The weak-lensing mass calibration out to redshift $\redshift\gtrsim1.1$ based on the HSC data sets was performed for a large sample of ($\gtrsim1700$) optically selected clusters over a footprint with an area of $\approx140$~deg$^2$ \citep{murata19}.
More recently, a weak-lensing analysis using the HSC data sets was carried out for an X-ray-selected sample of galaxy groups and clusters detected in the \XXL\ survey \citep{umetsu20}, clearly demonstrating the potential of the HSC weak-lensing data for the mass calibration of \eROSITA\ clusters.

To fully explore the survey capability of \eROSITA\ and its synergy with the HSC survey, the \eROSITA\ team began  the \eROSITA\ Final Equatorial Depth Survey (eFEDS) in a footprint with an area of $\approx140$~deg$^2$  significantly overlapping the HSC survey.
By design, the depth of the eFEDS reaches the average full depth of the eRASS in the equatorial area, thereby serving as a performance verification phase before the start of the main survey.
In this work, we make use of the latest HSC imaging (S20A) for the optical confirmation and redshift measurement of clusters selected in the eFEDS field, followed by the weak-lensing mass calibration using the three-year HSC data \citep[S19A][]{li21}, which are proprietary and will be released as an incremental data set of the third HSC Public Data Release.
The goal of this paper is to study various X-ray observable-to-mass-and-redshift scaling relations with the mass calibration using the HSC weak-lensing data, setting the stage for extending this analysis to the main eRASS work in the future.

The structure of this paper is as follows.
In Sect.~\ref{sec:data}, we describe the cluster sample and the data used in the analysis.
The full description of the weak-lensing analysis is given in Sect.~\ref{sec:wlanalysis}.
The simulation-based calibration of the weak-lensing mass and the observed count rate in X-rays is given in Sect.~\ref{sec:simulations}, followed by the modeling of scaling relations presented in Sect.~\ref{sec:xtom_modeling}.
The results and discussions are given in Sect.~\ref{sec:results}.
We discuss the potential systematic uncertainty in this work in Sect.~\ref{sec:sys}.
Finally, conclusions are made in Sect.~\ref{sec:conclusions}.
Throughout this paper, we assume a fiducial cosmology, namely a flat \LCDM\ cosmology with
$\OmegaM=0.3$,
a mean baryon density $\Omegab=0.05$,
the Hubble parameter $\Hnow = h\times100$\,km\,s\,$^{-1}$\,Mpc$^{-1}$ with
$h=0.7$, $\sigmaeight = 0.8$, and 
a spectral index of the primordial power spectrum $n_{\mathrm{s}}=0.95$.
While modeling the scaling relations, we vary the cosmological parameters around those of the fiducial cosmology, with the aim being to maintain the flexibility needed in a future cosmological analysis.
The cluster true mass \Mfiveoo\ is defined by a sphere with a radius of $\Rfiveoo$, such that the enclosed mass density is $500$ times the critical density $\rhocrit(\redshift)$ of the Universe at the cluster redshift.
All quoted errors represent the $68\percent$ confidence level (i.e., $1\sigma$) throughout this work, unless otherwise stated.
The notation $\mathcal{N}(x,y^2)$ ($\mathcal{U}(x,y)$) stands for a normal distribution with the mean $x$ and the standard deviation $y$ (a uniform distribution between $x$ and $y$).

%
%

\section{The cluster sample and data sets}
\label{sec:data}

In Sect.~\ref{sec:efeds}, a brief description of the eFEDS is given.
We then briefly introduce the Subaru HSC survey and the weak-lensing data in Sect.~\ref{sec:hsc}.
The cluster sample selected in the eFEDS is described in Sect.~\ref{sec:clustersample}, while their X-ray measurements are presented in Sect.~\ref{sec:xray_data}.

\subsection{The eROSITA Final Equatorial Depth Survey}
\label{sec:efeds}

The eFEDS is a small survey covering an area of $\approx140$~deg$^2$.
The depth of the eFEDS reaches an exposure time of $\approx2.2$~$k$sec, roughly the average full-depth of the eRASS, and therefore serves as a performance verification phase for \eROSITA\ science, by design.

The survey took place at an orbit around the L2 point in a scanning mode.
The field of view of \eROSITA\ is $\approx1$~deg$^2$.
The imaging is collected by seven  separate CCDs,  each with its own mirror, with good detector uniformity (no chip gaps).
The on-axis half-energy-width is $\approx18\arcsec$ at $1.49$~$k$eV, with an average of $\approx26\arcsec$ for the whole field of view \citep{predehl21}.
The imaging quality is excellent; the accuracy of a source location is better than $10\arcsec$ with a typical positional uncertainty at the level of $\approx4.6\arcsec$ ($1\sigma$).
The survey footprint consists of four scanning-mode units spanning a range of $\approx126^{\circ}$ to $\approx146^{\circ}$ ($\approx-3^{\circ}$ to $\approx+6^{\circ}$) in Right Ascension (Declination) in total.
The footprint of the eFEDS is visualized in Fig.~\ref{fig:footprint}.

The uniqueness of the eFEDS survey is that the footprint largely overlaps the HSC survey, which provides weak-lensing data products of the highest quality to date for a large sample of clusters.
This is especially true for galaxy clusters at high redshift ($\redshift\gtrsim0.7$), where a weak-lensing study of a large sample from the ground is only feasible using the HSC survey.
This provides a timely opportunity to statistically calibrate the mass of eROSITA-detected clusters out to redshifts beyond unity before the era of the \textit{Euclid} mission \citep{laureijs10} or the Legacy Survey of Space and Time \citep[LSST;][]{ivezic08} carried out by the Vera C. Rubin Observatory.
Therefore, the synergy between the eFEDS and the HSC survey presented in this paper lays the foundation for future work on this topic.

\subsection{The HSC survey and the weak-lensing data}
\label{sec:hsc}

The HSC survey \citep{aihara18a} is an imaging survey carried out as part of a Subaru Strategic Program with the goal being to map a sky area of 1100\,deg$^2$ through five broadband filters ($grizy$).
This is done using the wide-field camera Hyper Suprime-Cam \citep{miyazaki15,miyazaki18} installed on the 8.2\,m Subaru Telescope. 
The main goal of the HSC survey is to perform state-of-the-art weak-lensing studies, paving a way forward for science with the upcoming LSST.

The HSC survey comprises a three-layer imaging scheme: WIDE, DEEP, and UltraDEEP.
The $5\sigma$ limiting magnitudes of a $2\arcsec$ aperture in the WIDE layer are $26.5$~mag, $26.1$~mag, $25.9$~mag, $25.1$~mag, and $24.4$~mag for $g$-, $r$-, $i$-, $z$-, and $y$-band, respectively.
This represents the deepest imaging survey at the achieved area to date.
Moreover, the imaging quality is remarkable with a mean seeing of $0.58\arcsec$ at $i$-band \citep{mandelbaum18}.
With this high-quality imaging, together with a unique combination of depth and area, the data from the HSC survey are excellent for weak-lensing studies over a large sample of clusters.

In this work, we use the three-year shape catalog constructed from the second HSC Public Data Release (S19A) for weak-lensing studies.
This shape catalog is not released yet, and the full details are described in \cite{li21}.
In what follows, we provide a brief summary of the three-year shape catalog.
The shape measurement is carried out in the $i$-band imaging following the same methodology in constructing the first-year shape catalog (S16A) with a comprehensive description given in \cite{mandelbaum18}.
The shape measurement is rigorously calibrated against the image simulations \citep{mandelbaum18b}, such that the systematic uncertainty of the multiplicative bias meets the requirement of $\lesssim1.7\percent$.
Moreover, various null tests, as well as those on the map level \citep{oguri18b}, are all statistically consistent with zero.
Only galaxies that satisfy the weak-lensing flags with an $i$-band magnitude of smaller than $24.5$~mag are contained in the shape catalog.
With the high-quality of imaging, the source density in the HSC survey reaches $\approx22$ per arcmin$^2$.
It is worth mentioning that the same method of the shape measurement was used to construct the first-year weak-lensing catalog, which has been used to derive unbiased cosmic shears and cosmological constraints \citep{hikage19,hamana20}.
The three-year catalog covers an area of $\approx430$~deg$^2$, in which the footprint of the GAMA09 field significantly overlaps with the eFEDS.
As a result, the majority ($\approx72\percent$) of the eFEDS clusters are covered by the HSC shape catalog, from which the shear profile is derived.

It is worth mentioning that the biggest benefit of using the latest weak-lensing catalog instead of the first-year one is the increase in  the area coverage.
The number of eFEDS clusters covered by the S16A weak-lensing data is only $\approx50\percent$ of those currently covered by the S19A.
This implies an improvement of $\sqrt{2} -1 \approx 40\percent$ in terms of the signal-to-noise ratio of weak-lensing measurements from S16A to S19A.

\begin{figure}
\resizebox{0.5\textwidth}{!}{
\includegraphics[scale=1]{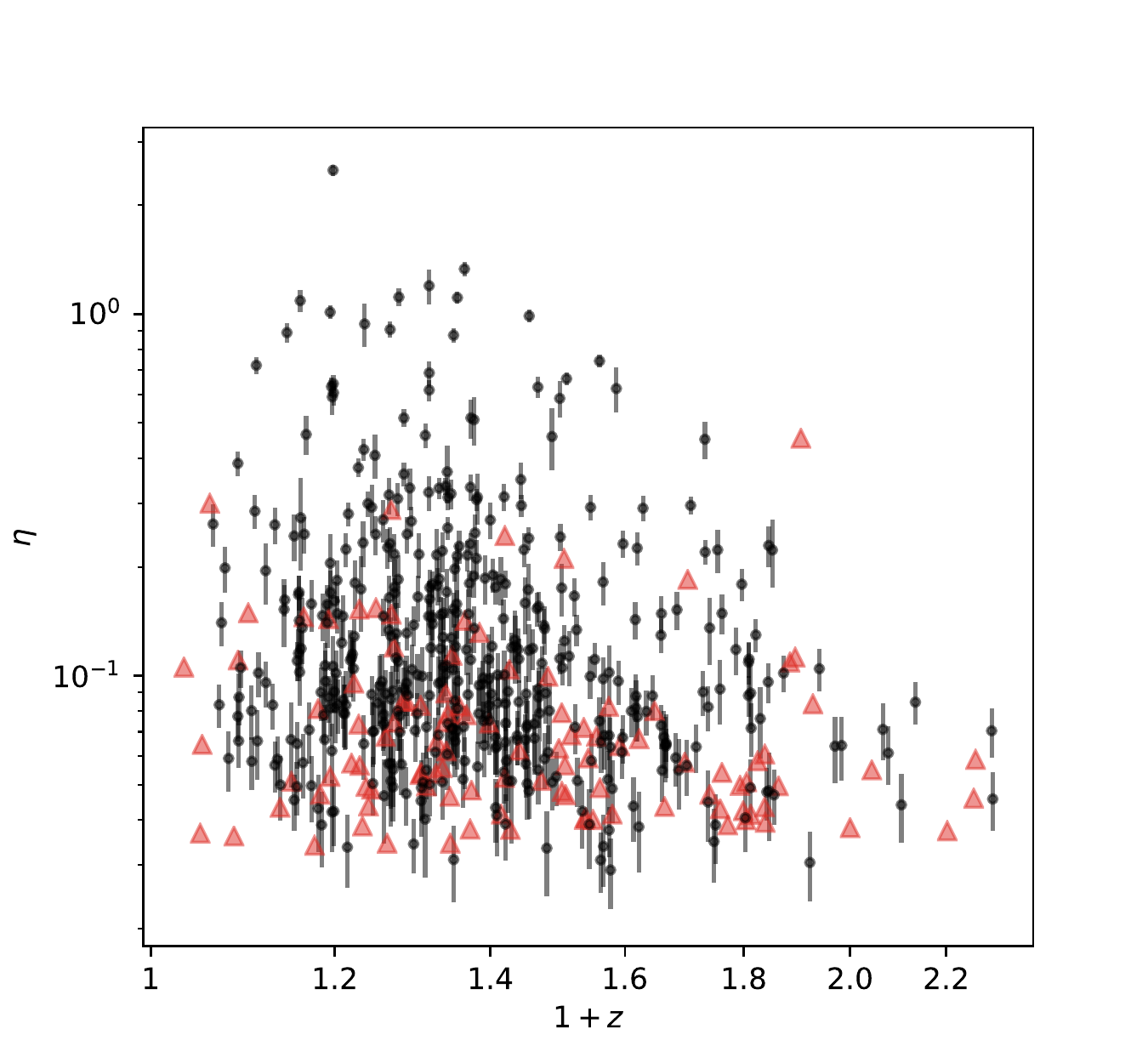}
}
\caption{
Observed count rate \rate\ (in the unit of counts per second) and the redshift \redshift\ of the 542 eFEDS clusters.
The $434$ secure clusters with $\fcont<0.2$ (see Sect.~\ref{sec:clustersample}) are shown as the circles, while the rest are marked by red triangles.
}
\label{fig:sample}
\end{figure}

\subsection{The eFEDS cluster sample}
\label{sec:clustersample}

The galaxy clusters in the eFEDS are identified by the source-detection algorithm in the \texttt{eSASS} pipeline.
The details of the X-ray source detection are given in \cite{brunner21}, to which we refer readers for a complete description.
In what follows, we give a brief summary of the cluster finding algorithm in the eFEDS.

The source detection of galaxy clusters is performed on the X-ray imaging at the energy band of $0.2-2.3$~$k$eV.
The source detection is performed with the \texttt{eSSASusers\_201009} software \citep{brunner21}. 
The tasks are based on the sliding-box algorithm, which searches for sources that are brighter than the calculated background in the X-ray images.
This process is iteratively performed to obtain a reliable estimation of background noise.
A list of X-ray sources is created after the iterations.
Then, the $\beta$-model \citep{cavaliere76} convolved with the derived point spread function is fitted to each source to derive the X-ray properties, including for example the detection likelihood $\mathcal{L}_{\mathrm{det}}$, the extent likelihood  $\mathcal{L}_{\mathrm{ext}}$, and the observed count rate \rate\ at $0.2-2.3$~$k$eV.
We note that the observed count rate \rate\ is estimated using the total photon counts evaluated by integrating the
best-fit model that depends on the detection radius, which is 
a characteristic scale to facilitate the detection of an extended source in the \texttt{eSASS} pipeline.
Thus, the observed count rate \rate\ is 
different from that estimated within the cluster radius \Rfiveoo.
We estimate the bias\footnote{We note that this bias refers to the difference between the observed count rate (returned by the \texttt{eSASS} pipeline) and the underlying true count rate enclosed by the cluster radius \Rfiveoo.} of the observed count rate \rate\ in Sect.~\ref{sec:etabias}.
Finally, the thresholds of $\mathcal{L}_{\mathrm{det}} > 5$ and $\mathcal{L}_{\mathrm{ext}} > 6$ are applied to create the catalog of cluster candidates, resulting in 542 systems in total in the eFEDS.
These 542 clusters and their X-ray properties are presented in \cite{liu21}.

The sample of 542 eFEDS clusters are then processed through the optical confirmation by running the Multi-Component Matched Filter (MCMF) algorithm \citep{klein18}.
The MCMF is run on the imaging from both the HSC S20A data set and the DESI Legacy Imaging Survey \citep{dey19}.
Full details of the optical confirmation are given in \cite{klein21}.
Briefly, the combination of the MCMF run on these two data sets yields, for each cluster, the photometric redshift estimate \zcl, the optical richness \rich, and the estimator \fcont\ of the probability that the cluster appears as a superposition of galaxies along the line of sight by chance.
The quantity of \fcont\ indicates the level of contamination in the optical confirmation.
For instance, a value of $\fcont=0.2$ means that there is a $20\percent$ chance of detecting a candidate with the derived richness \rich\ at the redshift \zcl\ in a random field \citep[see also][]{klein19}.
A value higher than this represents an even higher probability of a random superposition \citep{klein21}.
A cut on \fcont\ is equivalent to a reduction in the cluster contamination in the initial catalog, under the assumption that the contamination from X-rays and the optical is uncorrelated \citep[see also the discussion at length in][]{grandis20}.
The contamination from point sources in the initial catalog of eFEDS clusters is suggested to be at a level of $\approx20\percent$ using simulations \citep{liu21}.
Applying the cut of $\fcont<0.2$ results in a deduction of $18\percent$ (or $99$ clusters) in the sample size, which is consistent with the expectation from the simulation.
The residual contamination from nonextended sources is suggested to be at a level of $\approx4\percent$ with the cut of $\fcont<0.2$ \citep{klein19}, as suggested by running the MCMF algorithm on the sample of ROSAT clusters \citep{boller16}.
In this work, we apply the cut of $\fcont<0.2$ to the parent sample to construct a secure sample of $434$ clusters, which  is used to study their scaling relations.
The sample in the plane of the observed count rate \rate\ and redshift \redshift\ is presented in Fig.~\ref{fig:sample}.

In addition to the X-ray scaling relations, we also perform a modeling of the richness-to-mass-and-redshift (\rich--\Mfiveoo--\redshift) relation for the eFEDS clusters following the exact methodology described in Sect.~\ref{sec:wlsrmcalib}.
The results are shown in Appendix~\ref{sec:richness}.

\subsection{The X-ray measurements from the eFEDS}
\label{sec:xray_data}

In this work, we use five X-ray quantities as follow-up observables, which are the soft-band luminosity \Lx\ at the energy band of $0.5-2k$eV, the bolometric luminosity \Lb, the emission-weighted temperature \Tx, the gas mass \Mg, and the mass proxy \Yx\ of the ICM for each cluster.
These measurements will be fully presented in a forthcoming paper \citep{bahar21}, and the details of the X-ray analysis are given in \cite{ghirardini21} and \cite{liu21}.
We therefore only outline the important steps here.

We stress that these follow-up X-ray measurements (\Lx, \Lb, \Tx, \Mg, \Yx) are all estimated within the cluster radius ($<\Rfiveoo$) including the cluster core, because we do not have enough X-ray photons to derive these quantities with the core excised \citep{bahar21}.
To derive the cluster radius \Rfiveoo, the cluster mass needs to be known before the follow-up X-ray analysis.
In this work, the cluster mass \Mfiveoo\ is inferred from a joint modeling of the observed count rate and the weak-lensing shear profile (see Sect.~\ref{sec:wlmcalib}).
As the cluster mass based on the lensing modeling is still blinded at the time these X-ray observables are estimated, an early unblinding exclusively to the \eROSITA\ team is carried out to estimate the ``correct'' cluster radius \Rfiveoo\ (see more details in Sect.~\ref{sec:blinding}), which is then used as the aperture size to derive the X-ray observables.
After deriving the follow-up X-ray observables, the \eROSITA\ team delivers them to the HSC team for the modeling of the X-ray scaling relations.
It is important to note that the modeling of the follow-up X-ray scaling relations is solely led by the HSC team, which has no exchange of any mass-related information with the \eROSITA\ team after the early unblinding.
Therefore, the analysis of the X-ray scaling relations in this work is still performed blindly to minimize confirmation bias.

With the exception of \Tx, the X-ray follow-up measurements (\Lx, \Lb, and \Mg) are derived based on the fitting to the X-ray imaging extracted in the soft-energy band of $0.5-2k$eV.
The fitting model is centered on the X-ray center of each cluster, accounting for the sky and instrumental backgrounds.
The ICM density profile is parameterized by the 
\citet{vikhlinin06} model,
which is then projected onto the sky and fitted to the X-ray surface brightness profile.
In this way, the ICM mass \Mg\ is obtained by integrating the best-fit model to the cluster radius.
Meanwhile, the luminosities (\Lx\ and \Lb) are obtained from a conversion factor depending on the temperature and the cluster redshift.
The conversion factor is derived at each step of the Markov Chain Monte Carlo (MCMC) chain of the spectral analysis, and the full distribution of the conversion factor is used to compute the luminosity.
This effectively marginalizes the derived luminosity over the uncertainty arising from the spectral analysis, especially from the temperature measurements.
Moreover, this also allows us to estimate the luminosity even for clusters whose temperatures are poorly constrained.
We note that this conversion factor weakly depends on the temperature; there is only a factor of two difference between temperatures of 2~$k$eV and 30 $k$eV.

The temperature \Tx\ is obtained by the X-ray spectral analysis using the \texttt{XSPEC} \citep{arnaud96}.
The spectral analysis is challenging in the eFEDS observation because of the low number of X-ray photons.
The net number of X-ray counts obtained per cluster varies from $\approx5$ to $\approx1000$, 
with the first quartile, median, and the third quartile of $43$, $75$, and $147$, respectively.
Meanwhile, the signal-to-noise ratio (S/N) of such X-ray data sets extracted at the cluster radius \Rfiveoo\ spans a range of $0.02$ and $38.7$ with a median (mean) of $5.2$ ($6.5$).

Given the low photon counts, we are not able to robustly measure the temperature \Tx\ (thus \Yx) for all clusters.
In this work, we therefore restrict the modeling of the \Tx\ and \Yx\ scaling relations to a subsample of eFEDS clusters, for which the temperature can be reliably measured.
Specifically, the subsample is constructed by selecting clusters with a detection likelihood of $\mathcal{L}_{\mathrm{det}} > 50$ and an extent likelihood of $\mathcal{L}_{\mathrm{ext}} > 50$, resulting in $64$ clusters.
With these cuts on $\mathcal{L}_{\mathrm{det}}$ and $\mathcal{L}_{\mathrm{ext}}$, we construct a highly pure and complete subsample of eFEDS clusters with a median (mean) S/N of $13$ ($15$).
We assess the temperature measurement by comparing the temperatures obtained within \Rfiveoo\ to that within a fixed aperture with a physical radius of $500$~$k$pc, within which we can measure the temperature with the highest S/N for a maximal number of clusters \citep{liu21}. 
Given such a low number of X-ray photons, we do not expect significant differences between $\Tx(R<\Rfiveoo)$ and $\Tx(R<500~k\mathrm{pc})$.
As a result, we find that the difference between $\Tx(R<\Rfiveoo)$ and $\Tx(R<500~k\mathrm{pc})$ for the subsample with $\mathcal{L}_{\mathrm{det}} > 50$ and $\mathcal{L}_{\mathrm{ext}} > 50$ has an inverse-variance-weighted mean of $\left\langle\Tx(R<\Rfiveoo) - \Tx(R<500~k\mathrm{pc})\right\rangle = \left(0.06\pm0.36\right)~k\mathrm{eV}$,
showing no significant trend as a function of S/N.
This is consistent with \cite{giles16}, where they found no clear difference between the temperatures extracted within \Rfiveoo\ and $300$~$k$pc using the spectra with low counts.
We stress that the modeling of the X-ray scaling relations throughout this work is based on the temperature measurements $\Tx(R<\Rfiveoo)$, instead of $\Tx(R<500~k\mathrm{pc})$.

In summary, the modeling of the \Tx\ and \Yx\ scaling relations is performed based on the subsample of $64$ clusters with $\mathcal{L}_{\mathrm{det}} > 50$ and $\mathcal{L}_{\mathrm{ext}} > 50$, of which there are $47$ systems with weak-lensing measurements.
Meanwhile, we perform the modeling of other X-ray follow-up scaling relations (\Lx, \Lb, and \Mg) based on the full secure sample of $434$ eFEDS clusters, of which there are $313$ systems with weak-lensing measurements.

%
%

\section{Weak-lensing analysis}
\label{sec:wlanalysis}

In this section, we first provide a brief overview of the theory behind the weak-lensing analysis in Sect.~\ref{sec:basics}.
We then describe the blinding strategy in our weak-lensing analysis in Sect.~\ref{sec:blinding}.
Finally, we describe the weak-lensing analysis of the eFEDS clusters using the HSC data sets in Sects.~\ref{sec:source_selection} to \ref{sec:wlmodeling}.
Our goal is to extract the observed shear profiles for individual clusters and to quantify the systematic errors that will be accounted for in our forward-modeling approach.
We note that we largely follow the weak-lensing analysis as presented in \cite{umetsu20} and \cite{miyatake19}, to which we refer readers for more details.

\subsection{Weak-lensing basic theory}
\label{sec:basics}

A brief overview of weak gravitational lensing with emphasis on applications to galaxy clusters is given in this section.
We refer readers to \cite{bartelmann01}, \cite{hoekstra13}, and \cite{umetsu20b} for more details.

Cosmic structures deflect light rays, resulting in effects of gravitational lensing.
Galaxy clusters in the limit of the thin lens approximation act as a single lens embedded in a homogeneous universe, such that all sources behind clusters are lensed.
The  lensing strength of a source at redshift \zs\ arising from a galaxy cluster at redshift \zcl\ depends on the distances between the cluster, the observer, and the source.
This is characterized by the critical surface density \sigmacrit\ defined as
\begin{equation}
\label{eq:crit}
\sigmacrit\left(\zcl,\zs\right) = \frac{c^2}{4\pi G}\frac{
1
}{
D_{\mathrm{l}}\left(\zcl\right) \lensingeff\left(\zcl,\zs\right)
}
\, ,
\end{equation}
in which $G$ is Newton's constant, $c$ is the speed of light, $D_{\mathrm{l}}\left(\zcl\right)$ is the angular diameter distance of the cluster, and $\lensingeff$ is the lensing efficiency of the source,
\begin{equation}
\label{eq:beta}
\lensingeff\left(\zcl,\zs\right) = 
\begin{cases}
\frac{
D_{\mathrm{l},\mathrm{s}}\left(\zcl,\zs\right)
}{
D_{\mathrm{s}}\left(\zs\right)
} &~\mathrm{if}~\zs>\zcl \\ 
0 &~\mathrm{if}~\zs\leq\zcl 
\end{cases}
\, ,
\end{equation}
where $D_{\mathrm{l},\mathrm{s}}\left(\zcl,\zs\right)$ and $D_{\mathrm{s}}\left(\zs\right)$ are the angular diameter distances of the  cluster--source and observer--source pairs, respectively.

In terms of weak-shear effects, gravitational lensing arising from clusters distorts the imaging of background sources, resulting in a coherent distortion in the tangential direction around the centers of the clusters.
By measuring the shape of background sources, this effect can be quantified statistically by the  quantity of the ``reduced shear'' \gshear, defined as
\begin{equation}
\label{eq:reducedshear_def}
\gshear = \frac{\rshear}{1 - \kappa} \, ,
\end{equation}
where \rshear\ and $\kappa$ are the tangential shear and convergence of the cluster, respectively.
In an azimuthal average with respect to a cluster center, 
the tangential shear $\rshear\left(R\right)$ at the projected radius $R$ describes the differential surface mass density \deltaSigma\ of the cluster at $R$ with respect to the critical surface density inferred from the source, namely,
\begin{equation}
\label{eq:rshear_def}
\rshear\left(R\right) = \frac{
\deltaSigma\left(R\right)
}{
\sigmacrit
} \, .
\end{equation}
Meanwhile, the convergence $\kappa\left(R\right)$ is the ratio of the projected surface mass density $\Sigmam\left(R\right)$ to the critical surface density, as
\begin{equation}
\label{eq:kappa_def}
\kappa\left(R\right) = \frac{
\Sigmam\left(R\right)
}{
\sigmacrit
} \, .
\end{equation}
We note that $\deltaSigma\left(R\right) = \Sigmam\left(<R\right) - \Sigmam\left(R\right)$.

The shear profile, $\gshear\left(R\right)$, can then be statistically obtained by measuring the azimuthal average of tangential ellipticities over a large sample of background sources around the center of clusters (see Sect.~\ref{sec:gshear}). 
To infer the mass of a cluster from the observed shear profile, the redshift of sources must be known to compute the critical surface density \sigmacrit\ to interpret \rshear\ and $\kappa$ (see Sect.~\ref{sec:wlmodeling} for the modeling of \sigmacrit).
 
\subsection{Blinding analysis}
\label{sec:blinding}

We carry out the weak-lensing analysis in a blind fashion to avoid confirmation bias.
The blinding strategy is decided by the HSC weak-lensing team and has been widely used in cosmological analyses, such as those of \cite{hikage19} and \cite{hamana20}.
A full description of the blinding strategy is presented in \cite{hikage19}, to which we refer the reader for details. We summarize the blinding procedure below.

A two-level blinding strategy is used.
The first level of blinding is at the level of the catalog, while the second is at the level of the analysis itself.
In terms of the catalog level, we blind the true estimate of source shears by perturbing the multiplicative bias, as
\begin{equation}
\label{eq:multiplicative_bias_blinding}
m_{\mathrm{cat}, i} = m_{\mathrm{true}} + \dif m_{1,i} + \dif m_{2,i} \, ,
\end{equation}
where $m_{\mathrm{cat}, i}$ is the blinded multiplicative bias in the $i$-th shape catalog, $m_{\mathrm{true}}$ is the true estimate of the multiplicative bias, and $\dif m_{1,i}$ and $\dif m_{2,i}$ are two random variables for blinding.
While many HSC analyses are being carried out in parallel to this work, we prepare three blinded catalogs ($i\in\{0,1,2\}$) independently from other HSC analyses.
The random variables of $\dif m_{1,i}$ and $\dif m_{2,i}$ are encrypted and are different among these three blinded catalogs; the HSC analysis teams are also different.
The term $\dif m_{1,i}$ can only be decrypted by the leader of the analysis, and is removed before performing the analysis.
This term is needed to avoid an accidental unblinding by comparing the blinded catalogs of different analysis teams.
The term $\dif m_{2,i}$ can only be decrypted by a ``blinder-in-chief'', once we are ready to unblind the analysis.
The blinder-in-chief is not involved in the analysis and is not aware of the values of $\dif m_{2,i}$  until unblinding.
The value of $\dif m_{2,i}$ varies from $-0.1$ to $0.1$ randomly.
Only one blinded catalog among the three has $\dif m_{2,i}=0$, as the true catalog. 
The analysis team has to run the identical analysis on these three blinded catalogs, which is a computationally expensive element of our blinding strategy.
However, such a strategy comes with the advantage that an end-to-end rerun is not needed, once these catalogs are unblinded.
This completes the catalog-level of blinding.

For the analysis-level blinding, we never compare the results among three blinded catalogs until unblinding.
This avoids an automatic unblinding arising from an accidental comparison of the blinded catalogs with values of $\dif m_{2,i}$ all close to each other.
Moreover, for eFEDS clusters with available optical counterparts in the HSC survey, we never compare the mass estimates or shear profiles of common clusters between these two surveys.
Also, we never compare the mass distribution of eFEDS clusters with those from other cluster surveys until unblinding.
These strategies keep our weak-lensing analysis blinded.

It is worth mentioning that the first-year shape catalog of the HSC survey was already publicly available by the time we initialized this work.
However, we stress that we neither run the analysis using the public catalog, nor make any kind of comparisons between the blinded and public catalogs.

We unblind our analysis once the following criteria are met:
\begin{itemize}
\item The analysis codes pass validation tests using mock catalogs of at least ten times the size of the eFEDS sample, ensuring that the codes can recover the input parameters within statistical uncertainties.
\item The posteriors of parameters from the sampled chains are converged.
\item The best-fit models provide a good description of observed shear profiles.
\item The systematic errors arising from the miscentering on the final results are quantified.
\item The value $\dif m_{1,i}$ has been correctly subtracted from the blinded catalogs.
\item The selection bias due to resolution cuts and magnitude cuts has  been appropriately applied \citep[see][]{hikage19}.
\end{itemize}

It is important to stress that the measurement of the follow-up X-ray observables, i.e., \Lx, \Lb, \Mg, \Tx, and \Yx, is not performed in a blind analysis.
This is because the correct \Mfiveoo\ must be known to extract the X-ray spectra enclosed by the corresponding cluster radius \Rfiveoo.
Ideally, the X-ray spectral analysis should be performed in terms of profiles with fine radial bins around the cluster center, and we calculate the X-ray measurement at the corresponding cluster radius given a cluster mass at each iteration of the likelihood exploration.
However, the X-ray analysis, especially the extraction of spectra observed by \eROSITA, is extremely time-consuming and beyond what we can afford given the timescale. 
We therefore perform an early unblinding on the X-ray analysis in this work.
Specifically, the weak-lensing mass calibration (in Sect.~\ref{sec:wlmcalib}), which only involves the X-ray observable, namely the count rate \rate, is performed first in a completely blinded way as described above.
After the weak-lensing mass calibration passes the unblinding criteria, three blinded sets of cluster masses are produced based on the three blinded shape catalogs.
We then ask the blinder-in-chief to unblind the cluster mass privately and only to the \eROSITA\ team, such that they can extract the X-ray spectra within the ``correct'' \Rfiveoo.
A set of the X-ray observables extracted within the ``correct'' radius is then delivered to the HSC lensing team to perform the blind modeling of X-ray observable-to-mass-and-redshift relations (\Lx, \Lb, \Tx, \Mg, and \Yx, see Sect.~\ref{sec:wlsrmcalib}) for each blinded catalog.
After the early unblinding, we strictly forbid the exchange of information between the HSC  and \eROSITA\ teams by any means until the official unblinding takes place.
As the modeling of the scaling relations is performed solely by the lensing team, this ensures that the analysis in this work is still carried out blindly with an adjustment to minimize excessive measurements in X-rays.

\paragraph{Post-unblinding analysis}
Despite the careful treatment in the blinding analysis, we made changes to correct the discovered errors in a post-unblinding analysis. 
We refer readers to Appendix~\ref{sec:postblinding} for more details.
We find no significant changes after correcting the errors, and therefore the interpretations of this work are not affected.
We note that the post-unblinding analysis only affected the modeling of the follow-up X-ray scaling relations but not the weak-lensing mass calibration.

\subsection{Source selection}
\label{sec:source_selection}

We use the same source selection as in \cite{umetsu20} to select source galaxies behind each cluster.
Specifically, for a cluster at the redshift \zcl, the source selection is performed using the observed probability distribution of redshift, $P\left(\redshift\right)$, such that a galaxy is considered as a source galaxy if 
\begin{equation}
\label{sec:pzcut}
0.98 < \int_{{\redshift}_{\mathrm{min}}}^{\infty} P\left(\redshift\right)\dif\redshift~~\mathrm{and}~~{\redshift}_{\mathrm{MC}} < 2.5 \, ,
\end{equation}
where we set ${\redshift}_{\mathrm{min}} = \zcl + 0.2$, and ${\redshift}_{\mathrm{MC}}$ is the point estimator randomly sampled from $P\left(\redshift\right)$.
The contamination to lensing signals due to cluster members is suggested to be at the level of a few percent using our $P\left(\redshift\right)$-based source selection, as fully explored and quantified in \cite{medezinski18a}.
An independent examination of the cluster contamination in this work verifies that the level of contamination is at the level of $\lesssim3\percent$, on average (for more details, see Sect.~\ref{sec:contan}).

In this work, we use the photometric redshift estimated from the machine-learning-based code, \texttt{DEmP} \citep{hsieh14}, which has been widely used in obtaining not only redshifts but also other physical quantities, such as stellar mass \citep{lin17}.
Our source selection leads to source densities of 
$\approx15$, $\approx11$, $\approx6$, $\approx2$, and $\approx0.3$ galaxies per square arcmin for a cluster at a redshift of 
$\zcl=0.05$, $\zcl=0.25$, $\zcl=0.50$, $\zcl=0.78$, and $\zcl=1.1$, respectively.
At the median redshift of eFEDS clusters, $\ZPIV=0.35$, the source density reaches $\approx9.6$ galaxies per square arcmin.

\subsection{Tangential shear profiles}
\label{sec:gshear}

In what follows, we detail the procedure used to extract the observed tangential shear profile, as the lensing observable, for each cluster.
We largely follow the analysis presented in \cite{umetsu20} with one distinct difference: 
we measure the dimensionless tangential shear profile \gshear, as the quantity that can be directly observed in lensing observations without knowing the redshift of sources, while \cite{umetsu20} measured the differential surface mass density \deltaSigma.
The latter require prior knowledge of redshift and the distance-to-redshift relation with assumed cosmological parameters to 
convert \gshear\ to \deltaSigma.
This results in a difficulty in the lensing modeling in a cosmological analysis, where the cosmological parameters are varied and hence change the observable \deltaSigma.
Conversely, the tangential shear profile \gshear\ is directly observed and is invariant among different cosmology.
Therefore, the use of \gshear\ as a lensing observable enables a clean forward-modeling approach, which will be needed in a cosmological analysis in the future.

For each cluster at redshift \zcl, we derive the tangential shear profile \gshear\ in angular bins of clustercentric radius $\theta_{i}$,
\begin{equation}
\label{eq:gshear}
\gshear\left(\theta_{i}\right) = 
\frac{1
}{
2\mathfrak{R(\theta_{i})}\left(1 + \mathfrak{K(\theta_{i})}\right)
}
\frac{
\sum\limits_{s\in \theta_{i}} {\lensingw}_{s}  {\eplus}_{,s}  
}{
\sum\limits_{s\in \theta_{i}} {\lensingw}_{s}
}
\, ,
\end{equation}
where the subscript $s$ runs over the source sample that is selected according to the description in Sect.~\ref{sec:source_selection}; ${\lensingw}_{s}$ and ${\eplus}_{,s}$ are the lensing weight and tangential ellipticity of the $s$-th source galaxy, respectively; $\mathfrak{R}(\theta_{i})$ is the shear response \citep[see also][]{mandalbaum05b}; the factor $\left(1 + \mathfrak{K(\theta_{i})}\right)$ is the correction accounting for multiplicative shear bias, which is calibrated against the image simulations \citep{mandelbaum18,mandalbaum18b}.

The tangential ellipticity \eplus\ of a source galaxy around the cluster center is calculated as
\begin{equation}
\label{eq:eplus}
\eplus = -\cos(2\phi) e_{1} - \sin(2\phi) e_{2}
\, ,
\end{equation}
where $\left( e_{1}, e_{2} \right)$ is a two-component estimate of ellipticity measured in the cartesian coordinate of the sky defined in the HSC survey, and $\phi$ is the positional angle measured from the Right Ascension direction to the line connecting the cluster center and the source galaxy.
The lensing weight $\lensingw$ is evaluated as
\begin{equation}
\label{eq:lensing_weight}
\lensingw = \frac{1}{
{ \sigma_{\mathrm{e}} }^2 + { e_{\mathrm{rms}} }^2
}
\, ,
\end{equation}
where $\sigma_{\mathrm{e}}$ and $e_{\mathrm{rms}}$ are the measurement uncertainty and the root-mean-square estimate of the ellipticity per component, respectively.

The shear response $\mathfrak{R}(\theta_{i})$ in the angular bin $\theta_{i}$ is calculated as
\begin{equation}
\label{eq:response}
\mathfrak{R}(\theta_{i}) = 1 - \frac{
\sum\limits_{s\in \theta_{i}} {\lensingw}_{s} {{e_{\mathrm{rms}}}_{,s}}^2
}{
\sum\limits_{s\in \theta_{i}} {\lensingw}_{s}
}
\, ,
\end{equation}
where the subscript $s$ runs over the source sample in the angular bin $\theta_{i}$.
The factor $\left(1 + \mathfrak{K(\theta_{i})}\right)$ is calculated as
\begin{equation}
\label{eq:K}
1 + \mathfrak{K}(\theta_{i}) = \frac{
\sum\limits_{s\in \theta_{i}} {\lensingw}_{s} \left(1 + m_{s} \right)
}{
\sum\limits_{s\in \theta_{i}} {\lensingw}_{s}
}
\, ,
\end{equation}
where $m_{s}$ is the multiplicative bias of the $s$-th source galaxy.

The angular binning in clustercentric radii is defined as follows.
For each cluster, we first use a logarithmic binning between $0.2~\Mpch$ and $3.5~\Mpch$ with ten steps in physical units.
Then, the angular binning is obtained by dividing the physical radius by the angular diameter distance evaluated at the cluster redshift in the fiducial cosmology.
In this way, we ensure that roughly the same portion of the cluster profiles at different redshifts is probed.
The use of the angular-binned shear profile $\gshear\left(\theta\right)$ enables us to self-consistently calculate the cluster mass profile at any redshift-inferred distance in a forward-modeling approach where the cosmological parameters vary (see Sect.~\ref{sec:wlmodeling}).

For each cluster, we additionally compute the B-mode component of the shear profile $\gcros\left(\theta\right)$ according to Eq.~(\ref{eq:gshear}), after replacing $\eplus$ with $\ecros$, which is defined as
\begin{equation}
\label{eq:ecross}
\ecros =   \sin(2\phi) e_{1} -  \cos(2\phi) e_{2} 
\, .
\end{equation}
The azimuthally averaged cross-shear profile $\ecros\left(R\right)$ is expected to vanish if the signal is due to weak lensing,
therefore enabling a null-test for our lensing signals.
We verified that the stacked profile of $\gcros\left(\theta\right)$ in this work is indeed statistically consistent with zero.

\subsection{Covariance matrices}
\label{sec:covar}

For each cluster, we calculate the covariance matrix that will be used to model the observed \gshear, by following the prescription in \cite{umetsu20}.
Specifically, the covariance matrix $\mathbb{C}$ comprises a component characterizing the noise of shape measurements, $\mathbb{C}_{\mathrm{shape}}$, and another component $\mathbb{C}_{\mathrm{uLSS}}$ accounting for the uncertainty arising from uncorrelated large-scale structures around the cluster.
That is, 
\begin{equation}
\label{eq:covlensing}
\mathbb{C} = \mathbb{C}_{\mathrm{shape}} + \mathbb{C}_{\mathrm{uLSS}}
\, ,
\end{equation}
in which
\[
{\mathbb{C}_{\mathrm{shape}}}_{i,j} = \sigmashape^2(\theta_{i}) \delta_{i,j} \, ,
\]
where $\delta_{i,j}$ is a Kronecker delta function, and ${\sigmashape}^2(\theta_{i})$ in the angular bin $\theta_{i}$ is evaluated as
\begin{equation}
\label{eq:shapenoise}
\sigmashape^2(\theta_{i}) = \frac{
1
}{
4\mathfrak{R}^2(\theta_{i})\left(1 + \mathfrak{K}(\theta_{i})\right)^2 \sum\limits_{s\in \theta_{i}} {\lensingw}_{s}
}
\, .
\end{equation}

We compute $\mathbb{C}_{\mathrm{uLSS}}$ by following the prescription in Appendix~A of \cite{miyatake19} \citep[see also][]{hoekstra03}.
Specifically, for each cluster, we compute the nonlinear matter power spectrum $P^{\mathrm{NL}}_{\mathrm{m}}$ at the cluster redshift \zcl\   using the code \texttt{CAMB} \citep{lewis99} in order to derive the lensing power spectrum $C_{l}^{\kappa\kappa}$.
We use the full redshift distribution of the sources by stacking the observed redshift distributions of the source sample in order to infer the lensing weight function in calculating $C_{l}^{\kappa\kappa}$.
Finally, we compute $\mathbb{C}_{\mathrm{uLSS}}$ by integrating the product of $C_{l}^{\kappa\kappa}$ and two second-order Bessel functions associated with the radial binning of shear profiles.
This calculation of $\mathbb{C}_{\mathrm{uLSS}}$ is independently repeated for all clusters using the cosmological parameters fixed to the fiducial cosmology.

We note that the term $\mathbb{C}_{\mathrm{uLSS}}$ depends on the underlying cosmology.
Ideally, the variation in $\mathbb{C}_{\mathrm{uLSS}}$ due to a change of cosmological parameters needs to be taken into account in each iteration step of a forward modeling.
However, in this work, the covariance matrix is dominated by the shape noise in the fitting range of radii of interest.
Moreover, we adopt Gaussian priors on the cosmological parameters with the mean values used in the fiducial cosmology.
Hence, the cosmology-dependence of $\mathbb{C}_{\mathrm{uLSS}}$ is not expected to be a dominant factor in the final results of our work \citep[see][]{kodwani19}. 
Therefore, in this work we ignore the cosmological dependence of $\mathbb{C}_{\mathrm{uLSS}}$ so as to improve the speed of our calculation.
We leave the improvement of a cosmology-dependent $\mathbb{C}_{\mathrm{uLSS}}$ to future work.

We also note that the intrinsic variation of shear profiles at fixed cluster mass\footnote{This is characterized by the term $C^{\mathrm{int}}$ in Eq.~(16) of \cite{umetsu20}.} ---due to for example the triaxiality or concentration of halos--- is accounted for by modeling the intrinsic scatter of the weak-lensing mass bias (see Sect.~\ref{sec:bwl}).
Therefore, we do not include this term of intrinsic variation in Eq.~(\ref{eq:covlensing}).

\begin{figure*}
\centering
\resizebox{0.33\textwidth}{!}{
\includegraphics[scale=1]{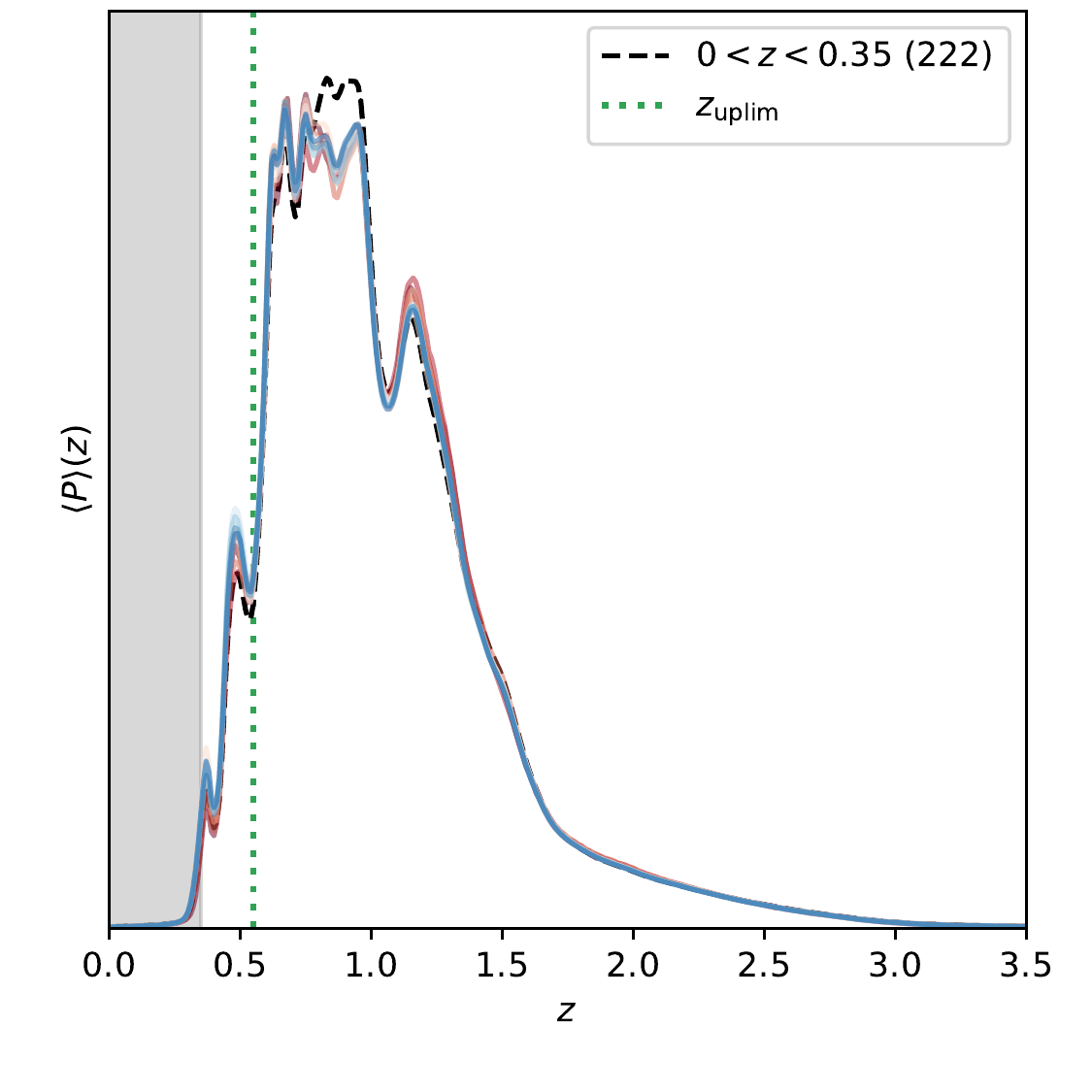}
}
\resizebox{0.33\textwidth}{!}{
\includegraphics[scale=1]{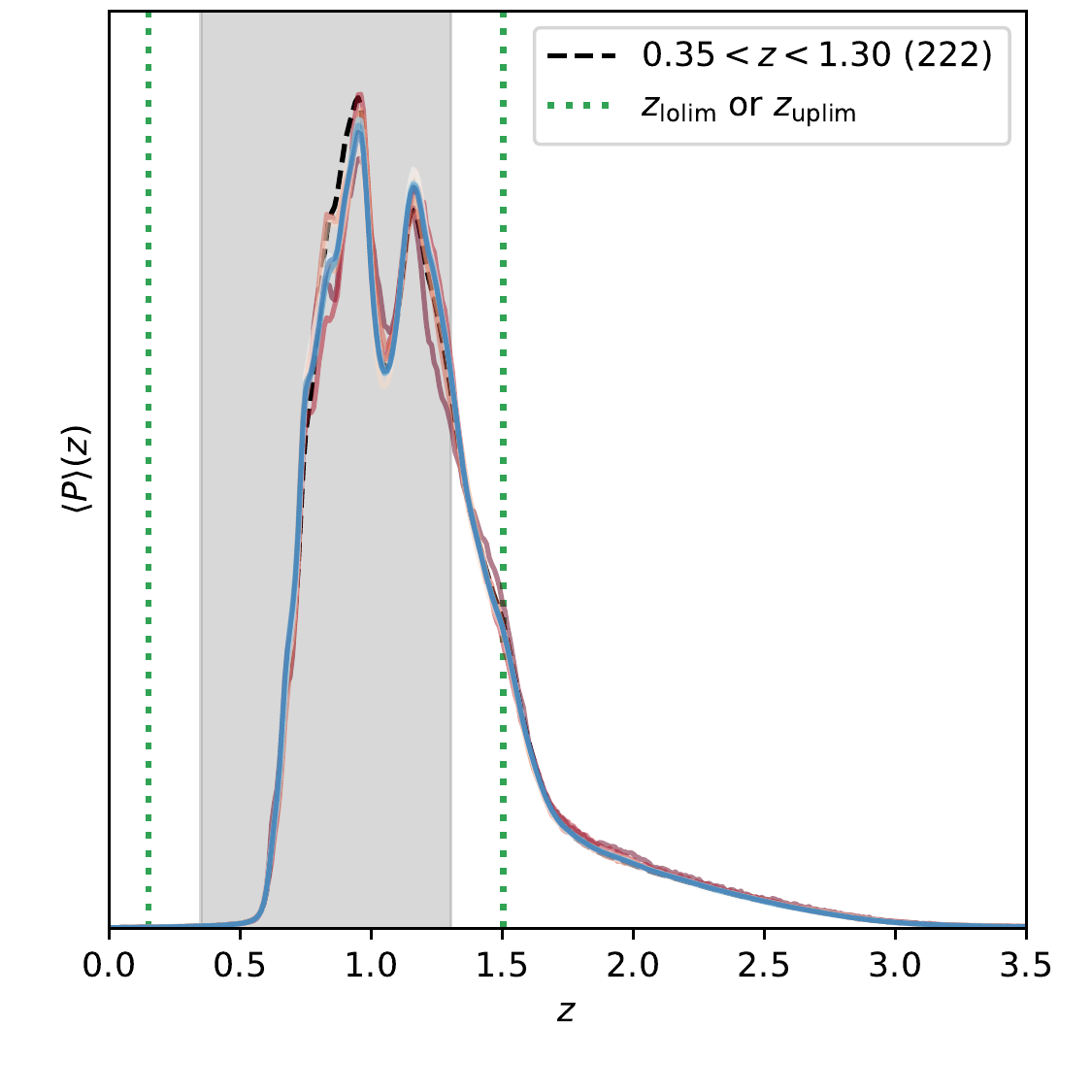}
}
\resizebox{0.33\textwidth}{!}{
\includegraphics[scale=1]{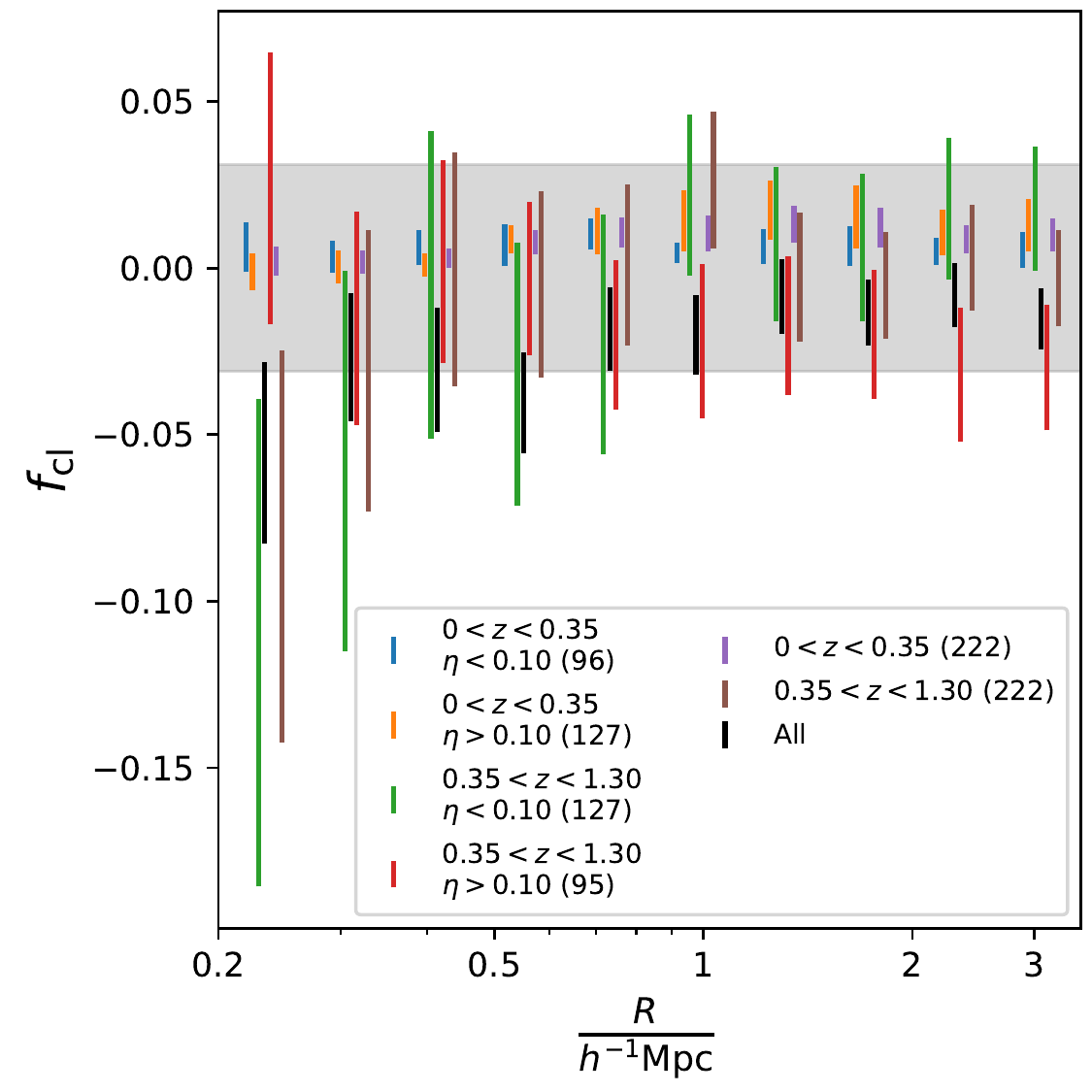}
}
\caption{
Comparison of the redshift distributions between cluster and random fields at low (left panel) and high (middle panel) redshifts.
In the left and middle panels, 
the color scheme ranging from dark red to blue indicates the scale of the clustercentric radius from the core to large radii, 
while the black dashed lines are the redshift distributions of random fields.
Meanwhile, the vertical gray regions indicate the redshift range of the cluster samples, while the upper and lower limits ($\redshift_{\mathrm{lolim}}$ and $\redshift_{\mathrm{uplim}}$) of the redshift interval in Eq.~(\ref{eq:fcl}) are shown by green dotted lines.
In the right panel, the profiles of cluster contamination in each subsample are shown with  the color scheme according to the sample binning.
The resulting profile of the total sample is in black.
The horizontal gray region indicates the conservative estimate of cluster contamination at a level of $\leq3\percent$ ($1\sigma$), which is used to calibrate the weak-lensing mass bias (see Sect.~\ref{sec:bwl}).
In these three panels, the numbers in the parentheses indicate the total numbers of the clusters used in the binning.
}
\label{fig:contam}
\end{figure*}

\subsection{Cluster member contamination}
\label{sec:contan}

Because of the uncertainty in determining redshifts, a sample of photometrically selected sources might contain galaxies that are not behind clusters, are therefore not lensed, and thus do not contribute to lensing signals.
The inclusion of these nonbackground galaxies dilutes lensing signals, and therefore introduces bias in inferred cluster masses.
For the contamination due to foreground galaxies, this bias can be statistically accounted for by properly weighting the lensing signal according to the observed redshift distribution $P\left(\redshift\right)$, with a calibration against random fields where no clusters are present.
However, cluster environments are extremely biased fields, of which the observed redshift distribution cannot be directly interpreted by those of random fields.
Moreover, the cluster contamination is expected to be more significant around the center of clusters, resulting in a radial dependence of bias in lensing signals.
This needs to be accounted for; otherwise significant bias will be introduced in the inferred mass \citep{medezinski18a}.

In this section, we examine the contamination of lensing signals due to cluster members by following the prescription in \cite{mcclintock19} \citep[see also][]{gruen14,melchior17,varga19,chiu20}.
Specifically, this method is based on a decomposition of the observed redshift distributions into a cluster component and the other from fore- and background (hereafter background, for simplicity).
By comparing the observed $P\left(\redshift\right)$ between the cluster and random fields, the contamination arising from cluster members can be quantified.
We stress that this methodology has been validated by simulations \citep{varga19}, showing that the underlying true cluster contamination can be recovered based on the observed $P\left(\redshift\right)$.

It is important to note that the source selection is different between \cite{mcclintock19} and this work:
these latter authors made use of all galaxies around clusters with available shape measurements, and properly assigned lensing weights to galaxies according to their observed redshift distributions, such that foreground galaxies and cluster members did not contribute to the lensing signals.
Then, they accounted for the residual contamination due to cluster members by quantifying the excess observed redshift distributions at the cluster redshift with respect to random fields.
In this work, on the other hand, we first minimize the cluster contamination by employing a stringent source selection based on the observed redshift distribution (see Sect.~\ref{sec:source_selection}); we then measure the lensing signal of these secure sources.
Consequently, our strategy leads to a clean sample with the cluster member contamination at the level of only a few percent \citep{medezinski18a,umetsu20}.
Importantly, our source selection still results in high densities of source galaxies thanks to the deep imaging of the HSC survey.

In this work, we conduct a nonparametric approach to quantify the cluster contamination by comparing the observed redshift distributions between the cluster and random fields.
In what follows, we express the clustercentric radius in the physical unit $R$ through the redshift--distance relation in the fiducial cosmology.
We assume that the observed redshift distribution $P\left(\redshift; R\right)$ of selected source galaxies at the radial bin $R$ can be written as
\begin{equation}
\label{eq:pzdecomp}
P\left(\redshift; R\right) = 
\left(1 - \fcl\left(R\right)\right)
P_{\mathrm{bkg}}\left(\redshift\right) + 
\fcl\left(R\right)
P_{\mathrm{cl}}\left(\redshift; R\right) \, ,
\end{equation}
where $\fcl\left(R\right)$ is the amount of cluster contamination at the radial bin $R$, $P_{\mathrm{bkg}}$ is the redshift distribution of galaxies at backgrounds, and $P_{\mathrm{cl}}$ is the redshift distribution of cluster members that lead to contamination.
By construction, we have $0\leq\fcl\left(R\right)\leq1$.
We can write Eq.~(\ref{eq:pzdecomp}) in the integral of redshift, 
\[
\int\dif\redshift P\left(\redshift; R\right) = 
\left(1 - \fcl\left(R\right)\right)
\int\dif\redshift P_{\mathrm{bkg}}\left(\redshift\right) + 
\fcl\left(R\right)
\int\dif\redshift P_{\mathrm{cl}}\left(\redshift; R\right) \, ,
\]
and then integrate over the redshift interval $\mathfrak{I}$, 
where the cluster component vanishes, and finally arrive at 
\begin{equation}
\label{eq:fcl}
\fcl\left(R\right) =  1 - 
\frac{
\int\limits_{\redshift\in\mathfrak{I}}\dif\redshift P\left(\redshift; R\right)
}{
\int\limits_{\redshift\in\mathfrak{I}}\dif\redshift P_{\mathrm{bkg}}\left(\redshift\right)
}
\, .
\end{equation}
In this way, the cluster contamination at the radial bin $R$ can be derived for a given set of observed $P\left(\redshift; R\right)$ and $P_{\mathrm{bkg}}\left(\redshift\right)$.

In practice, in the interest of more precise measurements, we estimate $\fcl\left(R\right)$ according to Eq.~(\ref{eq:fcl}) by stacking clusters in different redshift and count-rate bins.
Namely, the clusters are stacked in two redshift bins, low-redshift ($\redshift \leq 0.35$) and high-redshift ($\redshift>0.35$), and two count-rate bins, low-rate ($\rate<0.1$) and high-rate ($\rate>0.1$).
This leads to four subsamples in total based on the observed redshifts and count rates.
We note that we have tried a finer binning in redshift and find consistent results.
For each individual cluster, labeled $j$, we first derive the weighted probability distribution $P_{j}\left(\redshift;R_{i}\right)$ of redshift observed at the radial bin $R_{i}$ as
\begin{equation}
\label{eq:wpz}
P_{j}
\left(\redshift;R_{i}\right)
 = 
\frac{
\sum\limits_{s\in j, R_{i} } {\lensingw}_{s} P_{s}(\redshift)
}{
\sum\limits_{s\in j, R_{i} } {\lensingw}_{s}
} \, ,
\end{equation}
where $s$ runs over the source galaxies which are located at the radius bin $R_{i}$ around the cluster $j$.
In each subsample of clusters, we then derive the number-weighted average of redshift distributions observed at the radius bin $R_{i}$ as
\begin{equation}
\label{eq:average_wpz}
\left\langle
P
\right\rangle
\left(\redshift;R_{i}\right)
 = 
\frac{
\sum\limits_{j=1}^{N_{\mathrm{cl}}} P_{j}\left(\redshift;R_{i}\right) N_{\mathrm{gal}, j}\left(R_{i}\right)
}{
\sum\limits_{j=1}^{N_{\mathrm{cl}}} N_{\mathrm{gal}, j}\left(R_{i}\right)
} \, ,
\end{equation}
where $j$ runs over the subsample of $N_{\mathrm{cl}}$ clusters, and $N_{\mathrm{gal}, j}\left(R_{i}\right)$ is the number of observed galaxies at the radius bin $R_{i}$ around the cluster $j$.
We then define the redshift interval $\mathfrak{I}$ as 
$
\left\lbrace \redshift|\redshift < {\redshift}_{\mathrm{lolim}}\right\rbrace
\cup
\left\lbrace \redshift|\redshift > {\redshift}_{\mathrm{uplim}}\right\rbrace
$, where 
$
{\redshift}_{\mathrm{lolim}} = \mathtt{min}\left\lbrace{\redshift}_{j}\right\rbrace_{j=1}^{N_{\mathrm{cl}}} - \Delta\redshift
$
,
$
{\redshift}_{\mathrm{uplim}} = \mathtt{max}\left\lbrace{\redshift}_{j}\right\rbrace_{j=1}^{N_{\mathrm{cl}}} + \Delta\redshift
$,  
and ${\redshift}_{j}$ is the redshift of the $j$-th cluster.
We set $\Delta\redshift=0.2$, which is consistent with our source selection (see Sect.~\ref{sec:source_selection}).
We note that our results are not sensitive to the current choice of $\Delta\redshift=0.2$; 
we verified that the results with $\Delta\redshift=0.4$ are statistically consistent with those based on $\Delta\redshift=0.2$, although with much larger error bars.
In this way, we can estimate the numerator in Eq.~(\ref{eq:fcl}) by integrating 
$\left\langle P \right\rangle\left(\redshift;R_{i}\right)$ over the interval $\mathfrak{I}$.

Concomitently, we estimate $P_{\mathrm{bkg}}\left(\redshift\right)$ 
by repeating Eqs.~(\ref{eq:wpz}) and~(\ref{eq:average_wpz}) with the same scheme of radial binning and cluster redshift on a number of $N_{\mathrm{cl}}$ pointing on random fields.
This allows us to calculate the contribution arising from pure background sources in a fully consistent way as in cluster fields.
Finally, we compute $\fcl\left(R\right)$ in each subsample by following Eq.~(\ref{eq:fcl}).

The results of cluster contamination are shown in the right panel of Fig.~\ref{fig:contam}, where the contamination profiles of four subsamples are in blue, orange, green, and red.
These are all statistically consistent with zero contamination, with a mild exception that the subsample of low-\rate\ clusters at high redshift shows large variation but also with large error bars.
This is mainly due to noisy measurements of photo-\redshift\ for high-redshift sources.
We further repeat the procedure but without binning in \rate, resulting in two subsamples at low ($0<\redshift<0.35$) and high ($0.35<\redshift<1.20$) redshifts; their contamination profiles are shown in purple and brown, respectively.
No clear contamination is seen.
We additionally show the number-weighted average of redshift distributions (i.e., Eq.~(\ref{eq:average_wpz})) of these two subsamples in the left and middle panels, where the redder colors represent inner radii, together with those estimated from random fields in dashed lines.
As seen in these figures, they all show no signs of cluster contamination.
This is in perfect agreement with previous HSC work \citep{medezinski18a,medezinski18b,miyatake19,umetsu20}, demonstrating a highly pure source sample for weak-lensing studies.

Based on these results, we conclude that the level of cluster contamination in this work is at the level of $\lesssim3\percent$ ($1\sigma$), which is a conservative estimate, shown by the gray shaded region in the right panel of Fig.~\ref{fig:contam}.
This estimation of cluster contamination will be taken into account when deriving the weak-lensing mass bias (see Sect.~\ref{sec:bwl}).

\subsection{Cluster miscentering}
\label{sec:miscentering}

In this work, the cluster center is defined as the center of X-ray emissions.
However, X-ray centers
do not necessarily represent the true center of the total mass distribution of galaxy clusters.
The miscentering between the ``observed'' center and the ``true'' center causes bias in shear profiles with respect to perfectly centered clusters \citep{johnston07a,johnston07b}, especially at small clustercentric radii.
Consequently, the lensing modeling needs to account for this miscentering effect; otherwise bias in the inferred mass will be introduced.

It has been shown that miscentering has significant effects on the observed properties of optically selected clusters \citep[][]{biesiadzinski12,sehgal13,rozo14a}.
Therefore, extensive efforts have been made to quantify the offset between the optical center, usually defined by the BCG, and other center proxies, such as the peak of projected mass distributions inferred from lensing \citep{oguri10,zitrin12}, the second brightest cluster galaxy \citep{hoshino15}, and the ICM-based center defined in X-rays \citep{lin04b, mahdavi13, lauer14, zhang19} or at millimeter wavelength \citep{song12b, saro15,bleem20}.
It is worth mentioning that the center inferred from weak-lensing mass maps is largely affected by the noise arising from shape measurements \citep{dietrich12}, and therefore it is extremely difficult to determine the true center of overall cluster mass distributions from observations alone.
We note that it is possible to account for the modeling bias when centering on the weak-lensing mass peak  for clusters whose lensing signals are detected at high significance \citep{sommer21}, which is unfortunately not the case in our study.

In this work, we gauge the miscentering in the eFEDS clusters using the offset between the X-ray centers and the BCGs, which are used as the optical centers.
The BCGs are identified by the \texttt{MCMF} as the brightest galaxies that are consistent with the red-sequence prediction at the cluster redshift.
We note that the complex nature of BCGs may result in incorrect identification of the cluster center in the optical.
For example, the BCGs are suggested to scatter more from the red sequence than the typical bright galaxies \citep{burg14,kravtsov18}, and some BCGs may not be red in the group scale \citep{liu12}.
To assess the impact of   incorrectly identified optical centers, we re-derive the offset distribution (see below) by replacing the optical center with the peak of the galaxy density map for the clusters whose BCGs are significantly different from both the peak of galaxy density and X-ray maps.
We find that the resulting difference is negligible.

The offset distribution $P(x)$ between the X-ray and optical centers as a function of the dimensionless radius $x$ is characterized by a composite model \citep[see also][]{saro15,zhang19}:
\begin{equation}
\label{eq:miscentering}
P(x) = 
\left(1 - \fmis\right) \times \Pcen\left(x | \sigmacen,\alpha\right)
+ 
\fmis \times \Pmis\left(x | \sigmamis\right)
\, .
\end{equation}
In Eq.~(\ref{eq:miscentering}), we assume that the distribution $P(x)$ can be decomposed into two components:
one is the centering component characterized by a modified Rayleigh distribution $\Pcen(x|\sigmacen, \alpha)$, which reads
\begin{equation}
\label{eq:centering_distirbutions}
\Pcen(x|\sigmacen, \alpha) = 
\frac{
\frac{1}{\sigmacen}
\left(\frac{x}{\sigmacen}\right)^\alpha
\exp\left(-\frac{1}{2}\left(\frac{x}{\sigmacen}\right)^2\right)
}{
2^{\frac{\alpha-1}{2}}\Gamma\left(\frac{\alpha+1}{2}\right)
}
\, ,
\end{equation}
where $\Gamma$ is the gamma function; and the other is the miscentering component as modeled by a standard Rayleigh distribution,
\begin{equation}
\label{eq:miscentering_distirbutions}
\Pmis(x|\sigmamis) =
\frac{1}{\sigmamis}
\left(\frac{x}{\sigmamis}\right)
\exp\left(-\frac{1}{2}\left(\frac{x}{\sigmamis}\right)^2\right)
\, .
\end{equation}
These two components are weighted by the miscentering fraction \fmis\ in Eq.~(\ref{eq:miscentering}).
We note that the parameter $\alpha$ is required to be positive and characterizes the shape of the centering component:
The smaller $\alpha$ is the smaller radius where the peak of $\Pcen(x)$ occurs, and the steeper fall-off toward $x=0$.
Moreover, $\Pcen(x)$ reduces to a half-normal distribution and zero, as $\alpha$ approaches zero and $+\infty$, respectively.
Also, $\Pcen$ is identical to $\Pmis$ if $\alpha = 1$.

\begin{figure}
\resizebox{0.5\textwidth}{!}{
\includegraphics[scale=1]{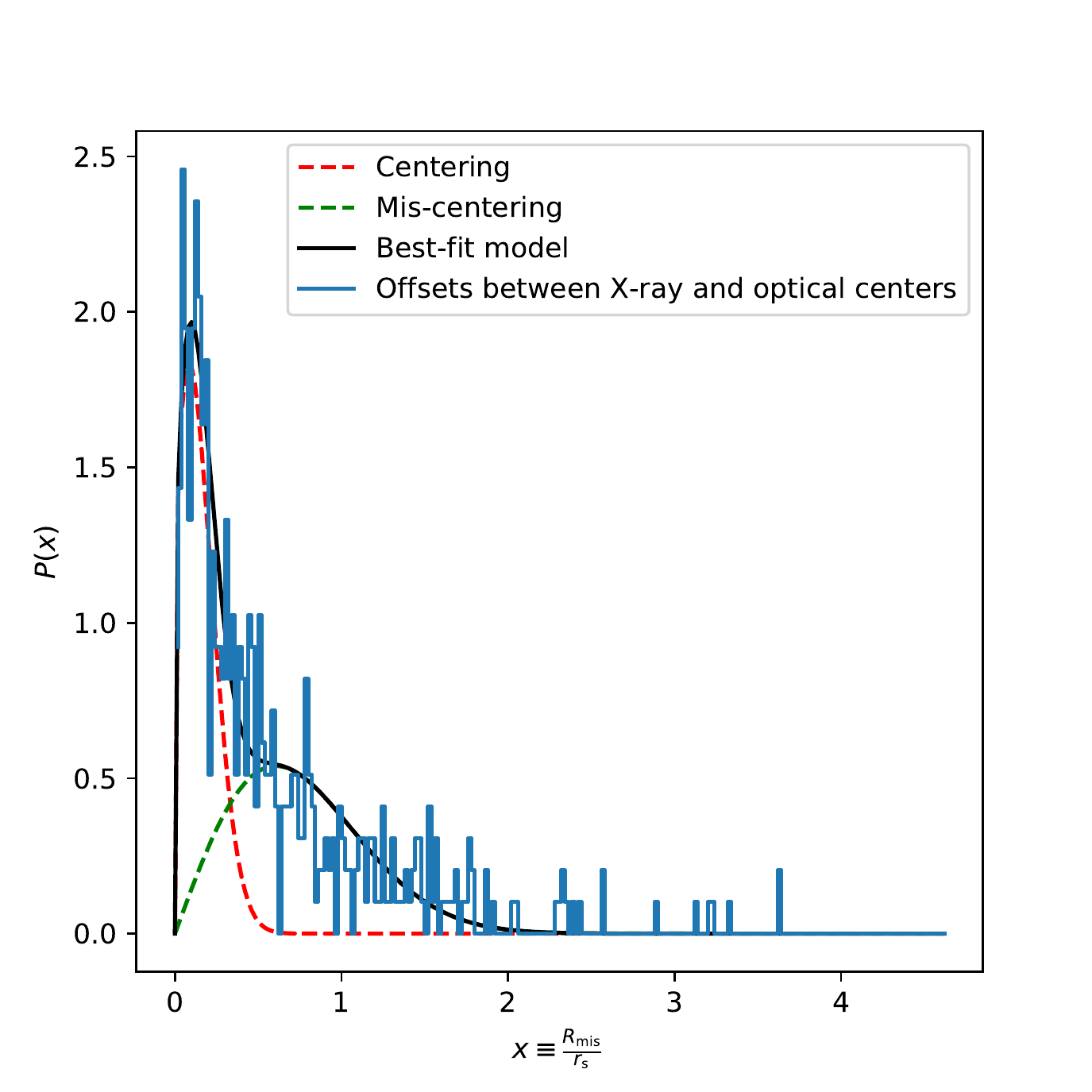}
}
\caption{
Distribution of offsets between the X-ray and optically defined centers of eFEDS clusters.
The blue histogram is the observed offset distribution in the unit of the scale radius.
The black solid line represents the best-fit model, i.e., Eq.~(\ref{eq:miscentering}), which consists of centering and miscentering components, as indicated by the red and green dashed lines, respectively.
}
\label{fig:miscentering}
\end{figure}

Our goal is to derive a universal offset distribution, based on which we model the miscentering of the shear profile for individual clusters. 
We define the dimensionless radius as $x \equiv R_{\mathrm{off}} / r_{\mathrm{s}}$, where $R_{\mathrm{off}}$ is the offset between the optical and X-ray centers, and $r_{\mathrm{s}}$ is the scale radius of a Navarro-Frenk-White \citep[hereafter NFW;][]{navarro1997} profile with a concentration parameter evaluated using the \cite{diemer15} fitting formula for a halo with the pivotal mass $\MPIV = 1.4\times10^{14}\Msunh$ at the pivotal redshift $\ZPIV = 0.35$.
For each cluster, the offset $R_{\mathrm{off}}$ is calculated first in the physical unit using the redshift--distance relation inferred in the fiducial cosmology.
We then convert $R_{\mathrm{off}}$ to $x$ using the fixed scale radius $r_{\mathrm{s}}$.
In this way, we effectively assume a universal mass and redshift for each cluster in deriving the offset distribution $P(x)$, as an ensemble behavior.
Later, when modeling the miscentering in Sect.~\ref{sec:wlmodeling}, we do vary the scale radius for a given weak-lensing mass according to the concentration-to-mass relation at each step of the likelihood exploration.
Ideally, different core radii based on individual cluster masses should be used in deriving the universal offset distribution, which could be achieved by an iterative process after obtaining the cluster mass.
However, we show below that the resulting posteriors of the cluster mass \Mfiveoo\ obtained with and without the modeling of the cluster core are consistent with each other (see Sect.~\ref{sec:mass_calibration_results}), suggesting that our modeling of the miscentering is sufficiently adequate in this work.

Figure~\ref{fig:miscentering} shows the offset distribution (blue histogram) and the best-fit miscentering model (black solid line) as a function of $x$.
The best-fit parameters for Eq.~(\ref{eq:miscentering}) are
\begin{multline}
\left(\fmis, \sigmacen, \alpha, \sigmamis\right) = \\
\left(
0.54\pm0.02, 0.17\pm0.01, 0.26\pm0.05, 0.61\pm0.03 
\right) \, .
\end{multline}

We stress the following particularities in modeling the miscentering of eFEDS clusters.
First, in this work we statistically correct for the miscentering for individual clusters assuming a universal distribution, $P(x)$, as directly determined from the data using Eq.~(\ref{eq:miscentering}).
For each cluster, this is done by taking an average of a data array representing a set of shear profiles with different offset radii $x$ weighted by the distribution $P(x)$ (see more details in Sect.~\ref{sec:wlmodeling}).
This approach could not be optimal on the basis of individual clusters; 
this is because a shear profile with average miscentering is not necessarily the best description of each cluster, of which the miscentering is just one realization from $P(x)$.
However, this approach is statistically correct, because we simultaneously model all observed shear profiles in an ensemble manner, in which the miscentering distribution follows the derived $P(x)$.
Moreover, we follow the identical modeling of the miscentering, including the distribution $P(x)$, in calibrating the weak-lensing mass bias using simulations (see Sect.~\ref{sec:bwl}), which ensures that we correctly infer the underlying true cluster mass in this framework.

Second, we determine the distribution $P(x)$ as the offset between the X-ray center and the BCG, while the required knowledge for modeling the miscentering is the offset between the X-ray  and ``true'' centers of a cluster potential.
That is, the distribution $P(x)$ contains the intrinsic miscentering of both the X-ray center and the BCG with respect to the true center, as well as their respective measurement uncertainties  at both wavelengths.
Among these, the dominant component is the intrinsic miscentering of the BCG with respect to the true center, because (1) the X-ray center is considered as a much better tracer of the total potential than other proxies \citep{lin04b, mahdavi13, lauer14, zhang19} and (2) 
the uncertainty on the measurement  of the optical center is much smaller than the characteristic miscentering radius. 
Therefore, the miscentering effect estimated in this work merely serves as an upper limit on X-ray miscentering.
It was suggested that X-ray centers could bias the mass estimate even when the modeling of the cluster core ($R<500$~$k$pc) is excluded \citep[][]{schrabback21}.
To accurately determine the intrinsic miscentering of eFEDS clusters, the most promising way is to utilize an end-to-end simulation suite, in which we compare the true center of halos with the X-ray center identified by the same cluster-finding algorithm.
A dedicated effort using simulations to determine the miscentering effect for eFEDS clusters is warranted, but we leave this for a future analysis.

To quantify the systematic errors arising from the modeling of miscentering, we compare the final results with and without the modeling of shear profiles in the cluster core ($R<0.5\Mpch$), that is including or excluding the three innermost radial bins, respectively.
We find that the miscentering effect is a subdominant factor in this work (see Sect.~\ref{sec:sys}).

\begin{figure}
\centering
\resizebox{0.5\textwidth}{!}{
\includegraphics[scale=1]{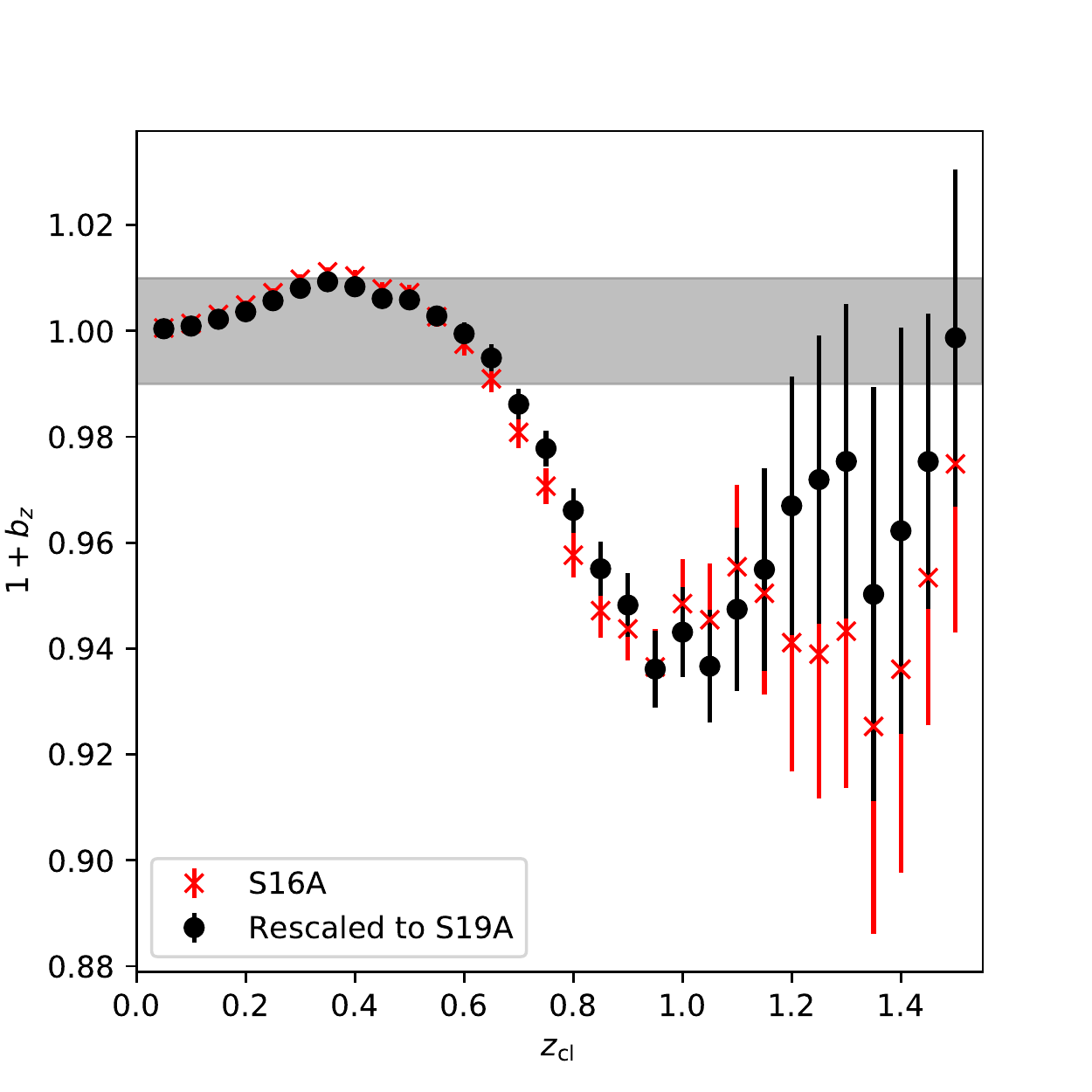}
}
\caption{
Photo-\redshift\ bias $\bz(\zcl)$ as a function of cluster redshift \zcl.
The result derived from the first-year HSC data is shown by the red crosses, which are re-scaled to the three-year HSC data shown by the black points based on the difference in the photo-\redshift\ between the data sets (see Sect.~\ref{sec:photozbias}).
We also show a reference at the level of $1\percent$ by the gray shaded region.
The resulting $\bz(\zcl)$ is applied to the calibration of weak-lensing mass bias for eFEDS clusters (see Sect.~\ref{sec:bwl}).
}
\label{fig:bz}
\end{figure}

\subsection{Photometric redshift bias}
\label{sec:photozbias}

The cluster mass is estimated from lensing signals with photo-\redshift, which is used to calculate the critical surface density \sigmacrit\ in converting \gshear\ to \deltaSigma\ (see Sect.~\ref{sec:basics}).
Therefore, a bias in photo-\redshift\ would ultimately bias the inferred cluster mass.

Given an observed shear profile, the bias $\bz(\zcl)$ in the inferred mass of a cluster at the redshift  \zcl\ due to the bias in photo-\redshift\ can be expressed in terms of the lensing efficiency \lensingeff, as
\begin{equation}
\label{eq:photozbias}
\frac{{\lensingeff}_{\mathrm{true}}}{{\lensingeff}_{\mathrm{obs}}}\left(\zcl\right) \equiv
1 + \bz(\zcl)
\, ,
\end{equation}
where ${\lensingeff}_{\mathrm{true}}$ and ${\lensingeff}_{\mathrm{obs}}$ are the lensing efficiency inferred by the true and observed redshifts, respectively.
In an ideal case, ${\lensingeff}_{\mathrm{true}}$ should be estimated from a sample of galaxies with redshifts secured by spectroscopic observations.
Moreover,  the distributions of the properties of this sample (e.g., of galaxy magnitudes or colors) should be consistent with those of the source sample, and should also be independent of the photo-\redshift\ calibration.
However, in practice this sample is difficult to obtain, especially for our source sample with the limiting depth of $i<24.5$~mag in the HSC survey.
Therefore, we make use of the galaxy sample observed in the COSMOS field \citep{ilbert09,laigle16} with high-quality photometric redshifts to assess the photo-\redshift\ bias in this work.

We follow the procedure detailed in \cite{miyatake19} to quantify the photo-\redshift\ bias $\bz(\zcl)$.
In what follows, we summarize the steps.
First, the photometric redshifts ${\redshift}_{\mathrm{COSMOS30}}$ estimated by the 30-band COSMOS photometry \citep{laigle16} are used to represent the true redshifts.
Second, as the COSMOS field is also observed by the HSC survey, we additionally prepare a catalog of the COSMOS field that is processed in the same configuration as the HSC WIDE layer, resulting in a HSC WIDE-depth version of the COSMOS catalog\footnote{We note that we only use a subset of galaxies in the COSMOS field that were not used to train the photo-\redshift\ of the source sample in the HSC survey. This HSC WIDE-depth version of the COSMOS catalog can be acquired at \url{https://hsc-release.mtk.nao.ac.jp/doc/index.php/s17a-wide-cosmos/}}.
We then estimate the photometric redshift of the galaxies in this HSC WIDE-depth COSMOS catalog in the same way as in the HSC survey.
This ensures the homogeneity of the photo-\redshift\ between the COSMOS reference catalog and the HSC data.
In this way, for each galaxy in the COSMOS field, we can assess the 30-band COSMOS photo-\redshift\ ${\redshift}_{\mathrm{COSMOS30}}$ as the true redshift, and the photometric redshift distribution $P_{\mathrm{HSC}}\left(\redshift\right)$ estimated by the HSC survey.
Moreover, we can select the galaxy sample in the COSMOS field in the same way as our source selection using the photometric redshift (see Sect.~\ref{sec:source_selection}).

Next, we employ a re-weighting technique \citep[for more details and some caveats of this method, see ][]{bonnett16,gruen17,hikage19}, such that the observed properties of the galaxy population in the COSMOS field match those of the HSC source sample.
Briefly, the HSC source sample is classified into cells of a self-organizing map \citep[SOM;][]{masters15} based on the $i$-band magnitude and four colors.
We then classify the COSMOS galaxies according to this SOM, and compute their new weights $w_{\mathrm{SOM}}$, such that the weighted distributions of the properties match those of the HSC sources \citep[see also][]{medezinski18a}.
Finally, we include the new weights $w_{\mathrm{SOM}}$ into the lensing weights $w$ in calculating the weighted distribution of redshift to infer the lensing efficiency.
In practice, for a cluster at redshift \zcl, we select the source sample, labeled $\mathsf{S}$, in the COSMOS field according to the description in Sect.~\ref{sec:source_selection}, and compute
\begin{equation}
\label{eq:photozbias_2}
{\lensingeff}_{\mathrm{true}/\mathrm{obs}}(\zcl) = 
\int
\lensingeff\left(\zcl,\redshift \right)
P_{\mathrm{true}/\mathrm{obs}}
\left({\redshift}\right)
\dif{\redshift}
\, ,
\end{equation}
where $P_{\mathrm{true}}(\redshift)$ is the distribution of true redshift ${\redshift}_{\mathrm{COSMOS30}}$ in the source sample $\mathsf{S}$, and 
\begin{equation}
\label{eq:photozbias_3}
P_{\mathrm{obs}}
\left(\redshift\right)
 = 
\frac{
\sum\limits_{s \in \mathsf{S}} {w_{\mathrm{SOM}}}_{,s}{\lensingw}_{s} {P_{\mathrm{HSC}}}_{,s}\left(\redshift\right)
}{
\sum\limits_{s \in \mathsf{S}} {w_{\mathrm{SOM}}}_{,s}{\lensingw}_{s}
} \, .
\end{equation}
Consequently, we compute the photo-\redshift\ bias $\bz(\zcl)$ following Eq.~(\ref{eq:photozbias}).

In practice, there is one additional factor that needs to be considered in this work:
the weighting factor $w_{\mathrm{SOM}}$ in the COSMOS catalog is specifically tuned to match the property of the sources observed in the first-year weak-lensing data (S16A), while the three-year (S19A) data are used in this work.
Therefore, the difference in the photo-\redshift\ between the first- and three-year data needs to be accounted for when calculating the photo-\redshift\ bias.
In this work, the difference in the photo-\redshift\ is taken into account by multiplying the ratio of the lensing efficiency between the S19 and S16 data sets.
Specifically, the ratio is calculated as
\[
R_{\beta}(\zcl) = \frac{
\left\langle\beta_{\mathrm{S16A}}(\zcl)\right\rangle
}{
\left\langle\beta_{\mathrm{S19A}}(\zcl)\right\rangle
} \, ,
\] 
where the $\beta_{\mathrm{S16A}}$ and $\beta_{\mathrm{S19A}}$ are the lensing efficiency calculated using the photo-\redshift\ from the S16A and S19A data sets, respectively, and the bracket $\left\langle\cdot\right\rangle$ stands for the mean for a selected sample of sources $\mathsf{S}$ given a cluster redshift \zcl.
In this way, the final photo-\redshift\ bias $b_{z,\mathrm{S19A}}$ used in this work is that from the S16A data set multiplying a re-scaling factor, namely
\begin{equation}
\label{eq:rescale_photoz}
1 + b_{z,\mathrm{S19A}}(\zcl) = R_{\beta}(\zcl) \times \left( 1 + b_{z,\mathrm{S16A}}(\zcl) \right) \, .
\end{equation}

Figure~\ref{fig:bz} shows the resulting $\bz(\zcl)$ as a function of cluster redshift \zcl.
As seen, the photo-\redshift\ bias on the cluster mass is estimated to be $\lesssim2\percent$ for clusters at redshift $\zcl\lesssim0.7$, and becomes $\lesssim6\percent$ for $\zcl\gtrsim0.8$.
There is no clear difference in the photo-\redshift\ bias between the S16A and S19A data sets, suggesting that their photo-\redshift\ performances are consistent with each other.
This is in close agreement with previous HSC studies, in which the mean photo-\redshift\ bias was estimated to be $\approx2\percent$ and $\approx0.9\percent$ for clusters selected in the ACTpol  \citep{miyazaki18} and XXL \citep{umetsu20} surveys, respectively.
We note that this resulting photo-\redshift\ bias $b_{z,\mathrm{S19A}}(\zcl)$ is included in deriving the weak-lensing mass bias for eFEDS clusters in Sect.~\ref{sec:bwl}.

\begin{figure*}
\resizebox{\textwidth}{!}{
\includegraphics[scale=1]{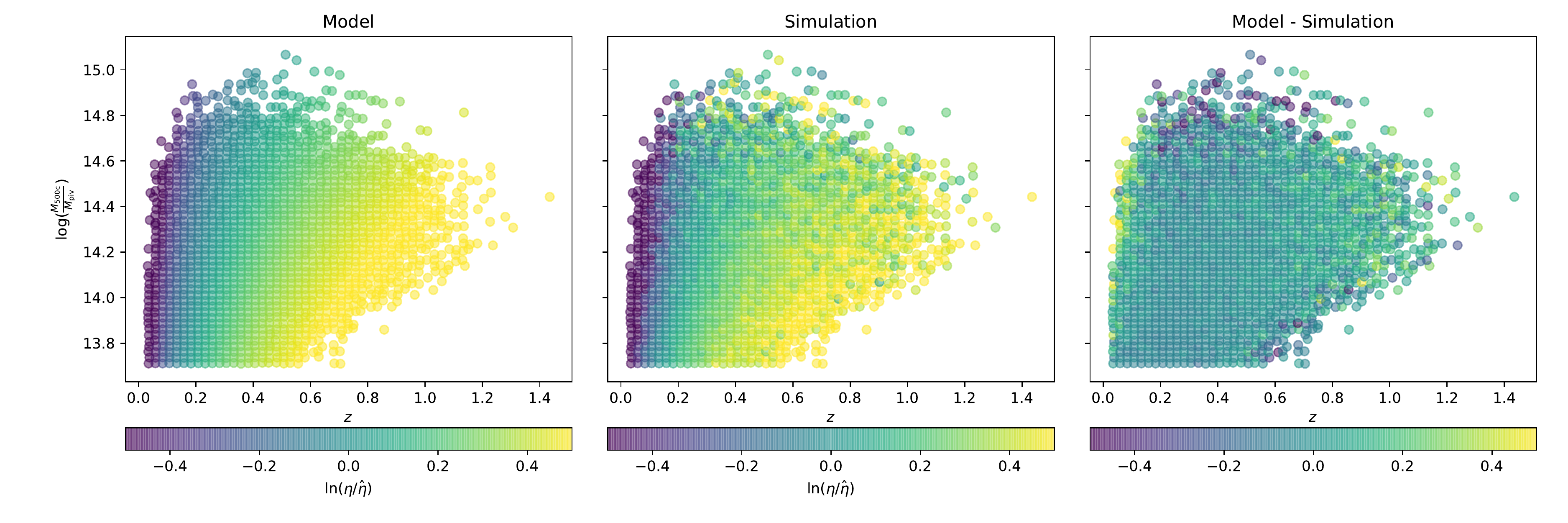}
}
\caption{
Bias \brate, which is defined as the ratio of the observed count rate \rate\ to the true count rate $\hat\rate$, from the simulation and the best-fit model (middle and left panels, respectively).
These are presented as a function of true  cluster mass \Mfiveoo\ at each redshift, color coded according to $\ln\left(\brate\right)$.
A clear gradient depending on both mass and redshift is seen.
The best-fit model, as described in Eq.~(\ref{eq:bias_rate}), provides a good description of the simulation, as suggested by their difference as shown in the right panel.
}
\label{fig:ratebias_sims}
\end{figure*}

\subsection{Modeling of shear profiles}
\label{sec:wlmodeling}

In this section, we describe the modeling of the observed shear profile of individual clusters, which is jointly fitted in the likelihood described in Sect.~\ref{sec:xtom_modeling}.

For an eFEDS cluster at redshift \zcl, there are three input observables used in the modeling: 
\begin{itemize}
\item the observed shear profile ${\gshear}\left(\theta\right)$ as a function of angular radius $\theta$, estimated in Eq.~(\ref{eq:gshear}),
\item the lensing covariance matrix $\mathbb{C}$, derived as Eq.~(\ref{eq:covlensing}), and
\item the observed redshift distribution $P\left(\redshift\right)$, estimated in Eq.~(\ref{eq:wpzbar}).
\end{itemize}
Given a weak-lensing mass \Mwl\ and a set of parameters 
$\mathbf{p}_{\mathrm{mis}}=\{\fmis, \sigmamis\}$
describing the miscentering distribution, together with a set of cosmological parameters 
$\mathbf{p}_{\mathrm{c}}$
that are used to calculate the redshift-inferred distance, the observed shear profile of an eFEDS cluster is modeled as \citep{seitz97}
\begin{equation}
\label{eq:lensingmodel}
\begin{aligned}
g_{+}^{\mathrm{mod}}\left(
\theta |\Mwl, \mathbf{p}_{\mathrm{mis}}, \mathbf{p}_{\mathrm{c}}
\right) = 
\frac{
{\gamma}_{\mathrm{mod}}\left(
\theta |\Mwl, \mathbf{p}_{\mathrm{mis}}, \mathbf{p}_{\mathrm{c}}
\right)
}{
1 - {\kappa}_{\mathrm{mod}}\left(
\theta |\Mwl, \mathbf{p}_{\mathrm{mis}}, \mathbf{p}_{\mathrm{c}}
\right)
}\times
\\
\left(
1 + {\kappa}_{\mathrm{mod}}\left(
\theta |\Mwl, \mathbf{p}_{\mathrm{mis}}, \mathbf{p}_{\mathrm{c}}
\right)
\left(
\frac{
\left\langle
{\lensingeff}^2
\left(
\mathbf{p}_{\mathrm{c}}
\right)
\right\rangle
}{
\left\langle
\lensingeff
\left(
\mathbf{p}_{\mathrm{c}}
\right)
\right\rangle^2
}
-1
\right)
\right)
\end{aligned}
\, ,
\end{equation}
in which 
\begin{equation}
\label{eq:beta_over_pz}
\left\langle
\lensingeff
\left(
\mathbf{p}_{\mathrm{c}}
\right)
\right\rangle
 = 
\int
\lensingeff\left(\zcl,\redshift|\mathbf{p}_{\mathrm{c}}\right)
P
\left(\redshift\right)
\dif\redshift
\, ,
\end{equation}
\begin{equation}
\label{eq:beta2_over_pz}
\left\langle
{\lensingeff}^2
\left(
\mathbf{p}_{\mathrm{c}}
\right)
\right\rangle
 = 
\int
{\lensingeff}^2\left(\zcl,\redshift|\mathbf{p}_{\mathrm{c}}\right)
P
\left(\redshift\right)
\dif\redshift
\, ,
\end{equation}
and the profiles ${\kappa}_{\mathrm{mod}}$ and ${\gamma}_{\mathrm{mod}}$ are calculated as
\begin{equation}
\label{eq:modelSigma}
{\kappa}_{\mathrm{mod}}
\left(
\theta|\Mwl, \mathbf{p}_{\mathrm{mis}}, \mathbf{p}_{\mathrm{c}}
\right) = 
\frac{
\Sigmam^{\mathrm{mod}}
\left(
\theta|\Mwl, \mathbf{p}_{\mathrm{mis}}, \mathbf{p}_{\mathrm{c}}
\right)
}
{
{\sigmacrit}
\left(
\mathbf{p}_{\mathrm{c}}
\right)
}
\, ,
\end{equation}
and
\begin{equation}
\label{eq:modeldeltaSigma}
{\gamma}_{\mathrm{mod}}
\left(
\theta|\Mwl, \mathbf{p}_{\mathrm{mis}}, \mathbf{p}_{\mathrm{c}}
\right) = 
\frac{
\deltaSigma^{\mathrm{mod}}
\left(
\theta|\Mwl, \mathbf{p}_{\mathrm{mis}}, \mathbf{p}_{\mathrm{c}}
\right)
}
{
{\sigmacrit}
\left(
\mathbf{p}_{\mathrm{c}}
\right)
}
\, ,
\end{equation}
where
\begin{equation}
\label{eq:modelsigmacrit}
\sigmacrit
\left(
\mathbf{p}_{\mathrm{c}}
\right)
= 
\frac{c^2}{4\pi G}\frac{
1
}{
D_{\mathrm{l}}
\left(
\zcl | \mathbf{p}_{\mathrm{c}}
\right) 
\left\langle
\lensingeff
\right\rangle
}
\, .
\end{equation}
Given that our source selection leads to a highly pure sample with cluster contamination that is consistent with zero without a clear radial trend (see Fig.~\ref{fig:contam}), we use a radially independent redshift distribution to calculate the critical surface density for all radii.
The observed redshift distribution $P\left(\redshift\right)$ used to calculate Eqs.~(\ref{eq:beta_over_pz}) and (\ref{eq:beta2_over_pz}) is estimated as
\begin{equation}
\label{eq:wpzbar}
P
\left(\redshift\right)
 = 
\frac{
\sum\limits_{s\in j } {\lensingw}_{s} P_{s}(\redshift)
}{
\sum\limits_{s\in j } {\lensingw}_{s}
} \, ,
\end{equation}
where the index $s$ runs over the sources with $R<3.5~\Mpch$ in the fiducial cosmology.

Given the set of cosmological parameters $\mathbf{p}_{\mathrm{c}}$, the calculation of $\Sigmam^{\mathrm{mod}}$ and $\deltaSigma^{\mathrm{mod}}$ is performed in the radial bins in physical units by converting $\theta$ to $R$, as
\begin{equation}
\label{eq:redshift-distance-relation}
R = D_{\mathrm{l}}\left(\zcl| \mathbf{p}_{\mathrm{c}}\right) \times \theta \, .
\end{equation}
This allows us to self-consistently compute the lensing profiles in every step of the likelihood exploration, given the variation of cosmological parameters.

The miscentering is accounted for in $\Sigmam^{\mathrm{mod}}$ (thus in $\deltaSigma^{\mathrm{mod}}$) as a composite model comprising a profile of a perfectly centered component and the profile of a miscentered component, weighted by a relative normalization associated with \fmis.
That is, 
\begin{align}
\label{eq:modelmiscentering}
\deltaSigma^{\mathrm{mod}}
&=
\left(1 - \fmis\right)
\deltaSigma^{\mathrm{cen}}
+
\fmis
\deltaSigma^{\mathrm{mis}}
\, .
\end{align}
The profile of $\Sigmam^{\mathrm{cen}}$ represents the surface mass density of a perfectly centered halo, which is evaluated using a NFW model. 
Meanwhile, the profile of $\Sigmam^{\mathrm{mis}}$ is modeled as an average of a set of NFW models with miscentering, weighted by a miscentering distribution $P\left(R_{\mathrm{mis}}\right)$, as
\begin{equation}
\label{eq:modelmiscentering_sigmam}
\Sigmam^{\mathrm{mis}}
\left(
R
\right)
 = 
 \int
\dif R_{\mathrm{mis}}
P\left(R_{\mathrm{mis}}\right)
\Sigmam^{\mathrm{cen}}
\left(
R|R_{\mathrm{mis}}
\right)
 \, ,
\end{equation}
where $\Sigmam^{\mathrm{cen}}\left(R|R_{\mathrm{mis}}\right)$ is the surface mass density of a halo with a central offset of $R_{\mathrm{mis}}$ azimuthally averaged over the positional angle $\phi$, expressed as \citep{yang06, johnston07a}
\begin{equation}
\label{eq:modelmiscentering_sigmam_2}
\Sigmam^{\mathrm{cen}}
\left(
R|R_{\mathrm{mis}}
\right)
 = 
 \int\limits_{0}^{2\pi}
 \frac{\dif\phi}
 {2\pi} 
\Sigmam^{\mathrm{cen}}
 \left(
 \sqrt{
 R^2 + R_{\mathrm{mis}}^2 + 2 R R_{\mathrm{mis}} \cos\phi
 }
 \right)
 \, .
\end{equation}

When estimating 
$\Sigmam^{\mathrm{cen}}$ and $\Sigmam^{\mathrm{mis}}$ in Eqs.~(\ref{eq:modelmiscentering}),
we fix the concentration parameter, given a weak-lensing mass \Mwl,
according to the concentration-to-mass relation from \cite{diemer15}.
With the concentration parameter, we calculate the corresponding scale radius $r_{\mathrm{s}}$ at the given \Mwl\  in order to convert the offset radius $R_{\mathrm{mis}}$ in Eq.~(\ref{eq:modelmiscentering_sigmam}) to the dimensionless radius $x$,
i.e., $x\equiv R_{\mathrm{mis}}/r_{\mathrm{s}}$.
In this way, we evaluate $P\left(R_{\mathrm{mis}}\right)$ at $R_{\mathrm{mis}}$ following the second component of the derived offset distribution $P\left(x\right)$, as in Eq.~(\ref{eq:miscentering}).
That is, we ignore the centering component,  $P_{\mathrm{cen}}$, in the offset distribution when calculating the profiles.
This is a reasonable approximation, because a miscentered shear profile with a miscentering distribution 
at the level of $\sigmacen\approx0.2$
in Eq.~(\ref{eq:centering_distirbutions}) is not significantly different from that of a perfectly centered model, $\Sigmam^{\mathrm{cen}}$, at the radial range of interest in this work.
By doing so,
we achieve a faster calculation of $\deltaSigma^{\mathrm{mod}}$ in Eq.~(\ref{eq:modelmiscentering}).

We note that we statistically account for the effect of miscentering in the modeling of observed shear profiles by assuming a universal distribution $P\left(R_{\mathrm{mis}}\right)$, as an average of all clusters.
This approach might not be optimal for individual clusters.
However, this is statistically correct in terms of accounting for the miscentering in an average behavior, which is the main focus of this work.
We refer readers to Sect.~\ref{sec:miscentering} for further discussion of this topic.

Finally, we compute the log probability of observing the shear profile $\gshear\left(\theta\right)$ given a weak-lensing mass \Mwl\ and a set of parameters,
$\mathbf{p} = \mathbf{p}_{\mathrm{mis}}\cup\mathbf{p}_{\mathrm{c}}$, 
while accounting for the covariance among radial bins, as
\begin{multline}
\label{eq:chi2}
\ln P\left(\gshear|\Mwl, \zcl, \mathbf{p}\right)
= \frac{-1}{2}
\times
\\
\left(
g_{+}^{\mathrm{mod}}\left(
\theta|\Mwl, \zcl, \mathbf{p}
\right)
 - \gshear\left(\theta\right)
\right)^{\mathrm{T}}
\bigcdot~
\mathbb{C}^{-1}
\bigcdot~
\\
\left(
g_{+}^{\mathrm{mod}}\left(
\theta|\Mwl, \zcl, \mathbf{p}
\right)
 - \gshear\left(\theta\right)
\right) + \mathrm{constant}
\,  .
\end{multline}
%

%
%

\section{Simulation calibrations}
\label{sec:simulations}

In this work, we use simulations to calibrate the relation between the observed and true underlying quantities.
These quantities are the X-ray count rate and the weak-lensing inferred mass, which are described in Sects.~\ref{sec:etabias} and~\ref{sec:bwl}, respectively.

\subsection{The X-ray count-rate bias}
\label{sec:etabias}

The observed count rate estimated by the \texttt{eSASS} pipeline is biased with respect to that enclosed by the cluster radius \Rfiveoo, which is referred to as the true count rate \truerate.
The main reason for this is that the observed count rate is estimated 
based on the best-fit model, which depends on the scale of extendedness \texttt{EXT}, which in turn is the best-fit core size that best facilitates the cluster detection and does not reflect the true underlying cluster core radius.
When constraining the true count-rate-to-mass-and-redshift relation, this bias needs to be accounted for.

In this work, we use large simulations to empirically calibrate the bias between the observed and true count rates at a given true cluster mass and redshift.
The simulation setup of point sources and clusters is fully described in \cite{comparat19} and \cite{comparat20}, respectively, to which we refer readers for more details.
In particular, the detail of the simulation scheme applied to eFEDS is described in \cite{liuteng21} and in Sect.~3 of \cite{liu21}, which is specifically for clusters.
For each simulated cluster, we first evaluate the true count rate \truerate\ enclosed by \Rfiveoo\  given the cluster true mass \Mfiveoo.
We note that the $K$-correction is included in calculating \truerate.
Meanwhile, we run the identical \texttt{eSASS} pipeline on the locations of simulated clusters to measure their observed count rate \rate.
In this way, we can derive the bias, 
\[
\brate \equiv \frac{\rate}{\truerate} \, , 
\]
as a function of cluster true mass and redshift.
We only use clusters with true counts larger than $40$ when calculating \brate, because the purity and completeness of the cluster sample drops significantly below this threshold.
The threshold at $40$ counts is motivated by the count distribution of eFEDS clusters: the minimum number of observed counts is $\approx40$ for the sample studied in this work.
Lowering this threshold would increase the false positive rate of the resulting cluster catalogs from the simulations.
A threshold of $30$ counts would increase the overall count-rate bias by $\approx4\percent$, which is still smaller than the modeling residual (see below) and hence is not a dominant factor in our analysis.
The results are shown in the middle panel of Fig.~\ref{fig:ratebias_sims}.

We empirically calibrate the bias by a power-law function in mass and redshift, as
\begin{multline}
\label{eq:bias_rate}
\brate\left(\Mfiveoo,\redshift\right) = 
\exp\left(\AF\right) \times
\left(\frac{\Mfiveoo}{\MPIV}\right)^{\BF + \deltaF\ln\left(\frac{\redshift}{\ZPIV}\right)} \left(\frac{\redshift}{\ZPIV}\right)^{\gammaF} \, ,
\end{multline}
with four parameters: $\left\lbrace\AF, \BF, \deltaF, \gammaF\right\rbrace$.
We use $\MPIV = 1.4\times10^{14}\Msunh$ and $\ZPIV = 0.35$ in Eq.~(\ref{eq:bias_rate}).
The best-fit model describing the bias is shown in the left panel of Fig.~\ref{fig:ratebias_sims}.
This model provides a good description of the simulated data, as suggested by their difference in the right panel of Fig.~\ref{fig:ratebias_sims}.
This difference, which is the modeling residual of the count-rate bias, can be well approximated by a Gaussian distribution with a standard deviation of $0.08$. 
This amount of modeling inaccuracy is absorbed into the Gaussian prior applied to the intrinsic scatter of observed count rates (see Sect.~\ref{sec:mcmc}).
We repeat the end-to-end procedure above for three different sets of simulations, which results in three sets of best-fit parameters.
These three sets of simulations differ from one another in terms of the underlying N-body dark-matter-only simulations, the area coverage, and the abundance of point-source populations (at a level of $25\percent$).
Exhaustive runs of various simulations spanning a complete list of configurations are too time-consuming and are beyond the scope of this work.
The goal of employing these different configurations is to gauge the inaccuracy of the derived count-rate bias as a systematic uncertainty that will be marginalized over in the analysis.
Therefore, we take the mean and standard error of the three sets of the best-fit parameters $\left\lbrace {\AF}_{,i}~, {\BF}_{,i}~, {\deltaF}_{,i}~, {\gammaF}_{,i}\right\rbrace_{i = 1, 2, 3}$ separately inferred from the three simulations as the priors that will be marginalized over.
The mean and standard error of these four parameters among the three simulations are
\begin{align}
\label{eq:ratebias_constraints}
\AF               &= 0.18     \pm 0.02    \, , \nonumber \\
\BF                &= -0.16   \pm 0.03   \, , \nonumber \\
\deltaF          &= -0.015 \pm 0.05   \, , \nonumber \\
\gammaF      &= 0.42     \pm 0.03   \, .
\end{align}
Finally, the constraints in Eq.~(\ref{eq:ratebias_constraints}) are used as the priors on these parameters in the modeling of scaling relations, effectively accounting for the bias in \rate.

It is worth mentioning that the synthetic clusters in the simulations are drawn from a library that is constructed based on real clusters, and that the resulting X-ray properties of the synthetic clusters have been shown to be in excellent agreement with the observed relations among luminosities, temperatures, and masses \citep{comparat20}.
Moreover, the halo shape, gas profile, and dynamical state of synthetic clusters are assigned to each simulated halo depending on the cluster mass and redshift, with dedicated treatments to capture the signature of cool-core clusters in terms of emissivity at cores \citep[see Section 3.2 in][]{comparat20}.
The resulting catalog of synthetic clusters has been shown to be in good agreement with the eFEDS clusters, and has been used to construct the selection function in \cite{liu21}.

It is important to stress that the modeling of the ``observed'' count rate \rate, as the main quantity of interest, is not affected by the resulting count-rate bias \brate.
This is because the parameters used to characterize the \rate--\Mfiveoo--\redshift\ relation are let free (see Eq.~(\ref{eq:countrate_to_mass}) in Sect.~\ref{sec:functionalforms}), such that a change in the resulting count-rate bias will be absorbed into the posteriors of the parameters given a fixed \rate.
As a result, the resulting count-rate bias will not change the overall modeling of the ``observed'' count-rate-to-mass-and-redshift relation, but only affects the inferred ``true'' count rate \truerate\ given a true cluster mass \Mfiveoo\ at a redshift \redshift.
However, including the modeling of the count-rate bias is still necessary, because it gives a flexible and complete functional form with which to empirically calibrate the  \rate--\Mfiveoo--\redshift\ relation.

\begin{figure}
\resizebox{0.5\textwidth}{!}{
\includegraphics[scale=1]{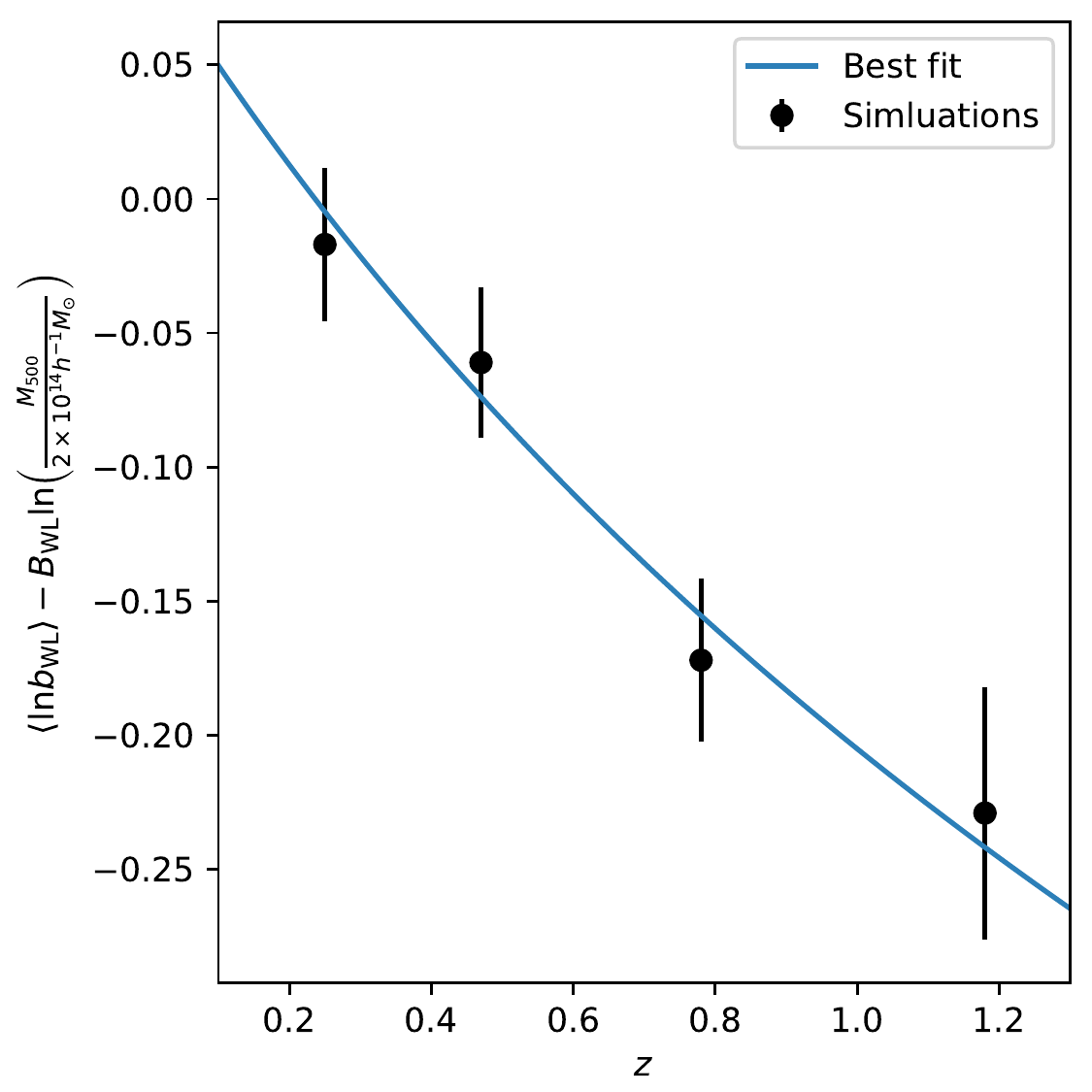}
}
\caption{
Best-fit weak-lensing mass bias \bwl\ normalized to the pivotal mass $2\times10^{14}\Msunh$ as a function of redshift.
This plot shows the results of the weak-lensing mass calibration including the modeling of the shear profiles at $R>0.2\Mpch$.
The results of the four snapshots in the hydrodynamics simulation are shown by the black points, while the best-fit redshift trend is indicated by the blue line.
We note that we account for the correlated uncertainties on the calibrated bias among the four snapshots of redshift when obtaining the best-fit relation.
}
\label{fig:grandisplot}
\end{figure}

\subsection{Weak-lensing mass bias}
\label{sec:bwl}

Weak lensing is expected to be the most direct way to probe the total mass of clusters, given that it does not rely on any assumption on the dynamical state of halos.
However, the weak-lensing inferred cluster mass could still be biased due to inaccurate assumptions \citep[e.g.,][]{pratt19,umetsu20}.
For example, clusters are triaxial \citep{clowe04,corless07,chiu18b}, such that the assumption of the spherical symmetry could introduce bias and additional scatter in the inferred mass.
Another example is that the density profile of clusters could deviate from the assumed functional form used in modeling due to the presence of correlated large-scale structures or substructures \citep{clowe04,king07}.
Moreover, the weak-lensing inferred mass depends on the fitting range of radii \citep{applegate14}, the accuracy of source photo-\redshift\ \citep{kelly14}, and the effect of miscentering \citep{rozo11b,vonderlinden14a}.
Therefore, the mass inferred from weak lensing is not free from bias and needs to be calibrated.

In this work, we calibrate the weak-lensing inferred mass by simulations, similarly to the methodology in \cite{schrabback18}, \cite{dietrich19}, and \cite{sommer21}.
We note that a similar approach was also adopted for shear-selected clusters \citep{chen20}.
The goal is to calibrate both the weak-lensing mass bias \bwl\ as a function of  true cluster mass and redshift, and the intrinsic scatter at fixed mass.
To properly account for the various systematic and statistical effects impacting weak-lensing mass measurements, we follow the method presented by \cite{grandis21}. 
We use the Magneticum\footnote{http://www.magneticum.org/index.html} \citep{dolag21}, a suite of hydrodynamical structure formation simulations, to create a library of halo density contrasts at different halo masses and redshifts. 
We then use the systematic and statistical properties of our lensing survey described in Sect.~\ref{sec:wlanalysis} to transform the density contrasts into synthetic shear profiles. 
These properties include (1) the multiplicative bias as a function of the source redshift, (2) the redshift distribution of selected sources at a given cluster redshift, (3) the cluster member contamination\footnote{Because there is no clear cluster contamination in this work (see Sect.~\ref{sec:contan}), we set the upper bound of the cluster contamination to $6\percent$ at a level of $2\sigma$ in the inner most bin, following a projected NFW model toward large radii.}, (4) the miscentering distribution of the eFEDS clusters, and (5) the photo-\redshift\ bias of lensing sources observed in the three-year weak-lensing data.
The shear profiles are then drawn including the scatter due to uncorrelated large-scale structures, as described in Eq.~(\ref{eq:covlensing}).
Finally, these shear profiles are fitted with our shear model following the identical procedure described in Sect.~\ref{sec:wlmodeling} to extract the weak-lensing mass \Mwl.
We note that the same radial binning is used in deriving the \Mwl\ from both synthetic shear profiles and those from the HSC data. 
Given the true halo mass \Mfiveoo\ from the simulation, we then fit a weak-lensing mass to true halo mass relation at a given cluster redshift by a log-normal distribution, resulting in a set of best-fit mean and variance.
The posteriors on these parameters are derived by drawing independent realizations of the shear library while varying the weak-lensing parameters, as described in \cite{grandis21}.

We perform the same procedure above at four different cluster redshifts: $0.25$, $0.47$, $0.78$, and $1.18$.
With the results at these four snapshots of redshift, we perform a joint fit of the weak-lensing mass bias as a function of the cluster true mass and redshift, which is described as
\begin{multline}
\label{eq:bwl}
\left\langle\ln\bwl\left(\Mfiveoo, \redshift\right)\right\rangle \equiv 
\left\langle\ln\left(\frac{\Mwl}{\Mfiveoo}\right)\right\rangle =\\
\ln \Awl + B_{\mathrm{WL}}\times \ln\left(\frac{\Mfiveoo}{2\times10^{14}\Msunh}\right) +
\gamma_{\mathrm{WL}} \times \ln\left(\frac{1 + \redshift}{1 + 0.6}\right)
\, ,
\end{multline}
with log-normal intrinsic scatter \sigmawl\ at fixed mass,
\begin{equation}
\label{eq:bwl_sigma}
\sigmawl \equiv \left(\mathrm{Var}\left[\ln\left(\bwl|\Mfiveoo\right)\right] \right)^{\frac{1}{2}}
\, .
\end{equation}
The resulting constraints on the parameters are
\begin{align}
\label{eq:wlbias_constraints}
\Awl               &= 0.896   \pm 0.076    \, , \nonumber \\
\Bwl                &= -0.055 \pm 0.021   \, , \nonumber \\
\gammawl      &= -0.426 \pm 0.087   \, , \nonumber \\
\sigmawl        &= 0.296   \pm 0.047 \, .
\end{align}
We note that we do not observe significant redshift- or mass-dependent scatter  \sigmawl\ in the weak-lensing mass bias; we therefore use the same  \sigmawl\ for all clusters in this work.
The weak-lensing mass bias obtained from the simulations and the best-fit relation as a function of redshift are shown in Fig.~\ref{fig:grandisplot}, in which we can see that the model provides a good description of the resulting \bwl.

We also repeat the calibration procedure for the modeling without the cluster core (i.e., excluding the three inner most bins), for which the resulting constraints are
\begin{align}
\label{eq:wlbias_constraints_nocore}
\Awl               &= 0.903   \pm 0.057    \, , \nonumber \\
\Bwl                &= -0.057 \pm 0.022   \, , \nonumber \\
\gammawl      &= -0.474 \pm 0.080   \, , \nonumber \\
\sigmawl        &= 0.238   \pm 0.032  \, .
\end{align}
These constraints are used as the priors on the parameters of the weak-lensing-mass-bias-to-mass relation, which accounts for the bias of weak-lensing masses in the likelihood modeling.

\begin{table}
\centering
\caption{
Summary of the priors used in the modeling (see Sect.~\ref{sec:mcmc} for more details).
The first and second columns represent the name of and the prior on the parameters, respectively. 
The former and latter priors on the parameters of the \Mwl--\Mfiveoo--\redshift\ relation describe the cases of modeling the lensing profile with $R>0.2\Mpch$ and $R>0.5\Mpch$, respectively.
}
\label{tab:priors}
\resizebox{!}{0.4\textheight}{
\begin{tabular}{ll}
\hline\hline
Parameter &Prior \\[3pt]
\hline
\multicolumn{2}{c}{The \Mwl--\Mfiveoo--\redshift\ relation} \\
\multicolumn{2}{c}{Eq.~(\ref{eq:bwl}) } \\
\hline
\Awl         &$\mathcal{N}(0.896,0.076^2)$  or $\mathcal{N}(0.903,0.057^2)$    \\[3pt]
\Bwl          &$\mathcal{N}(-0.055,0.021^2)$ or $\mathcal{N}(-0.057,0.022^2)$  \\[3pt]
\gammawl &$\mathcal{N}(-0.426,0.087^2)$ or $\mathcal{N}(-0.474,0.08^2)$   \\[3pt]
\sigmawl   &$\mathcal{N}(0.293,0.047^2)$  or $\mathcal{N}(0.238,0.032^2)$   \\[3pt]
\hline
\multicolumn{2}{c}{The \rate--\Mfiveoo--\redshift\ relation} \\
\multicolumn{2}{c}{Eq.~(\ref{eq:countrate_to_mass}) with $C_{\mathrm{SS},\rate} = 2$ } \\[3pt]
\hline
\Aeta           &$\mathcal{U}(0,0.5)$              \\[3pt]
\Beta           &$\mathcal{U}(0,5)$                 \\[3pt]
\deltaeta      &$\mathcal{U}(-3,3)$               \\[3pt]
\gammaeta  &$\mathcal{U}(-3,3)$               \\[3pt]
\sigmaeta    &$\mathcal{N}(0.3, 0.08^2)$    and $\mathcal{U}(0.05, 0.8)$ \\[3pt]
\AF             &$\mathcal{N}(0.18,0.02^2)$    \\[3pt]
\BF             &$\mathcal{N}(-0.16,0.03^2)$   \\[3pt]
\deltaF       &$\mathcal{N}(-0.015,0.05^2)$ \\[3pt]
\gammaF   &$\mathcal{N}(0.42,0.03^2)$     \\[3pt]
\hline
\multicolumn{2}{c}{The \Xlabel--\Mfiveoo--\redshift\ relation} \\
\multicolumn{2}{c}{Eq.~(\ref{eq:x_to_mass}) } \\[3pt]
\hline
$B_{\Xlabel}$               &$\mathcal{U}(0,5)$  \\[3pt]
$\delta_{\Xlabel}$        &$\mathcal{U}(-3,3)$  \\[3pt]
$\gamma_{\Xlabel}$    &$\mathcal{U}(-3,3)$  \\[3pt]
$\sigma_{\Xlabel}$      &$\mathcal{U}(0.05, 0.8)$ \\[3pt]
\hline
\multicolumn{2}{c}{$\Xlabel = \Lx$ with $C_{\mathrm{SS},\Lx} = 2$}\\[3pt]
$\frac{A_{\Lx}}{\mathrm{ergs}}$             &$\mathcal{U}(10^{41},10^{46})$ \\[3pt]
\hline
\multicolumn{2}{c}{$\Xlabel = \Lb$ with $C_{\mathrm{SS},\Lb} = 7/3$}\\[3pt]
$\frac{A_{\Lb}}{\mathrm{ergs}}$             &$\mathcal{U}(10^{41},10^{44})$  \\[3pt]
\hline
\multicolumn{2}{c}{$\Xlabel = \Tx$ with $C_{\mathrm{SS},\Tx} = 2/3$}\\[3pt]
$\frac{A_{\Tx}}{k\mathrm{eV}}$            &$\mathcal{U}(0.5,25)$  \\[3pt]
\hline
\multicolumn{2}{c}{$\Xlabel = \Mg$ with $C_{\mathrm{SS},\Mg} = 0$} \\[3pt]
$\frac{A_{\Mg}}{\Msun}$            &$\mathcal{U}(5\times10^{11},5\times10^{13})$  \\[3pt]
\hline
\multicolumn{2}{c}{$\Xlabel = \Yx$ with $C_{\mathrm{SS},\Yx} = 2/3$}\\[3pt]
$\frac{A_{\Yx}}{k\mathrm{eV}\cdot\Msun}$            &$\mathcal{U}(10^{12},10^{14})$  \\[3pt]
\hline
\multicolumn{2}{c}{Correlated scatter} \\
\hline
$\rho_{\mathrm{WL},\mathcal{X}}$, $\rho_{\mathrm{WL},\rate}$,$\rho_{\mathcal{X},\rate}$ & $\mathcal{U}(-0.9,0.9)$ \\[3pt]
\hline
\multicolumn{2}{c}{Miscentering} \\
\hline
\fmis         &$\mathcal{N}(0.54,0.02^2)$  \\[3pt]
\sigmamis &$\mathcal{N}(0.61,0.03^2)$  \\[3pt]
\hline
\multicolumn{2}{c}{Cosmological parameters} \\
\hline
\OmegaM     &$\mathcal{N}(0.3,0.016^2)$  \\[3pt]
\sigmaeight &$\mathcal{N}(0.8,0.014^2)$   \\[3pt]
\Hnow         &$\mathcal{N}(70,5.6^2)$        \\[3pt]
\hline\hline
\end{tabular}
}
\end{table}
%

%
%

\section{Modeling of X-ray-observable-to-mass-and-redshift relations}
\label{sec:xtom_modeling}

In this section, we describe the forward modeling of X-ray observables as functions of the cluster mass and redshift.
Given the X-ray observables and weak-lensing shear profiles of a sample of eFEDS clusters, our goal is to model the X-ray-observable-to-mass-and-redshift relations by simultaneously fitting these observables on the basis of individual clusters.

We perform two types of modeling.
First, we model the count rate \rate\ as a function of cluster mass and redshift together with the mass calibration, using the observed shear profiles from the HSC data.
Second, we 
model the follow-up X-ray observables (\Lx, \Lb, \Tx, \Mg, and \Yx) as a function of cluster mass and redshift, jointly with the modeling of the count rate and the weak-lensing mass calibration.
These two types of modeling are described in Sects.~\ref{sec:wlmcalib} and ~\ref{sec:wlsrmcalib}, respectively.

It is important to stress that these fitting frameworks account for Malmquist bias arising from the selection function of eFEDS clusters, and for \citet{eddington13} bias due to the steep gradient of the halo mass function.
We have fully verified and extensively used these modeling strategies in previous work \citep{bocquet15,liu15a,chiu16c,chiu18a,bulbul19,bocquet19,chiu20,chiu20b,schrabback21}, to which we refer readers for more details.

Although the eFEDS clusters are selected by both the detection likelihood and extent likelihood, instead of the observed count rate \rate\ as the main X-ray proxy used in this work, we note that our strategy to model the scaling relation still provides the unbiased result, as demonstrated in \cite{grandis20}.

It is important to stress that we do not impose any additional cut on the follow-up observables, including the weak-lensing shear profile, such that the selection of the cluster sample does not depend on them and is only determined by the selection used in constructing the initial cluster catalog.
Therefore, the likelihoods presenting the weak-lensing mass calibration and the modeling of the X-ray scaling relations (in Sects.~\ref{sec:wlmcalib}~and~\ref{sec:wlsrmcalib}, respectively) are complete and free from the selection bias.
Moreover, the inclusion of the modeling of the correlated scatter among the X-ray follow-up observable, the weak-lensing mass, and the count rate accounts for the possible selection biases in the X-ray scaling relations.

By using mock catalogs that are at least ten times larger than the eFEDS sample, we verified that the modeling codes can recover the input parameters within statistical uncertainties that are $\sqrt{10}\approx3$ times smaller than those expected for the real sample.
The mock catalogs are generated following the procedure in \cite{grandis19}, and we perform the modeling on them in an identical way to that used on the real data.
This ensures that the results in this work are robust and unbiased.

\subsection{Functional forms of X-ray scaling relations}
\label{sec:functionalforms}

We first summarize the functional forms used to describe the X-ray observable-to-mass-and-redshift relations.
The adopted form follows that in \cite{bulbul19}, in which there are five parameters describing the behavior of an observable as a function of mass and redshift.
Specifically, for an X-ray observable labeled \Xlabel, we have
\begin{itemize}
\item $A_{\Xlabel}$ describing the normalization at the pivotal mass \MPIV\ and the pivotal redshift \ZPIV;
\item $B_{\Xlabel}$ describing the overall power-law index in mass;
\item $\gamma_{\Xlabel}$ describing the deviation from the self-similar redshift evolution \citep{kaiser1986} of the observable \Xlabel;
\item $\delta_{\Xlabel}$ characterizing the redshift-dependent power-law index in the cluster mass;
\item $\sigma_{\Xlabel}$ describing the log-normal scatter of \Xlabel\ at fixed mass.
\end{itemize}
In this way, the X-ray observable-to-mass-and-redshift (\Xlabel--\Mfiveoo--\redshift) relation reads,
\begin{multline}
\label{eq:x_to_mass}
\left\langle\ln\mathcal{X}|\Mfiveoo\right\rangle 
= \ln A_{\mathcal{X}} + \\
\left[ B_{\mathcal{X}} + \delta_{\mathcal{X}}\ln\left(\frac{1 + \redshift}{1 + \ZPIV}\right) \right] \times
\ln\left(\frac{\Mfiveoo}{\MPIV}\right) +
\\
C_{\mathrm{SS},\mathcal{X}} \times \ln\left(\frac{\Ez}{\Ezpiv}\right) + 
\gamma_{\mathcal{X}} \times 
\ln \left(\frac{1 + \redshift}{1 + \ZPIV}\right)
\, ,
\end{multline}
with log-normal intrinsic scatter at fixed mass of 
\begin{equation}
\label{eq:intrinsic_scatter_x}
\sigma_{\mathcal{X}} \equiv \left(\mathrm{Var}\left[\ln\mathcal{X}|\Mfiveoo \right] \right)^{\frac{1}{2}} \, ,
\end{equation}
where $C_{\mathrm{SS},\mathcal{X}}$ is the power-law index of the observable $\mathcal{X}$ in redshift predicted by the self-similar model.
In this work, we have five follow-up X-ray observables, namely $\mathcal{X} = \{\Lx, \Lb, \Mg, \Tx,\Yx\}$.
We model the \Xlabel--\Mfiveoo--\redshift\ relation separately for each X-ray observable.
The self-similar prediction of $C_{\mathrm{SS},\mathcal{X}}$ for each of them is listed in Table~\ref{tab:priors}.

For the count-rate-to-mass-and-redshift (\rate--\Mfiveoo--\redshift) relation, we additionally include two factors: 
 the scaling depending on the luminosity distance $D_{\mathrm{L}}$ to the redshift of clusters \citep{grandis19}, and  the count-rate bias $\brate$ calibrated by the simulations (see Sect.~\ref{sec:etabias}).
The resulting \rate--\Mfiveoo--\redshift\ relation reads,
\begin{multline}
\label{eq:countrate_to_mass}
\left\langle\ln\rate|\Mfiveoo\right\rangle 
= \ln \Aeta + \\
\left[ \Beta + \deltaeta\ln\left(\frac{1 + \redshift}{1 + \ZPIV}\right) \right] \times
\ln\left(\frac{\Mfiveoo}{\MPIV}\right) +
\\
C_{\mathrm{SS},\rate} \times \ln\left(\frac{\Ez}{\Ezpiv}\right) + 
\gammaeta \times 
\ln \left(\frac{1 + \redshift}{1 + \ZPIV}\right)
\\
-2\times\ln\left(\frac{D_{\mathrm{L}}\left(\redshift\right)}{D_{\mathrm{L}}\left(\ZPIV\right)}\right)
+ \ln\left( \brate\left(\Mfiveoo, \redshift\right) \right) 
\, ,
\end{multline}
with log-normal intrinsic scatter at fixed mass of 
\begin{equation}
\label{eq:intrinsic_scatter_rate}
\sigmaeta \equiv \left(\mathrm{Var}\left[\ln \rate|\Mfiveoo\right]\right)^{\frac{1}{2}} \, ,
\end{equation}
Throughout this work, except for \Tx\ and \Yx, we use the pivotal mass $\MPIV=1.4\times10^{14}\Msunh$ and the pivotal redshift $\ZPIV=0.35$ as the median values of our eFEDS cluster sample.
As the scaling relations of \Tx\ and \Yx\ are derived based on the subsample (see Sect.~\ref{sec:xray_data}) with a relatively high mass scale, we instead use the pivotal mass $\MPIV=2.5\times10^{14}\Msunh$ for $\Xlabel \in \left\lbrace\Tx, \Yx\right\rbrace$.

\subsection{Weak-lensing mass calibration}
\label{sec:wlmcalib}

Our goal is to model the observed X-ray count rate \rate\ of eFEDS clusters as a function of cluster mass and redshift with the weak-lensing calibration using the observed shear profile of individual clusters.

Specifically, given a cluster at redshift \zcl\ with an observed X-ray count rate \rate\ and a shear profile \gshear, we compute the probability of observing \gshear\ given \rate\ and a set of the parameter $\mathbf{p}$, namely,
\begin{align}
\label{eq:mcalib_single}
P\left(\gshear|\rate,\zcl,\mathbf{p}\right) &=
\frac{
N\left(\gshear, \rate|\zcl,\mathbf{p}\right)
}{
N\left(\rate|\zcl,\mathbf{p}\right)
} \nonumber \\
& =
\frac{
\int
P\left(\gshear,\rate|\Mfiveoo,\zcl,\mathbf{p}\right)
n\left(\Mfiveoo,\zcl\right)\dif\Mfiveoo
}{
\int
P\left(\rate|\Mfiveoo,\zcl,\mathbf{p}\right)
n\left(\Mfiveoo,\zcl\right)\dif\Mfiveoo
}
\, ,
\end{align}
where $N\left(\rate|\zcl,\mathbf{p}\right)$ and $N\left(\gshear, \rate|\zcl,\mathbf{p}\right)$ denote the numbers of clusters observed with the count rate \rate\ and the observable set $\{ \gshear, \rate\}$, respectively; 
$n\left(\Mfiveoo,\zcl\right)$ is the halo mass function evaluated at the true cluster mass \Mfiveoo\ at the cluster redshift \zcl\ using the \cite{bocquet16} fitting formula with the inclusion of baryons;
$P\left(\rate|\Mfiveoo,\zcl,\mathbf{p}\right)$ and 
$P\left(\gshear,\rate|\Mfiveoo,\zcl,\mathbf{p}\right)$ describe the probabilities of observing \rate\ and $\{ \gshear, \rate\}$, respectively, for the cluster with the true mass \Mfiveoo\ at the redshift \zcl\ given the parameter set $\mathbf{p}$.
In Eq.~(\ref{eq:mcalib_single}), the inclusion of the halo mass function is needed to account for the \citet{eddington13} bias.
We note that the difference in the final results obtained between the dark-matter-only and hydro-simulation mass functions is expected to be negligible given the current sample size \citep{bocquet16,castro21}.

The probability $P\left(\rate|\Mfiveoo,\zcl,\mathbf{p}\right)$ in the denominator of Eq.~(\ref{eq:mcalib_single}) is calculated as
\begin{multline}
\label{eq:mcalib_denominator}
P\left(\rate|\Mfiveoo,\zcl,\mathbf{p}\right) =
\int\dif\tilde\rate
P\left(\rate|\tilde\rate\right)
P\left(\tilde\rate|\Mfiveoo,\zcl,\mathbf{p}\right)
\, ,
\end{multline}
where 
the term $P\left(\tilde\rate|\Mfiveoo,\zcl,\mathbf{p}\right)$ describes the log-normal distribution of the pristine count rate $\tilde\rate$ with the mean following Eq.~(\ref{eq:countrate_to_mass}) and the intrinsic scatter \sigmaeta.
This characterizes the intrinsic distribution of the pristine count rate while including the observed bias \brate\ calibrated by the simulations. 
The term $P\left(\rate|\tilde\rate\right)$ accounts for the measurement uncertainty of \rate\
with respect to $\tilde\rate$.

The probability $P\left(\gshear,\rate|\Mfiveoo,\zcl,\mathbf{p}\right)$ reads
\begin{multline}
\label{eq:mcalib_num}
P\left(\gshear,\rate|\Mfiveoo,\zcl,\mathbf{p}\right) = \\
\int\int\dif\tilde\rate\dif\Mwl
P\left(\gshear|\Mwl\right) 
P\left(\rate|\tilde\rate\right)
P\left(\Mwl, \tilde\rate|\Mfiveoo,\zcl,\mathbf{p}\right) 
\, ,
\end{multline}
in which the term $P\left(\gshear|\Mwl\right)$ accounts for the measurement uncertainty of the observed \gshear\ given the weak-lensing mass \Mwl, which is evaluated using Eq.~(\ref{eq:chi2}).
The term $P\left(\Mwl, \tilde\rate|\Mfiveoo,\zcl,\mathbf{p}\right)$ then describes a joint distribution of \Mwl\ and $\tilde\rate$ characterized by both the intrinsic scatter of these two
quantities and the correlated scatter between them given the underlying true cluster mass \Mfiveoo.
That is, $P\left(\Mwl, \tilde\rate|\Mfiveoo,\zcl,\mathbf{p}\right)$ follows a multivariate log-normal distribution of $\left(\Mwl, \tilde\rate\right)$ characterized by an intrinsic covariance matrix, 
\begin{equation}
\label{eq:lognormal_covar}
\Sigma_{\Mwl, \rate} = 
\begin{pmatrix}
\sigma_{\mathrm{WL}}^2 & \sigma_{\mathrm{WL}} \sigmaeta \rho_{\mathrm{WL},\rate} \\
\sigma_{\mathrm{WL}} \sigmaeta \rho_{\mathrm{WL},\rate} & \sigmaeta^2
\end{pmatrix}
\, ,
\end{equation}
where $\rho_{\mathrm{WL},\rate}$ is the correlation coefficient between \Mwl\ and $\hat{\rate}$.

In this way, the joint likelihood of a sample comprising \Ncl\ eFEDS clusters, given the set of observables $\{ {\gshear}_{,i}, \rate_{i}\}_{i=1}^{\Ncl}$, is calculated as
\begin{equation}
\label{eq:like_mcalib}
\mathcal{L} \left(\mathbf{p}\right)
= \prod\limits_{i=1}^{\Ncl}
P\left({\gshear}_{,i}|\rate_{i}, \mathbf{p}\right)
\, .
\end{equation}
In this framework, we have a parameter set $\mathbf{p}$ consisting of 19 free parameters, that is, 
\begin{align}
\label{eq:parameter_mcalib}
\mathbf{p} &= 
\mathbf{p}_{\rate}\cup
\mathbf{p}_{\mathrm{F}}\cup
\mathbf{p}_{\mathrm{WL}}\cup
\mathbf{p}_{\mathrm{mis}}
\cup\{\rho_{\mathrm{WL},\rate}\}
\cup\mathbf{p}_{\mathrm{c}}
\, , \nonumber \\
\mathbf{p}_{\rate} &= \{\Aeta, \Beta, \deltaeta, \gammaeta, \sigmaeta\} \, , \nonumber \\
\mathbf{p}_{\mathrm{F}} &= \{\AF, \BF, \gammaF, \sigmaF\} \, , \nonumber \\
\mathbf{p}_{\mathrm{WL}} &= \{\Awl, \Bwl, \gammawl, \sigmawl\} \, , \nonumber \\
\mathbf{p}_{\mathrm{mis}} &= \{\fmis, \sigmamis\} \, , \nonumber \\
\mathbf{p}_{\mathrm{c}} &= \{\OmegaM, \Hnow, \sigmaeight\}, \,
\end{align}
where  $\mathbf{p}_{\rate}$ describes the \rate--\Mfiveoo--\redshift\ relation as in Eqs.~(\ref{eq:countrate_to_mass});
$\mathbf{p}_{\mathrm{F}}$ and $\mathbf{p}_{\mathrm{WL}}$ describe the \brate--\Mfiveoo--\redshift\ and \Mwl--\Mfiveoo--\redshift\  relations in  Eqs.~(\ref{eq:bias_rate})~and~(\ref{eq:bwl}), respectively; and 
$\mathbf{p}_{\mathrm{mis}}$ characterizes the miscentering distribution as in Eq.~(\ref{eq:miscentering});
$\mathbf{p}_{\mathrm{c}}$ consists of the cosmological parameters, i.e., \OmegaM, \Hnow, and \sigmaeight, which we vary in the modeling below.

\begin{figure*}
\resizebox{\textwidth}{!}{
\includegraphics[scale=1]{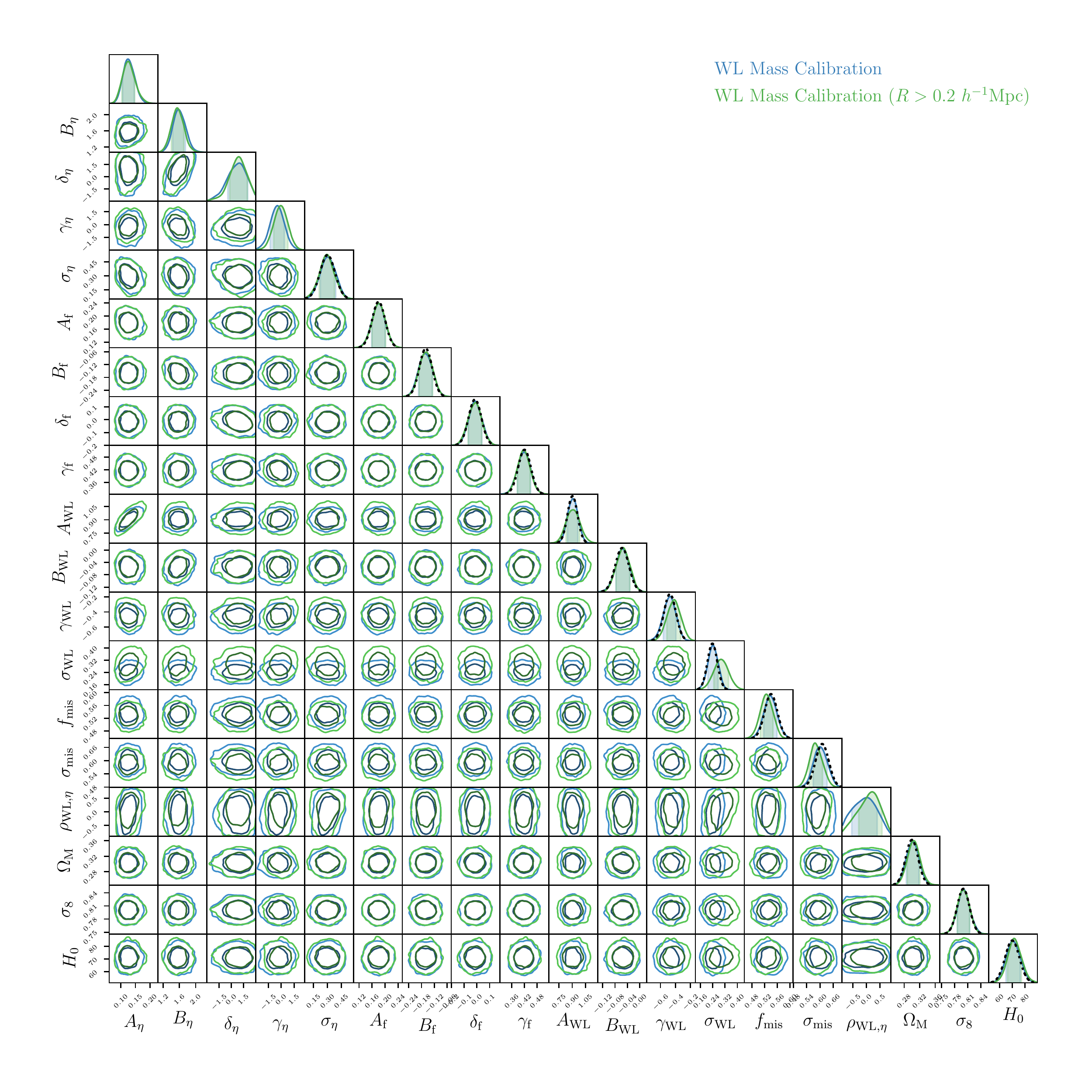}
}
\caption{
Parameter constraints in the joint modeling of X-ray count rates and weak-lensing shear profiles.
The results from the modeling of shear profiles at $R>0.2$~\Mpch\ and $R>0.5$~\Mpch\ are shown in green and blue, respectively.
The marginalized posteriors of the parameters are presented in the diagonal subplots, where the priors are indicated by the dotted lines.
The off-diagonal subplots contain the correlations between the parameters.
The contours display the $1\sigma$ and $2\sigma$ confidence levels.
We only show the priors in the case of fitting shear profiles at $R>0.5$~\Mpch, for clarity.
As seen, there is no clear evidence of a difference in the results between the modeling of shear profiles at $R>0.2$~\Mpch\ and $R>0.5$~\Mpch.
}
\label{fig:chain_mcalib}
\end{figure*}
\begin{figure*}
\centering
\resizebox{\textwidth}{!}{
\includegraphics[scale=1]{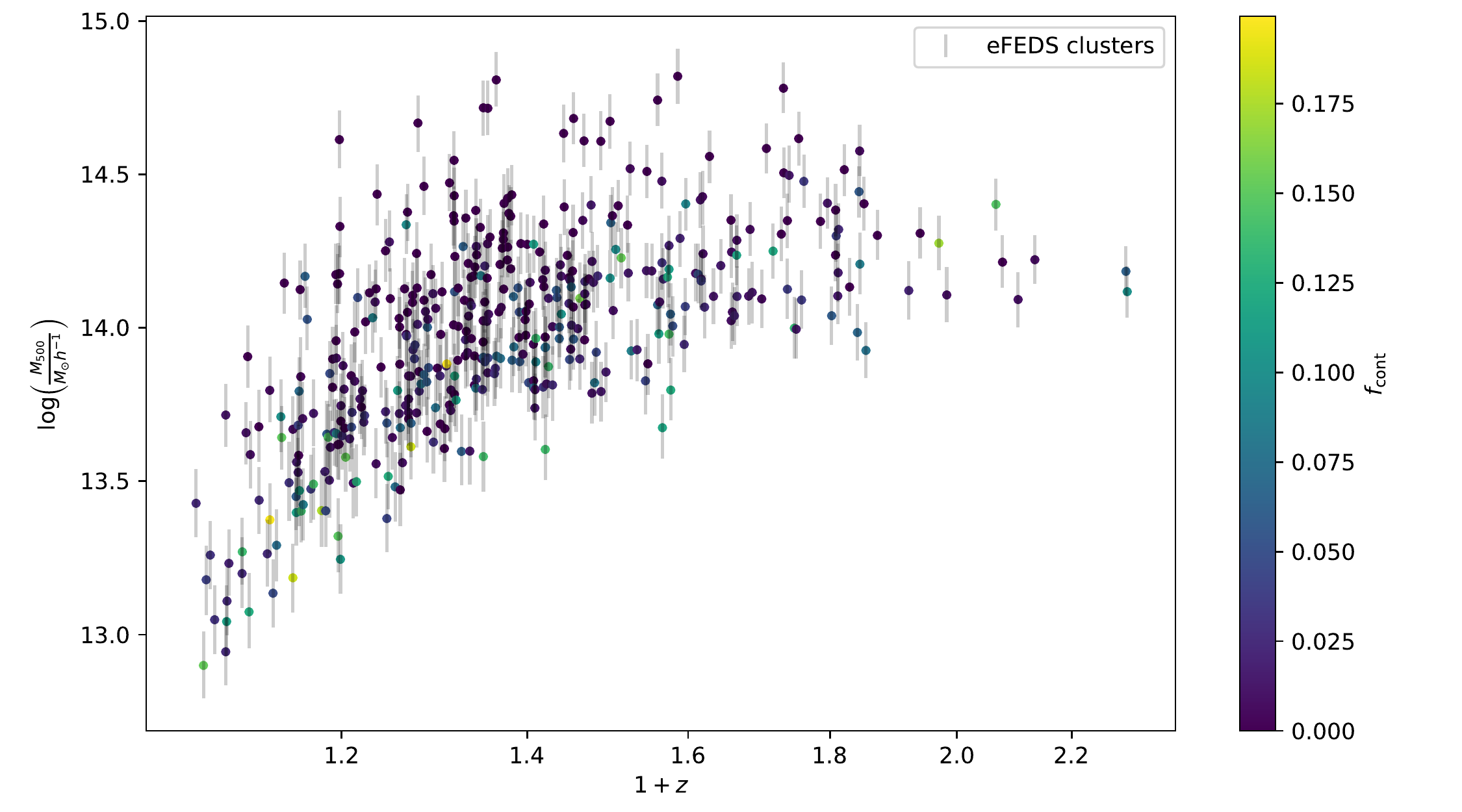}
}
\caption{
Cluster mass \Mfiveoo\ and redshift \redshift\ of each cluster color coded by the quantity \fcont.
The mass is randomly sampled from the posterior of \Mfiveoo\ as an ensemble mass (see Eq.~(\ref{eq:mass_post})).
}
\label{fig:mz}
\end{figure*}
\begin{figure*}
\centering
\resizebox{0.48\textwidth}{!}{
\includegraphics[scale=1]{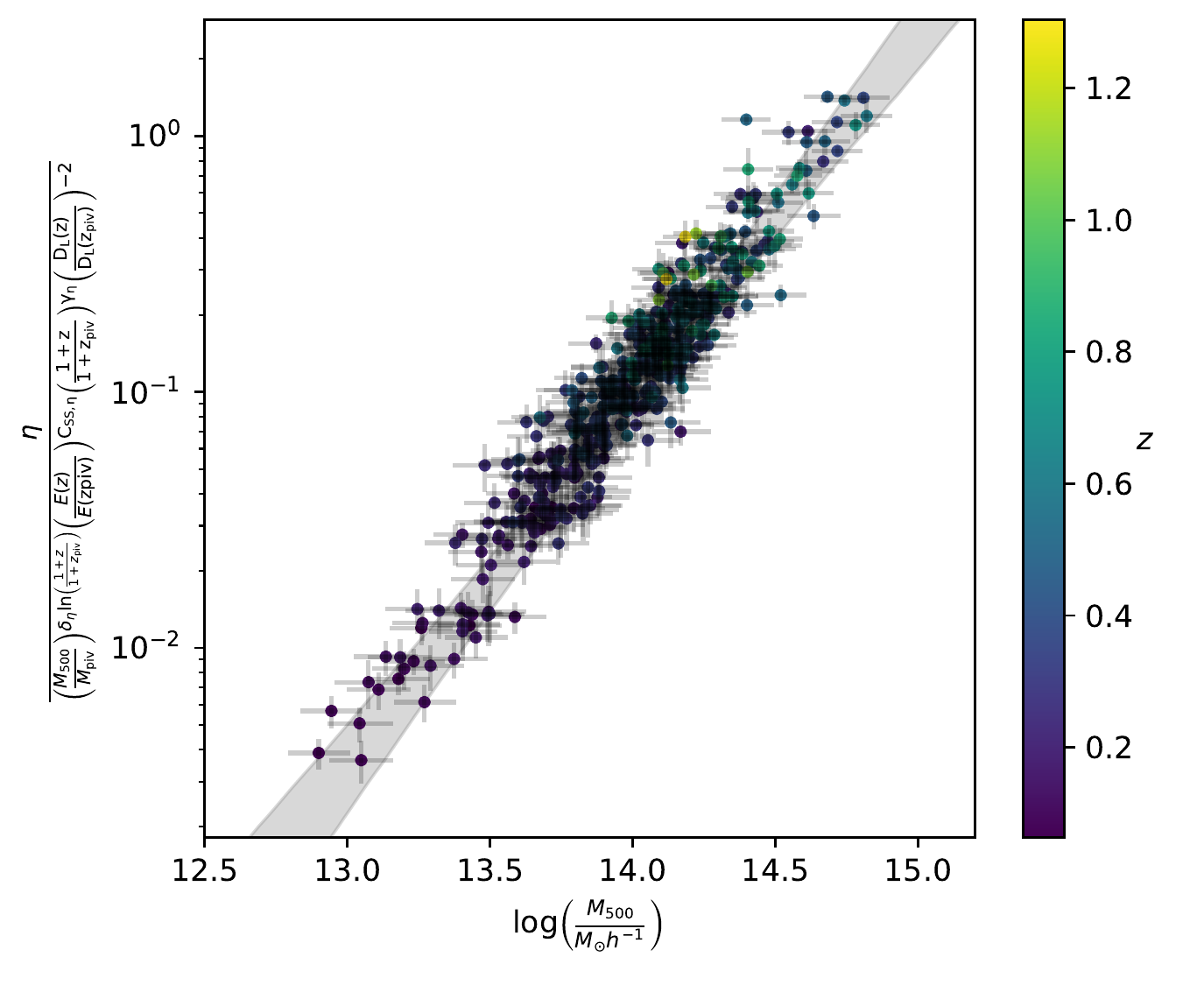}
}
\resizebox{0.48\textwidth}{!}{
\includegraphics[scale=1]{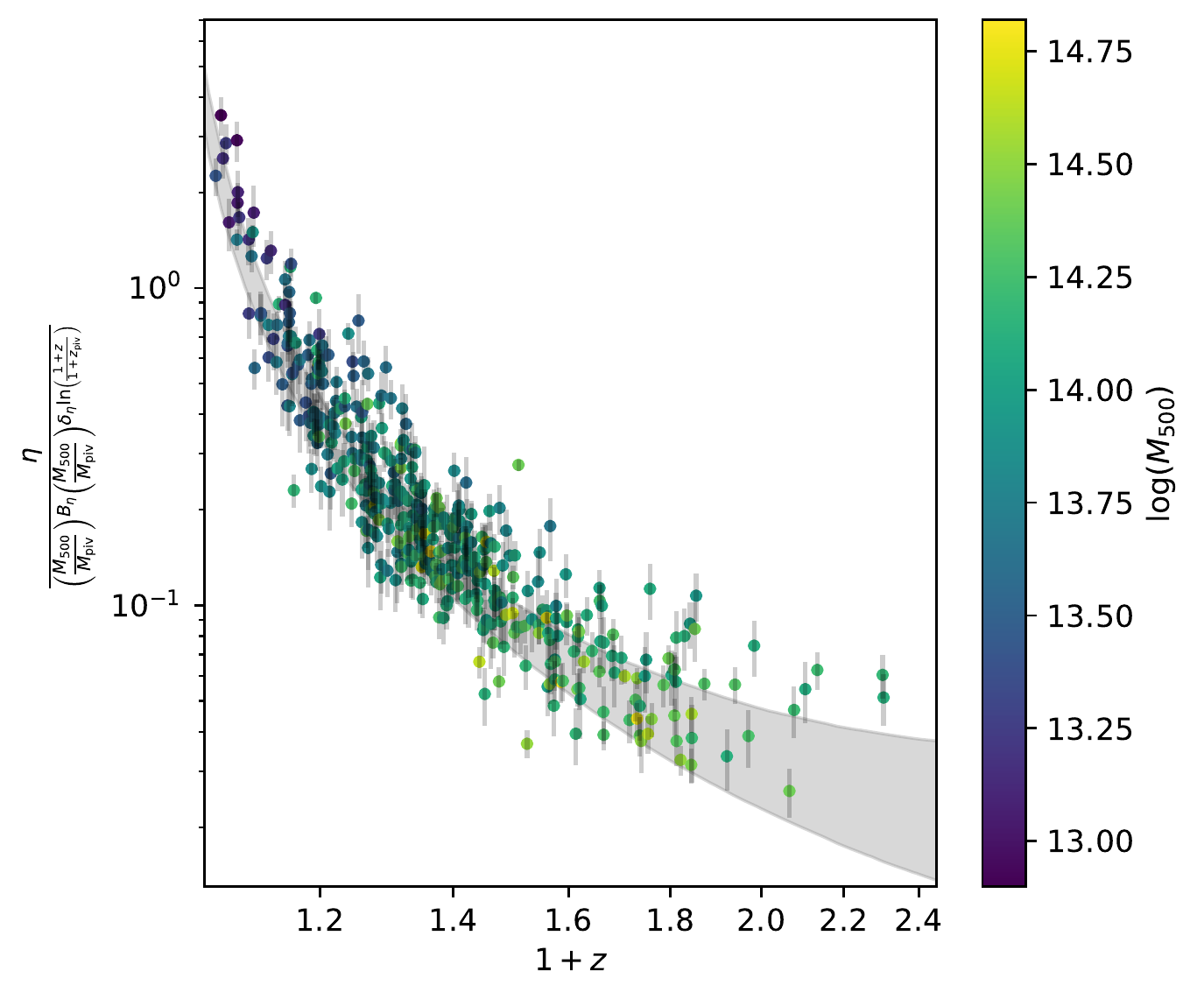}
}
\caption{
Observed count rate \rate\ as a function of the cluster mass \Mfiveoo\ (left panel) and redshift \redshift\ (right panel).
The circles in the left and right panels represent the measurements of \rate\ color-coded by the cluster redshift and mass, respectively.
The gray shaded regions are the best-fit models.
In the left (right) panel, we normalize the observed count rate to that at the pivotal redshift $\ZPIV = 0.35$ (pivotal mass $\MPIV = 1.4\times10^{14}\Msunh$).
}
\label{fig:cmz}
\end{figure*}

\subsection{Joint modeling of X-ray scaling relations}
\label{sec:wlsrmcalib}

A joint modeling of an X-ray observable \Xlabel\ scaling relation and the weak-lensing mass calibration (as described in Sect.~\ref{sec:wlmcalib}) is described as follows.
Given an eFEDS cluster at redshift \zcl\ with an available set of observables $\{\Xlabel, \gshear, \rate\}$, we evaluate the probability of observing the cluster with \Xlabel\ and \gshear, given \rate\ and the set of parameters $\mathbf{p}$.
That is, 
\begin{align}
\label{eq:like_sr_single}
P\left(\mathcal{X},\gshear|\rate, \zcl, \mathbf{p}\right) &=
\frac{
 N\left(\mathcal{X},\gshear,\rate|\zcl, \mathbf{p}\right) 
}{
 N\left(\rate|\zcl, \mathbf{p}\right) 
}
\, \nonumber \\
&= 
\frac{
\int
P\left(\mathcal{X}, \gshear,\rate|\Mfiveoo,\zcl,\mathbf{p}\right)
n\left(\Mfiveoo,\zcl\right)\dif\Mfiveoo
}{
\int
P\left(\rate|\Mfiveoo,\zcl,\mathbf{p}\right)
n\left(\Mfiveoo,\zcl\right)\dif\Mfiveoo
}
\, ,
\end{align}
where $P\left(\mathcal{X}, \gshear,\rate|\Mfiveoo,\zcl,\mathbf{p}\right)$ is the joint probability of observing the cluster with the observables of $\{\mathcal{X}, \gshear, \rate\}$ at the true cluster mass \Mfiveoo\ and redshift \zcl, given the parameter $\mathbf{p}$.
This term is calculated as
\begin{multline}
\label{eq:xsrmcalib_num}
P\left(\mathcal{X}, \gshear,\rate|\Mfiveoo,\zcl,\mathbf{p}\right)
= \int\int\int\dif\tilde\rate\dif\Mwl\dif\tilde\Xlabel
\\
P\left(\mathcal{X}|\tilde\Xlabel\right)
P\left(\gshear|\Mwl\right) 
P\left(\rate|\tilde\rate\right) 
P\left(\tilde\Xlabel, \Mwl, \tilde\rate|\Mfiveoo,\zcl,\mathbf{p}\right)
\, .
\end{multline}

In an analogy to Eq.~(\ref{eq:mcalib_num}), the term $P\left(\tilde\Xlabel, \Mwl, \tilde\rate|\Mfiveoo,\zcl,\mathbf{p}\right)$ describes the intrinsic distribution of $\{ \tilde\Xlabel , \Mwl, \tilde\rate\}$ given the  
true cluster mass \Mfiveoo, assuming a multivariate log-normal distribution characterized by an intrinsic covariance matrix, 
\begin{equation}
\label{eq:lognormal_covar_3params}
\Sigma_{\Mwl, \mathcal{X}, \rate} = 
\begin{pmatrix}
\sigma_{\mathrm{WL}}^2 
& \sigma_{\mathrm{WL}} \sigma_{\mathcal{X}} \rho_{\mathrm{WL},\mathcal{X}} 
& \sigma_{\mathrm{WL}} \sigmaeta \rho_{\mathrm{WL},\rate} \\
\sigma_{\mathrm{WL}} \sigma_{\mathcal{X}} \rho_{\mathrm{WL},\mathcal{X}} 
& \sigma_{\mathcal{X}}^2
& \sigma_{\mathcal{X}} \sigmaeta \rho_{\mathcal{X},\rate}  \\
\sigma_{\mathrm{WL}} \sigmaeta \rho_{\mathrm{WL},\rate}
& \sigma_{\mathcal{X}} \sigmaeta \rho_{\mathcal{X},\rate}
& \sigmaeta^2
\end{pmatrix}
\, ,
\end{equation}
where $\rho_{\mathrm{X}, \mathrm{Y}}$ denotes the correlation coefficient  between the observables $\mathrm{X}$ and $\mathrm{Y}$.
The term $P\left(\Xlabel|\tilde\Xlabel\right)$ evaluates the probability of observing \Xlabel\ with the measurement uncertainty given the intrinsic value of $\tilde\Xlabel$ predicted at the true cluster mass \Mfiveoo.

If a cluster is not covered by the HSC survey (thus does not have an observed weak-lensing shear profile), we evaluate the probability of observing this cluster with the X-ray observable \Xlabel\ given the count rate \rate, namely,
\begin{align}
\label{eq:sr_only_x}
P\left(\Xlabel|\rate,\zcl,\mathbf{p}\right) &=
\frac{
N\left(\Xlabel, \rate|\zcl,\mathbf{p}\right)
}{
N\left(\rate|\zcl,\mathbf{p}\right)
} \nonumber \\
& =
\frac{
\int
P\left(\Xlabel,\rate|\Mfiveoo,\zcl,\mathbf{p}\right)
n\left(\Mfiveoo,\zcl\right)\dif\Mfiveoo
}{
\int
P\left(\rate|\Mfiveoo,\zcl,\mathbf{p}\right)
n\left(\Mfiveoo,\zcl\right)\dif\Mfiveoo
}
\, .
\end{align}
We note that Eq.~(\ref{eq:sr_only_x}) is identical to Eq.~(\ref{eq:mcalib_single}) with a substitution of \gshear\ with \Xlabel.
Accordingly, the probability $P\left(\Xlabel,\rate|\Mfiveoo,\zcl,\mathbf{p}\right)$ is calculated as
\begin{multline}
\label{eq:sr_only_x_num}
P\left(\Xlabel,\rate|\Mfiveoo,\zcl,\mathbf{p}\right) =\\
\int\int
\dif\tilde\rate
\dif\tilde\Xlabel 
P\left(\Xlabel|\tilde\Xlabel\right) 
P\left(\rate|\tilde\rate\right)
P\left(\tilde\Xlabel, \tilde\rate|\Mfiveoo,\zcl,\mathbf{p}\right) 
\, ,
\end{multline}
where $P\left(\tilde\Xlabel, \tilde\rate|\Mfiveoo,\zcl,\mathbf{p}\right) $ describes the intrinsic distribution of $\{\tilde\Xlabel, \tilde\rate\}$ characterized as a 
two-dimensional log-normal distribution with a covariance matrix of
\begin{equation}
\label{eq:sr_only_x_lognormal_covar}
\Sigma_{\Xlabel, \rate} = 
\begin{pmatrix}
\sigma_{\Xlabel}^2 & \sigma_{\Xlabel} \sigmaeta \rho_{\Xlabel,\rate} \\
\sigma_{\Xlabel} \sigmaeta \rho_{\Xlabel,\rate} & \sigmaeta^2
\end{pmatrix}
\, .
\end{equation}

If a cluster 
does not have the X-ray observable \Xlabel\ but simply a weak-lensing shear profile,
the likelihood reduces to the pure mass calibration, as in Eq.~(\ref{eq:mcalib_single}).

Finally, the likelihood of the joint modeling given a sample of eFEDS clusters is calculated as
\begin{multline}
\label{eq:like_sr}
\mathcal{L} \left(\mathbf{p}\right)
= \prod\limits_{i=1}
P\left(\Xlabel_{i}, {\gshear}_{,i}|\rate_{i}, \mathbf{p}\right) 
\\
\times
\prod\limits_{j=1}
P\left({\gshear}_{,j}|\rate_{j}, \mathbf{p}\right)
\times
\prod\limits_{k=1}
P\left(\Xlabel_{k}|\rate_{k}, \mathbf{p}\right)
\, ,
\end{multline}
where $i$ runs over the clusters with the available observables of $\{\Xlabel, \gshear \rate \}$,
$j$ runs over the clusters with only observed shear profiles but \Xlabel, 
and $k$ runs over the clusters with only \Xlabel\ but \gshear.
In total, we have a parameter set $\mathbf{p}$ consisting of 26 free parameters, i.e., 
\begin{align}
\label{eq:parameter_xsrmcalib}
\mathbf{p} &= 
\mathbf{p}_{\rate}\cup
\mathbf{p}_{\mathrm{F}}\cup
\mathbf{p}_{\mathcal{X}}\cup
\mathbf{p}_{\mathrm{WL}}\cup
\mathbf{p}_{\mathrm{mis}}\cup
\mathbf{p}_{\rho}\cup
\mathbf{p}_{\mathrm{c}}
\, , \nonumber \\
\mathbf{p}_{\mathcal{X}} &= \{ A_{\mathcal{X}},B_{\mathcal{X}}, \delta_{\mathcal{X}}, \gamma_{\mathcal{X}}, \sigma_{\mathcal{X}}  \} \, , \nonumber \\
\mathbf{p}_{\rho} &= \{\rho_{\mathrm{WL},\mathcal{X}}, \rho_{\mathrm{WL},\rate}, \rho_{\mathcal{X},\rate}\}, 
\,
\end{align}
where  $\mathbf{p}_{\rate}$, $\mathbf{p}_{\mathrm{F}}$, $\mathbf{p}_{\mathrm{WL}}$, $\mathbf{p}_{\mathrm{mis}}$, and $\mathbf{p}_{\mathrm{c}}$ are defined in Eq.~(\ref{eq:parameter_mcalib});
$\mathbf{p}_{\mathcal{X}}$ describes the $\mathcal{X}$--\Mfiveoo--\redshift\ relation;
$\mathbf{p}_{\rho}$ denotes the intrinsic correlation coefficients among the three observables.

It is worth noting that we expect a degeneracy between the parameters of the scaling relations obtained in a forward modeling.
This is because we are constraining the scaling relations between the
true cluster mass \Mfiveoo\ and the multiple observables $\left\lbrace\Xlabel, \rate, \gshear\right\rbrace$, where the latter is fixed in the modeling.
In this way, the posteriors of \Mfiveoo\ and the scaling relation parameters are simultaneously obtained by exploring the likelihood space. 
As a result, a change in the posterior of the normalization of one relation, for example,  will coherently change those of the other scaling relations given a set of fixed observables.
This is especially true for the parameters between the \rate--\Mfiveoo--\redshift\ and \Xlabel--\Mfiveoo--\redshift\ relations, because the degeneracy with the weak-lensing mass will be broken by applying informative priors on the parameters of the \Mwl--\Mfiveoo--\redshift\ relation (see Sect.~\ref{sec:bwl}).
We see below that the strongest degeneracy occurs in the case of $\Xlabel = \Mg$ (see Sect.~\ref{sec:mg_results}).
Despite the degeneracy, we stress that the inclusion of the \rate--\Mfiveoo--\redshift\ relation in modeling the \Xlabel--\Mfiveoo--\redshift\ relations is still needed, because the count rate is used as the X-ray mass proxy.
By including the \rate--\Mfiveoo--\redshift\ relation, we fully account for the parameter degeneracy in constraining the \Xlabel--\Mfiveoo--\redshift\ relation.

It is also important to make one remark:
As in Eq.~(\ref{eq:like_sr}), we use all available $313$ shear profiles in modeling the \Xlabel--\Mfiveoo--\redshift\ relations, such that the constraining power from the weak-lensing mass calibration essentially remains the same for all cases of $\Xlabel = \left\lbrace\Lx, \Lb, \Tx, \Mg, \Yx\right\rbrace$.
Even in the modeling of \Tx\ and \Yx\ scaling relations, where only $64$ clusters are present with temperature measurements, the relative errors on the normalization ($A_{\Tx}$ and $A_{\Yx}$) are much better than what would be obtained with the first term in
Eq.~(\ref{eq:like_sr}) alone, which is due to the fact that we have fully used the available data sets in two surveys.
The numbers of clusters used in each modeling with the combination of the data sets are presented in the last three columns of Table~\ref{tab:model_params}.

\subsection{Statistical inference}
\label{sec:mcmc}

Our statistical inference can be written in a generic form as follows.
Letting $\mathbf{p}$ be a set of parameters describing the modeling and $\mathcal{D}$  a data vector, the posterior $P(\mathbf{p} | \mathcal{D})$ of $\mathbf{p}$ given the observed data can be written, through Bayes' theorem, as
\begin{equation}
\label{eq:posterior}
P(\mathbf{p} | \mathcal{D}) \propto \mathcal{L}(\mathcal{D}| \mathbf{p}) \cdot\mathcal{P(\mathbf{p})} \, ,
\end{equation}
where $\mathcal{L}(\mathcal{D}| \mathbf{p})$ is the likelihood of observing $\mathcal{D}$ given the parameter $\mathbf{p}$, and $\mathcal{P(\mathbf{p})}$ is the prior on $\mathbf{p}$.
The parameter space is explored by the Affine Invariant Markov Chain Monte Carlo (MCMC) algorithm, which is carried out by \texttt{emcee} \citep{foreman13,foreman19}.

The following priors are adopted in the modeling:
The constraints on the parameters describing the bias in the observed count rate and the weak-lensing mass, as in Eqs.~(\ref{eq:bias_rate})~and~(\ref{eq:bwl}), respectively, are used as the priors on these parameters.
This effectively accounts for the bias in the observed quantities using the simulation calibration (see Sect.~\ref{sec:simulations}).
When constraining the \rate--\Mfiveoo--\redshift\ relation, we adopt a Gaussian prior on the log-normal scatter \sigmaeta\ derived from the simulation.
Specifically, this includes two factors: 
First, the intrinsic scatter of the true count rate within each mass and redshift bin in the middle panel of Fig.~\ref{fig:ratebias_sims} is quantified at a level of $\approx22\percent$.
Second, the scatter of the residual in $\ln\brate$ between the model and the simulated data is at a level of $\approx8\percent$, as evaluated using all mass and redshift bins in the right panel of Fig.~\ref{fig:ratebias_sims}.
These two lead to the intrinsic scatter of \rate\ at a level of $\approx30\percent$ in total.
As a result, we use a Gaussian prior, $\mathcal{N}(0.3, 0.08^2)$ with a dispersion of $0.08$ on \sigmaeta.
We note that the scatter of the modeling residual at a level of $8\percent$ is used as the dispersion of the Gaussian prior as a conservative estimate.

When modeling other X-ray observable-to-mass-and-redshift relations, we adopt uniform priors on the parameters. 
The Gaussian priors on the parameters, \fmis\ and \sigmamis, are adopted following the constraints of the miscentering, as quantified in Sect.~\ref{sec:miscentering}.
We adopt the Gaussian priors on the cosmological parameters, i.e., \OmegaM, \sigmaeight, and \Hnow, based on the forecast from \cite{grandis19} if including the ground-based weak-lensing mass calibration.
The adopted priors are summarized in Table~\ref{tab:priors}.

\begin{table*}
    \centering
    \caption{
    Parameter constraints on the X-ray scaling relations.
    The first (second) row contains the result of the weak-lensing mass calibration, which is to simultaneously model the count rate \rate\ and the shear profile \gshear\ in a radial range of $R>0.5\Mpch$ ($R>0.2\Mpch$).
    The following rows present the joint modeling of the follow-up X-ray scaling relation and the weak-lensing mass calibration.
    The first column records the type of modeling.
    The second to sixth columns contain the best-fit parameters describing the \rate--\Mfiveoo--\redshift\ relation.
    The seventh to eleventh show the parameter constraints of the target \Xlabel--\Mfiveoo--\redshift\ relation.
    The twelfth to fourteenth columns record the correlation coefficient in the intrinsic scatter among the underlying mass proxies.
    In each modeling, the number $N_{\rate,\mathrm{WL}}$ ($N_{\Xlabel, \rate,\mathrm{WL}}$, $N_{\Xlabel, \rate}$) of clusters used with the observable set of $\left\lbrace\rate, \gshear\right\rbrace$ ($\left\lbrace\Xlabel, \rate, \gshear\right\rbrace$, $\left\lbrace\Xlabel, \rate\right\rbrace$) is shown in the fifteenth, sixteenth, and seventeenth columns.
    }
    \label{tab:model_params}
    \resizebox{\textwidth}{!}{
    \begin{tabular}{ccccccccccccccccc}
        \hline
                Model & $A_{\eta}$ & $B_{\eta}$ & $\delta_{\eta}$ & $\gamma_{\eta}$ & $\sigma_{\eta}$ & $A_{\Xlabel}$ & $B_{\Xlabel}$ & $\delta_{\Xlabel}$ & $\gamma_{\Xlabel}$ & $\sigma_{\Xlabel}$ & $\rho_{\rm{WL},\Xlabel}$ & $\rho_{\rm{WL},\eta}$ & $\rho_{\Xlabel,\eta}$
                & $N_{\rate,\gshear}$
                & $N_{\Xlabel, \rate, \gshear}$
                & $N_{\Xlabel, \rate}$                      \\[3pt] 
                \hline
                WL Mass Calibration & $0.124^{+0.022}_{-0.019}$ & $1.58^{+0.17}_{-0.14}$ & $1.0^{+1.0}_{-1.4}$ & $-0.44^{+0.81}_{-0.85}$ & $0.301^{+0.089}_{-0.078}$ & -- & -- & -- & -- & -- & -- & $0.01^{+0.39}_{-0.55}$ & -- & $313$ & -- & -- \\ [3pt]
                WL Mass Calibration ($R > 0.2~h^{-1}\rm{Mpc}$) & $0.124^{+0.023}_{-0.021}$ & $1.54^{+0.17}_{-0.14}$ & $0.9^{+1.0}_{-1.1}$ & $-0.04^{+0.82}_{-0.83}$ & $0.294^{+0.081}_{-0.083}$ & -- & -- & -- & -- & -- & -- & $0.22^{+0.36}_{-0.50}$ & -- & $313$ & -- & -- \\ [3pt]
                The $L_{\rm{X}}$--$M_{500\rm{c}}$--$z$ Relation + WL Mass Calibration & $0.124^{+0.021}_{-0.019}$ & $1.50^{+0.15}_{-0.14}$ & $0.3^{+1.0}_{-1.3}$ & $-0.50^{+0.74}_{-0.84}$ & $0.337^{+0.054}_{-0.069}$ & $3.36^{+0.53}_{-0.49}$ & $1.44^{+0.14}_{-0.13}$ & $-0.07^{+1.26}_{-0.79}$ & $-0.51^{+0.93}_{-0.75}$ & $0.120^{+0.138}_{-0.060}$ & $0.24^{+0.38}_{-0.67}$ & $-0.21^{+0.41}_{-0.31}$ & $0.35^{+0.29}_{-0.45}$ & $0$ & $313$ & $121$ \\ [3pt]
                The $L_{\rm{b}}$--$M_{500\rm{c}}$--$z$ Relation + WL Mass Calibration & $0.127^{+0.022}_{-0.018}$ & $1.55^{+0.16}_{-0.14}$ & $0.2^{+1.3}_{-1.0}$ & $-0.67^{+0.99}_{-0.80}$ & $0.305^{+0.062}_{-0.059}$ & $9.2^{+1.6}_{-1.3}$ & $1.59\pm 0.14$ & $0.2^{+1.3}_{-1.1}$ & $-0.45^{+1.00}_{-0.86}$ & $0.102^{+0.143}_{-0.043}$ & $0.61^{+0.26}_{-0.58}$ & $-0.31^{+0.40}_{-0.34}$ & $0.38^{+0.28}_{-0.67}$ & $0$ & $313$ & $121$ \\ [3pt]
                The $T_{\rm{X}}$--$M_{500\rm{c}}$--$z$ Relation + WL Mass Calibration & $0.130^{+0.020}_{-0.021}$ & $1.61^{+0.17}_{-0.14}$ & $0.6^{+1.2}_{-1.1}$ & $-0.61^{+0.88}_{-0.90}$ & $0.288^{+0.067}_{-0.089}$ & $3.27^{+0.26}_{-0.31}$ & $0.65\pm 0.11$ & $-0.02^{+0.66}_{-0.70}$ & $-1.03^{+0.54}_{-0.75}$ & $0.069^{+0.061}_{-0.014}$ & $0.47^{+0.36}_{-0.59}$ & $0.00^{+0.27}_{-0.57}$ & $0.65^{+0.22}_{-0.59}$ & $266$ & $47$ & $17$ \\ [3pt]
                The $M_{\rm{g}}$--$M_{500\rm{c}}$--$z$ Relation + WL Mass Calibration & $0.114^{+0.019}_{-0.017}$ & $1.49^{+0.15}_{-0.16}$ & $1.56^{+0.85}_{-1.12}$ & $0.16^{+0.80}_{-0.85}$ & $0.361^{+0.044}_{-0.061}$ & $1.08\pm 0.13$ & $1.190^{+0.099}_{-0.118}$ & $0.40^{+0.78}_{-0.70}$ & $0.32^{+0.59}_{-0.61}$ & $0.074^{+0.063}_{-0.019}$ & $0.08^{+0.60}_{-0.43}$ & $-0.30^{+0.32}_{-0.38}$ & $0.63^{+0.23}_{-0.49}$ & $0$ & $313$ & $121$ \\ [3pt]
                The $Y_{\rm{X}}$--$M_{500\rm{c}}$--$z$ Relation + WL Mass Calibration & $0.128^{+0.020}_{-0.015}$ & $1.60^{+0.18}_{-0.14}$ & $0.73^{+1.39}_{-0.79}$ & $-0.36^{+0.75}_{-0.81}$ & $0.224^{+0.070}_{-0.068}$ & $6.9^{+1.5}_{-1.2}$ & $1.78^{+0.24}_{-0.20}$ & $-1.1^{+1.9}_{-1.2}$ & $-1.43^{+1.20}_{-0.93}$ & $0.106^{+0.171}_{-0.047}$ & $0.12^{+0.53}_{-0.50}$ & $-0.44^{+0.50}_{-0.36}$ & $0.794^{+0.079}_{-0.344}$ & $266$ & $47$ & $17$ \\ [3pt]
                \hline
    \end{tabular}
    }
\end{table*}
%

%
%

\section{Results and Discussion}
\label{sec:results}

We first present and discuss the mass calibration in Sect.~\ref{sec:mass_calibration_results}; we then turn to the constraint of the follow-up X-ray scaling relations in Sects.~\ref{sec:lm_results}~to~\ref{sec:ym_results}.

\subsection{The \rate--\Mfiveoo--\redshift\ relation and the cluster mass \Mfiveoo}
\label{sec:mass_calibration_results}

The count rate-to-mass-and-redshift (\rate--\Mfiveoo--\redshift) relation is obtained by simultaneously modeling the observed count rate \rate\ and shear profile \gshear, as detailed in Sect.~\ref{sec:wlmcalib}.
The parameter constraints are shown in Fig.~\ref{fig:chain_mcalib} and listed in Table~\ref{tab:model_params}.
We only present the parameters of the \rate--\Mfiveoo--\redshift\ relation and correlated scatter in Table~\ref{tab:model_params}, as others are largely following the adopted Gaussian priors.

As clearly seen in Fig.~\ref{fig:chain_mcalib}, there is no statistical difference between the modeling of shear profiles at $R>0.2\Mpch$ and $R>0.5\Mpch$.
This consistency is not trivial: 
as we adopt the different simulation-calibrated priors on the parameters of $\{\Awl, \Bwl, \gammawl, \sigmawl\}$ 
for the cases of $R>0.2\Mpch$ and $R>0.5\Mpch$ (see Sect.~\ref{sec:bwl}),
the consistent results suggest that the simulation-calibrated \Mwl--\Mfiveoo--\redshift\ relation accurately accounts for the weak-lensing mass bias, especially at the cluster core.
Hereafter, we focus on the results of the modeling with $R>0.5\Mpch$, as the default analysis of the mass calibration in this work.

The resulting count-rate-to-mass-and-redshift relation is obtained as
\begin{multline}
\label{eq:cm}
\left\langle\ln\left( \frac{\rate}{ \mathrm{counts}/\mathrm{sec} } \Bigg|\Mfiveoo \right)\right\rangle 
= \ln \left( \ansAeta \right) + \\
\left[ \left( \ansBeta \right) + \left( \ansdeltaeta \right) \ln\left(\frac{1 + \redshift}{1 + \ZPIV}\right) \right] \times
\ln\left(\frac{\Mfiveoo}{\MPIV}\right) \\
+ 2 \times \ln\left(\frac{\Ez}{\Ezpiv}\right)
-2\times\ln\left(\frac{D_{\mathrm{L}}\left(\redshift\right)}{D_{\mathrm{L}}\left(\ZPIV\right)}\right) \\
+ \left( \ansgammaeta \right) \times 
\ln \left(\frac{1 + \redshift}{1 + \ZPIV}\right)
+ \ln\left( \brate(\Mfiveoo, \redshift) \right)
\, ,
\end{multline}
with the log-normal scatter of $\sigmaeta = \anssigmaeta$.
The last term, $\ln\left( \brate(\Mfiveoo, \redshift) \right)$, characterizes the bias in the observed count rate with respect to the true count rate through the simulation calibration, as described in Sect.~\ref{sec:etabias}.
By subtracting $\ln\left( \brate(\Mfiveoo, \redshift) \right)$ from Eq.~(\ref{eq:cm}), the remaining terms describe the true count-rate-to-mass-and-redshift relation.

With the derived \rate--\Mfiveoo--\redshift\ relation, we estimate the lensing-calibrated posterior of the cluster mass \Mfiveoo\ for each cluster at the redshift \redshift\ given the observed count rate \rate, as
\begin{equation}
\label{eq:mass_post}
P(\Mfiveoo | \rate, \redshift, \mathbf{p}) \propto P(\rate | \Mfiveoo, \redshift, \mathbf{p})~P(\Mfiveoo | \redshift, \mathbf{p}) \, ,
\end{equation}
where $\mathbf{p}$ is the best-fit parameter of the modeling, and $P(\Mfiveoo | \redshift, \mathbf{p})$ is the normalized mass function evaluated at the cluster redshift.
We then randomly sample a mass estimate from the resulting posterior as the ensemble mass for individual clusters \citep[see more applications in][]{bocquet19,chiu20b}.
In this way, the individual cluster mass is statistically inferred from the ensemble population modeling.
The same method but with a different terminology, the ``mass forecasting'', has been  used to estimate the mass of clusters based on the X-ray temperature  in the \XXL\ survey \citep{umetsu20}.
We show the ensemble mass of each cluster in Fig.~\ref{fig:mz}.
The eFEDS sample spans a mass range between $\approx10^{13}\Msunh$ and $\approx10^{14.8}\Msunh$ in terms of \Mfiveoo, of which low-mass systems ($\Mfiveoo\lesssim10^{14}\Msunh$) are mostly at low redshift ($\redshift\lesssim0.4$).
The typical uncertainty in the cluster mass is at a level of $\approx27\percent$ ($\approx32\percent$, $\approx24\percent$) for a cluster with $\Mfiveoo\approx10^{14}\Msunh$ ($\approx10^{13.6}\Msunh$, $\approx10^{14.4}\Msunh$).
The mass \Mfiveoo\ of the eFEDS clusters is presented in Appendix~\ref{sec:clustermasstable}.

We show the observed count rate \rate\ as a function of the cluster mass \Mfiveoo\ and redshift \redshift, together with this best-fit model, in Fig.~\ref{fig:cmz}.
When showing the mass-trend in the left panel, we divide the count rate by the redshift-dependent quantities at the cluster redshift.
In this way, the count rate \rate\ is re-normalized to the pivotal redshift.
Similarly, we re-normalize the count rate to the pivotal mass when showing the redshift trend in the right panel of Fig.~\ref{fig:cmz}.
As seen, the best-fit model provides a good description of the eFEDS clusters.
Based on the result, the true count rate observed by \eROSITA\ is \ansAeta\ counts per second for a cluster with the pivotal mass $\MPIV=1.4\times10^{14}\Msunh$ at the pivotal redshift $\ZPIV = 0.35$.
Moreover, the true count rate scales with mass and redshift as $\appropto{\Mfiveoo}^{\ansBeta}$ and $\appropto (1 + \redshift)^{\ansgammaeta}$, respectively, without significant cross-scaling ($\deltaeta = \ansdeltaeta$).
We stress that there is no significant deviation from the expected redshift trend of the true count rate ($\gammaeta = \ansgammaeta$), which is consistent with what we observe in other X-ray scaling relations for the eFEDS sample.
We continue our discussion of this perspective in Sects.~\ref{sec:lm_results}~to~\ref{sec:ym_results}.

For both the lensing modeling with and without the cluster core, the correlated scatter $\rho_{\mathrm{WL},\rate}$ between the count rate and weak-lensing mass is constrained as statistically consistent with zero.
This suggests that the true count rate and weak-lensing-inferred mass are nearly independent of each other.
This is in line with the picture that the source of the intrinsic scatter in X-rays and in the optical is not strongly correlated.
The former is related to various activities in the ICM caused by either the output of energetic sources or the thermal emission of clusters, while the latter is largely attributed to the halo orientation, concentration, and the projection of line-of-sight structures.

\begin{figure}
\centering
\resizebox{0.5\textwidth}{!}{
\includegraphics[scale=1]{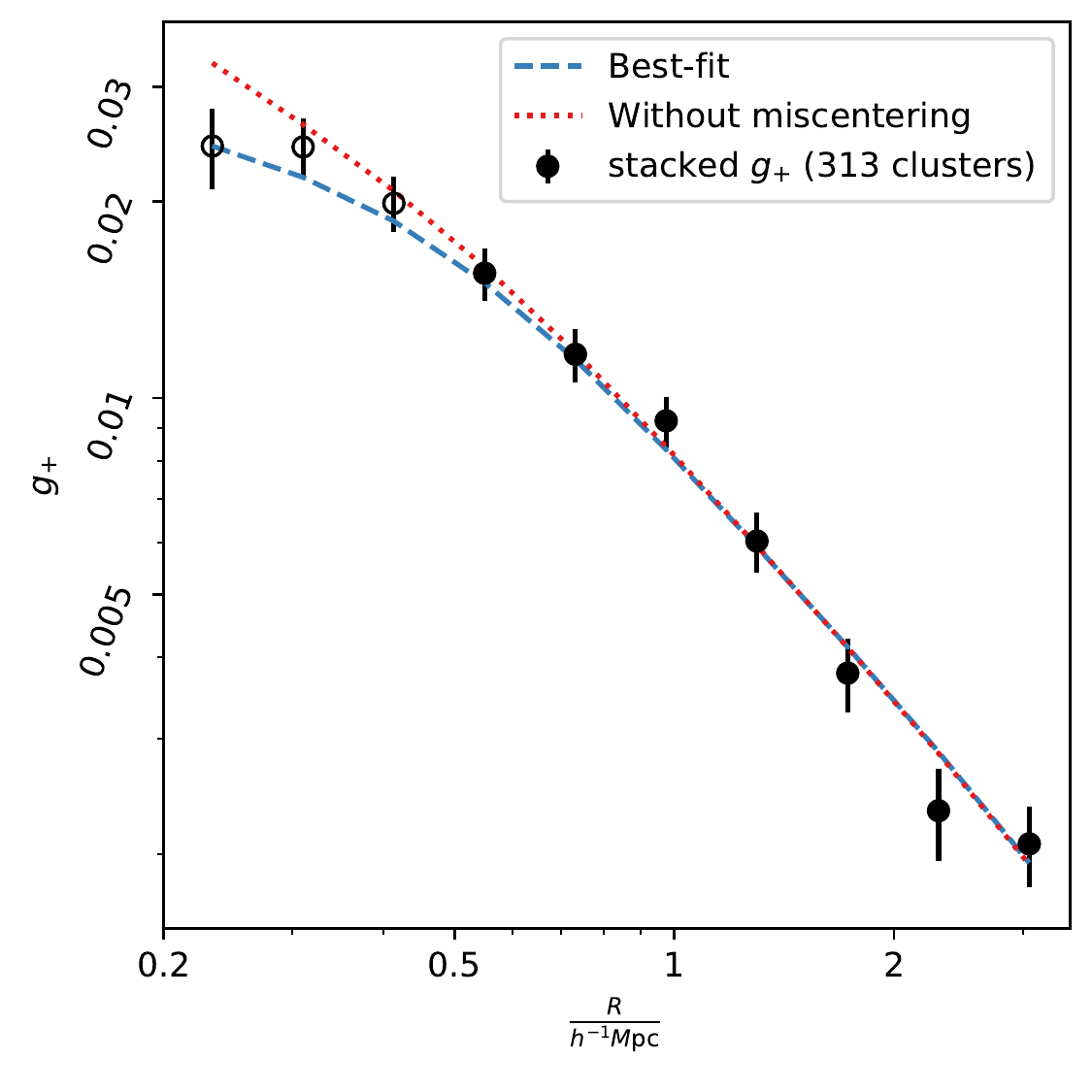}
}
\caption{
Stacked weak-lensing shear profile (black points), and the best-fit lensing models with (blue-dashed line) and without (red-dotted line) the miscentering component.
The open circles indicate the inner regime of $R < 0.5~\Mpch$, which are not used in the fiducial lensing model in this work.
}
\label{fig:stacked_shear}
\end{figure}

Finally, we show the stacked shear profile and the best-fit models in Fig.~\ref{fig:stacked_shear}.
To derive the stacked profile, we properly weight the lensing shear profiles \gshear\ of individual clusters at each radial bin.
We use the diagonal term\footnote{In the case that off-diagonal terms are large, the exact form including the correlation among radial bins must be used, as Eq.~(136) in \cite{umetsu20b}. In this work, the off-diagonal terms in the lensing covariance matrix are small, and we stack the measurements only for the purpose of visualizations. Thus, we use the diagonal approximation when producing Fig.~\ref{fig:stacked_shear}.} of the lensing covariance matrix (see Eq.~(\ref{eq:covlensing})) as the radius-dependent weight for each cluster.
We repeat the same stacking procedure for both observed shear profiles and those predicted by the best-fit parameters, as shown by the black points and blue-dashed line, respectively.
We additionally derive the best-fit model without the miscentering by setting $\fmis = 0$, as shown by the red dotted line.
It is seen that the best-fit model with the miscentering provides a good description of the stacked shear profile, and that the miscentered model comes into agreement with that of perfectly centered sources at $R\gtrsim0.5\Mpch$.
This picture is also supported by the consistent constraints between the lensing modeling with and without the cluster core.
However, we note that the best-fit miscentered model (blue-dashed) is slightly lower than the stacked shear profile (open circles) in the inner radial range $R<0.5~\Mpch$ at a level of $\lesssim1\sigma$, indicating  that the miscentering of the X-ray centers is over-corrected.
This is not surprising, as the modeling of the miscentering is based on the offset distribution between the X-ray and optical centers, 
instead of the former and the true center, 
in which case the offset distribution is prone to the miscentering of the optical centers (see more discussions in Sect.~\ref{sec:miscentering}).
We stress that the modeling of the miscentering is subdominant to our results, and that the use of simulations to calibrate the X-ray miscentering is clearly needed in a future work.

\begin{figure*}
\resizebox{\textwidth}{!}{
\includegraphics[scale=1]{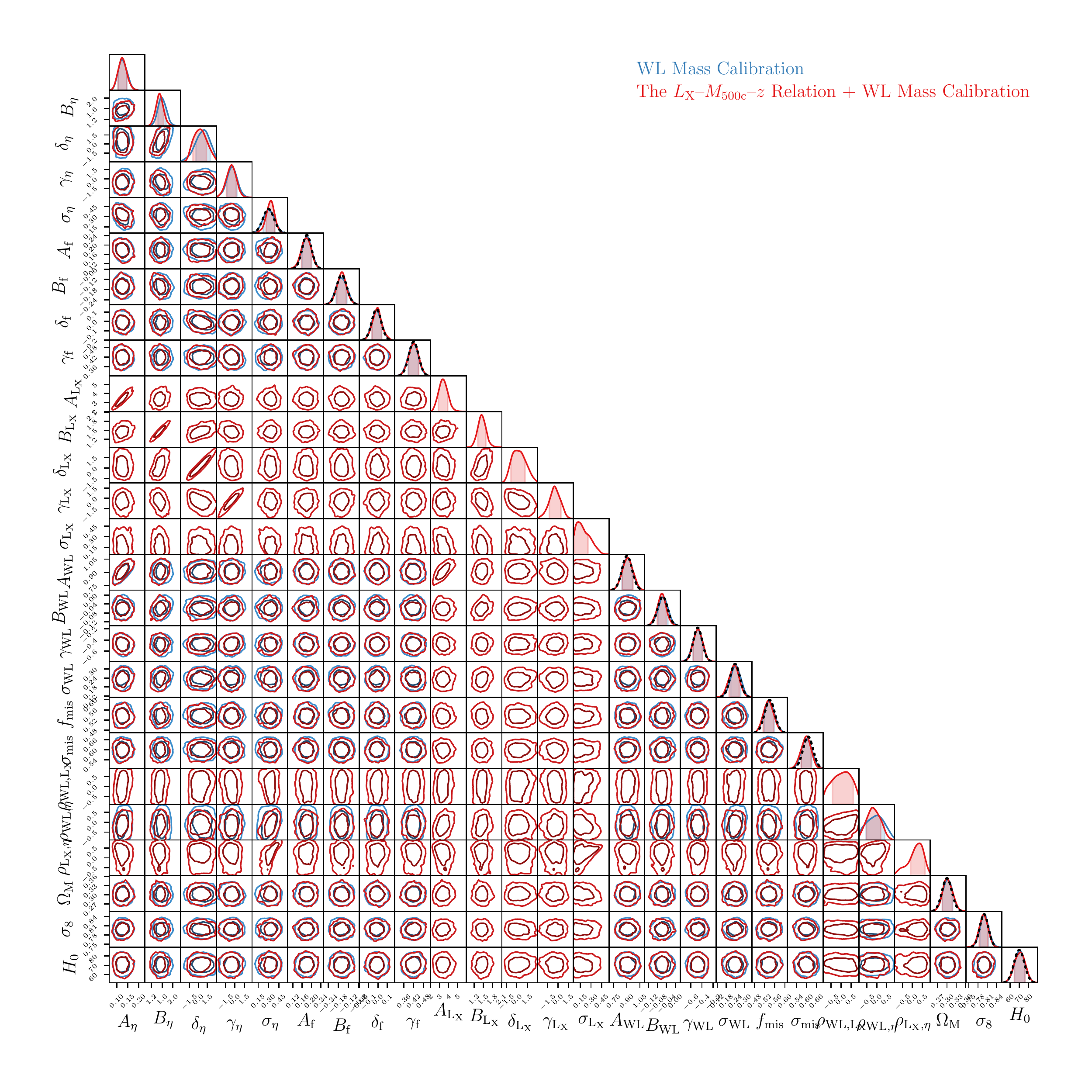}
}
\caption{
Parameter constraints in the joint modeling of the \Lx--\Mfiveoo--\redshift\ relation and the weak-lensing mass calibration.
The weak-lensing mass calibration consists of the modeling of X-ray count rates and shear profiles at $R>0.5$~\Mpch.
The result of the weak-lensing mass calibration is shown in blue, while that of the joint modeling with  the \Lx--\Mfiveoo--\redshift\ relation is shown in red.
The marginalized posteriors of the parameters and the correlation between them  are presented in the same way as in Fig.~\ref{fig:chain_mcalib}.
The adopted priors are indicated by the dotted lines in the diagonal subplots.
}
\label{fig:chain_lxmcalib}
\end{figure*}
\begin{figure*}
\resizebox{\textwidth}{!}{
\includegraphics[scale=1]{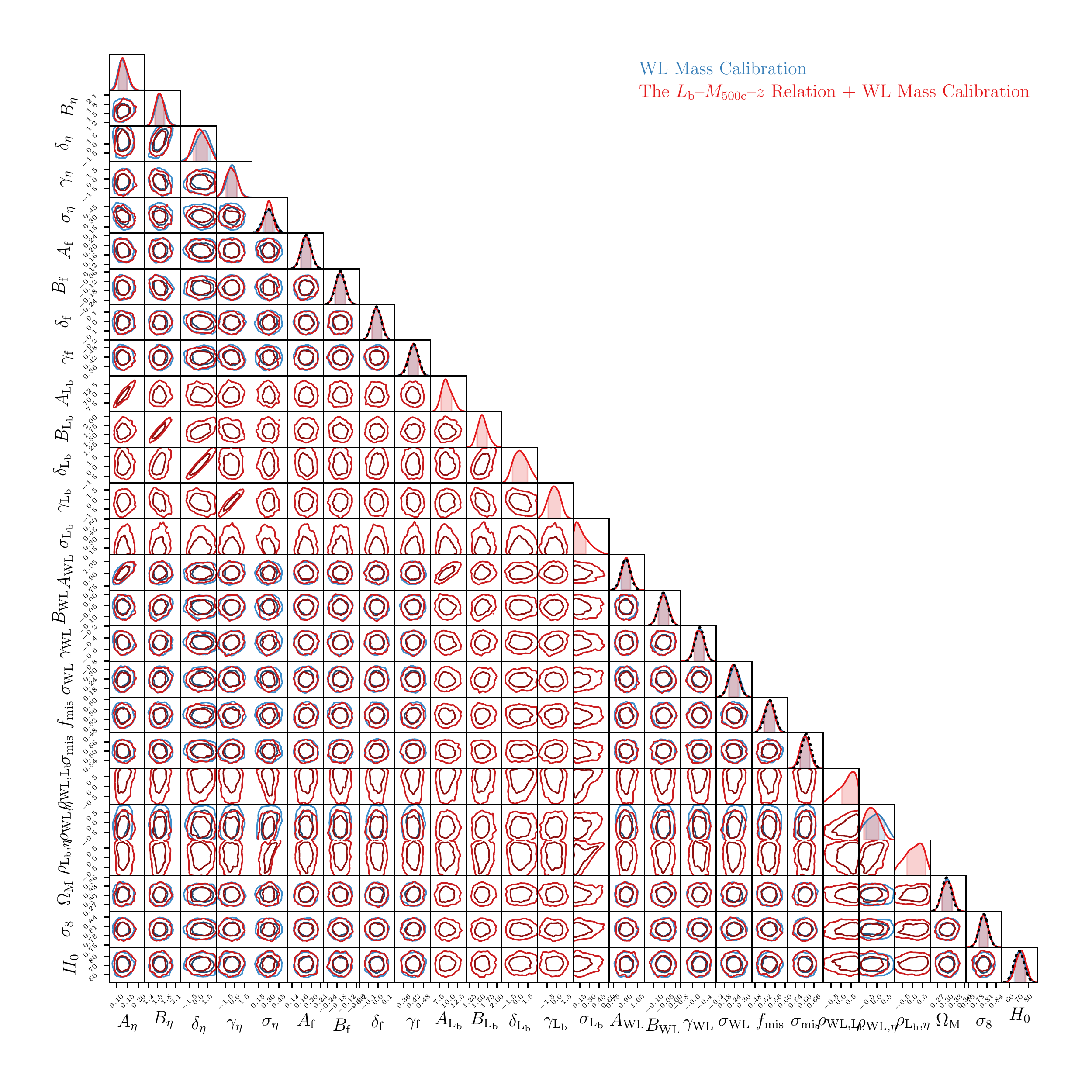}
}
\caption{
Parameter constraints in the joint modeling of the \Lb--\Mfiveoo--\redshift\ relation and the weak-lensing mass calibration presented in an identical way to those shown in Fig.~\ref{fig:chain_lxmcalib}.
}
\label{fig:chain_lbmcalib}
\end{figure*}

\subsection{The \Lx--\Mfiveoo--\redshift\ relation}
\label{sec:lm_results}

The luminosity-to-mass-and-redshift relation has been intensively studied \citep[e.g.,][]{reiprich02, maughan07, vikhlinin09a, arnaud10} and has been shown to have larger scatter at fixed mass than other X-ray mass proxies.
This scatter is suggested to be related to complex mergers \citep{ritchie02,poole08} or the nature of cluster cores, for example, the presence of a cool core \citep{pratt09}.
Despite the large scatter, the X-ray luminosity is relatively easy to obtain with a high signal-to-noise ratio, and thus becomes a convenient mass proxy for clusters with only a few tens of photon counts.
Moreover, the luminosity has a direct link to the observed count rate \rate, which is used as the mass proxy in this work.
Therefore, the luminosity provides straightforward access to the physical properties of eFEDS clusters.

The luminosity-to-mass-and-redshift relations predicted from the self-similar model \citep{kaiser1986} are
\[
\Lx \propto \Mfiveoo \times E(\redshift)^{2} \, ,
\]
and
\[
\Lb \propto {\Mfiveoo}^{\frac{4}{3}} \times E(\redshift)^{\frac{7}{3}} \, ,
\]
for the soft-band and bolometric luminosities, respectively.
We refer readers to \cite{boehringer12} and \cite{giodini13} for a more recent and complete discussion of the self-similar scaling among X-ray observables.
Here, we assume that the emissivity in the soft band does not depend on the temperature 
\citep[see Fig.~1 in][]{mohr99}, such that the mass and redshift-dependence of the temperature does not enter into the soft-band luminosity-to-mass-and-redshift relation.

The parameter constraints from the joint modeling of  the \Lx--\Mfiveoo--\redshift\ (\Lb--\Mfiveoo--\redshift) relation and the weak-lensing mass calibration are shown in Fig.~\ref{fig:chain_lxmcalib} (Fig.~\ref{fig:chain_lbmcalib}).
The results are listed in Table~\ref{tab:model_params}.
In this work, the \Lx--\Mfiveoo--\redshift\ relation of the eFEDS sample is derived as
\begin{multline}
\label{eq:lm}
\left\langle\ln\left(\frac{ \Lx  }{ \mathrm{ergs} } \Bigg| \Mfiveoo \right) \right\rangle 
= \ln \left( \ansAlx \right) + \ln \left( 10^{43} \right) + \\
\left[ \left( \ansBlx \right) + \left( \ansdeltalx \right) \ln\left(\frac{1 + \redshift}{1 + \ZPIV}\right) \right] \times
\ln\left(\frac{\Mfiveoo}{\MPIV}\right) \\
 + 2 \times \ln\left(\frac{\Ez}{\Ezpiv}\right) + 
\left( \ansgammalx \right)
 \times \ln \left(\frac{1 + \redshift}{1 + \ZPIV}\right),
\end{multline}
with the log-normal scatter of $\sigmalx = \anssigmalx$ at fixed mass.
Meanwhile, the \Lb--\Mfiveoo--\redshift\ relation is derived as
\begin{multline}
\label{eq:lbm}
\left\langle\ln\left( \frac{ \Lb }{\mathrm{ergs} } \Bigg| \Mfiveoo \right) \right\rangle 
= \ln \left( \ansAlb \right) + \ln \left( 10^{43} \right) + \\
\left[ \left( \ansBlb \right) + \left( \ansdeltalb \right) \ln\left(\frac{1 + \redshift}{1 + \ZPIV}\right) \right] \times
\ln\left(\frac{\Mfiveoo}{\MPIV}\right) \\
 +\frac{7}{3} \times \ln\left(\frac{\Ez}{\Ezpiv}\right) + 
\left( \ansgammalb \right)
 \times \ln \left(\frac{1 + \redshift}{1 + \ZPIV}\right),
\end{multline}
with the log-normal scatter of $\sigmalb = \anssigmalb$.
The soft-band (bolometric) luminosity is constrained as $\ansAlx \times 10^{43}$ ($\ansAlb \times 10^{43}$)~ergs at the pivotal mass $\MPIV=1.4\times10^{14}\Msunh$ at the pivotal redshift $\ZPIV = 0.35$.
The intrinsic scatter of the soft-band ($\sigmalx = \anssigmalx$) and bolometric ($\sigmalb = \anssigmalb$) luminosity are statistically consistent with each other.
We note that the asymmetric posterior of the scatter is mostly due to the lower bound of the imposed prior.

\begin{figure*}
\centering
\resizebox{0.48\textwidth}{!}{
\includegraphics[scale=1]{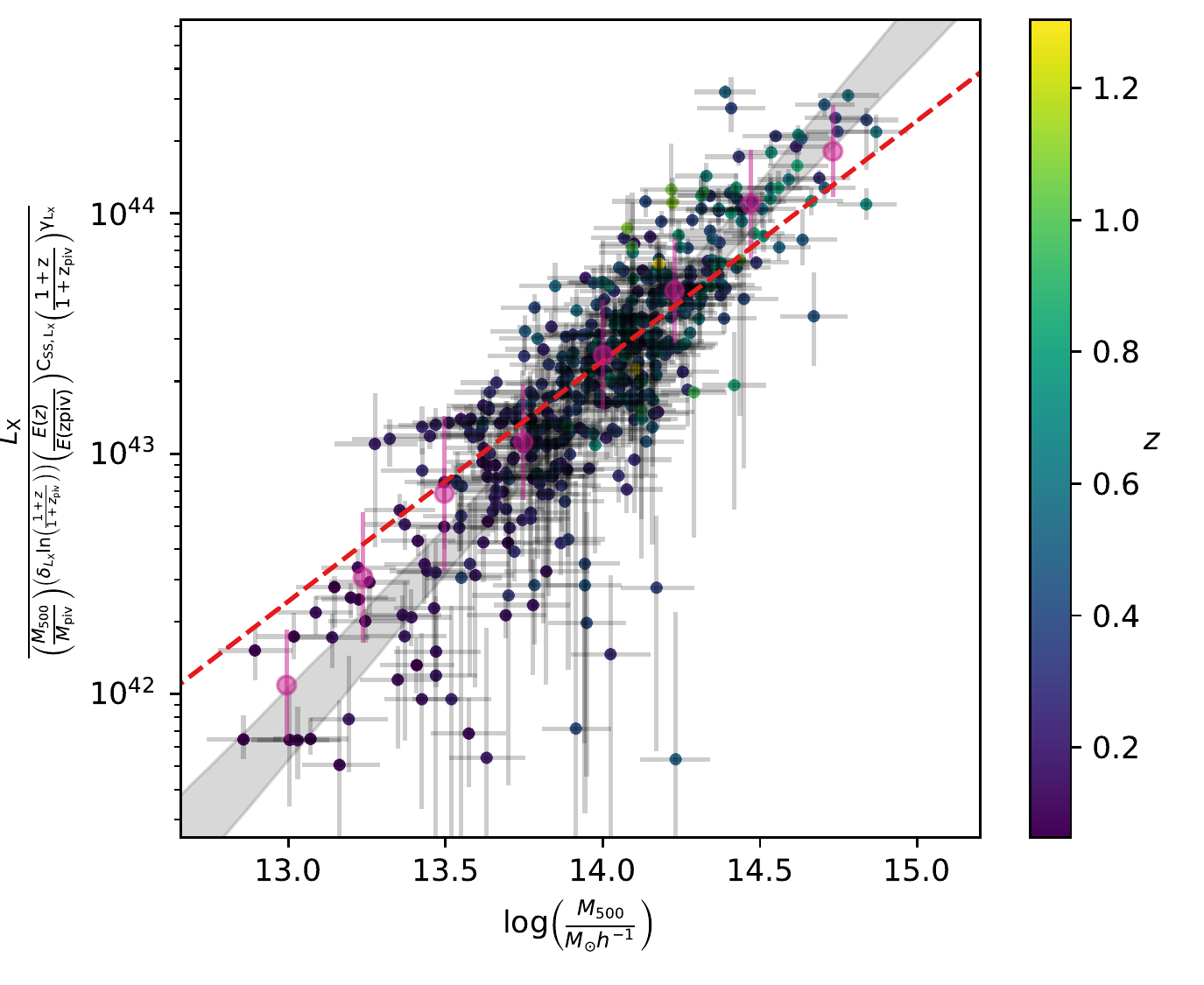}
}
\resizebox{0.48\textwidth}{!}{
\includegraphics[scale=1]{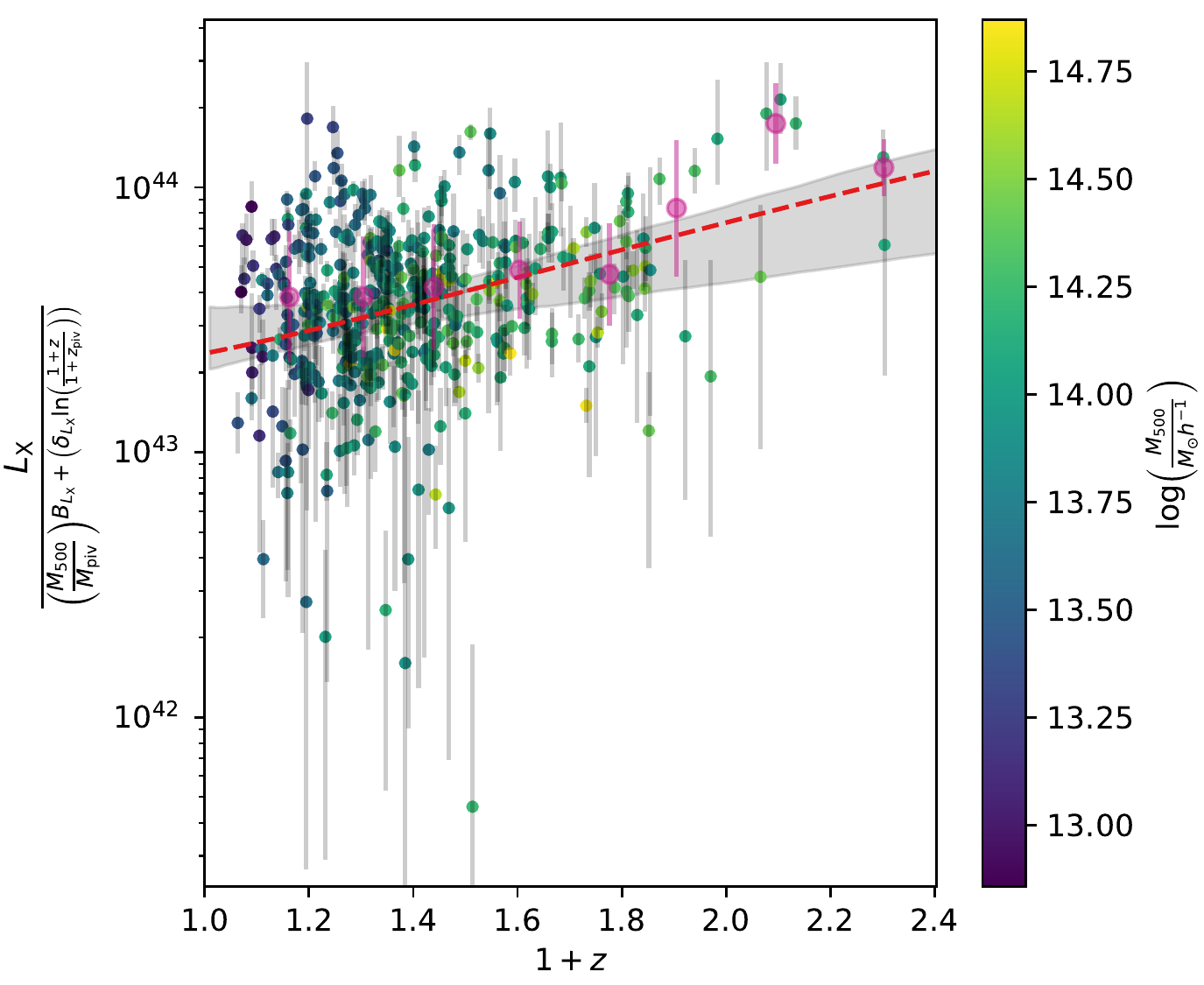}
}
\caption{
Observed soft-band luminosity \Lx\ as a function of the cluster mass \Mfiveoo\ (left panel) and redshift \redshift\ (right panel).
When showing the mass (redshift) trend in the left (right) panel, we re-normalize the observed quantity to the pivotal redshift $\ZPIV = 0.35$ (mass $\MPIV = 1.4\times10^{14}\Msunh$) as done in a similar manner in order to  produce Fig.~\ref{fig:cmz}.
The best-fit model and observed clusters are shown in the same way as in Fig.~\ref{fig:cmz}.
The pink circles indicate the weighted mean of the eFEDS sample.
}
\label{fig:lmz}
\end{figure*}

\begin{figure*}
\centering
\resizebox{0.48\textwidth}{!}{
\includegraphics[scale=1]{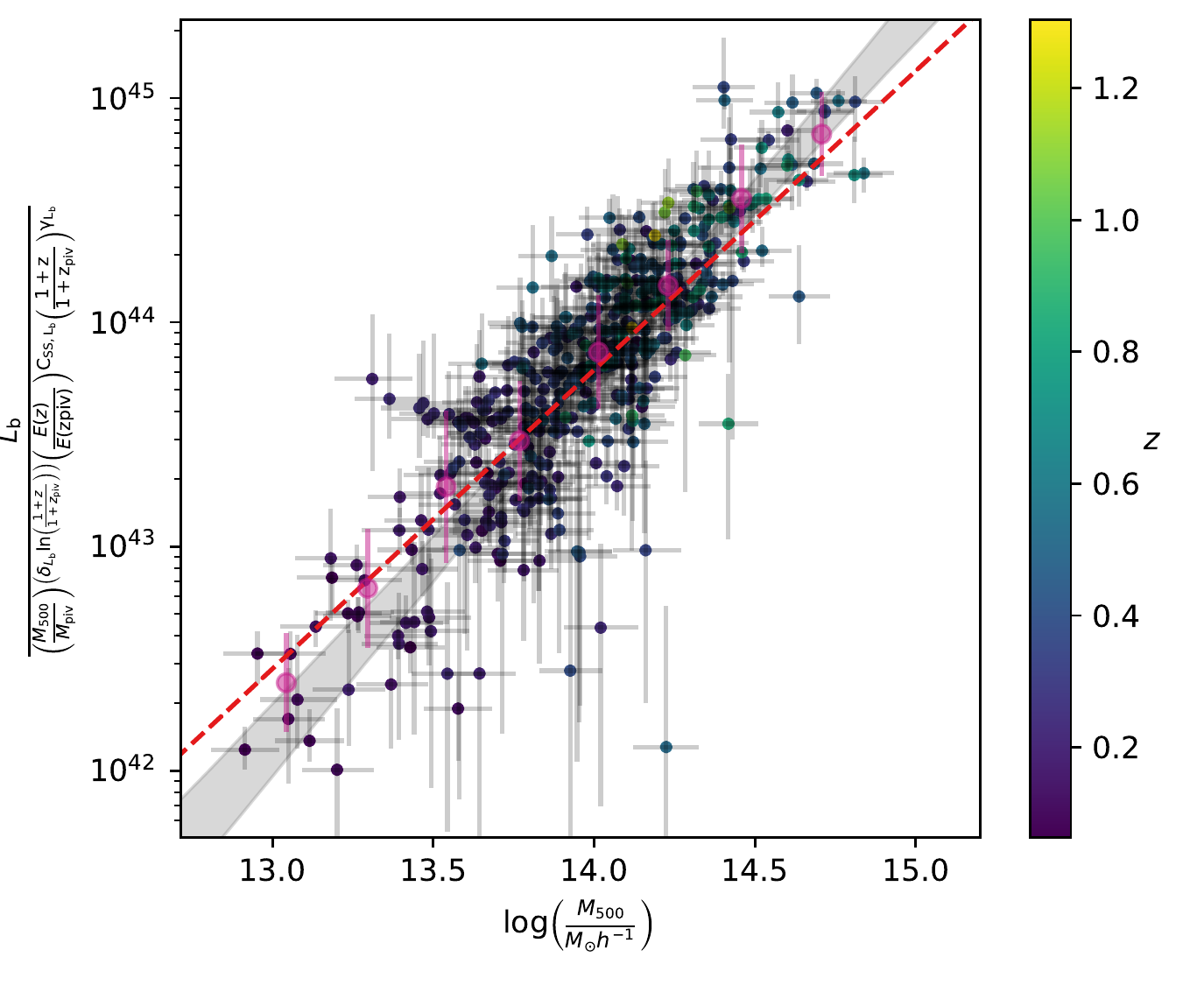}
}
\resizebox{0.48\textwidth}{!}{
\includegraphics[scale=1]{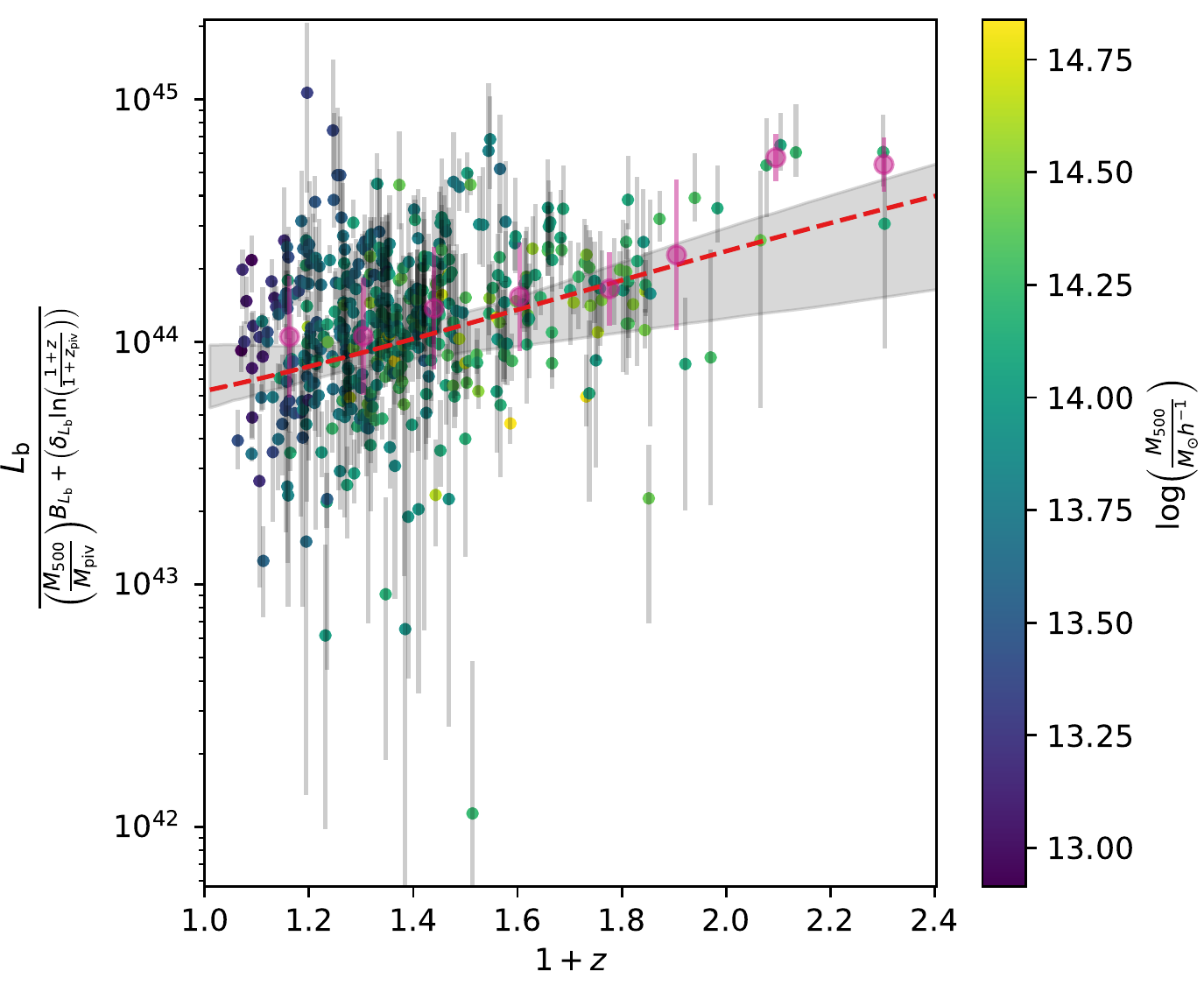}
}
\caption{
Observed bolometric luminosity \Lb\ as a function of the cluster mass \Mfiveoo\ (left panel) and redshift \redshift\ (right panel).
This plot is produced and shown in the same way as in Fig.~\ref{fig:lmz}.
}
\label{fig:lbmz}
\end{figure*}

We discuss our results as follows.
First, a mass-trend slope that is steeper than (statistically consistent with) the self-similar prediction is obtained for the \Lx--\Mfiveoo--\redshift\ (\Lb--\Mfiveoo--\redshift) relation with $\Blx = \ansBlx$ ($\Blb = \ansBlb$), 
corresponding to a significance level of 
$\approx3.4\sigma$ ($\approx1.7\sigma$).
Second, there is no significant deviation in the redshift trend ($\gammalx = \ansgammalx$ and $\gammalb = \ansgammalb$) from the self-similar model ($\gamma = 0$).
Finally, no sign of a redshift-dependent mass trend ($\deltalx = \ansdeltalx$ and $\deltalb = \ansdeltalb$) is revealed for the eFEDS sample.
This picture can be better visualized in Figs.~\ref{fig:lmz}~and~\ref{fig:lbmz} for the  \Lx--\Mfiveoo--\redshift\ and \Lb--\Mfiveoo--\redshift\ relations, respectively.
In Figs.~\ref{fig:lmz}~and~\ref{fig:lbmz}, we also re-normalize the luminosity to the pivotal mass and redshift, respectively, as done in a similar way in Fig.~\ref{fig:cmz}. 
As seen in Fig.~\ref{fig:lbmz}, the mass and redshift trends of \Lb\ are statistically consistent with the self-similar predictions.
For \Lx\ in Fig.~\ref{fig:lmz}, the mass-trend slope of the best-fit model (gray shaded region) is steeper than that predicted by the self-similar model (red dashed line) in the left panel, while the redshift trends of them show good consistency
in the right panel.
These are all in broad agreement with the results of \cite{bulbul19}, despite the fact that their sample of clusters is selected by the SZE signature observed by the SPT.
We stress that the SPT sample of \cite{bulbul19} only probes massive clusters with $\Mfiveoo\gtrsim10^{14.3}\Msunh$, while the eFEDS sample includes low-mass systems with $\Mfiveoo\lesssim10^{14}\Msunh$.
That is, this picture still holds down to a group scale for luminosity--mass--redshift relations.
Our result for \Lx\ suggests that (1) the mass-trend slope of eFEDS clusters is steeper than the self-similar prediction down to a mass scale of $\Mfiveoo\approx10^{13}\Msunh$, while the redshift trend is in good agreement with the self-similarity out to redshift $\redshift\approx1.3$, and that (2) there is no significant difference in the luminosity--mass--redshift relation between the eROSITA-selected and the SZE-selected clusters.

In terms of correlated scatter, we find no clear correlation between the count rate and the luminosity ($\rho_{\Lx,\rate}=\ansrholxeta$ and $\rho_{\Lb,\rate}=\ansrholbeta$; see also Figs.~\ref{fig:chain_lxmcalib}~and~\ref{fig:chain_lbmcalib}).
Meanwhile, no clear correlation is seen between the luminosity and the weak-lensing mass ($\rho_{\mathrm{WL},\Lx} = \ansrhowllx$ and $\rho_{\mathrm{WL},\Lb} = \ansrhowllb$).
No correlation is shown between the count rate and weak-lensing mass either, as consistent with the result in Sect.~\ref{sec:wlmcalib}.
We note that we do expect positive $\rho_{\Lx,\rate}$ and $\rho_{\Lb,\rate}$, given a direct relation between the count rate and luminosity.
However, the constraining power based on the eFEDS sample is generally not strong enough to precisely constrain the intrinsic correlated scatter.
A larger sample based on the first-year \eROSITA\ survey will significantly improve these constraints in a future work.


\subsubsection*{Comparisons with the literature}

\begin{figure*}
\centering
\resizebox{0.48\textwidth}{!}{
\includegraphics[scale=1]{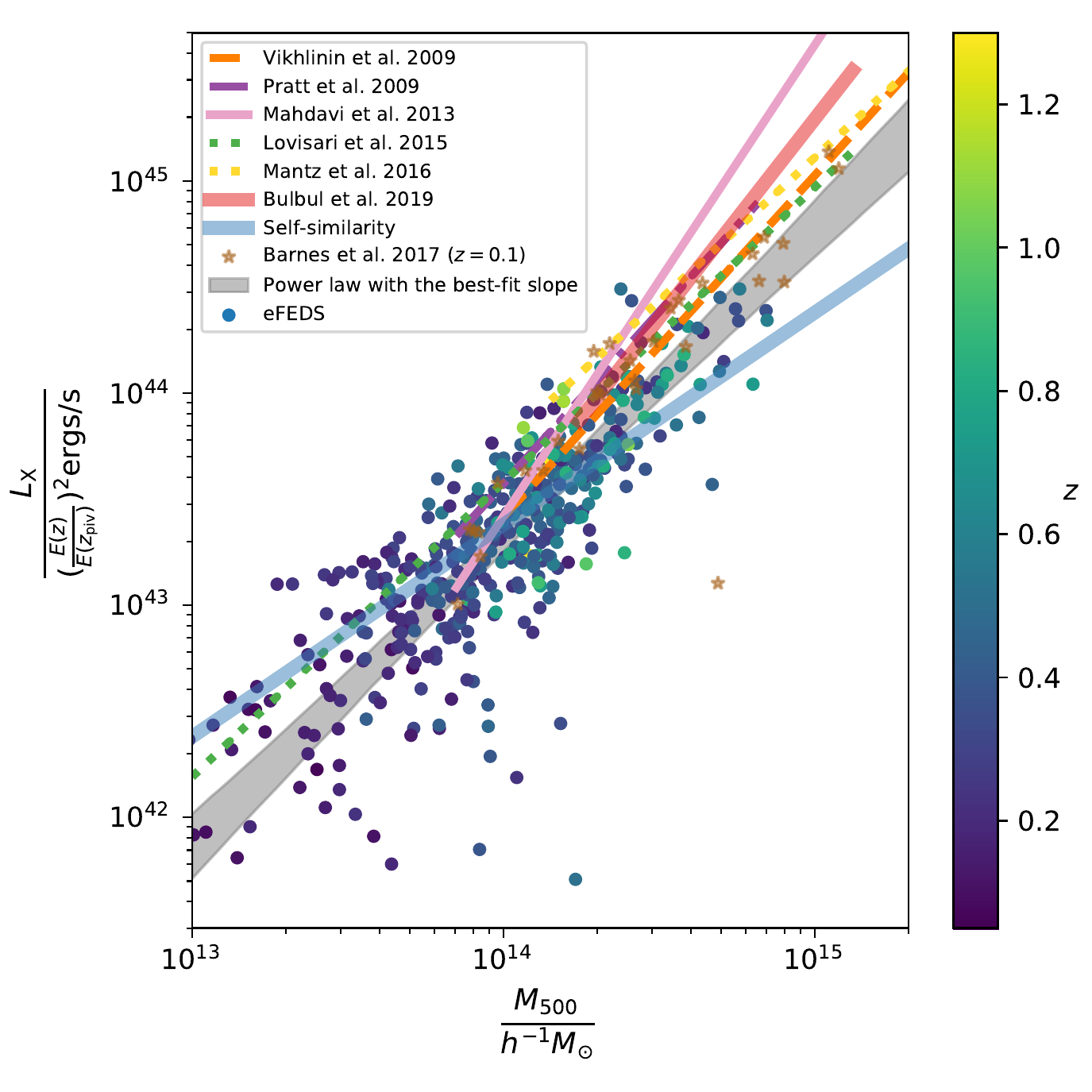}
}
\resizebox{0.48\textwidth}{!}{
\includegraphics[scale=1]{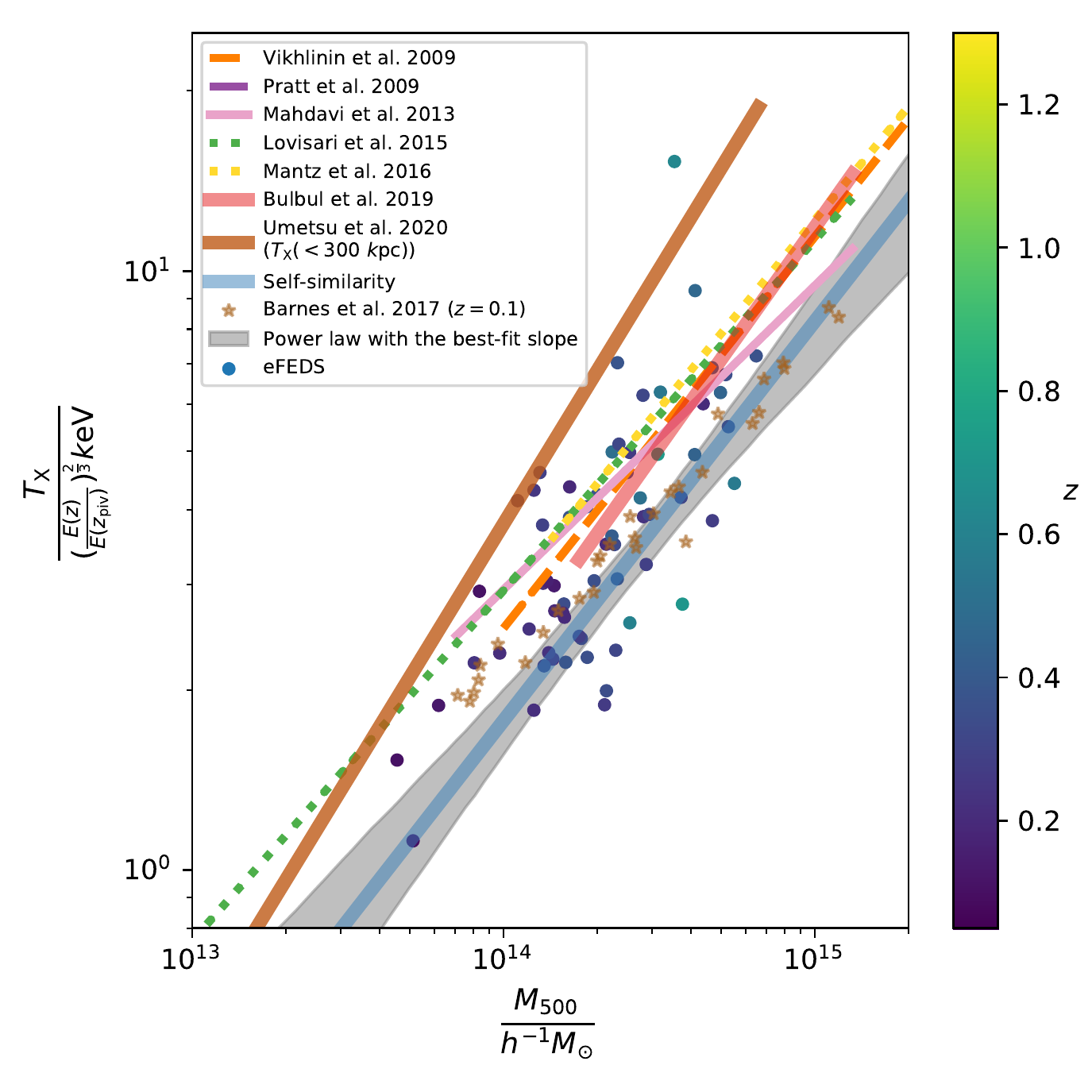}
}
\resizebox{0.48\textwidth}{!}{
\includegraphics[scale=1]{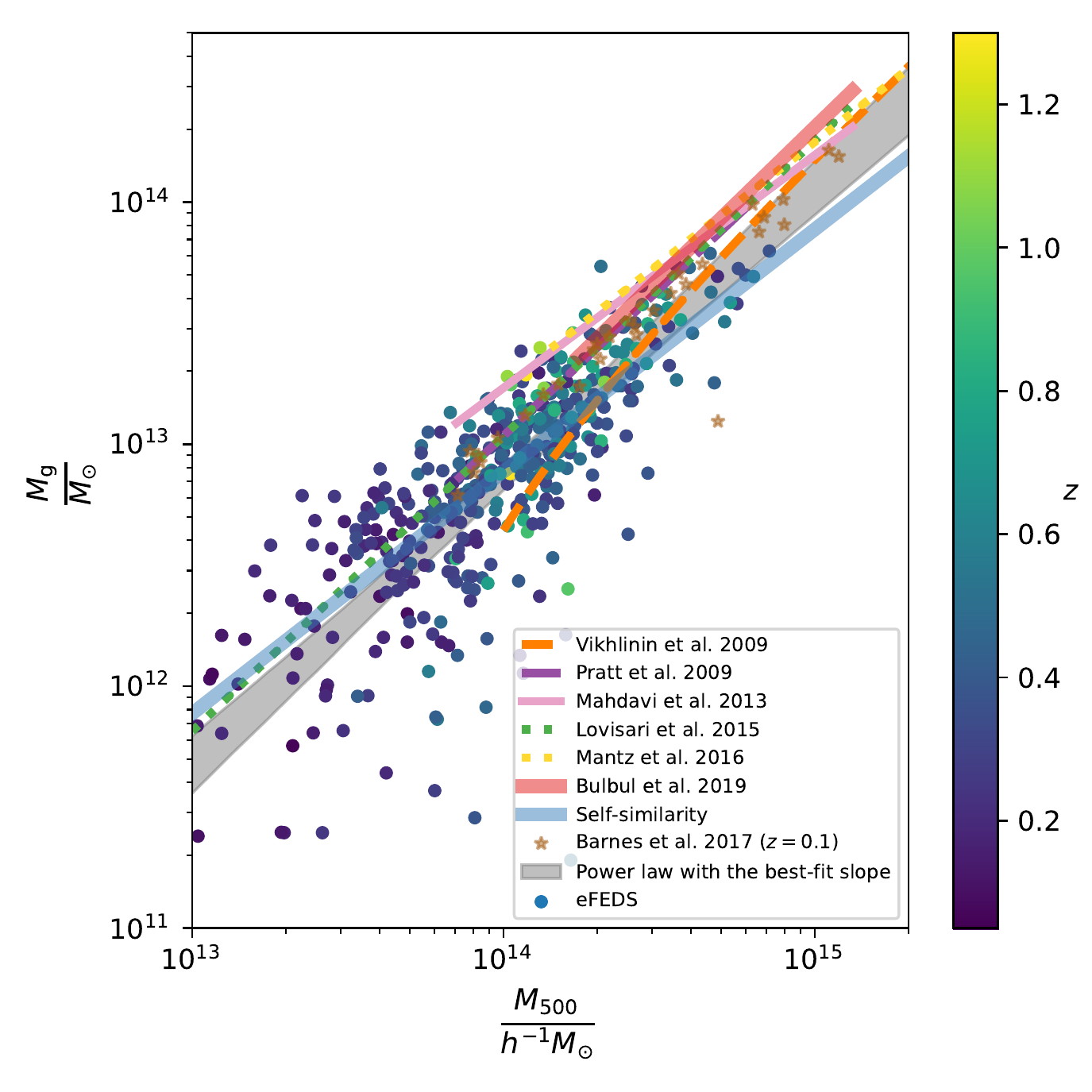}
}
\resizebox{0.48\textwidth}{!}{
\includegraphics[scale=1]{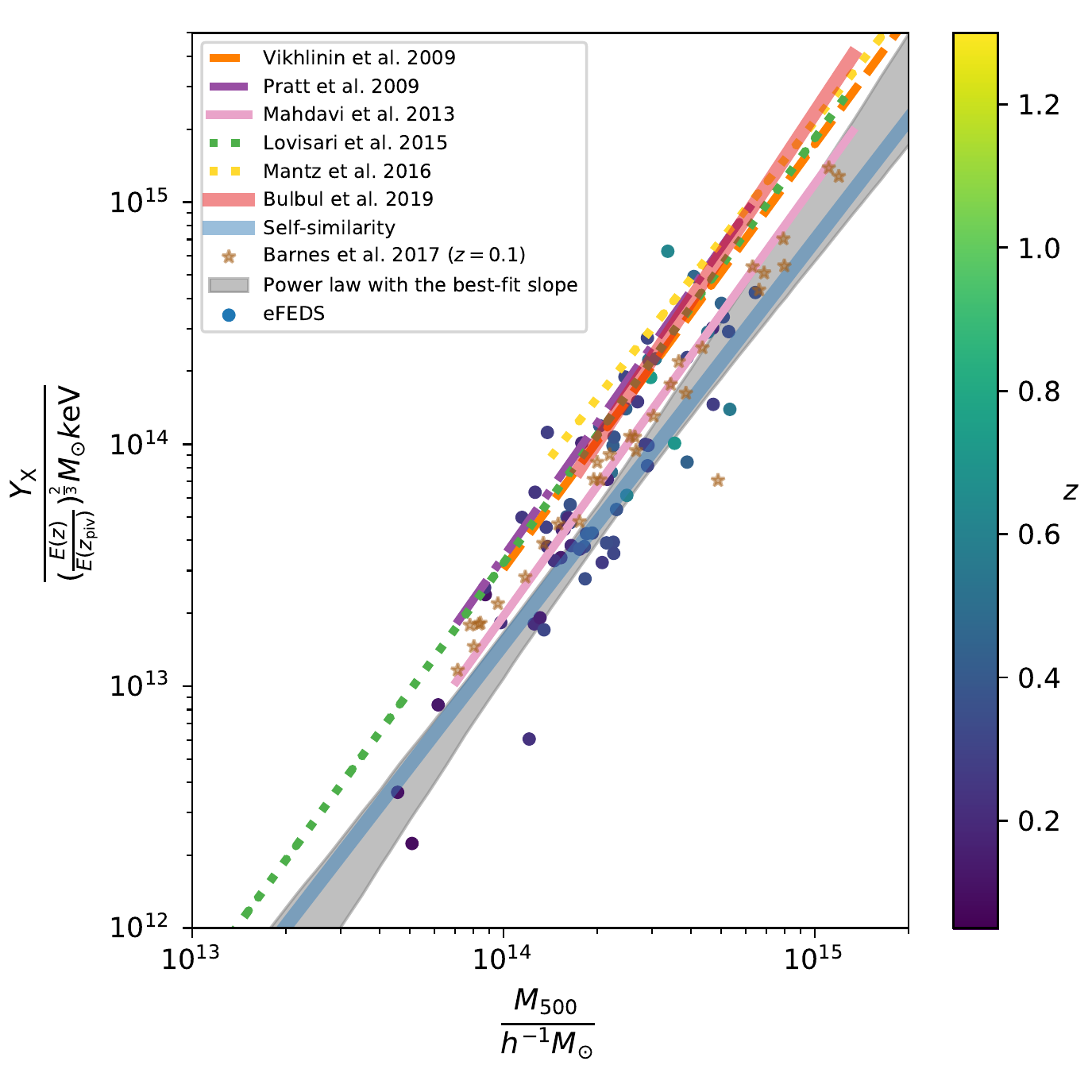}
}
\caption{
Comparison between the eFEDS sample and the results from the literature for \Lx\ (upper-left), \Tx\ (upper-right), \Mg\ (lower-left), and \Yx\ (lower-right).
The eFEDS clusters are shown by the circles color coded by the redshift.
The 68\percent\ confidence levels of the best-fit models at the pivotal redshift $\ZPIV=0.35$ are plotted as the gray shaded regions.
The self-similar prediction with the normalization anchored to the best-fit value of the eFEDS sample is shown by the blue line, while the results from the literature are indicated by different colors.
The simulated clusters \citep{barnes17} are marked as the golden stars.
We apply a correction factor to the cluster mass from the literature to account for the systematic difference in \Mfiveoo, following Appendix~\ref{sec:correctionM500}.
}
\label{fig:comparisons}
\end{figure*}

The \Lx--\Mfiveoo--\redshift\ relation of the eFEDS sample has a mass trend ($\Blx = \ansBlx$) that is broadly consistent with the literature, including 
\cite{pratt09} who found a slope of $1.62\pm0.12$ using an X-ray-selected sample of clusters (REXCESS) at redshift $\redshift\approx0.1$,
\cite{vikhlinin09a} with a slope of $1.61\pm0.14$ from a cluster sample out to $\redshift\approx0.5$ using high-quality \CHANDRA\ data;
\cite{andersson11} who derived a slope of $1.16\pm0.20$ using a sample of 15 SPT-selected clusters with the redshift scaling fixed to $E(\redshift)^{1.85}$; 
\cite{lovisari15} in which a slope of $1.39\pm0.05$ was obtained from a sample of local clusters and groups down to a mass scale of $\approx10^{13}\Msun$; 
\cite{mantz16b} where a slope of $1.35\pm0.06$ is obtained for a sample of $\approx200$ clusters selected from the \textit{ROSAT} survey;
and \cite{bulbul19} who derived a slope of $1.93^{+0.19}_{-0.18}$ using 59 SPT-selected clusters out to redshift $\redshift\approx1.3$ observed with the \XMMNEWTON\ telescope.
\cite{schellenberger17} obtained a slope of $1.35\pm0.07$ for an X-ray-flux-limited sample based on a joint data set from \CHANDRA\ and \ROSAT.
Except for \cite{bulbul19}, which has a steeper slope than ours at a level of $\approx2\sigma$, the resulting mass trend of the soft-band luminosity of the eFEDS sample is in agreement with the literature at a level of $\lesssim1.5\sigma$.
We show the comparison of \Lx\ with the literature in the upper-left panel of Fig.~\ref{fig:comparisons}, in which we additionally include the result of the simulated clusters (golden stars) from the C-Eagle cosmological hydrodynamical simulation \citep{barnes17} and the self-similar model (blue line) with the normalization anchored to the best-fit value of the eFEDS sample.
As seen, the mass scaling of the eFEDS sample is consistent with that found by these latter authors, including the simulated clusters, and shows a steeper mass trend than the self-similar model ($B = 1$) at a level of $\approx3\sigma$.

In terms of the bolometric luminosity \Lb, we find that the resulting slope of the eFEDS sample ($\Blb = \ansBlb$) is shallower than that from \cite{pratt09} and \cite{bulbul19}, with $1.90\pm0.11$ and $2.12^{+0.23}_{-0.18}$, at a level of $\approx1.6\sigma$ and $\approx2.2\sigma$, respectively.
It is worth mentioning that a mass scaling of $\Lb\propto{\Mfiveoo}^{1.52^{+0.10}_{-0.09}}$ was obtained by combining heterogeneous samples in \cite{reichert11}, which is in good agreement with our results.

We constrain the deviation of \Lx\ and \Lb\ from the self-similar redshift trend as $\gammalx = \ansgammalx$ and $\gammalb = \ansgammalb$, respectively.
This implies that the redshift trend of \Lx\ and \Lb\ closely follows the self-similar model, which is in excellent agreement with the result from the SPT-selected massive clusters out to a redshift of $\redshift\approx1.3$ \citep{bulbul19}.
A consistent result is also reported in \cite{mantz16b} with $\gammalx^{-0.65\pm0.38}$ for the soft-band luminosity.
Moreover, \cite{vikhlinin09a} found a redshift scaling of $\Lx\propto E(\redshift)^{1.85\pm0.42}$, which is also consistent with the self-similarity ($C_{\mathrm{SS},\Lx} = 2$).
By homogenizing the different samples from the literature, \cite{reichert11} derived a redshift scaling\footnote{
These authors obtained $\Mfiveoo\propto {\Ez}^{-0.93^{+0.62}_{-0.12}}$. For the redshift scaling of \Lb\ given a cluster mass, the slope is obtained by multiplying $\frac{-4}{3}$ to $-0.93^{+0.62}_{-0.12}$.
}
of $\Lb\propto {\Ez}^{1.24^{+0.16}_{-0.82}}$ out to a high redshift $\redshift\approx1.4$, which is significantly shallower than the self-similar prediction and the eFEDS results.

The log-normal scatter of the soft-band (bolometric) luminosity is constrained as $\sigmalx = \anssigmalx$ ($\sigmalb = \anssigmalb$) at fixed mass.
This is in broad agreement with, but mildly smaller than, the scatter of $0.3$ in the soft-band luminosity derived from the simulated clusters in \cite{barnes17}.
\cite{lovisari15} and \cite{vikhlinin09a} derived the scatter of $0.245$ and $0.396$, respectively, which are both mildly higher than ours in the soft-band luminosity.
Meanwhile, a high value of $0.43\pm0.03$ was derived in \cite{mantz16b} for the scatter of the soft-band luminosity. 
An interesting result from the FABLE simulation \citep{henden19} suggests that the scatter of bolometric luminosity decreases with redshift from $\approx0.25$ at redshift $\redshift\approx0$ to $\approx0.13$ at redshift $\redshift\approx1$.
Based on these comparisons, we conclude that the scatter of the X-ray luminosity in the eFEDS sample is in broad agreement with the literature.

\begin{figure*}
\resizebox{\textwidth}{!}{
\includegraphics[scale=1]{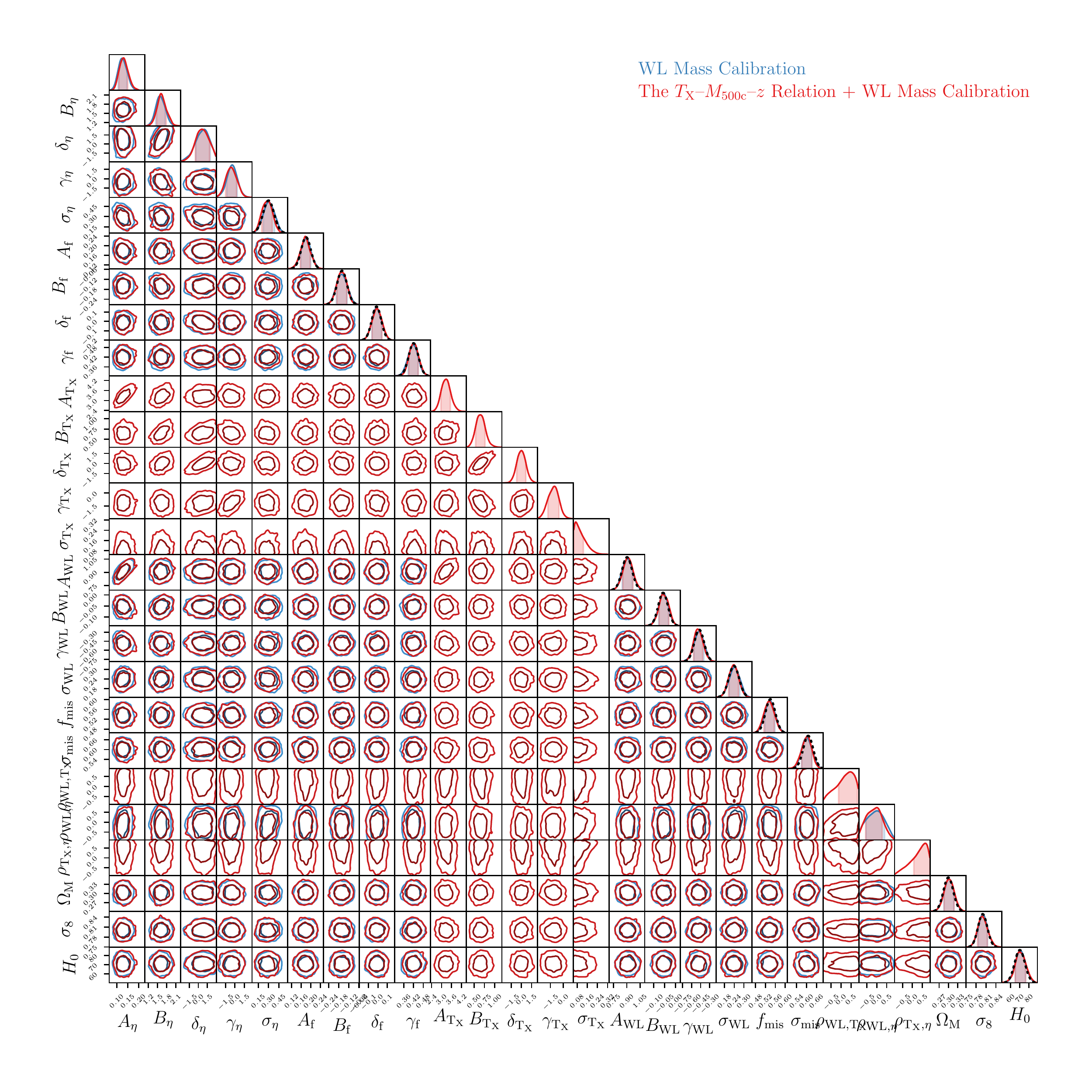}
}
\caption{
Parameter constraints in the joint modeling of the \Tx--\Mfiveoo--\redshift\ relation and the weak-lensing mass calibration, presented in the same way as in Fig.~\ref{fig:chain_lxmcalib}.
}
\label{fig:chain_txmcalib}
\end{figure*}
\begin{figure*}
\centering
\resizebox{0.48\textwidth}{!}{
\includegraphics[scale=1]{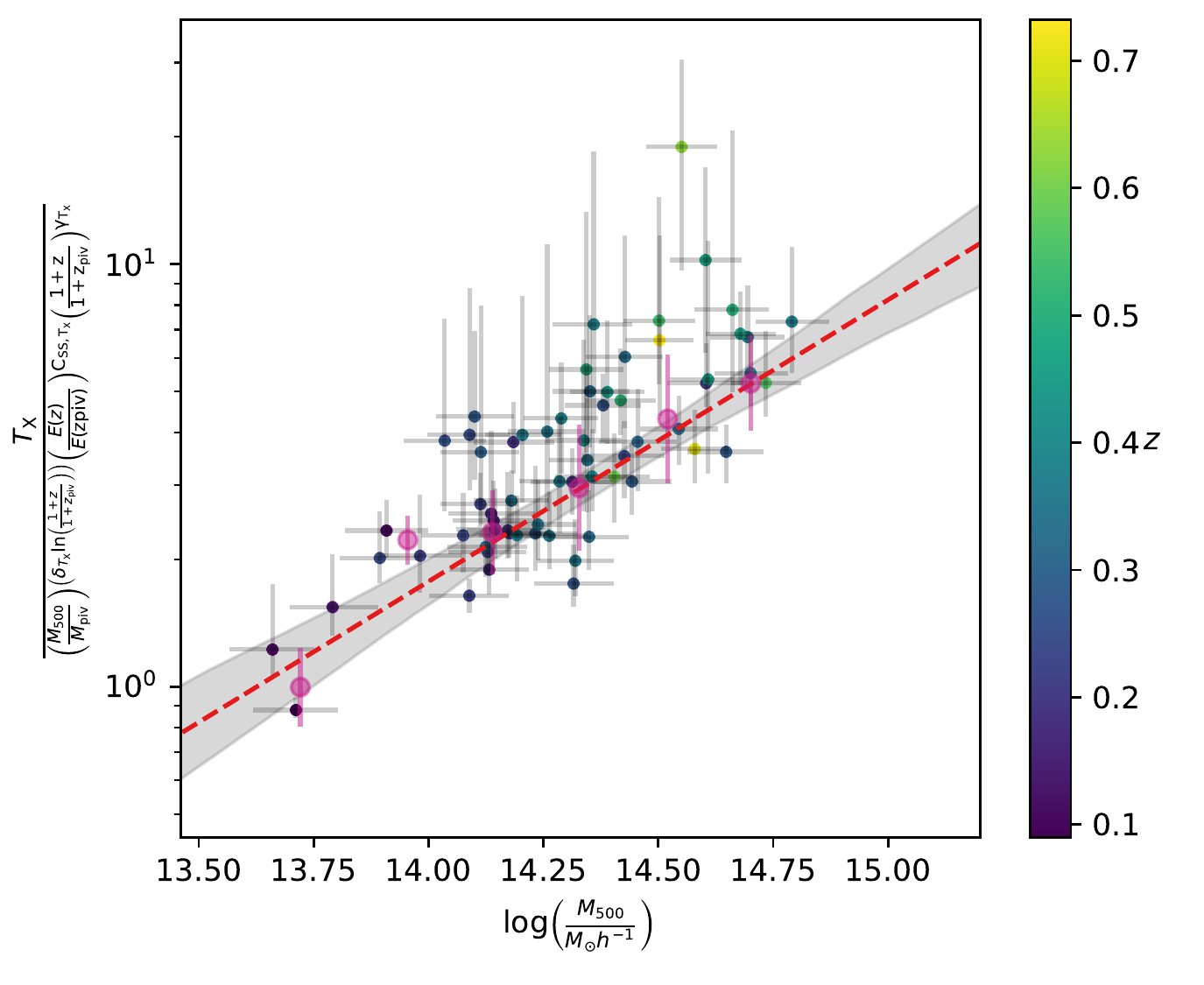}
}
\resizebox{0.48\textwidth}{!}{
\includegraphics[scale=1]{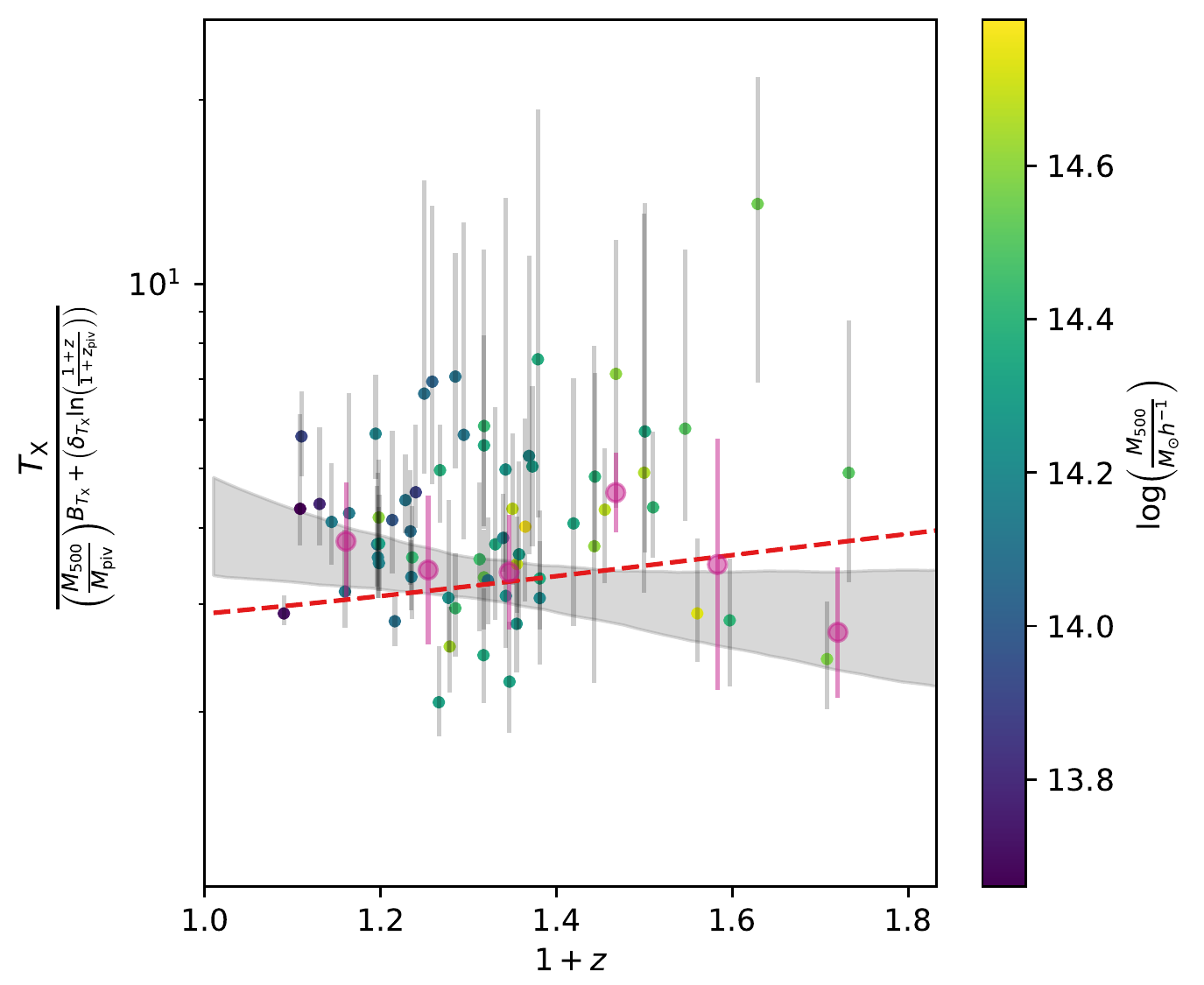}
}
\caption{
Observed X-ray temperature \Tx\ as a function of the cluster mass \Mfiveoo\ (left panel) and redshift \redshift\ (right panel).
This plot is produced and shown in the same way as in Fig.~\ref{fig:lmz}.
}
\label{fig:tmz}
\end{figure*}

\subsection{The \Tx--\Mfiveoo--\redshift\ relation}
\label{sec:tx_results}

The self-similar prediction of the \Tx--\Mfiveoo--\redshift\ relation reads
\[
\Tx \propto {\Mfiveoo}^{\frac{2}{3}} E(\redshift)^{\frac{2}{3}} \, ,
\]
which is obtained by assuming that galaxy clusters reach the virial condition ($M / R \sim T$) with the conversion between the cluster mass and radius ($\Mfiveoo \propto {\Rfiveoo}^3 \rhocrit$).
The X-ray temperature was considered as a reliable mass proxy in early studies of galaxy clusters \citep{smith79,mohr97,mohr99,ikebe02}; therefore the temperature-to-mass relation was intensively investigated using the X-ray hydrostatic mass \citep[e.g.,][]{finoguenov01,vikhlinin06} before the lensing mass became available \citep[e.g.,][]{lieu16}.
In practice, galaxy clusters are not fully virialized and contain nonthermal emissions \citep{ohara06}, resulting in scatter at fixed mass and deviation from self-similarity.
Thus, studying the scaling relation of the temperature sheds light on this perspective.

For the eFEDS sample, the \Tx--\Mfiveoo--\redshift\ relation is constrained as
\begin{multline}
\label{eq:tm}
\left\langle\ln\left( \frac{ \Tx }{ k\mathrm{eV}  } \Bigg|\Mfiveoo \right) \right\rangle 
= \ln \left(\ansAt \right) + \\
\left[ \left( \ansBt \right) + \left( \ansdeltat \right) \ln\left(\frac{1 + \redshift}{1 + \ZPIV}\right) \right] \times
\ln\left(\frac{\Mfiveoo}{\MPIV}\right) \\
 +\frac{2}{3} \times \ln\left(\frac{\Ez}{\Ezpiv}\right) + 
\left( \ansgammat \right)
 \times \ln \left(\frac{1 + \redshift}{1 + \ZPIV}\right)
 \, ,
\end{multline}
with log-normal scatter of $\sigmat = \anssigmat$ at fixed mass.
The parameter constraints of the \Tx--\Mfiveoo--\redshift\ relation are shown in Fig.~\ref{fig:chain_txmcalib} and are listed in Table~\ref{tab:model_params}.
At the pivotal mass $\MPIV = 2.5\times10^{14}\Msunh$ and the pivotal redshift $\ZPIV = 0.35$, the temperature of the eFEDS sample is constrained as $\At = \ansAt$~$k$eV.

In terms of mass and redshift trends, the resulting \Tx--\Mfiveoo--\redshift\ relation shows (1) a mass trend ($\Bt = \ansBt$) that is in good agreement with the self-similarity ($B = 2/3$), (2) a redshift trend ($\gammat = \ansgammat$) that is shallower than, but still statistically consistent with, the self-similar behavior at a significance of $1.7\sigma$, and (3) no clear redshift-dependent mass slope with $\deltat = \ansdeltat$.
The  results of a self-similar mass trend and a redshift scaling shallower than the self-similar prediction are not consistent
with the picture we see in the scaling relation of the luminosity in Sect.~\ref{sec:lm_results}, where we find a steeper mass trend than the self-similar slope with a nearly zero deviation from the self-similar redshift scaling.
This picture can be visualized in Fig.~\ref{fig:tmz}, where the mass and redshift trends of the temperature are presented in the left and right panels, respectively.

A possible explanation to the shallower redshift trend could be as follows.
In the right panel of Fig.~\ref{fig:tmz}, it is seen that the \Tx\ measurements above $\approx4$~$k$eV contain larger error bars and higher scatter than the low-temperature regime.
This is not surprising, given that the capability of the \eROSITA\ telescope to measure high temperatures ($\gtrsim5$~$k$eV) is limited by the reduction of the sensitivity to the high-energy band at $>3$~$k$eV \citep{liu21}.
Consequently, the noisy measurements decrease the constraining power of the high-temperature eFEDS clusters, which are mainly the high-mass systems at high redshift.
As a result, the modeling of the \Tx--\Mfiveoo--\redshift\ relation is weighted toward the low-mass eFEDS clusters or groups at low redshift ($\redshift\lesssim0.3$).
The shallow redshift trend could also be attributed to either the fact that the redshift-dependence of \Tx\ is intrinsically shallower than the self-similar prediction at a group scale,
or that the temperature of the eFEDS clusters with the low mass ($\Mfiveoo\lesssim10^{14}\Msunh$) at low redshift ($\redshift\lesssim0.3$) is higher than the full population on average.
Currently, the redshift trend is only marginally shallower than the self-similar prediction at a level of $\approx1.7\sigma$.
However, in the future,  deep follow-up observations from either \XMMNEWTON\ or \CHANDRA\ telescopes are clearly warranted for a thorough investigation of the \Tx--\Mfiveoo--\redshift\ relation for \eROSITA-detected clusters.

The log-normal intrinsic scatter of \Tx\ for the eFEDS sample is constrained as $\sigmat=\anssigmat$, with a correlation coefficient of $\rho_{\Tx,\rate} = \ansrhoteta$ and $\rho_{\mathrm{WL},\Tx} =\ansrhowlt$ with the count rate and weak-lensing mass, respectively.
Again, no clear correlated scatter between the count rate and weak-lensing mass is seen ($\rho_{\mathrm{WL},\rate} = \ansrhowletat$) in the joint modeling of the \Tx--\Mfiveoo--\redshift\ relation and the mass calibration.

\subsubsection*{Comparisons with the literature}

We compare the \Tx--\Mfiveoo--\redshift\ relation of the eFEDS sample with the literature as follows.
In \cite{arnaud07}, the core-excised temperature is constrained to have a mass-dependent scaling with a slope of $0.58\pm0.03$ using a sample of ten nearby clusters; this slope becomes self-similar as $0.67\pm0.07$ if restricting the sample to clusters with temperature $\Tx\gtrsim3$~$k$eV \citep{arnaud05}. 
The mass trend from \cite{vikhlinin09a} was derived with a self-similar slope of $0.65\pm0.04$ for the core-excised temperature-to-mass relation using a sample of clusters with $10^{14}\lesssim\frac{\Mfiveoo}{\Msunh}\lesssim10^{15}$ out to redshift $\redshift\approx0.5$.
Based on a combined and local sample from the literature, \cite{reichert11} obtained a scaling of $\Tx\propto{\Mfiveoo}^{0.57\pm0.03}$, which is statistically consistent with the eFEDS result (within $1\sigma$).
In \cite{mahdavi13}, the core-excised temperature-to-mass relation was derived as $0.51^{+0.42}_{-0.16}$ ($0.70^{+0.11}_{-0.08}$) with respect to the weak-lensing (X-ray hydrostatic) mass using a sample of $50$ massive clusters in the redshift range of $0.15<\redshift<0.55$.
The core-excised temperature-to-mass relation was found to be $0.58\pm0.01$ by \cite{lovisari15}, which is shallower than the self-similar prediction at high significance.
In \cite{mantz16b}, the mass trend was found to be $0.63\pm0.03$ for ROSAT-selected clusters in the range $10^{14}\lesssim\frac{\Mfiveoo}{\Msunh}\lesssim10^{15}$.
Using a sample of 59 SPT clusters with $\Mfiveoo\gtrsim3\times10^{14}\Msunh$ in the redshift range $0.2\lesssim\redshift\lesssim1.3$, \cite{bulbul19} determined the mass trend as $0.79^{+0.09}_{-0.10}$ and $0.81^{+0.09}_{-0.08}$ for core-included and core-excised temperature, respectively.
The mass-trend slope of the temperature measured within an aperture with a radius of $300$~$k$pc was derived as $0.56^{+0.12}_{-0.10}$ for a sample of galaxy clusters and groups selected in the \XXL\ survey \citep{lieu16,peirre16}; this slope became $0.60^{+0.04}_{-0.05}$ when combining the sample of galaxy groups from the COSMOS field \citep{kettula13} and the sample of massive clusters from \cite{mahdavi13}.
A highly self-similar redshift trend of the temperature, $\Tx\propto{\Ez}^{0.69\pm0.05}$, was also derived in \cite{reichert11}, which is consistent with the eFEDS result.
The latest result from the \XXL\ survey using the S16A HSC lensing data delivered a constraint on the mass trend of $\Tx(<300k\mathrm{pc})$ as $0.85\pm0.31$ and $0.75\pm0.27$ when the redshift scaling is a free parameter and fixed to the self-similar prediction, respectively \citep{umetsu20}.
In summary, the resulting mass trend of the eFEDS sample ($\Bt = \ansBt$) is statistically consistent with the literature at a level of $\lesssim1.5\sigma$.
The upper-right panel of Fig.~\ref{fig:comparisons} contains the comparison, showing good consistency between the eFEDS sample and the previous work.
In addition, we also show the simulated clusters from the C-Eagle cosmological simulation \citep{barnes17} as the golden stars, which nicely follow the trend of the eFEDS sample.

In terms of the redshift trend, the result of the eFEDS sample shows a mild deviation ($\gammat = \ansgammat$) from the self-similar behavior ($\Tx\propto {E(z)}^{\frac{2}{3}}$) at a level of $\approx1.7\sigma$. 
This redshift trend is consistent with the results of \cite{bulbul19}, who obtained a scaling as $\propto {E(z)}^{\frac{2}{3}} (1 + \redshift)^{-0.22^{+0.29}_{-0.35}}$ for core-included temperature out to redshift $\redshift\approx1.3$.
A similar picture is also suggested for the core-excised temperature in \cite{bulbul19}, in which the redshift scaling is constrained as ${E(z)}^{\frac{2}{3}} (1 + \redshift)^{-0.30^{+0.27}_{-0.28}}$.
The latest result from the \XXL\ survey \citep{umetsu20} obtained a redshift scaling of $\propto {E(\redshift)}^{0.18\pm0.66}$ for the temperature estimated in an aperture with a radius of $300$~$k$pc, which is much shallower than but still statistically consistent with the self-similar behavior, given the large error bar.
In summary, the eFEDS sample shows a  redshift scaling that is mildly shallower than the self-similar prediction but is still statistically consistent with the previous work.

The log-normal intrinsic scatter of \Tx\ is constrained as $\sigmat=\anssigmat$ for the eFEDS sample, which is in broad agreement with the literature:
In \cite{bulbul19}, the scatter of the core-excised and core-included temperature was constrained as $0.13^{+0.05}_{-0.04}$ and $0.18\pm0.04$, respectively.
\cite{mantz16b} derived an intrinsic scatter of $0.161\pm0.019$ for the \ROSAT\ sample.
The sample of simulated clusters from \cite{barnes17} suggests a scatter of $0.14$ for the core-included temperature.
The intrinsic scatter of the core-excised temperature at fixed mass\footnote{We make use of Eq.~(4) in \cite{evrard14} to convert the intrinsic scatter of the mass at fixed temperature to that of the temperature at fixed mass.} was suggested to lie in the range between $\approx0.10$ and $\approx0.16$ by \cite{vikhlinin09a}.
A value of $0.23^{+0.17}_{-0.10}$ for the intrinsic scatter of \Tx\ was suggested by \cite{mahdavi13}.
Similarly, the intrinsic scatter was obtained as $0.30^{+0.12}_{-0.10}$ for the \XXL\ clusters, and was reduced to $0.18^{+0.13}_{-0.11}$ if excluding disturbed clusters \citep{lieu16}.
The simulation work in \cite{henden19} suggested the intrinsic scatter to be at a level of $\approx0.23,$ slightly decreasing toward high redshift.

\begin{figure*}
\resizebox{\textwidth}{!}{
\includegraphics[scale=1]{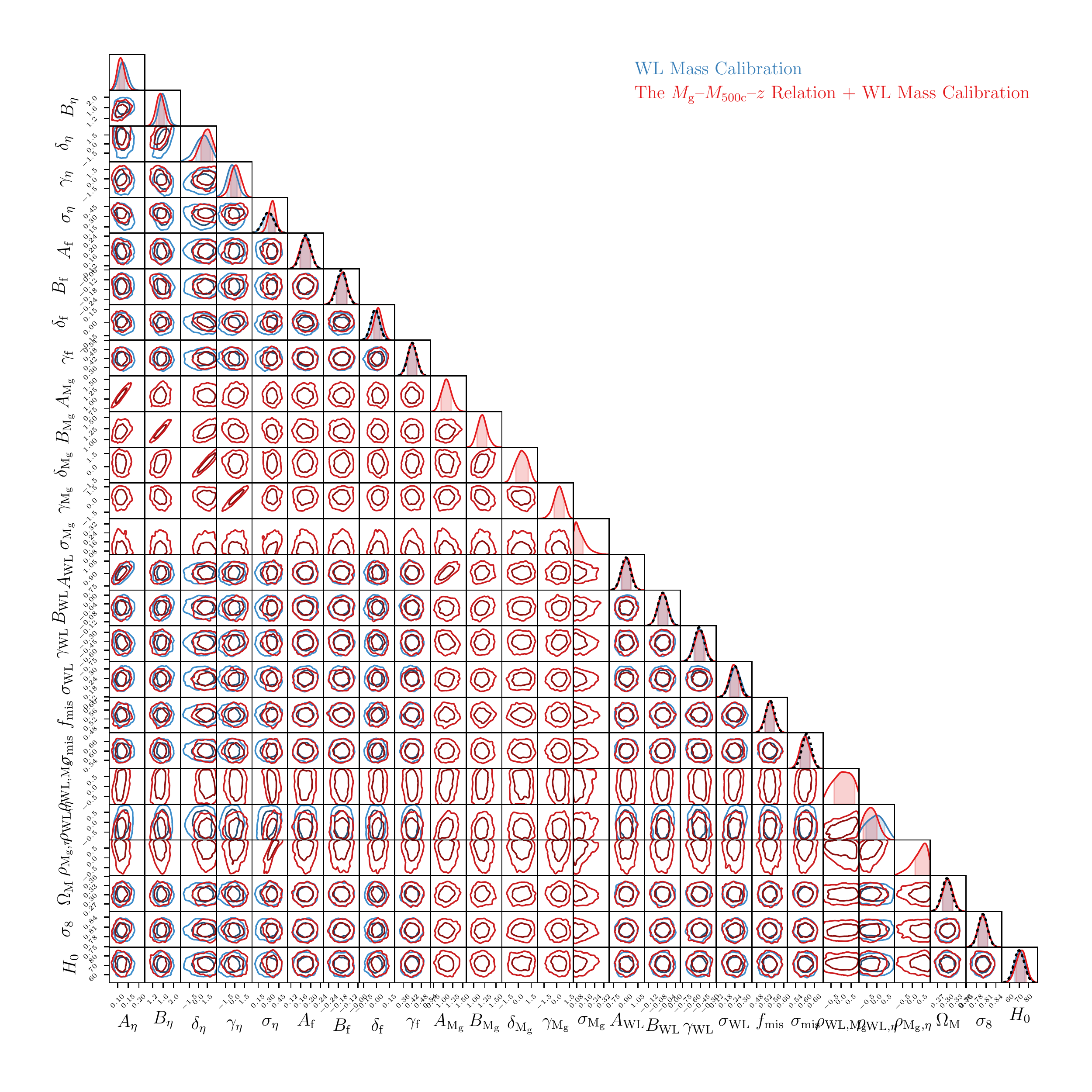}
}
\caption{
Parameter constraints in the joint modeling of the \Mg--\Mfiveoo--\redshift\ relation and the weak-lensing mass calibration, presented in the same way as in Fig.~\ref{fig:chain_lxmcalib}.
}
\label{fig:chain_mgmcalib}
\end{figure*}
\begin{figure*}
\centering
\resizebox{0.48\textwidth}{!}{
\includegraphics[scale=1]{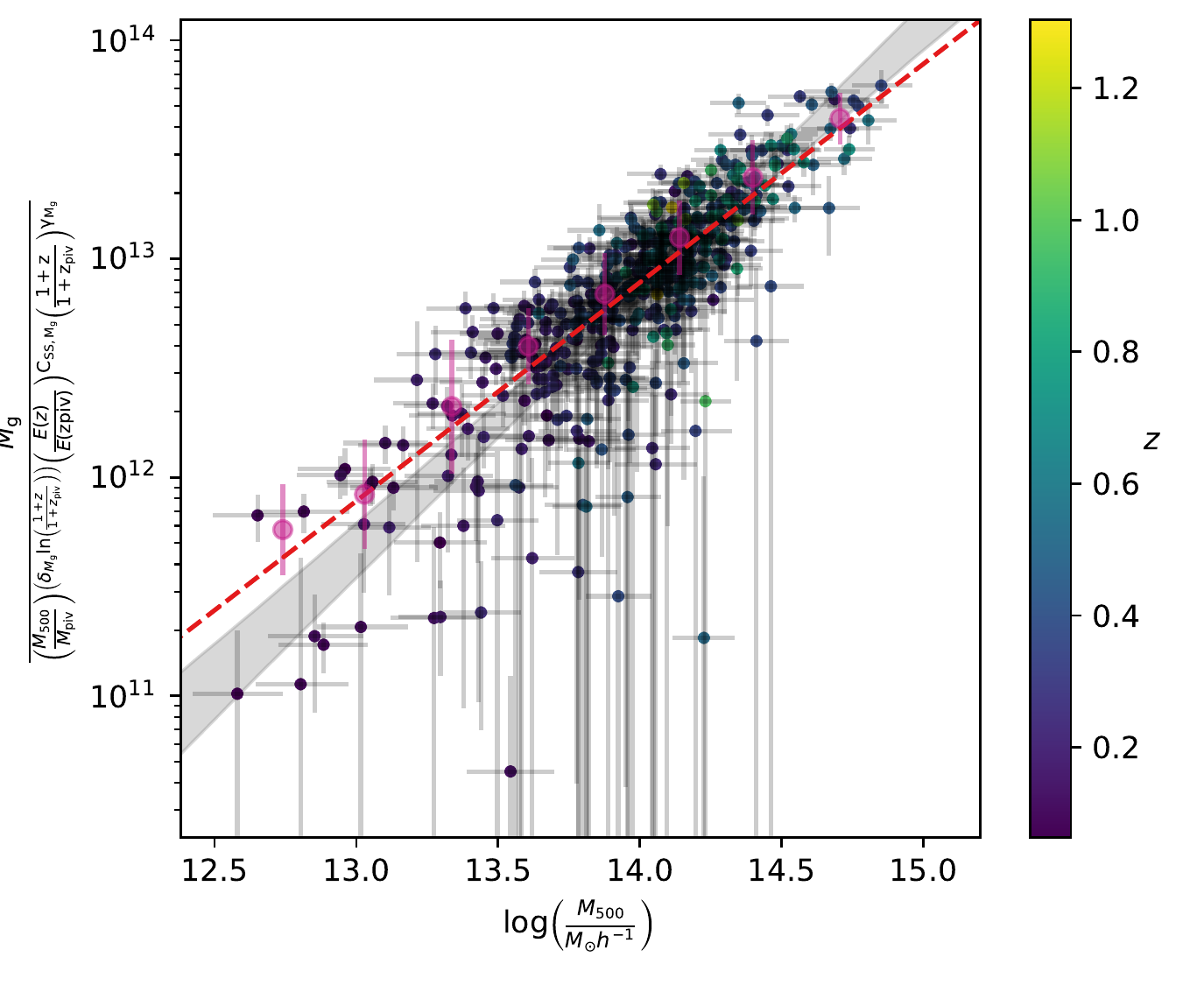}
}
\resizebox{0.48\textwidth}{!}{
\includegraphics[scale=1]{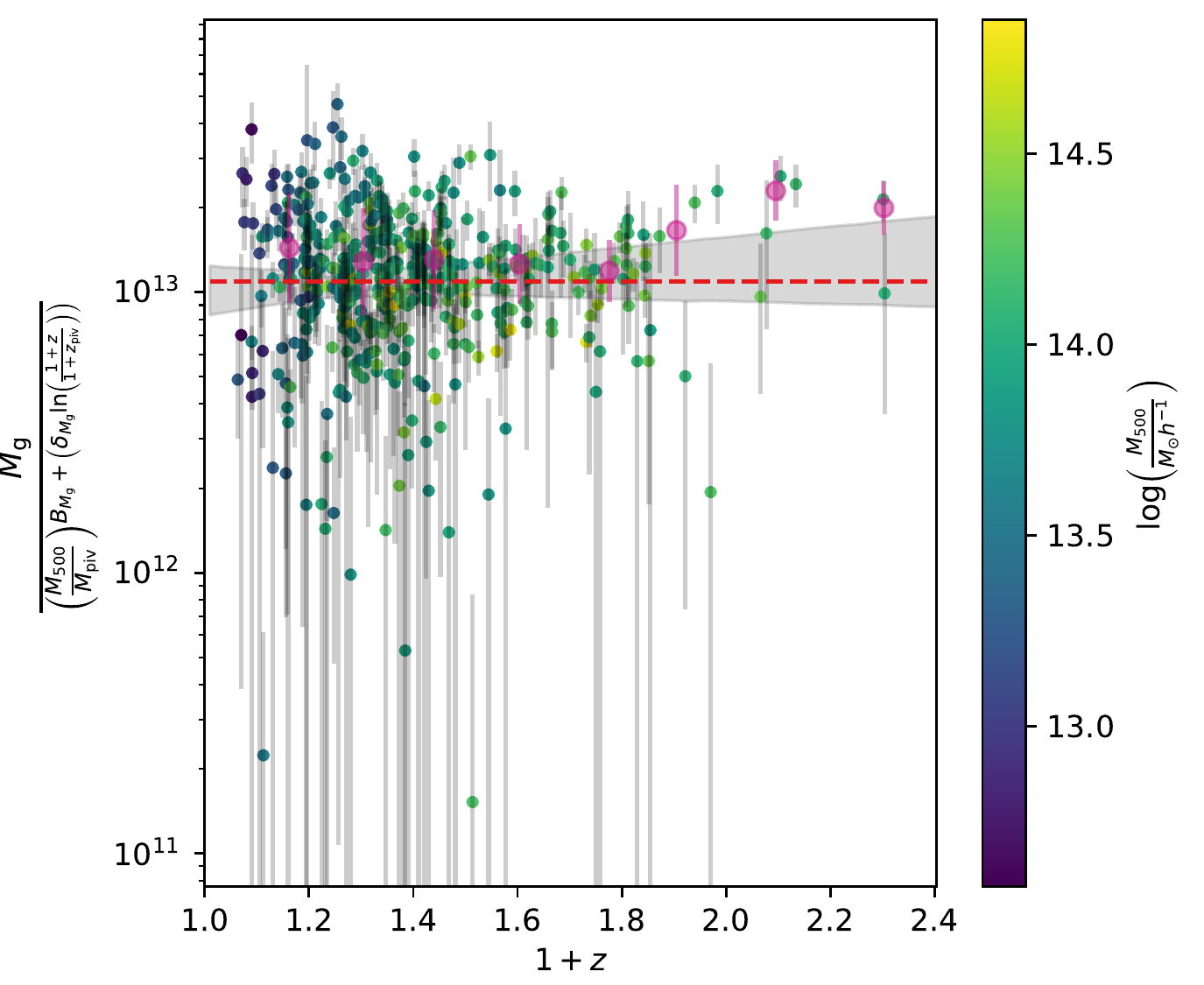}
}
\caption{
Observed ICM mass \Mg\ as a function of the cluster mass \Mfiveoo\ (left panel) and redshift \redshift\ (right panel).
This plot is produced and shown in the same way as in Fig.~\ref{fig:lmz}.
}
\label{fig:mgmz}
\end{figure*}

\subsection{The \Mg--\Mfiveoo--\redshift\ relation}
\label{sec:mg_results}

The self-similar model predicts the \Mg--\Mfiveoo--\redshift\ relation as
\[
\Mg \propto \Mfiveoo \, ,
\]
meaning that the ICM mass fraction, $\frac{\Mg}{\Mfiveoo}$, is invariant among clusters at any redshift.
This has an important implication: Galaxy clusters are considered as a ``closed box'', inside which baryons cannot escape from the potential, meaning that the baryon fraction of clusters is equivalent to that of the Universe.
In practice, the deviation from the universal baryon fraction implies an effect from baryonic physics \citep{lin03b} or the assembly history of galaxy clusters \citep{chiu16c,chiu18a}.

The \Mg--\Mfiveoo--\redshift\ relation of the eFEDS sample is derived as
\begin{multline}
\label{eq:mgm}
\left\langle\ln\left( \frac{ \Mg }{ \Msun } \Bigg| \Mfiveoo \right) \right\rangle 
= \ln \left( \ansAmg \right) + \ln \left( 10^{13} \right) + \\
\left[ \left( \ansBmg \right) + \left( \ansdeltamg \right) \ln\left(\frac{1 + \redshift}{1 + \ZPIV}\right) \right] \times
\ln\left(\frac{\Mfiveoo}{\MPIV}\right) \\
+ \left( \ansgammamg \right)
 \times \ln \left(\frac{1 + \redshift}{1 + \ZPIV}\right) \, ,
\end{multline}
with a log-normal scatter of $\sigmamg = \anssigmamg$ at fixed mass.
The parameter constraints 
are shown in Fig.~\ref{fig:chain_mgmcalib} and are listed in Table~\ref{tab:model_params}.
The ICM mass of the eFEDS clusters is constrained to be $\ansAmg \times 10^{13} \Msun$, corresponding to an ICM mass fraction of $\approx0.053\pm0.007$ (with $h = 0.7$) at the pivotal mass $\MPIV = 1.4\times10^{14}\Msunh$ and redshift $\ZPIV = 0.35$.

Based on these latter results, the scaling trends in the mass ($\Bmg = \ansBmg$) and redshift ($\gammamg = \ansgammamg$) are both steeper than, but still statistically consistent with, the self-similar predictions of $B = 1$ and $\gamma = 0$ at significance levels of $\lesssim1\sigma$.
Moreover, no redshift-dependent mass trend at a significant level is seen with $\deltamg = \ansdeltamg$.
That is, our results for the eFEDS clusters suggest that the ICM mass \Mg\ largely follows the self-similar prediction in both the mass and redshift trends with only mild deviations.

The log-normal intrinsic scatter of \Mg\ at fixed cluster mass is constrained as $\sigmamg = \anssigmamg$ with a mildly positive correlation coefficient $\rho_{\Mg,\rate} = \ansrhomgeta$ with that from the count rate \rate\ at a level of $\approx1\sigma$.
This is in line with expectations, because the ICM mass and count rate are both derived mainly based on an integral of the X-ray surface brightness profile, as a pair of highly correlated quantities.
Meanwhile, there is no significant correlated scatter between the ICM mass and the weak-lensing mass ($\rho_{\mathrm{WL},\Mg}=\ansrhowlmg$) and between the count rate and weak-lensing mass ($\rho_{\mathrm{WL},\rate} = \ansrhowletamg$), as consistent with other follow-up X-ray scaling relations.

We make an additional remark.
Although the ICM mass has low intrinsic scatter at fixed mass and can be directly calculated based on the best-fit model of X-ray imaging, it is a less ideal mass proxy than the luminosity.
This is because prior knowledge of the underlying cluster mass is needed to properly calculate the integrated ICM mass enclosed by \Rfiveoo, given that $\Mfiveoo(<r)\appropto r$ for a large radius $r$.
On the other hand, the luminosity is dominated by the cluster core and hence does not strongly rely on a prior on the cluster radius, despite the large scatter.

\vspace{-0.35cm}
\subsubsection*{Comparisons with the literature}

We first compare the mass trend of \Mg\ between the eFEDS sample and the literature:
The mass trend of the \Mg--\Mfiveoo--\redshift\ relation was constrained as $1.25\pm0.06$ by \cite{arnaud07} based on a sample of ten nearby clusters.
In \cite{vikhlinin09a}, the scaling of the ICM mass fraction is constrained as $\frac{\Mg}{\Mfiveoo}\propto{0.037\pm0.006\ln\Mfiveoo}$, corresponding to a mass trend\footnote{This slope is obtained by re-fitting the data from \cite{vikhlinin09a} in the functional form of $\frac{\Mg}{\Mfiveoo} = B \left(\Mfiveoo\right)$. See also Sect.~5.2 in \cite{chiu18a}.} as $\Mg\propto{\Mfiveoo}^{1.15\pm0.02}$.
Based on a representative sample of  local clusters selected by the X-ray luminosity, \cite{pratt09} obtained a mass-trend slope of $1.21\pm0.03$ in a mass range of $10^{14}\Msunh \lesssim \Mfiveoo \lesssim 10^{15}\Msunh$.
\cite{zhang12} derived a mass scaling of $\Mg\propto {\Mfiveoo}^{1.38\pm0.36}$ using a sample of 19 nearby clusters with masses of $3\times10^{14}\Msunh \lesssim \Mfiveoo \lesssim 10^{15}\Msunh$.
\cite{mahdavi13} constrained the mass trend with a self-similar slope of $1.04\pm0.1$ for massive clusters with a temperature above $3$~$k$eV.
A slope of $1.22\pm0.04$ was derived in \cite{lovisari15} across the mass range of $10^{13}\Msunh\lesssim\Mfiveoo\lesssim10^{15}\Msunh$.
A mass-trend slope of $1.007\pm0.012$, which is highly consistent with self-similarity, was also suggested by \cite{mantz16b} using the \ROSAT-selected clusters.
The 100 brightest galaxy clusters in the \XXL\ survey with $\Mfiveoo\gtrsim2\times10^{13}\Msunh$ resulted in a slope of $1.21^{+0.11}_{-0.10}$ \citep{eckert16}.
The ICM mass of SPT-selected clusters at a mass range of $\Mfiveoo\gtrsim3\times{10}^{14}\Msunh$ was constrained as $\Mg\propto{\Mfiveoo}^{1.33\pm0.09}$ and $\Mg\propto {\Mfiveoo}^{1.26^{+0.12}_{-0.09}}$ in \cite{chiu18a} and \cite{bulbul19}, respectively.
In summary, the mass trend of \Mg\ obtained from the eFEDS sample shows broad consistency with the previous work and no clear tendency to deviate from the self-similar behavior ($B = 1$). 
We plot these comparisons in the lower-left panel of Fig.~\ref{fig:comparisons}, where the simulated clusters (golden stars) from \cite{barnes17} nicely show an extrapolation from the eFEDS sample toward the high-mass end.As there is no significant redshift-dependent mass scaling ($\deltamg = \ansdeltamg$) of the eFEDS sample, our result suggests that this picture still holds out to redshift $\redshift\approx1.3$.

With the capability of \eROSITA\ in discovering high-redshift clusters, we are able to constrain the redshift trend of the \Mg--\Mfiveoo--\redshift\ relation and make comparisons with previous work.
The redshift trend of \Mg\ in the eFEDS sample is constrained to have a slope of $\gammamg = \ansgammamg$, showing 
no clear deviation from self-similarity ($\gamma = 0$) out to redshift $\redshift\approx1.3$ (see Fig.~\ref{fig:mgmz}). 
Based on the sample from the SPT, the redshift scaling of \Mg\ with respect to the pivotal mass was constrained as $\Mg\propto(1+\redshift)^{-0.15\pm0.22}$  and $\Mg\propto(1+\redshift)^{0.18^{+0.30}_{-0.31}}$ in \cite{chiu18a} and \cite{bulbul19}, respectively.
A steep mass trend as $\Mg\propto {E(\redshift)}^{1.76\pm1.22}$ with a large error bar was derived in \cite{sereno20} based on the \XXL\ sample.
Briefly, the ICM mass of the eFEDS sample shows a consistent picture with previous work, namely no significant deviation from self-similarity in the redshift trend is suggested.

For the eFEDS sample, the intrinsic scatter of the ICM mass at fixed cluster mass is constrained as $\sigmamg = \anssigmamg$, where the asymmetric error bars are attributed to the lower bound of the adopted prior.
This result is consistent with the SPT sample of massive clusters ($\Mfiveoo\gtrsim3\times10^{14}\Msunh$) out to high redshift $\redshift\approx1.3$ with the scatter of $0.11\pm0.02$ \citep{chiu18a} and $0.10^{+0.05}_{-0.07}$ \citep{bulbul19} using the \CHANDRA\ and \XMMNEWTON\ telescopes, respectively.
\cite{pratt09} derived the scatter of the ICM mass fraction at fixed cluster mass as $0.12\pm0.03$.
The simulation results from \cite{henden19} suggest the scatter of $0.12^{+0.03}_{-0.02}$ at redshift $\redshift = 0$, decreasing to $\approx0.05$ at redshift $\redshift\gtrsim1$.
A value of $0.13$ was obtained for the intrinsic scatter of \Mg\ in \cite{arnaud07} using the conversion formula in \cite{evrard14}.
Similarly, the intrinsic scatter of \Mg\ at a level of $0.14\pm0.06$ was obtained in \cite{mahdavi13}.
The simulations in \cite{henden19} suggest a decreasing intrinsic scatter at the level of $\approx0.27$ at redshift $\redshift\approx0$ toward a value of $\approx0.12$ at redshift $\redshift\gtrsim1$. 
Based on these comparisons, we conclude that the eFEDS sample has intrinsic scatter of \Mg\ that is broadly consistent with that found in previous studies, suggesting that the ICM mass can be used as a mass proxy that shows little scatter.

\begin{figure*}
\resizebox{\textwidth}{!}{
\includegraphics[scale=1]{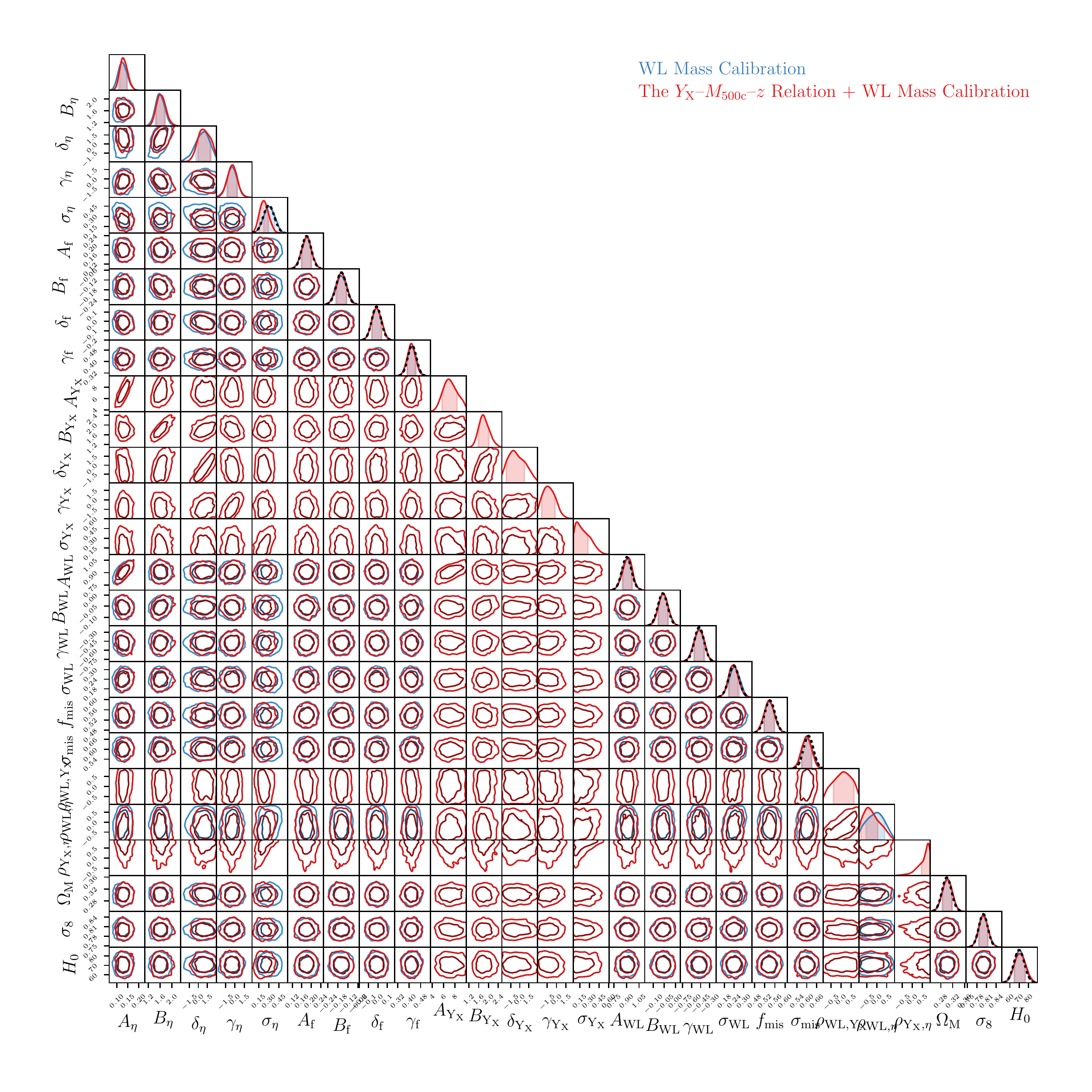}
}
\caption{
Parameter constraints in the joint modeling of the \Yx--\Mfiveoo--\redshift\ relation and the weak-lensing mass calibration, presented in the same way as in Fig.~\ref{fig:chain_lxmcalib}.
}
\label{fig:chain_yxmcalib}
\end{figure*}
\begin{figure*}
\centering
\resizebox{0.48\textwidth}{!}{
\includegraphics[scale=1]{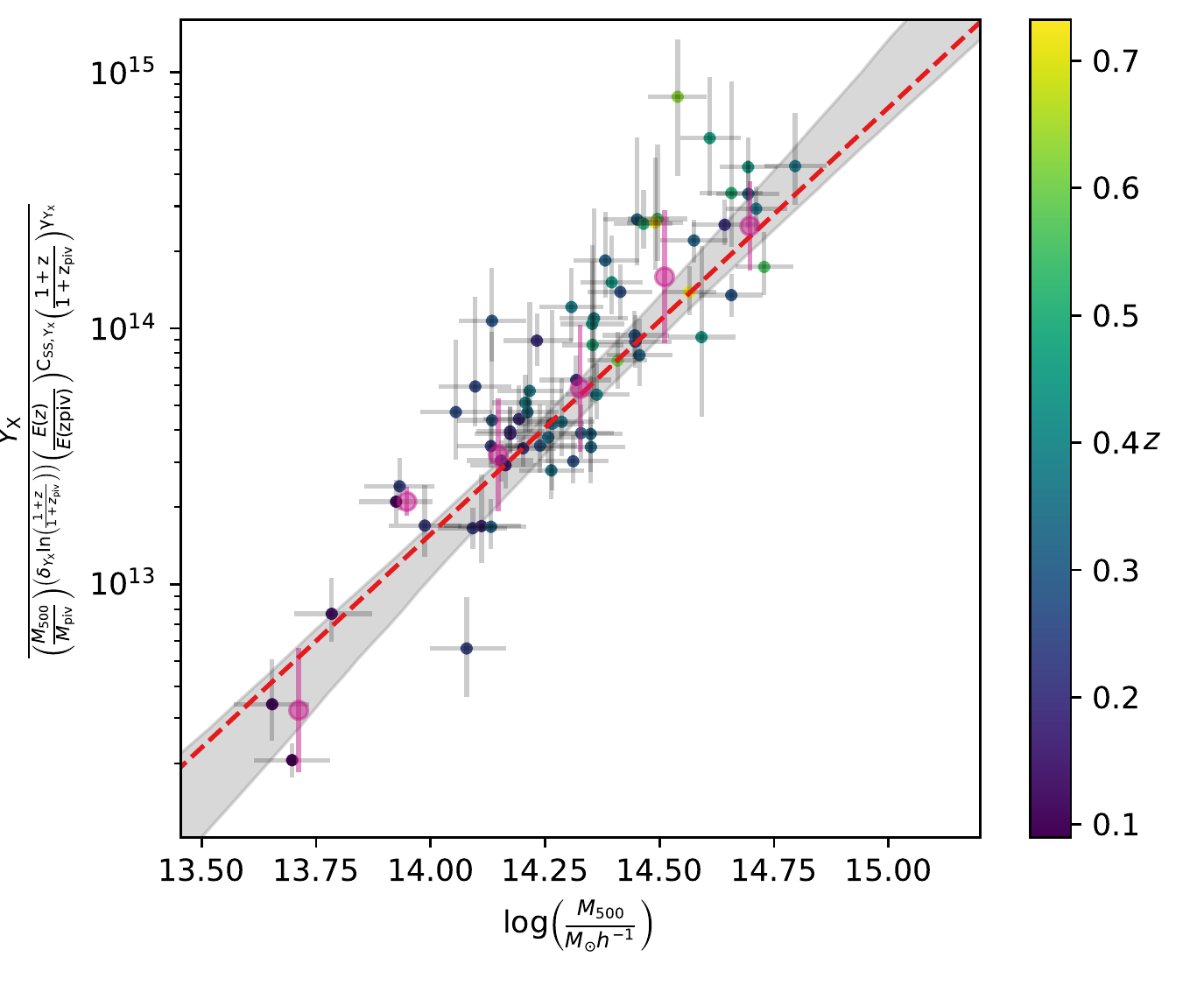}
}
\resizebox{0.48\textwidth}{!}{
\includegraphics[scale=1]{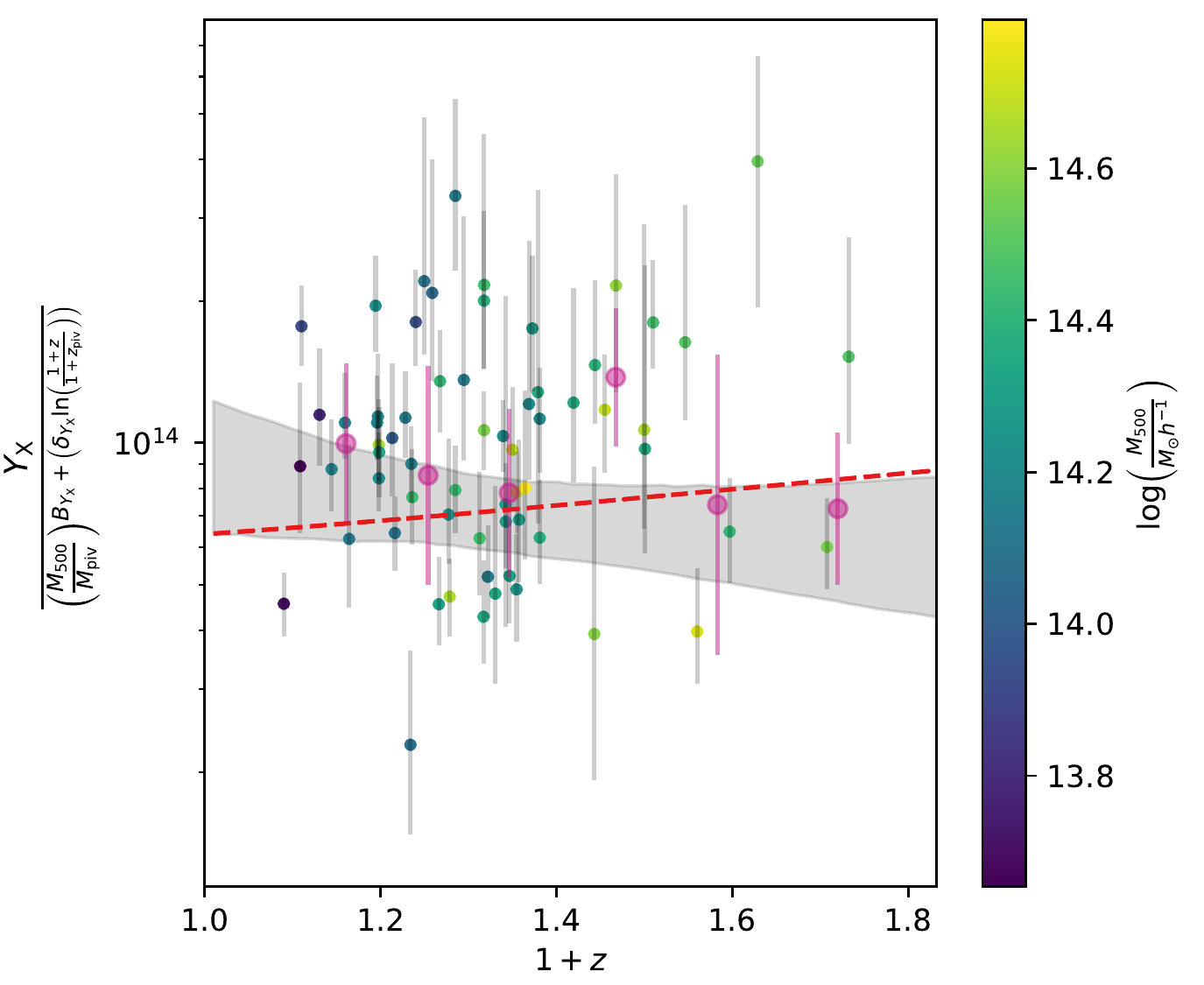}
}
\caption{
Observed \Yx\ as a function of the cluster mass \Mfiveoo\ (left panel) and redshift \redshift\ (right panel).
This plot is produced and shown in the same way as in Fig.~\ref{fig:lmz}.
}
\label{fig:ymz}
\end{figure*}

\subsection{The \Yx--\Mfiveoo--\redshift\ relation}
\label{sec:ym_results}

The mass proxy, \Yx, is defined as a product of the temperature \Tx\ and the ICM mass \Mg, namely, $\Yx \equiv \Tx\Mg$.
The mass proxy \Yx\ was first proposed in \cite{kravtsov06a} and is demonstrated to be insensitive to baryonic feedback \citep[see also][]{nagai07}.
It is found to show remarkably low scatter and is widely used to infer the mass of clusters via X-rays \citep{vikhlinin09a,arnaud10}.
In addition, this quantity can be directly connected to the SZE observable, providing a window to study galaxy clusters in synergy with X-ray and SZE \citep{bonamente08,andersson11,benson13}.
The \Yx--\Mfiveoo--\redshift\ relation has a self-similar scaling as
\[
\Yx \propto {\Mfiveoo}^{\frac{5}{3}} E(z)^{\frac{2}{3}} \, ,
\]
which is directly derived from the self-similar behavior of \Tx\ and \Mg\ by definition.

In this work, the \Yx--\Mfiveoo--\redshift\ relation is derived as
\begin{multline}
\label{eq:ym}
\left\langle\ln\left( \frac{ \Yx }{ \Msun\times k\mathrm{eV} } \Bigg|\Mfiveoo \right) \right\rangle 
= \ln \left( \ansAy \right) + \ln \left( 10^{13}\right) + \\
\left[ \left( \ansBy \right) + \left( \ansdeltay \right) \ln\left(\frac{1 + \redshift}{1 + \ZPIV}\right) \right] \times
\ln\left(\frac{\Mfiveoo}{\MPIV}\right) \\
 +\frac{2}{3} \times \ln\left(\frac{\Ez}{\Ezpiv}\right) + 
\left( \ansgammay \right)
 \times \ln \left(\frac{1 + \redshift}{1 + \ZPIV}\right)
 \, ,
\end{multline}
with log-normal scatter of $\sigmay = \anssigmay$.
The mass proxy \Yx\ of the eFEDS sample is constrained as $\ansAy \times 10^{13}\Msun~k\mathrm{eV}$ at the pivotal mass $\MPIV=2.5\times10^{14}\Msunh$ and the pivotal redshift $\ZPIV=0.35$.
We show the parameter constraints in Fig.~\ref{fig:chain_yxmcalib} with the results listed in Table~\ref{tab:model_params}.
The mass and redshift trends of \Yx\ for the eFEDS sample are shown in Fig.~\ref{fig:ymz}, following the normalization scheme in producing Fig.~\ref{fig:lmz}.

Based on our results, we find a self-similar mass trend ($\By = \ansBy$)
and no significant deviation ($\gammay = \ansgammay$) from self-similarity ($\Yx\propto{E(\redshift)}^{\frac{2}{3}}$) in the redshift scaling.
Both the mass and redshift scaling are better visualized in Fig.~\ref{fig:ymz}.
In addition, no significant redshift-dependent mass trend is seen ($\deltay = \ansdeltay$), suggesting that the mass scaling holds out to high redshift $\redshift\approx1.3$.
The self-similar slopes in both the mass and redshift scaling are expected, as a combined result of the \Tx--\Mfiveoo--\redshift\ and \Mg--\Mfiveoo--\redshift\ relations.
We note that the \Yx\ measurements are more noisy and show greater scatter above the best-fit model (the gray area in Fig.~\ref{fig:ymz}), 
which is clearly due to the higher uncertainty at the high-temperature end  (see Sect.~\ref{sec:tx_results}).

The intrinsic scatter of \Yx\ is constrained as $\sigmay = \anssigmay$ with a correlation coefficient of $\rho_{\mathrm{WL},\Yx} = \ansrhowly$ which is consistent with zero.
Interestingly, the correlation between the intrinsic scatter of  \Yx\ and \rate\ is constrained to be positive ($\rho_{\mathrm{\Yx},\rate} = \ansrhoyeta$) at a significance level that is higher than those of $\rho_{\mathrm{\Tx},\rate}$ and $\rho_{\mathrm{\Mg},\rate}$.
This is expected, because both 
$\rho_{\mathrm{\Tx},\rate}$ and $\rho_{\mathrm{\Mg},\rate}$ are also constrained to be mildly positive.

\subsubsection*{Comparisons with the literature}

Similarly to the X-ray scaling relations discussed in Sects.~\ref{sec:lm_results}~to~\ref{sec:mg_results}, we first compare the mass trend ($\By = \ansBy$) of the \Yx--\Mfiveoo--\redshift\ relation of the eFEDS sample with the results of other studies:
In a pilot study, \cite{arnaud07} derived a slope of $1.82\pm0.09$ based on a sample of ten nearby clusters while fixing the redshift scaling to the self-similarity ($\Yx\propto{E(\redshift)}^{\frac{2}{3}}$).
\cite{vikhlinin09a} constrained the slope of the mass trend as $1.75^{+0.10}_{-0.09}$ for \ROSAT-selected clusters out to $\redshift\approx0.5$, assuming $\Yx\propto{E(\redshift)}^{\frac{2}{3}}$.
A mass-trend slope of $1.79^{+0.26}_{-0.20}$ using a sample of 50 massive clusters was derived in \cite{mahdavi13} with the fixed redshift scaling of $E(\redshift)^{\frac{2}{3}}$.
Later, \cite{lovisari15} derived a slope of $1.75\pm0.03$ using a joint sample of galaxy groups and massive clusters with temperatures above $3$~$k$eV. 
The \ROSAT-selected sample was used in \cite{mantz16b} to constrain the \Yx--\Mfiveoo--\redshift\ relation, resulting in a slope of $1.63\pm0.04$.
Using a sample of 59 massive ($\Mfiveoo\gtrsim3\times10^{14}\Msunh$) clusters selected in the SPT, the mass scaling of \Yx\ was constrained to be $2.02^{+0.16}_{-0.17}$ out to redshift $\redshift\approx1.3$.
We compare these results with that of the eFEDS sample in the lower-right panel of Fig.~\ref{fig:comparisons}, where we additionally show the simulated clusters (golden stars) from \cite{barnes17}.
As seen, the eFEDS sample has the mass scaling of \Yx\ that is consistent with previous work.

As an interesting comparison, our result of $\By = \ansBy$ is also in good agreement with the mass trend of the integrated pressure $Y_{\mathrm{SZ}}$ inferred from the SZE, which was derived with a slope of $1.67\pm0.29$ by \cite{andersson11} using 15 SPT clusters with the redshift scaling fixed to $E(\redshift)^{\frac{2}{3}}$.

In terms of the redshift trend, the eFEDS sample shows a scaling as $\Yx\propto{E(\redshift)}^{\frac{2}{3}}(1 + \redshift)^{\ansgammay}$.
Our result is consistent with that of 
\cite{bulbul19}, in which \Yx\ was derived with the scaling as $\Yx\propto{E(\redshift)}^{\frac{2}{3}}(1 + \redshift)^{-0.17^{+0.47}_{-0.50}}$, showing no significant deviation from self-similarity.
A cosmological analysis based on the abundance of SPT-selected clusters resulted in a slope of $0.310^{+0.209}_{-0.140}$ \citep{bocquet19}, which is statistically consistent with our result.
Based on these comparisons, the mass proxy \Yx\ of the eFEDS sample shows a self-similar redshift trend, as consistent with previous work.

The log-normal intrinsic scatter of \Yx\ at fixed cluster mass is constrained as $\sigmay = \anssigmay$, which is in good agreement with the previous literature:
The \ROSAT-sample from \cite{mantz16b} resulted in the scatter of $0.185\pm0.016$.
A joint analysis of cluster abundance and the integrated pressure from both X-rays and SZE resulted in the intrinsic scatter of $0.184^{+0.087}_{-0.089}$ for the SPT-selected clusters \citep{bocquet19}.
The sample of simulated clusters from \cite{barnes17} suggested the scatter of $0.23$ for the \Yx--\Mfiveoo--\redshift\ relation.
A similar value of $\approx0.23$ was also suggested by the simulations from \cite{henden19} at redshift $\redshift\approx0$, decreasing to $\approx0.15$ for clusters at redshift $\redshift\gtrsim1.0$.
The intrinsic scatter in the core-included and core-excised \Yx\ of the SPT clusters at fixed mass was constrained as $0.16^{+0.07}_{-0.10}$ and $0.13^{+0.07}_{-0.08}$, respectively \citep{bulbul19}.
These results, together with that from the eFEDS, imply that \Yx\ serves as a low scattering proxy.

%
%

\section{Systematic errors}
\label{sec:sys}

The key and novel ingredients in this work are the simulation-based calibrations of the observed count rate \rate\ and the weak-lensing mass \Mwl, as described in Sect.~\ref{sec:simulations}.
The calibrations are obtained by performing the end-to-end analysis on the simulations with properties that are as realistic as those observed, resulting in an empirical relation between the observed quantity and the underlying true mass.
As a result, a systematic change in the simulation calibrations  results in a systematic uncertainty of the scaling relation (and hence the cluster mass).
Therefore, in this work we effectively marginalize over the systematic uncertainty by applying the informative priors, which are obtained from the simulation-based calibrations, on the parameters of Eqs.~(\ref{eq:bias_rate})~and~(\ref{eq:bwl}).
That is, the systematic errors of the simulation calibration are marginalized over in our results.

One of the remaining systematic errors is from whether or not the cluster core is included in the modeling of weak-lensing shear profiles.
As seen in Table~\ref{tab:model_params}, including the cluster core results in a negligible difference in the parameters of the \rate--\Mfiveoo--\redshift\ relation compared to the current uncertainty.
This, again, is supported by our end-to-end calibration of the weak-lensing mass bias presented in Sect.~\ref{sec:bwl}.
Quantitatively, the cluster mass would decrease by $\approx0.01$~dex if the cluster core were include ; this is a subdominant systematic uncertainty in this work.

A source of potential systematic error could be the photo-\redshift\ bias in the 30-band COSMOS sample, if indeed there is one.
In this work, the photo-\redshift\ from the 30-band COSMOS catalog \citep{laigle16} is used as the true redshift.
Hence, a photo-\redshift\ bias in the COSMOS catalog would result in a biased lensing signal; this is especially true for the faint-end regime \citep[see also][]{schrabback10}.
This issue, the reliability of the photo-\redshift\ in the COSMOS catalog, was also thoroughly discussed in a recent cosmic shear analysis \citep[see Section 8 in][]{hildebrandt20}, in which the authors also estimated how the outlier fraction of the photo-\redshift\ may affect the mean redshift of the lensing sources.
Although the accuracy and the outlier fraction of the COSMOS photo-\redshift\ are estimated to be less than $\approx1\percent$ for relatively bright galaxies \citep[$i\lesssim21$~mag;][]{laigle16}, the outlier fraction is quantified to be at a level of $\approx6\percent$ for galaxies with an $i$-band magnitude of between $23$ and $24$.
However, it is unclear and hard to quantify how the outlier fraction of the COSMOS catalog affects the cluster mass estimate without a representative spectroscopic sample out to high redshift.
Moreover, our source selection effectively removes source galaxies with low photo-\redshift, which further affects the systematic impact arising from catastrophic photo-\redshift\ outliers.
Here, we make an assumption that the galaxies with available spectroscopic redshifts\footnote{
Specifically, we make use of the spectroscopic redshifts estimated from
3DHST \citep{skelton14,momcheva16}, 
C3R2 \citep{masters19},
DEIMOS \citep{hasinger18}, 
FMOS-COSMOS \citep{silverman15}, 
LEGA-C \citep{straatman18}, 
PRIMUS \citep{coil11,cool13}, 
SDSS \citep{alam15,paris18}, 
and zCOSMOS \citep{lilly09}.
}
in the COSMOS field provide a representative sample of those at the faint end and/or the high redshift, and try to estimate the bias arising from the inaccuracy in the COSMOS photo-\redshift\ \citep[see also][]{schrabback18,schrabback18b,raihan20}.
Specifically, we re-calculate Eq.~(\ref{eq:photozbias}) based on the COSMOS galaxy sample with available spectroscopic redshifts, and replace the observed and true redshifts with the COSMOS 30-band photo-\redshift\ and the spectroscopic redshift when evaluating Eq.~(\ref{eq:photozbias_2}), respectively.
The resulting photo-\redshift\ bias $\bz\left(\zcl\right)$, which indicates the bias arising from the inaccuracy of the COSMOS 30-band photo-\redshift, shows a behavior that is generally consistent with zero at $\redshift\lesssim0.6$ with a weak trend that slightly increases to a level of $\approx1\percent$ at $\redshift\approx0.8$ and then decreases toward a level of $\approx-3\percent$ at $\redshift\approx1.2$, despite with large error bars.
This result suggests that the inaccuracy of the COSMOS photo-\redshift\ may introduce systematic bias to the cluster mass (primarily at high redshift, $\redshift\gtrsim0.8$) at a level of $\lesssim3\percent$, assuming that the available spectroscopic sample in the COSMOS is complete.
For a study with a significantly larger sample size from the \eROSITA\ All-Sky Survey, a more detailed investigation on the photo-\redshift\ bias using an alternative method \citep[e.g., the clustering-based calibration;][]{gatti20} is clearly recommended.

Another source of systematic error comes from the fact that we ignore the correlation between the uncertainty in the follow-up X-ray quantities and that of 
the cluster mass.
Because we perform the early unblinding (see Sect.~\ref{sec:blinding}) to self-consistently estimate the X-ray properties extracted at the  ``fixed'' cluster radius \Rfiveoo, a change in the cluster mass \Mfiveoo\ will lead to a change in \Rfiveoo\ and the follow-up X-ray observables.
We gauge this correlated uncertainty as follows.
In this work, the statistical uncertainty of each cluster mass, which is estimated as the dispersion of the resulting posterior of \Mfiveoo\ (see Sect.~\ref{sec:mass_calibration_results}), is at the level of $\approx32\percent$, $\approx27\percent$, and $\approx23\percent$ for a cluster with a mass of $\approx10^{13.6}\Msunh$, $\approx10^{14}\Msunh$, and $\approx10^{14.4}\Msunh$, corresponding to a change in the radius at a level of $\approx10\percent$, $\approx9\percent$, and $\approx8\percent$, respectively.
The ICM mass \Mg\ is more sensitive to the cluster radius among the X-ray observables, because it is a direct integration of the surface brightness profile up to \Rfiveoo.
Assuming that the ICM density distribution $n_{\mathrm{e}}$ follows a $\beta$-model \citep{cavaliere76} with a typical value of $\beta=2/3$, the integrated profile of the ICM  is  $\propto\int n_{\mathrm{e}} r^2 \dif r \appropto r$.
That is, the change in \Rfiveoo\ leads to roughly the same amount of uncertainty in \Mg.
Meanwhile, a typical measurement uncertainty of the ICM mass is on the order of $\approx50\percent$, $\approx42\percent$, and $\approx28\percent$ for an eFEDS cluster with a mass of $\approx10^{13.6}\Msunh$, $\approx10^{14}\Msunh$, and $\approx10^{14.4}\Msunh$, respectively.
In comparison, the radius-induced statistical uncertainty is much smaller than the current measurement uncertainty and is thus negligible.
In terms of the systematic uncertainty in the cluster mass, which is expected to be at the level of $\approx7.6\percent$ as inferred from the weak-lensing mass bias calibration (see Eq.~(\ref{eq:wlbias_constraints}) in Sect.~\ref{sec:bwl}), 
the corresponding change in \Rfiveoo\ is at the level of $\approx7.6\percent/3\approx2.5\percent$.
On the other hand, the uncertainties on the mean \Mg\ at the mass bins centering at $\log\left(\frac{\Mfiveoo}{\Msunh}\right)=13.6$, $14$, and $14.4$ with a bin width of $0.2$~dex are approximately $50\percent/\sqrt{30}\approx9\percent$, $42\percent/\sqrt{90}\approx4\percent$, and $28\percent/\sqrt{64}\approx3.5\percent$, respectively, where the numbers in the square root are the numbers of the clusters in the bins.
That is, the change due to a systematic uncertainty in the cluster mass is still subdominant compared to the current measurement uncertainty.
Therefore, we conclude that the uncertainty due to the change in \Mfiveoo\ is expected to be subdominant for the modeling of the X-ray scaling relations. 
We note that the luminosity, which is also estimated based on the surface brightness profile, is dominated by the cluster core\footnote{$\Lx\propto\int n_{\mathrm{e}}^2 \sqrt{T} r^2 \dif r \appropto C - \frac{1}{r}$, with $C$ up to a constant. Here, we assume that the temperature is insensitive to the cluster radius.} 
and is thus less sensitive to the cluster radius.
An improvement overcoming these systematic errors is to use an iterative way to estimate the X-ray observable from a profile in each step of the likelihood exploration \citep[e.g.,][]{benson13,bocquet19}, but we defer this to a future work.

%
%

\section{Conclusions}
\label{sec:conclusions}

We present the first weak-lensing mass calibration of \eROSITA-detected clusters selected in the eFEDS survey using the three-year weak-lensing shape catalog (S19A) from the HSC survey.
We use a sample of $434$ optically confirmed clusters at redshift $0.01\lesssim\redshift\lesssim1.3$ (median $0.35$) with contamination from point sources at the level of $\approx2\percent$.
The uniform coverage of the HSC survey enables us to extract the lensing shear profile for the majority ($\approx70\percent$, or $313$ clusters) of the sample.

We first perform the weak-lensing mass calibration by simultaneously modeling the observed count rate \rate, as the X-ray mass proxy, and the shear profile \gshear\ on the basis of individual clusters through a Bayesian population modeling of the count-rate-to-mass-and-redshift (\rate--\Mfiveoo--\redshift) and weak-lensing-mass-to-mass-and-redshift (\Mwl--\Mfiveoo--\redshift) relations, respectively.
The weak-lensing mass \Mwl, which is obtained by fitting a mass model to \gshear, is carefully calibrated against cosmological hydrodynamic simulations that are exclusively constructed according to the observed lensing properties, including the multiplicative bias of shape measurements as a function of source redshift, the shape noise of  \gshear, the uncertainty arising from uncorrelated large-scale structures, the cluster member contamination, the miscentering of the X-ray center, the redshift distribution of sources, the bias from the photo-\redshift, the choice of the radial bins in the fitting, and the inaccurate assumption about the halo mass distribution.
In this way, we empirically establish the connection between the lensing observable \gshear\ and \Mfiveoo\ at the cluster redshift. 
Similarly, we calibrate the bias in the observed count rate \rate\ with respect to the true count rate through a calibration by running the end-to-end \eROSITA\ analysis pipeline on large simulations, enabling us to robustly connect \rate\ to \Mfiveoo.
With these simulation-based calibrations, we calibrate the \rate--\Mfiveoo--\redshift\ and \Mwl--\Mfiveoo--\redshift\ relations by simultaneously modeling 
the individual \rate\ and \gshear\ in a forward-modeling framework.
As a result, the count-rate-inferred and lensing-calibrated posterior of \Mfiveoo\ for each cluster is obtained.

The resulting mass of the eFEDS sample spans a range of $10^{13}\Msunh\lesssim\Mfiveoo\lesssim10^{15}\Msunh$ with a median value of $\Mfiveoo\approx10^{14}\Msunh$, of which the low-mass clusters ($\Mfiveoo\lesssim10^{14}\Msunh$) are mainly located at low redshift ($\redshift\lesssim0.4$). 
The characteristic uncertainty of the cluster mass, which is estimated as the dispersion of the mass posterior $P(\Mfiveoo)$, is at the level of $\approx27\percent$ for a cluster with $\Mfiveoo\approx10^{14}\Msunh$.
The mass estimates are not sensitive to the weak-lensing modeling of the cluster core; excluding the radial range of $0.2\Mpch\lesssim R \lesssim0.5\Mpch$ results in a negligible change ($\approx0.01$~dex) in the cluster mass.
Currently, the statistical uncertainty of \Aeta, the normalization of the X-ray-mass-proxy-to-mass-and-redshift relation, is constrained to be at the level of 
$\approx20\percent$.
If the weak-lensing mass calibration can be controlled with a subpercent accuracy, the uncertainty on \Aeta\ is expected to be constrained with a precision of $\approx1.8\percent$ to first order in the final \eROSITA\ survey, assuming that half of the clusters ($\approx100,000/2=50,000$) have weak-lensing follow-ups.
The mass calibration shows that, as expected, \eROSITA\ is able to reliably detect clusters down to much lower mass scales than any currently existing or planned SZ survey.

Leveraging the weak-lensing mass calibration, we further explore the relationships between cluster mass, redshift, and the follow-up X-ray observables, including the rest-frame soft-band (bolometric) luminosity \Lx\ (\Lb), the core-included temperature \Tx, the gas mass \Mg, and the mass proxy \Yx.
These X-ray observables can be well described by a power-law function of the cluster mass and redshift with log-normal intrinsic scatter at fixed mass.
We make use of the full combination of the data from both the \eROSITA\ and HSC surveys to significantly tighten the constraints on the X-ray observable relations, even for those observables, for example, \Tx, which cannot be estimated for any of the detected clusters with high signal-to-noise ratios.
Except for \Lx, the scaling relations of the follow-up X-ray observables show 
\begin{itemize}
\item a mass trend that is statistically consistent with the self-similar prediction at a level of 
$\lesssim1.7\sigma$,
\item no significant deviation from self-similarity at the level of 
$\lesssim1.7\sigma$
in terms of redshift scaling, and 
\item no clear sign of a redshift-dependent power-law index in the cluster mass.
\end{itemize}
Meanwhile, we find that the \Lx--\Mfiveoo--\redshift\ relation has a steeper mass trend than the self-similar prediction at the level of $\approx3\sigma$ but a redshift scaling that is in good agreement with self-similarity.
This suggests that the eFEDS clusters, which are dominated by low-mass clusters or galaxy groups, show X-ray scaling relations that are largely consistent with self-similar predictions in terms of mass and redshift scaling.
A generally steeper mass dependence of the luminosity clearly indicates that nongravitational processes play an important role in low-mass systems at the scale of galaxy groups, which is the main population studied in this work.
The energy input from nongravitational processes lowers the gas content within \Rfiveoo\ and hence decreases the X-ray luminosity; 
moreover, this effect is mass-dependent and is more prominent at a low-mass scale, resulting in a steeper slope seen in the luminosity-to-mass relation.
This picture is also in line with the \Mg--\Mfiveoo--\redshift\ relation of the eFEDS sample with a steeper mass trend than the self-similar prediction, despite the low significance due to large error bars.

We compare the mass and redshift trends of the X-ray scaling relations derived from the eFEDS sample with those from the previous work and find good consistency at a level of  $\lesssim2\sigma$.
This work extends the study of the X-ray scaling relations to low-mass clusters over a wide range of redshift using the \eROSITA-selected sample, which confers advantages over SZ-based selections.
It is interesting to mention the comparison with \cite{bulbul19}, the study of X-ray scaling relations based on the largest SZ-selected sample out to $\redshift\approx1.3$.
Our work probes a similar range of redshift but a mass scale that is much lower than that of these latter authors at the scale of massive clusters ($\Mfiveoo\approx4.5\times10^{14}\Msunh$).
A significantly steeper mass dependence than the self-similar prediction was found by \cite{bulbul19} for all the X-ray observables (\Lx, \Lb, \Tx, \Mg, and \Yx) at the level of $\gtrsim3\sigma$.
Despite broad consistency ($\approx2\sigma$) with their results, our eFEDS sample generally shows a mass scaling that is shallower than theirs and is statistically consistent with the self-similar predictions at $\lesssim2\sigma$. 
This may suggest a broken power law \citep{lebrun17} of the X-ray observables in the transition between groups and clusters, or a possible mass- and redshift-dependent systematic bias  in the cluster mass and/or X-ray observables between \cite{bulbul19} and this work.
Deep X-ray follow-ups of a large sample selected in the \eROSITA\ All-Sky Survey will shed light on this subject.

We find that the gas mass \Mg\ shows the smallest intrinsic scatter compared to other follow-up observables, suggesting that the ICM mass-based mass proxy may be most useful for clusters with $\approx100$ X-ray photons observed by large surveys.

This work clearly demonstrates the capability of the HSC weak-lensing data set in calibrating the mass of a sizable sample of galaxy clusters and groups detected by \eROSITA\ out to high redshift ($\redshift\approx1.3$).
This work not only presents the weak-lensing mass calibration of \eROSITA-detected clusters in synergy with a wide and deep lensing survey, but also lays the foundation for the cluster cosmology in a combination of the revolutionary \eROSITA\ sample and the state-of-the-art lensing data from the HSC.

%
%

\section*{Acknowledgments}
\label{sec:acknowledgments}

The authors thank the referee for constructive comments that lead to improvements of this paper.
I-Non Chiu thanks Hung-Hsu Chan and Chia-Ying Lin for useful discussions that lead to improvements of this paper. 
This work is supported in part by the national science foundation of China (Nos. 11833005,  11890691, 11890692, 11621303, 11890693, 11421303), 111 project No. B20019 and Shanghai Natural Science Foundation, grant No. 15ZR1446700, 19ZR1466800. 
This work made use of the Gravity Supercomputer at the Department of Astronomy, Shanghai Jiao Tong University.
This work is supported in part by the Ministry of Science and Technology of Taiwan (grant MOST 106-2628-M-001-003-MY3 and MOST
109-2112-M-001-018-MY3) and by Academia Sinica (grant AS-IA-107-M01).
This work made use of the computing resources in the National Center for High-Performance Computing (NCHC) in Taiwan.

This work is based on data from \eROSITA, the soft X-ray instrument aboard SRG, a joint Russian-German science mission supported by the Russian Space Agency (Roskosmos), in the interests of the Russian Academy of Sciences represented by its Space Research Institute (IKI), and the Deutsches Zentrum f{\"{u}}r Luft- und Raumfahrt (DLR). The SRG spacecraft was built by Lavochkin Association (NPOL) and its subcontractors, and is operated by NPOL with support from the Max Planck Institute for Extraterrestrial Physics (MPE).

The development and construction of the \eROSITA\ X-ray instrument was led by MPE, with contributions from the Dr. Karl Remeis Observatory Bamberg \& ECAP (FAU Erlangen-Nuernberg), the University of Hamburg Observatory, the Leibniz Institute for Astrophysics Potsdam (AIP), and the Institute for Astronomy and Astrophysics of the University of T{\"{u}}bingen, with the support of DLR and the Max Planck Society. The Argelander Institute for Astronomy of the University of Bonn and the Ludwig Maximilians Universit{\"{a}}t Munich also participated in the science preparation for \eROSITA.

The \eROSITA\ data shown here were processed using the \texttt{eSASS/NRTA} software system developed by the German \eROSITA\ consortium.

The Hyper Suprime-Cam (HSC) collaboration includes the astronomical communities of Japan and Taiwan, and Princeton University.  The HSC instrumentation and software were developed by the National Astronomical Observatory of Japan (NAOJ), the Kavli Institute for the Physics and Mathematics of the Universe (Kavli IPMU), the University of Tokyo, the High Energy Accelerator Research Organization (KEK), the Academia Sinica Institute for Astronomy and Astrophysics in Taiwan (ASIAA), and Princeton University.  Funding was contributed by the FIRST program from the Japanese Cabinet Office, the Ministry of Education, Culture, Sports, Science and Technology (MEXT), the Japan Society for the Promotion of Science (JSPS), Japan Science and Technology Agency  (JST), the Toray Science  Foundation, NAOJ, Kavli IPMU, KEK, ASIAA, and Princeton University.

This paper makes use of software developed for the Legacy Survey of Space and Time carried out by the Vera C. Rubin Observatory. We thank the LSST Project for making their code available as free software at \url{http://dm.lsst.org}.

This paper is based in part on data collected at the Subaru Telescope and retrieved from the HSC data archive system, which is operated by Subaru Telescope and Astronomy Data Center (ADC) at NAOJ. Data analysis was in part carried out with the cooperation of Center for Computational Astrophysics (CfCA), NAOJ.
We are honored and grateful for the opportunity of observing the Universe from Maunakea, which has the cultural, historical and natural significance in Hawaii.

The Pan-STARRS1 Surveys (PS1) and the PS1 public science archive have been made possible through contributions by the Institute for Astronomy, the University of Hawaii, the Pan-STARRS Project Office, the Max Planck Society and its participating institutes, the Max Planck Institute for Astronomy, Heidelberg, and the Max Planck Institute for Extraterrestrial Physics, Garching, The Johns Hopkins University, Durham University, the University of Edinburgh, the Queen's University Belfast, the Harvard-Smithsonian Center for Astrophysics, the Las Cumbres Observatory Global Telescope Network Incorporated, the National Central University of Taiwan, the Space Telescope Science Institute, the National Aeronautics and Space Administration under grant No. NNX08AR22G issued through the Planetary Science Division of the NASA Science Mission Directorate, the National Science Foundation grant No. AST-1238877, the University of Maryland, Eotvos Lorand University (ELTE), the Los Alamos National Laboratory, and the Gordon and Betty Moore Foundation.
This work was supported in part by World Premier International Research Center Initiative (WPI Initiative), MEXT, Japan, and JSPS KAKENHI Grant Numbers JP19KK0076, JP20H00181, and JP20H05856.

This work is possible because of the efforts in the LSST \citep{juric17,ivezic19} and PS1 \citep{chambers16, schlafly12, tonry12, magnier13}, and in the HSC \citep{aihara18a} developments including the deep imaging of the COSMOS field \citep{tanaka17}, the on-site quality-assurance system \citep{furusawa18}, the Hyper Suprime-Cam \citep{miyazaki15, miyazaki18, komiyama18}, the design of the filters \citep{kawanomoto18},  the data pipeline \citep{bosch18}, the design of bright-star masks \citep{coupon18}, the characterization of the photometry by the code \texttt{Synpipe} \citep{huang18}, the photometric redshift estimation \citep{tanaka18}, the shear calibration \citep{mandelbaum18}, and the public data releases \citep{aihara18b, aihara19}.

This work made use of the IPython package \citep{PER-GRA:2007}, \texttt{SciPy} \citep{virtanen_scipy}, \texttt{TOPCAT}, an interactive graphical viewer and editor for tabular data \citep{topcat1,topcat2}, \texttt{matplotlib}, a Python library for publication quality graphics \citep{Hunter:2007}, \texttt{Astropy}, a community-developed core Python package for Astronomy \citep{2013A&A...558A..33A}, \texttt{NumPy} \citep{van2011numpy}. 
This work made use of \texttt{Pathos} \citep{pathos} in parallel computing. 
This work made use of \citet{bocquet16b} and \cite{hinton2016} for producing the corner plots for the parameter constraints.
The code \texttt{Colossus} \citep{diemer18} is heavily used to calculate cosmology-dependent quantities in this work.

%
%

\section*{Data Availability}
\label{sec:data_availability}

The \eROSITA\ X-ray detection parameters are publicly available in \cite{liu21} and the redshifts are provided in \cite{klein21}.
The X-ray observables of the eFEDS sample will be made publicly available in \cite{bahar21}.
The HSC weak-lensing shape catalog will be made publicly available following the plan of the HSC Public Data Releases.
The other data products underlying this article will be shared upon a reasonable request to the corresponding author.

%
%

\bibliographystyle{aa}
\bibliography{literature}

\begin{thebibliography}{220}
\expandafter\ifx\csname natexlab\endcsname\relax\def\natexlab#1{#1}\fi

\bibitem[{{Adami} {et~al.}(2018){Adami}, {Giles}, {Koulouridis}, {Pacaud},
  {Caretta}, {Pierre}, {Eckert}, {Ramos-Ceja}, {Gastaldello}, {Fotopoulou},
  {Guglielmo}, {Lidman}, {Sadibekova}, {Iovino}, {Maughan}, {Chiappetti},
  {Alis}, {Altieri}, {Baldry}, {Bottini}, {Birkinshaw}, {Bremer}, {Brown},
  {Cucciati}, {Driver}, {Elmer}, {Ettori}, {Evrard}, {Faccioli}, {Granett},
  {Grootes}, {Guzzo}, {Hopkins}, {Horellou}, {Lef{\`e}vre}, {Liske}, {Malek},
  {Marulli}, {Maurogordato}, {Owers}, {Paltani}, {Poggianti}, {Polletta},
  {Plionis}, {Pollo}, {Pompei}, {Ponman}, {Rapetti}, {Ricci}, {Robotham},
  {Tuffs}, {Tasca}, {Valtchanov}, {Vergani}, {Wagner}, {Willis}, \& {XXL
  Consortium}}]{adami18}
{Adami}, C., {Giles}, P., {Koulouridis}, E., {et~al.} 2018, \aap, 620, A5

\bibitem[{{Aihara} {et~al.}(2019){Aihara}, {AlSayyad}, {Ando}, {Armstrong},
  {Bosch}, {Egami}, {Furusawa}, {Furusawa}, {Goulding}, {Harikane}, {Hikage},
  {Ho}, {Hsieh}, {Huang}, {Ikeda}, {Imanishi}, {Ito}, {Iwata}, {Jaelani},
  {Kakuma}, {Kawana}, {Kikuta}, {Kobayashi}, {Koike}, {Komiyama}, {Li},
  {Liang}, {Lin}, {Luo}, {Lupton}, {Lust}, {MacArthur}, {Matsuoka}, {Mineo},
  {Miyatake}, {Miyazaki}, {More}, {Murata}, {Namiki}, {Nishizawa}, {Oguri},
  {Okabe}, {Okamoto}, {Okura}, {Ono}, {Onodera}, {Onoue}, {Osato}, {Ouchi},
  {Shibuya}, {Strauss}, {Sugiyama}, {Suto}, {Takada}, {Takagi}, {Takata},
  {Takita}, {Tanaka}, {Terai}, {Toba}, {Uchiyama}, {Utsumi}, {Wang}, {Wang}, \&
  {Yamada}}]{aihara19}
{Aihara}, H., {AlSayyad}, Y., {Ando}, M., {et~al.} 2019, \pasj, 71, 114

\bibitem[{{Aihara} {et~al.}(2018{\natexlab{a}}){Aihara}, {Arimoto},
  {Armstrong}, {Arnouts}, {Bahcall}, {Bickerton}, {Bosch}, {Bundy}, {Capak},
  {Chan}, {Chiba}, {Coupon}, {Egami}, {Enoki}, {Finet}, {Fujimori}, {Fujimoto},
  {Furusawa}, {Furusawa}, {Goto}, {Goulding}, {Greco}, {Greene}, {Gunn},
  {Hamana}, {Harikane}, {Hashimoto}, {Hattori}, {Hayashi}, {Hayashi},
  {He{\l}miniak}, {Higuchi}, {Hikage}, {Ho}, {Hsieh}, {Huang}, {Huang},
  {Ikeda}, {Imanishi}, {Inoue}, {Iwasawa}, {Iwata}, {Jaelani}, {Jian},
  {Kamata}, {Karoji}, {Kashikawa}, {Katayama}, {Kawanomoto}, {Kayo}, {Koda},
  {Koike}, {Kojima}, {Komiyama}, {Konno}, {Koshida}, {Koyama}, {Kusakabe},
  {Leauthaud}, {Lee}, {Lin}, {Lin}, {Lupton}, {Mandelbaum}, {Matsuoka},
  {Medezinski}, {Mineo}, {Miyama}, {Miyatake}, {Miyazaki}, {Momose}, {More},
  {More}, {Moritani}, {Moriya}, {Morokuma}, {Mukae}, {Murata}, {Murayama},
  {Nagao}, {Nakata}, {Niida}, {Niikura}, {Nishizawa}, {Obuchi}, {Oguri},
  {Oishi}, {Okabe}, {Okamoto}, {Okura}, {Ono}, {Onodera}, {Onoue}, {Osato},
  {Ouchi}, {Price}, {Pyo}, {Sako}, {Sawicki}, {Shibuya}, {Shimasaku},
  {Shimono}, {Shirasaki}, {Silverman}, {Simet}, {Speagle}, {Spergel},
  {Strauss}, {Sugahara}, {Sugiyama}, {Suto}, {Suyu}, {Suzuki}, {Tait},
  {Takada}, {Takata}, {Tamura}, {Tanaka}, {Tanaka}, {Tanaka}, {Tanaka},
  {Terai}, {Terashima}, {Toba}, {Tominaga}, {Toshikawa}, {Turner}, {Uchida},
  {Uchiyama}, {Umetsu}, {Uraguchi}, {Urata}, {Usuda}, {Utsumi}, {Wang}, {Wang},
  {Wong}, {Yabe}, {Yamada}, {Yamanoi}, {Yasuda}, {Yeh}, {Yonehara}, \&
  {Yuma}}]{aihara18a}
{Aihara}, H., {Arimoto}, N., {Armstrong}, R., {et~al.} 2018{\natexlab{a}},
  \pasj, 70, S4

\bibitem[{{Aihara} {et~al.}(2018{\natexlab{b}}){Aihara}, {Armstrong},
  {Bickerton}, {Bosch}, {Coupon}, {Furusawa}, {Hayashi}, {Ikeda}, {Kamata},
  {Karoji}, {Kawanomoto}, {Koike}, {Komiyama}, {Lang}, {Lupton}, {Mineo},
  {Miyatake}, {Miyazaki}, {Morokuma}, {Obuchi}, {Oishi}, {Okura}, {Price},
  {Takata}, {Tanaka}, {Tanaka}, {Tanaka}, {Uchida}, {Uraguchi}, {Utsumi},
  {Wang}, {Yamada}, {Yamanoi}, {Yasuda}, {Arimoto}, {Chiba}, {Finet},
  {Fujimori}, {Fujimoto}, {Furusawa}, {Goto}, {Goulding}, {Gunn}, {Harikane},
  {Hattori}, {Hayashi}, {He{\l}miniak}, {Higuchi}, {Hikage}, {Ho}, {Hsieh},
  {Huang}, {Huang}, {Imanishi}, {Iwata}, {Jaelani}, {Jian}, {Kashikawa},
  {Katayama}, {Kojima}, {Konno}, {Koshida}, {Kusakabe}, {Leauthaud}, {Lee},
  {Lin}, {Lin}, {Mandelbaum}, {Matsuoka}, {Medezinski}, {Miyama}, {Momose},
  {More}, {More}, {Mukae}, {Murata}, {Murayama}, {Nagao}, {Nakata}, {Niida},
  {Niikura}, {Nishizawa}, {Oguri}, {Okabe}, {Ono}, {Onodera}, {Onoue}, {Ouchi},
  {Pyo}, {Shibuya}, {Shimasaku}, {Simet}, {Speagle}, {Spergel}, {Strauss},
  {Sugahara}, {Sugiyama}, {Suto}, {Suzuki}, {Tait}, {Takada}, {Terai}, {Toba},
  {Turner}, {Uchiyama}, {Umetsu}, {Urata}, {Usuda}, {Yeh}, \&
  {Yuma}}]{aihara18b}
{Aihara}, H., {Armstrong}, R., {Bickerton}, S., {et~al.} 2018{\natexlab{b}},
  \pasj, 70, S8

\bibitem[{{Alam} {et~al.}(2015){Alam}, {Albareti}, {Allende Prieto}, {Anders},
  {Anderson}, {Anderton}, {Andrews}, {Armengaud}, {Aubourg}, {Bailey}, {Basu},
  {Bautista}, {Beaton}, {Beers}, {Bender}, {Berlind}, {Beutler}, {Bhardwaj},
  {Bird}, {Bizyaev}, {Blake}, {Blanton}, {Blomqvist}, {Bochanski}, {Bolton},
  {Bovy}, {Shelden Bradley}, {Brandt}, {Brauer}, {Brinkmann}, {Brown},
  {Brownstein}, {Burden}, {Burtin}, {Busca}, {Cai}, {Capozzi}, {Carnero
  Rosell}, {Carr}, {Carrera}, {Chambers}, {Chaplin}, {Chen}, {Chiappini},
  {Chojnowski}, {Chuang}, {Clerc}, {Comparat}, {Covey}, {Croft}, {Cuesta},
  {Cunha}, {da Costa}, {Da Rio}, {Davenport}, {Dawson}, {De Lee}, {Delubac},
  {Deshpande}, {Dhital}, {Dutra-Ferreira}, {Dwelly}, {Ealet}, {Ebelke},
  {Edmondson}, {Eisenstein}, {Ellsworth}, {Elsworth}, {Epstein}, {Eracleous},
  {Escoffier}, {Esposito}, {Evans}, {Fan}, {Fern{\'a}ndez-Alvar}, {Feuillet},
  {Filiz Ak}, {Finley}, {Finoguenov}, {Flaherty}, {Fleming}, {Font-Ribera},
  {Foster}, {Frinchaboy}, {Galbraith-Frew}, {Garc{\'\i}a},
  {Garc{\'\i}a-Hern{\'a}ndez}, {Garc{\'\i}a P{\'e}rez}, {Gaulme}, {Ge},
  {G{\'e}nova-Santos}, {Georgakakis}, {Ghezzi}, {Gillespie}, {Girardi},
  {Goddard}, {Gontcho}, {Gonz{\'a}lez Hern{\'a}ndez}, {Grebel}, {Green},
  {Grieb}, {Grieves}, {Gunn}, {Guo}, {Harding}, {Hasselquist}, {Hawley},
  {Hayden}, {Hearty}, {Hekker}, {Ho}, {Hogg}, {Holley-Bockelmann}, {Holtzman},
  {Honscheid}, {Huber}, {Huehnerhoff}, {Ivans}, {Jiang}, {Johnson},
  {Kinemuchi}, {Kirkby}, {Kitaura}, {Klaene}, {Knapp}, {Kneib}, {Koenig},
  {Lam}, {Lan}, {Lang}, {Laurent}, {Le Goff}, {Leauthaud}, {Lee}, {Lee},
  {Licquia}, {Liu}, {Long}, {L{\'o}pez-Corredoira}, {Lorenzo-Oliveira},
  {Lucatello}, {Lundgren}, {Lupton}, {Mack}, {Mahadevan}, {Maia}, {Majewski},
  {Malanushenko}, {Malanushenko}, {Manchado}, {Manera}, {Mao}, {Maraston},
  {Marchwinski}, {Margala}, {Martell}, {Martig}, {Masters}, {Mathur},
  {McBride}, {McGehee}, {McGreer}, {McMahon}, {M{\'e}nard}, {Menzel},
  {Merloni}, {M{\'e}sz{\'a}ros}, {Miller}, {Miralda-Escud{\'e}}, {Miyatake},
  {Montero-Dorta}, {More}, {Morganson}, {Morice-Atkinson}, {Morrison},
  {Mosser}, {Muna}, {Myers}, {Nand ra}, {Newman}, {Neyrinck}, {Nguyen},
  {Nichol}, {Nidever}, {Noterdaeme}, {Nuza}, {O'Connell}, {O'Connell},
  {O'Connell}, {Ogando}, {Olmstead}, {Oravetz}, {Oravetz}, {Osumi}, {Owen},
  {Padgett}, {Padmanabhan}, {Paegert}, {Palanque-Delabrouille}, {Pan},
  {Parejko}, {P{\^a}ris}, {Park}, {Pattarakijwanich}, {Pellejero-Ibanez},
  {Pepper}, {Percival}, {P{\'e}rez-Fournon}, {Ṕrez-Ra`fols}, {Petitjean},
  {Pieri}, {Pinsonneault}, {Porto de Mello}, {Prada}, {Prakash},
  {Price-Whelan}, {Protopapas}, {Raddick}, {Rahman}, {Reid}, {Rich}, {Rix},
  {Robin}, {Rockosi}, {Rodrigues}, {Rodr{\'\i}guez-Torres}, {Roe}, {Ross},
  {Ross}, {Rossi}, {Ruan}, {Rubi{\~n}o-Mart{\'\i}n}, {Rykoff},
  {Salazar-Albornoz}, {Salvato}, {Samushia}, {S{\'a}nchez}, {Santiago},
  {Sayres}, {Schiavon}, {Schlegel}, {Schmidt}, {Schneider}, {Schultheis},
  {Schwope}, {Sc{\'o}ccola}, {Scott}, {Sellgren}, {Seo}, {Serenelli}, {Shane},
  {Shen}, {Shetrone}, {Shu}, {Silva Aguirre}, {Sivarani}, {Skrutskie},
  {Slosar}, {Smith}, {Sobreira}, {Souto}, {Stassun}, {Steinmetz}, {Stello},
  {Strauss}, {Streblyanska}, {Suzuki}, {Swanson}, {Tan}, {Tayar}, {Terrien},
  {Thakar}, {Thomas}, {Thomas}, {Thompson}, {Tinker}, {Tojeiro}, {Troup},
  {Vargas-Maga{\~n}a}, {Vazquez}, {Verde}, {Viel}, {Vogt}, {Wake}, {Wang},
  {Weaver}, {Weinberg}, {Weiner}, {White}, {Wilson}, {Wisniewski},
  {Wood-Vasey}, {Ye`che}, {York}, {Zakamska}, {Zamora}, {Zasowski}, {Zehavi},
  {Zhao}, {Zheng}, {Zhou}, {Zhou}, {Zou}, \& {Zhu}}]{alam15}
{Alam}, S., {Albareti}, F.~D., {Allende Prieto}, C., {et~al.} 2015, \apjs, 219,
  12

\bibitem[{Andersson {et~al.}(2011)Andersson, Benson, Ade, Aird, Armstrong,
  Bautz, Bleem, Brodwin, Carlstrom, Chang, Crawford, Crites, de~Haan, Desai,
  Dobbs, Dudley, Foley, Forman, Garmire, George, Gladders, Halverson, High,
  Holder, Holzapfel, Hrubes, Jones, Joy, Keisler, Knox, Lee, Leitch, Lueker,
  Marrone, McMahon, Mehl, Meyer, Mohr, Montroy, Murray, Padin, Plagge, Pryke,
  Reichardt, Rest, Ruel, Ruhl, Schaffer, Shaw, Shirokoff, Song, Spieler,
  Stalder, Staniszewski, Stark, Stubbs, Vanderlinde, Vieira, Vikhlinin,
  Williamson, Yang, Zahn, \& Zenteno}]{andersson11}
Andersson, K., Benson, B., Ade, P., {et~al.} 2011, \apj, 738, 48

\bibitem[{Applegate {et~al.}(2014)Applegate, von~der Linden, Kelly, Allen,
  Allen, Burchat, Burke, Ebeling, Mantz, \& Morris}]{applegate14}
Applegate, D., von~der Linden, A., Kelly, P., {et~al.} 2014, \mnras, 439, 48

\bibitem[{{Arnaud}(1996)}]{arnaud96}
{Arnaud}, K.~A. 1996, in Astronomical Society of the Pacific Conference Series,
  Vol. 101, Astronomical Data Analysis Software and Systems V, ed. G.~H.
  {Jacoby} \& J.~{Barnes}, 17

\bibitem[{Arnaud {et~al.}(2005)Arnaud, Pointecouteau, \& Pratt}]{arnaud05}
Arnaud, M., Pointecouteau, E., \& Pratt, G. 2005, \aap, 441, 893

\bibitem[{Arnaud {et~al.}(2007)Arnaud, Pointecouteau, \& Pratt}]{arnaud07}
Arnaud, M., Pointecouteau, E., \& Pratt, G. 2007, \aap, 474, L37

\bibitem[{Arnaud {et~al.}(2010)Arnaud, Pratt, Piffaretti, B{\"{o}}hringer,
  Croston, \& Pointecouteau}]{arnaud10}
Arnaud, M., Pratt, G., Piffaretti, R., {et~al.} 2010, \aap, 517, A92+

\bibitem[{{Astropy Collaboration} {et~al.}(2013){Astropy Collaboration},
  {Robitaille}, {Tollerud}, {Greenfield}, {Droettboom}, {Bray}, {Aldcroft},
  {Davis}, {Ginsburg}, {Price-Whelan}, {Kerzendorf}, {Conley}, {Crighton},
  {Barbary}, {Muna}, {Ferguson}, {Grollier}, {Parikh}, {Nair}, {Unther},
  {Deil}, {Woillez}, {Conseil}, {Kramer}, {Turner}, {Singer}, {Fox}, {Weaver},
  {Zabalza}, {Edwards}, {Azalee Bostroem}, {Burke}, {Casey}, {Crawford},
  {Dencheva}, {Ely}, {Jenness}, {Labrie}, {Lim}, {Pierfederici}, {Pontzen},
  {Ptak}, {Refsdal}, {Servillat}, \& {Streicher}}]{2013A&A...558A..33A}
{Astropy Collaboration}, {Robitaille}, T.~P., {Tollerud}, E.~J., {et~al.} 2013,
  \aap, 558, A33

\bibitem[{{Bahar} {et~al.}(in preparation){Bahar}, {Name}, \& et~al.}]{bahar21}
{Bahar}, N., {Name}, N., \& et~al. in preparation

\bibitem[{{Barnes} {et~al.}(2017){Barnes}, {Kay}, {Bah{\'e}}, {Dalla Vecchia},
  {McCarthy}, {Schaye}, {Bower}, {Jenkins}, {Thomas}, {Schaller}, {Crain},
  {Theuns}, \& {White}}]{barnes17}
{Barnes}, D.~J., {Kay}, S.~T., {Bah{\'e}}, Y.~M., {et~al.} 2017, \mnras, 471,
  1088

\bibitem[{Bartelmann \& Schneider(2001)}]{bartelmann01}
Bartelmann, M. \& Schneider, P. 2001, \physrep, 340, 291

\bibitem[{Benson {et~al.}(2013)Benson, de~Haan, Dudley, Reichardt, Aird,
  Andersson, Armstrong, Ashby, Bautz, Bayliss, Bazin, Bleem, Brodwin,
  Carlstrom, Chang, Cho, Clocchiatti, Crawford, Crites, Desai, Dobbs, Foley,
  Forman, George, Gladders, Gonzalez, Halverson, Harrington, High, Holder,
  Holzapfel, Hoover, Hrubes, Jones, Joy, Keisler, Knox, Lee, Leitch, Liu,
  Lueker, Luong-Van, Mantz, Marrone, McDonald, McMahon, Mehl, Meyer, Mocanu,
  Mohr, Montroy, Murray, Natoli, Padin, Plagge, Pryke, Rest, Ruel, Ruhl,
  Saliwanchik, Saro, Sayre, Schaffer, Shaw, Shirokoff, Song, Spieler, Stalder,
  Staniszewski, Stark, Story, Stubbs, Suhada, van Engelen, Vanderlinde, Vieira,
  Vikhlinin, Williamson, Zahn, \& Zenteno}]{benson13}
Benson, B., de~Haan, T., Dudley, J., {et~al.} 2013, \apj, 763, 147

\bibitem[{{Biesiadzinski} {et~al.}(2012){Biesiadzinski}, {McMahon}, {Miller},
  {Nord}, \& {Shaw}}]{biesiadzinski12}
{Biesiadzinski}, T., {McMahon}, J., {Miller}, C.~J., {Nord}, B., \& {Shaw}, L.
  2012, \apj, 757, 1

\bibitem[{Bleem {et~al.}(2015)Bleem, Stalder, de~Haan, Aird, Allen, Applegate,
  Ashby, Bautz, Bayliss, Benson, Bocquet, Brodwin, Carlstrom, Chang, Chiu, Cho,
  Clocchiatti, Crawford, Crites, Desai, Dietrich, Dobbs, Foley, Forman, George,
  Gladders, Gonzalez, Halverson, Hennig, Hoekstra, Holder, Holzapfel, Hrubes,
  Jones, Keisler, Knox, Lee, Leitch, Liu, Lueker, Luong-Van, Mantz, Marrone,
  McDonald, McMahon, Meyer, Mocanu, Mohr, Murray, Padin, Pryke, Reichardt,
  Rest, Ruel, Ruhl, Saliwanchik, Saro, Sayre, Schaffer, Schrabback, Shirokoff,
  Song, Spieler, Stanford, Staniszewski, Stark, Story, Stubbs, Vanderlinde,
  Vieira, Vikhlinin, Williamson, Zahn, \& Zenteno}]{bleem15}
Bleem, L., Stalder, B., de~Haan, T., {et~al.} 2015, \apjs, 216, 27

\bibitem[{{Bleem} {et~al.}(2020){Bleem}, {Bocquet}, {Stalder}, {Gladders},
  {Ade}, {Allen}, {Anderson}, {Annis}, {Ashby}, {Austermann}, {Avila}, {Avva},
  {Bayliss}, {Beall}, {Bechtol}, {Bender}, {Benson}, {Bertin}, {Bianchini},
  {Blake}, {Brodwin}, {Brooks}, {Buckley-Geer}, {Burke}, {Carlstrom}, {Rosell},
  {Carrasco Kind}, {Carretero}, {Chang}, {Chiang}, {Citron}, {Moran},
  {Costanzi}, {Crawford}, {Crites}, {da Costa}, {de Haan}, {De Vicente},
  {Desai}, {Diehl}, {Dietrich}, {Dobbs}, {Eifler}, {Everett}, {Flaugher},
  {Floyd}, {Frieman}, {Gallicchio}, {Garc{\'\i}a-Bellido}, {George}, {Gerdes},
  {Gilbert}, {Gruen}, {Gruendl}, {Gschwend}, {Gupta}, {Gutierrez}, {Halverson},
  {Harrington}, {Henning}, {Heymans}, {Holder}, {Hollowood}, {Holzapfel},
  {Honscheid}, {Hrubes}, {Huang}, {Hubmayr}, {Irwin}, {James}, {Jeltema},
  {Joudaki}, {Khullar}, {Klein}, {Knox}, {Kuropatkin}, {Lee}, {Li}, {Lidman},
  {Lowitz}, {MacCrann}, {Mahler}, {Maia}, {Marshall}, {McDonald}, {McMahon},
  {Melchior}, {Menanteau}, {Meyer}, {Miquel}, {Mocanu}, {Mohr}, {Montgomery},
  {Nadolski}, {Natoli}, {Nibarger}, {Noble}, {Novosad}, {Padin}, {Palmese},
  {Parkinson}, {Patil}, {Paz-Chinch{\'o}n}, {Plazas}, {Pryke}, {Ramachandra},
  {Reichardt}, {Remolina Gonz{\'a}lez}, {Romer}, {Roodman}, {Ruhl}, {Rykoff},
  {Saliwanchik}, {Sanchez}, {Saro}, {Sayre}, {Schaffer}, {Schrabback},
  {Serrano}, {Sharon}, {Sievers}, {Smecher}, {Smith}, {Soares-Santos}, {Stark},
  {Story}, {Suchyta}, {Tarle}, {Tucker}, {Vanderlinde}, {Veach}, {Vieira},
  {Wang}, {Weller}, {Whitehorn}, {Wu}, {Yefremenko}, \& {Zhang}}]{bleem20}
{Bleem}, L.~E., {Bocquet}, S., {Stalder}, B., {et~al.} 2020, \apjs, 247, 25

\bibitem[{{Bocquet} \& {Carter}(2016)}]{bocquet16b}
{Bocquet}, S. \& {Carter}, F.~W. 2016, 1

\bibitem[{{Bocquet} {et~al.}(2019){Bocquet}, {Dietrich}, {Schrabback}, {Bleem},
  {Klein}, {Allen}, {Applegate}, {Ashby}, {Bautz}, {Bayliss}, {Benson},
  {Brodwin}, {Bulbul}, {Canning}, {Capasso}, {Carlstrom}, {Chang}, {Chiu},
  {Cho}, {Clocchiatti}, {Crawford}, {Crites}, {de Haan}, {Desai}, {Dobbs},
  {Foley}, {Forman}, {Garmire}, {George}, {Gladders}, {Gonzalez}, {Grandis},
  {Gupta}, {Halverson}, {Hlavacek-Larrondo}, {Hoekstra}, {Holder}, {Holzapfel},
  {Hou}, {Hrubes}, {Huang}, {Jones}, {Khullar}, {Knox}, {Kraft}, {Lee}, {von
  der Linden}, {Luong-Van}, {Mantz}, {Marrone}, {McDonald}, {McMahon}, {Meyer},
  {Mocanu}, {Mohr}, {Morris}, {Padin}, {Patil}, {Pryke}, {Rapetti},
  {Reichardt}, {Rest}, {Ruhl}, {Saliwanchik}, {Saro}, {Sayre}, {Schaffer},
  {Shirokoff}, {Stalder}, {Stanford}, {Staniszewski}, {Stark}, {Story},
  {Strazzullo}, {Stubbs}, {Vanderlinde}, {Vieira}, {Vikhlinin}, {Williamson},
  \& {Zenteno}}]{bocquet19}
{Bocquet}, S., {Dietrich}, J.~P., {Schrabback}, T., {et~al.} 2019, \apj, 878,
  55

\bibitem[{Bocquet {et~al.}(2016)Bocquet, Saro, Dolag, \& Mohr}]{bocquet16}
Bocquet, S., Saro, A., Dolag, K., \& Mohr, J. 2016, \mnras, 456, 2361

\bibitem[{Bocquet {et~al.}(2015)Bocquet, Saro, Mohr, Aird, Ashby, Bautz,
  Bayliss, Bazin, Benson, Bleem, Brodwin, Carlstrom, Chang, Chiu, Cho,
  Clocchiatti, Crawford, Crites, Desai, de~Haan, Dietrich, Dobbs, Foley,
  Forman, Gangkofner, George, Gladders, Gonzalez, Halverson, Hennig,
  Hlavacek-Larrondo, Holder, Holzapfel, Hrubes, Jones, Keisler, Knox, Lee,
  Leitch, Liu, Lueker, Luong-Van, Marrone, McDonald, McMahon, Meyer, Mocanu,
  Murray, Padin, Pryke, Reichardt, Rest, Ruel, Ruhl, Saliwanchik, Sayre,
  Schaffer, Shirokoff, Spieler, Stalder, Stanford, Staniszewski, Stark, Story,
  Stubbs, Vanderlinde, Vieira, Vikhlinin, Williamson, Zahn, \&
  Zenteno}]{bocquet15}
Bocquet, S., Saro, A., Mohr, J., {et~al.} 2015, \apj, 799, 214

\bibitem[{{B{\"o}hringer} {et~al.}(2012){B{\"o}hringer}, {Dolag}, \&
  {Chon}}]{boehringer12}
{B{\"o}hringer}, H., {Dolag}, K., \& {Chon}, G. 2012, \aap, 539, A120

\bibitem[{B{\"{o}}hringer {et~al.}(2004)B{\"{o}}hringer, Schuecker, Guzzo,
  Collins, Voges, Cruddace, Ortiz-Gil, Chincarini, {De Grandi}, Edge,
  MacGillivray, Neumann, Schindler, \& Shaver}]{bohringer04}
B{\"{o}}hringer, H., Schuecker, P., Guzzo, L., {et~al.} 2004, \aap, 425, 367

\bibitem[{{B{\"o}hringer} {et~al.}(2001){B{\"o}hringer}, {Schuecker}, {Guzzo},
  {Collins}, {Voges}, {Schindler}, {Neumann}, {Cruddace}, {De Grandi},
  {Chincarini}, {Edge}, {MacGillivray}, \& {Shaver}}]{bohringer01}
{B{\"o}hringer}, H., {Schuecker}, P., {Guzzo}, L., {et~al.} 2001, \aap, 369,
  826

\bibitem[{{Boller} {et~al.}(2016){Boller}, {Freyberg}, {Tr{\"u}mper}, {Haberl},
  {Voges}, \& {Nandra}}]{boller16}
{Boller}, T., {Freyberg}, M.~J., {Tr{\"u}mper}, J., {et~al.} 2016, \aap, 588,
  A103

\bibitem[{Bonamente {et~al.}(2008)Bonamente, Joy, LaRoque, Carlstrom, Nagai, \&
  Marrone}]{bonamente08}
Bonamente, M., Joy, M., LaRoque, S., {et~al.} 2008, \apj, 675, 106

\bibitem[{{Bonnett} {et~al.}(2016){Bonnett}, {Troxel}, {Hartley}, {Amara},
  {Leistedt}, {Becker}, {Bernstein}, {Bridle}, {Bruderer}, {Busha}, {Carrasco
  Kind}, {Childress}, {Castander}, {Chang}, {Crocce}, {Davis}, {Eifler},
  {Frieman}, {Gangkofner}, {Gaztanaga}, {Glazebrook}, {Gruen}, {Kacprzak},
  {King}, {Kwan}, {Lahav}, {Lewis}, {Lidman}, {Lin}, {MacCrann}, {Miquel},
  {O'Neill}, {Palmese}, {Peiris}, {Refregier}, {Rozo}, {Rykoff}, {Sadeh},
  {S{\'a}nchez}, {Sheldon}, {Uddin}, {Wechsler}, {Zuntz}, {Abbott}, {Abdalla},
  {Allam}, {Armstrong}, {Banerji}, {Bauer}, {Benoit-L{\'e}vy}, {Bertin},
  {Brooks}, {Buckley-Geer}, {Burke}, {Capozzi}, {Carnero Rosell}, {Carretero},
  {Cunha}, {D'Andrea}, {da Costa}, {DePoy}, {Desai}, {Diehl}, {Dietrich},
  {Doel}, {Fausti Neto}, {Fernandez}, {Flaugher}, {Fosalba}, {Gerdes},
  {Gruendl}, {Honscheid}, {Jain}, {James}, {Jarvis}, {Kim}, {Kuehn},
  {Kuropatkin}, {Li}, {Lima}, {Maia}, {March}, {Marshall}, {Martini},
  {Melchior}, {Miller}, {Neilsen}, {Nichol}, {Nord}, {Ogando}, {Plazas},
  {Reil}, {Romer}, {Roodman}, {Sako}, {Sanchez}, {Santiago}, {Smith},
  {Soares-Santos}, {Sobreira}, {Suchyta}, {Swanson}, {Tarle}, {Thaler},
  {Thomas}, {Vikram}, {Walker}, \& {Dark Energy Survey
  Collaboration}}]{bonnett16}
{Bonnett}, C., {Troxel}, M.~A., {Hartley}, W., {et~al.} 2016, \prd, 94, 042005

\bibitem[{{Borm} {et~al.}(2014){Borm}, {Reiprich}, {Mohammed}, \&
  {Lovisari}}]{borm14}
{Borm}, K., {Reiprich}, T.~H., {Mohammed}, I., \& {Lovisari}, L. 2014, \aap,
  567, A65

\bibitem[{{Bosch} {et~al.}(2018){Bosch}, {Armstrong}, {Bickerton}, {Furusawa},
  {Ikeda}, {Koike}, {Lupton}, {Mineo}, {Price}, {Takata}, {Tanaka}, {Yasuda},
  {AlSayyad}, {Becker}, {Coulton}, {Coupon}, {Garmilla}, {Huang}, {Krughoff},
  {Lang}, {Leauthaud}, {Lim}, {Lust}, {MacArthur}, {Mandelbaum}, {Miyatake},
  {Miyazaki}, {Murata}, {More}, {Okura}, {Owen}, {Swinbank}, {Strauss},
  {Yamada}, \& {Yamanoi}}]{bosch18}
{Bosch}, J., {Armstrong}, R., {Bickerton}, S., {et~al.} 2018, \pasj, 70, S5

\bibitem[{{Brunner} {et~al.}(2021){Brunner}, {Liu}, {Lamer}, {Georgakakis},
  {Merloni}, {Brusa}, {Bulbul}, {Dennerl}, {Friedrich}, {Liu}, {Maitra},
  {Nandra}, {Ramos-Ceja}, {Sanders}, {Stewart}, {Boller}, {Buchner}, {Clerc},
  {Comparat}, {Dwelly}, {Eckert}, {Finoguenov}, {Freyberg}, {Ghirardini},
  {Gueguen}, {Haberl}, {Kreykenbohm}, {Krumpe}, {Osterhage}, {Pacaud},
  {Predehl}, {Reiprich}, {Robrade}, {Salvato}, {Santangelo}, {Schrabback},
  {Schwope}, \& {Wilms}}]{brunner21}
{Brunner}, H., {Liu}, T., {Lamer}, G., {et~al.} 2021, arXiv e-prints,
  arXiv:2106.14517

\bibitem[{{Bulbul} {et~al.}(2019){Bulbul}, {Chiu}, {Mohr}, {McDonald},
  {Benson}, {Bautz}, {Bayliss}, {Bleem}, {Brodwin}, {Bocquet}, {Capasso},
  {Dietrich}, {Forman}, {Hlavacek-Larrondo}, {Holzapfel}, {Khullar}, {Klein},
  {Kraft}, {Miller}, {Reichardt}, {Saro}, {Sharon}, {Stalder}, {Schrabback}, \&
  {Stanford}}]{bulbul19}
{Bulbul}, E., {Chiu}, I.~N., {Mohr}, J.~J., {et~al.} 2019, \apj, 871, 50

\bibitem[{{Castro} {et~al.}(2021){Castro}, {Borgani}, {Dolag}, {Marra},
  {Quartin}, {Saro}, \& {Sefusatti}}]{castro21}
{Castro}, T., {Borgani}, S., {Dolag}, K., {et~al.} 2021, \mnras, 500, 2316

\bibitem[{{Cavaliere} \& {Fusco-Femiano}(1976)}]{cavaliere76}
{Cavaliere}, A. \& {Fusco-Femiano}, R. 1976, \aap, 500, 95

\bibitem[{{Chambers} {et~al.}(2016){Chambers}, {Magnier}, {Metcalfe},
  {Flewelling}, {Huber}, {Waters}, {Denneau}, {Draper}, {Farrow}, {Finkbeiner},
  {Holmberg}, {Koppenhoefer}, {Price}, {Rest}, {Saglia}, {Schlafly}, {Smartt},
  {Sweeney}, {Wainscoat}, {Burgett}, {Chastel}, {Grav}, {Heasley}, {Hodapp},
  {Jedicke}, {Kaiser}, {Kudritzki}, {Luppino}, {Lupton}, {Monet}, {Morgan},
  {Onaka}, {Shiao}, {Stubbs}, {Tonry}, {White}, {Ba{\~n}ados}, {Bell},
  {Bender}, {Bernard}, {Boegner}, {Boffi}, {Botticella}, {Calamida},
  {Casertano}, {Chen}, {Chen}, {Cole}, {Deacon}, {Frenk}, {Fitzsimmons},
  {Gezari}, {Gibbs}, {Goessl}, {Goggia}, {Gourgue}, {Goldman}, {Grant},
  {Grebel}, {Hambly}, {Hasinger}, {Heavens}, {Heckman}, {Henderson}, {Henning},
  {Holman}, {Hopp}, {Ip}, {Isani}, {Jackson}, {Keyes}, {Koekemoer}, {Kotak},
  {Le}, {Liska}, {Long}, {Lucey}, {Liu}, {Martin}, {Masci}, {McLean}, {Mindel},
  {Misra}, {Morganson}, {Murphy}, {Obaika}, {Narayan}, {Nieto-Santisteban},
  {Norberg}, {Peacock}, {Pier}, {Postman}, {Primak}, {Rae}, {Rai}, {Riess},
  {Riffeser}, {Rix}, {R{\"o}ser}, {Russel}, {Rutz}, {Schilbach}, {Schultz},
  {Scolnic}, {Strolger}, {Szalay}, {Seitz}, {Small}, {Smith}, {Soderblom},
  {Taylor}, {Thomson}, {Taylor}, {Thakar}, {Thiel}, {Thilker}, {Unger},
  {Urata}, {Valenti}, {Wagner}, {Walder}, {Walter}, {Watters}, {Werner},
  {Wood-Vasey}, \& {Wyse}}]{chambers16}
{Chambers}, K.~C., {Magnier}, E.~A., {Metcalfe}, N., {et~al.} 2016, arXiv
  e-prints, arXiv:1612.05560

\bibitem[{{Chen} {et~al.}(2020){Chen}, {Oguri}, {Lin}, \& {Miyazaki}}]{chen20}
{Chen}, K.-F., {Oguri}, M., {Lin}, Y.-T., \& {Miyazaki}, S. 2020, \apj, 891,
  139

\bibitem[{Chiu {et~al.}(2016{\natexlab{a}})Chiu, Mohr, McDonald, Bocquet,
  Ashby, Bayliss, Benson, Bleem, Brodwin, Desai, Dietrich, Forman, Gangkofner,
  Gonzalez, Hennig, Liu, Reichardt, Saro, Stalder, Stanford, Song, Schrabback,
  {\v{S}}uhada, Strazzullo, \& Zenteno}]{chiu16a}
Chiu, I., Mohr, J., McDonald, M., {et~al.} 2016{\natexlab{a}}, \mnras, 455, 258

\bibitem[{{Chiu} {et~al.}(2018{\natexlab{a}}){Chiu}, {Mohr}, {McDonald},
  {Bocquet}, {Desai}, {Klein}, {Israel}, {Ashby}, {Stanford}, {Benson},
  {Brodwin}, {Abbott}, {Abdalla}, {Allam}, {Annis}, {Bayliss},
  {Benoit-L{\'e}vy}, {Bertin}, {Bleem}, {Brooks}, {Buckley-Geer}, {Bulbul},
  {Capasso}, {Carlstrom}, {Rosell}, {Carretero}, {Castander}, {Cunha},
  {D'Andrea}, {da Costa}, {Davis}, {Diehl}, {Dietrich}, {Doel},
  {Drlica-Wagner}, {Eifler}, {Evrard}, {Flaugher}, {Garc{\'{\i}}a-Bellido},
  {Garmire}, {Gaztanaga}, {Gerdes}, {Gonzalez}, {Gruen}, {Gruendl}, {Gschwend},
  {Gupta}, {Gutierrez}, {Hlavacek-L}, {Honscheid}, {James}, {Jeltema}, {Kraft},
  {Krause}, {Kuehn}, {Kuhlmann}, {Kuropatkin}, {Lahav}, {Lima}, {Maia},
  {Marshall}, {Melchior}, {Menanteau}, {Miquel}, {Murray}, {Nord}, {Ogando},
  {Plazas}, {Rapetti}, {Reichardt}, {Romer}, {Roodman}, {Sanchez}, {Saro},
  {Scarpine}, {Schindler}, {Schubnell}, {Sharon}, {Smith}, {Smith},
  {Soares-Santos}, {Sobreira}, {Stalder}, {Stern}, {Strazzullo}, {Suchyta},
  {Swanson}, {Tarle}, {Vikram}, {Walker}, {Weller}, \& {Zhang}}]{chiu18a}
{Chiu}, I., {Mohr}, J.~J., {McDonald}, M., {et~al.} 2018{\natexlab{a}}, \mnras,
  478, 3072

\bibitem[{Chiu {et~al.}(2016{\natexlab{b}})Chiu, Saro, Mohr, Desai, Bocquet,
  Capasso, Gangkofner, Gupta, \& Liu}]{chiu16c}
Chiu, I., Saro, A., Mohr, J., {et~al.} 2016{\natexlab{b}}, \mnras, 458, 379

\bibitem[{{Chiu} {et~al.}(2020{\natexlab{a}}){Chiu}, {Okumura}, {Oguri},
  {Agrawal}, {Umetsu}, \& {Lin}}]{chiu20b}
{Chiu}, I.~N., {Okumura}, T., {Oguri}, M., {et~al.} 2020{\natexlab{a}}, \mnras,
  498, 2030

\bibitem[{{Chiu} {et~al.}(2020{\natexlab{b}}){Chiu}, {Umetsu}, {Murata},
  {Medezinski}, \& {Oguri}}]{chiu20}
{Chiu}, I.~N., {Umetsu}, K., {Murata}, R., {Medezinski}, E., \& {Oguri}, M.
  2020{\natexlab{b}}, \mnras, 495, 428

\bibitem[{{Chiu} {et~al.}(2018{\natexlab{b}}){Chiu}, {Umetsu}, {Sereno},
  {Ettori}, {Meneghetti}, {Merten}, {Sayers}, \& {Zitrin}}]{chiu18b}
{Chiu}, I.-N., {Umetsu}, K., {Sereno}, M., {et~al.} 2018{\natexlab{b}}, \apj,
  860, 126

\bibitem[{{Clowe} {et~al.}(2004){Clowe}, {De Lucia}, \& {King}}]{clowe04}
{Clowe}, D., {De Lucia}, G., \& {King}, L. 2004, \mnras, 350, 1038

\bibitem[{{Coil} {et~al.}(2011){Coil}, {Blanton}, {Burles}, {Cool},
  {Eisenstein}, {Moustakas}, {Wong}, {Zhu}, {Aird}, {Bernstein}, {Bolton}, \&
  {Hogg}}]{coil11}
{Coil}, A.~L., {Blanton}, M.~R., {Burles}, S.~M., {et~al.} 2011, \apj, 741, 8

\bibitem[{{Comparat} {et~al.}(2020){Comparat}, {Eckert}, {Finoguenov},
  {Schmidt}, {Sanders}, {Nagai}, {Lau}, {K�fer}, {Pacaud}, {Clerc},
  {Reiprich}, {Bulbul}, {Chitham}, {Chiang}, {Ghirardini}, {Gonzalez-Perez},
  {Gozaliasl}, {Fitzpatrick}, {Klypin}, {Merloni}, {Nandra}, {Liu}, {Prada},
  {Ramos-Ceja}, {Salvato}, {Seppi}, {Tempel}, \& {Yepes}}]{comparat20}
{Comparat}, J., {Eckert}, D., {Finoguenov}, A., {et~al.} 2020, The Open Journal
  of Astrophysics, 3, 13

\bibitem[{{Comparat} {et~al.}(2019){Comparat}, {Merloni}, {Salvato}, {Nandra},
  {Boller}, {Georgakakis}, {Finoguenov}, {Dwelly}, {Buchner}, {Del Moro},
  {Clerc}, {Wang}, {Zhao}, {Prada}, {Yepes}, {Brusa}, {Krumpe}, \&
  {Liu}}]{comparat19}
{Comparat}, J., {Merloni}, A., {Salvato}, M., {et~al.} 2019, \mnras, 487, 2005

\bibitem[{{Cool} {et~al.}(2013){Cool}, {Moustakas}, {Blanton}, {Burles},
  {Coil}, {Eisenstein}, {Wong}, {Zhu}, {Aird}, {Bernstein}, {Bolton}, {Hogg},
  \& {Mendez}}]{cool13}
{Cool}, R.~J., {Moustakas}, J., {Blanton}, M.~R., {et~al.} 2013, \apj, 767, 118

\bibitem[{{Corless} \& {King}(2007)}]{corless07}
{Corless}, V.~L. \& {King}, L.~J. 2007, \mnras, 380, 149

\bibitem[{{Costanzi} {et~al.}(2019{\natexlab{a}}){Costanzi}, {Rozo}, {Rykoff},
  {Farahi}, {Jeltema}, {Evrard}, {Mantz}, {Gruen}, {Mandelbaum}, {DeRose},
  {McClintock}, {Varga}, {Zhang}, {Weller}, {Wechsler}, \&
  {Aguena}}]{costanzi19}
{Costanzi}, M., {Rozo}, E., {Rykoff}, E.~S., {et~al.} 2019{\natexlab{a}},
  \mnras, 482, 490

\bibitem[{{Costanzi} {et~al.}(2019{\natexlab{b}}){Costanzi}, {Rozo}, {Simet},
  {Zhang}, {Evrard}, {Mantz}, {Rykoff}, {Jeltema}, {Gruen}, {Allen},
  {McClintock}, {Romer}, {von der Linden}, {Farahi}, {DeRose}, {Varga},
  {Weller}, {Giles}, {Hollowood}, {Bhargava}, {Bermeo-Hernandez}, {Chen},
  {Abbott}, {Abdalla}, {Avila}, {Bechtol}, {Brooks}, {Buckley-Geer}, {Burke},
  {Rosell}, {Kind}, {Carretero}, {Crocce}, {Cunha}, {da Costa}, {Davis}, {De
  Vicente}, {Diehl}, {Dietrich}, {Doel}, {Eifler}, {Estrada}, {Flaugher},
  {Fosalba}, {Frieman}, {Garc{\'\i}a-Bellido}, {Gaztanaga}, {Gerdes},
  {Giannantonio}, {Gruendl}, {Gschwend}, {Gutierrez}, {Hartley}, {Honscheid},
  {Hoyle}, {James}, {Krause}, {Kuehn}, {Kuropatkin}, {Lima}, {Lin}, {Maia},
  {March}, {Marshall}, {Martini}, {Menanteau}, {Miller}, {Miquel}, {Mohr},
  {Ogando}, {Plazas}, {Roodman}, {Sanchez}, {Scarpine}, {Schindler},
  {Schubnell}, {Serrano}, {Sevilla-Noarbe}, {Sheldon}, {Smith},
  {Soares-Santos}, {Sobreira}, {Suchyta}, {Swanson}, {Tarle}, {Thomas}, \&
  {Wechsler}}]{costanzi18}
{Costanzi}, M., {Rozo}, E., {Simet}, M., {et~al.} 2019{\natexlab{b}}, \mnras,
  488, 4779

\bibitem[{{Costanzi} {et~al.}(2021){Costanzi}, {Saro}, {Bocquet}, {Abbott},
  {Aguena}, {Allam}, {Amara}, {Annis}, {Avila}, {Bacon}, {Benson}, {Bhargava},
  {Brooks}, {Buckley-Geer}, {Burke}, {Carnero Rosell}, {Carrasco Kind},
  {Carretero}, {Choi}, {da Costa}, {Pereira}, {De Vicente}, {Desai}, {Diehl},
  {Dietrich}, {Doel}, {Eifler}, {Everett}, {Ferrero}, {Fert{\'e}}, {Flaugher},
  {Fosalba}, {Frieman}, {Garc{\'\i}a-Bellido}, {Gaztanaga}, {Gerdes},
  {Giannantonio}, {Giles}, {Grandis}, {Gruen}, {Gruendl}, {Gupta}, {Gutierrez},
  {Hartley}, {Hinton}, {Hollowood}, {Honscheid}, {James}, {Jeltema}, {Krause},
  {Kuehn}, {Kuropatkin}, {Lahav}, {Lima}, {MacCrann}, {Maia}, {Marshall},
  {Menanteau}, {Miquel}, {Mohr}, {Morgan}, {Myles}, {Ogando}, {Palmese},
  {Paz-Chinch{\'o}n}, {Plazas}, {Rapetti}, {Reichardt}, {Romer}, {Roodman},
  {Ruppin}, {Salvati}, {Samuroff}, {Sanchez}, {Scarpine}, {Serrano},
  {Sevilla-Noarbe}, {Singh}, {Smith}, {Soares-Santos}, {Stark}, {Suchyta},
  {Swanson}, {Tarle}, {Thomas}, {To}, {Tucker}, {Varga}, {Wechsler}, {Zhang},
  {DES}, \& {SPT Collaborations}}]{costanzi21}
{Costanzi}, M., {Saro}, A., {Bocquet}, S., {et~al.} 2021, \prd, 103, 043522

\bibitem[{{Coupon} {et~al.}(2018){Coupon}, {Czakon}, {Bosch}, {Komiyama},
  {Medezinski}, {Miyazaki}, \& {Oguri}}]{coupon18}
{Coupon}, J., {Czakon}, N., {Bosch}, J., {et~al.} 2018, \pasj, 70, S7

\bibitem[{de~Haan {et~al.}(2016)de~Haan, Benson, Bleem, Allen, Applegate,
  Ashby, Bautz, Bayliss, Bocquet, Brodwin, Carlstrom, Chang, Chiu, Cho,
  Clocchiatti, Crawford, Crites, Desai, Dietrich, Dobbs, Doucouliagos, Foley,
  Forman, Garmire, George, Gladders, Gonzalez, Gupta, Halverson,
  Hlavacek-Larrondo, Hoekstra, Holder, Holzapfel, Hou, Hrubes, Huang, Jones,
  Keisler, Knox, Lee, Leitch, von~der Linden, Luong-Van, Mantz, Marrone,
  McDonald, McMahon, Meyer, Mocanu, Mohr, Murray, Padin, Pryke, Rapetti,
  Reichardt, Rest, Ruel, Ruhl, Saliwanchik, Saro, Sayre, Schaffer, Schrabback,
  Shirokoff, Song, Spieler, Stalder, Stanford, Staniszewski, Stark, Story,
  Stubbs, Vanderlinde, Vieira, Vikhlinin, Williamson, \& Zenteno}]{deHaan16}
de~Haan, T., Benson, B., Bleem, L., {et~al.} 2016, \apj, 832, 95

\bibitem[{{DES Collaboration} {et~al.}(2020){DES Collaboration}, {Abbott},
  {Aguena}, {Alarcon}, {Allam}, {Allen}, {Annis}, {Avila}, {Bacon}, {Bermeo},
  {Bernstein}, {Bertin}, {Bhargava}, {Bocquet}, {Brooks}, {Brout},
  {Buckley-Geer}, {Burke}, {Carnero Rosell}, {Carrasco Kind}, {Carretero},
  {Castander}, {Cawthon}, {Chang}, {Chen}, {Choi}, {Costanzi}, {Crocce}, {da
  Costa}, {Davis}, {De Vicente}, {DeRose}, {Desai}, {Diehl}, {Dietrich},
  {Dodelson}, {Doel}, {Drlica-Wagner}, {Eckert}, {Eifler}, {Elvin-Poole},
  {Estrada}, {Everett}, {Evrard}, {Farahi}, {Ferrero}, {Flaugher}, {Fosalba},
  {Frieman}, {Garcia-Bellido}, {Gatti}, {Gaztanaga}, {Gerdes}, {Giannantonio},
  {Giles}, {Grandis}, {Gruen}, {Gruendl}, {Gschwend}, {Gutierrez}, {Hartley},
  {Hinton}, {Hollowood}, {Honscheid}, {Hoyle}, {Huterer}, {James}, {Jarvis},
  {Jeltema}, {Johnson}, {Kent}, {Krause}, {Kron}, {Kuehn}, {Kuropatkin},
  {Lahav}, {Li}, {Lidman}, {Lima}, {Lin}, {MacCrann}, {Maia}, {Mantz},
  {Marshall}, {Martini}, {Mayers}, {Melchior}, {Mena}, {Menanteau}, {Miquel},
  {Mohr}, {Nichol}, {Nord}, {Ogando}, {Palmese}, {Paz-Chinchon}, {Plazas
  Malag{\'o}n}, {Prat}, {Rau}, {Romer}, {Roodman}, {Rooney}, {Rozo}, {Rykoff},
  {Sako}, {Samuroff}, {Sanchez}, {Saro}, {Scarpine}, {Schubnell}, {Scolnic},
  {Serrano}, {Sevilla}, {Sheldon}, {Smith}, {Suchyta}, {Swanson}, {Tarle},
  {Thomas}, {To}, {Troxel}, {Tucker}, {Varga}, {von der Linden}, {Walker},
  {Wechsler}, {Weller}, {Wilkinson}, {Wu}, {Yanny}, {Zhang}, \&
  {Zuntz}}]{desclustercosmology20}
{DES Collaboration}, {Abbott}, T., {Aguena}, M., {et~al.} 2020, arXiv e-prints,
  arXiv:2002.11124

\bibitem[{{Dey} {et~al.}(2019){Dey}, {Schlegel}, {Lang}, {Blum}, {Burleigh},
  {Fan}, {Findlay}, {Finkbeiner}, {Herrera}, {Juneau}, {Landriau}, {Levi},
  {McGreer}, {Meisner}, {Myers}, {Moustakas}, {Nugent}, {Patej}, {Schlafly},
  {Walker}, {Valdes}, {Weaver}, {Y{\`e}che}, {Zou}, {Zhou}, {Abareshi},
  {Abbott}, {Abolfathi}, {Aguilera}, {Alam}, {Allen}, {Alvarez}, {Annis},
  {Ansarinejad}, {Aubert}, {Beechert}, {Bell}, {BenZvi}, {Beutler}, {Bielby},
  {Bolton}, {Brice{\~n}o}, {Buckley-Geer}, {Butler}, {Calamida}, {Carlberg},
  {Carter}, {Casas}, {Castander}, {Choi}, {Comparat}, {Cukanovaite}, {Delubac},
  {DeVries}, {Dey}, {Dhungana}, {Dickinson}, {Ding}, {Donaldson}, {Duan},
  {Duckworth}, {Eftekharzadeh}, {Eisenstein}, {Etourneau}, {Fagrelius},
  {Farihi}, {Fitzpatrick}, {Font-Ribera}, {Fulmer}, {G{\"a}nsicke},
  {Gaztanaga}, {George}, {Gerdes}, {Gontcho}, {Gorgoni}, {Green}, {Guy},
  {Harmer}, {Hernandez}, {Honscheid}, {Huang}, {James}, {Jannuzi}, {Jiang},
  {Joyce}, {Karcher}, {Karkar}, {Kehoe}, {Kneib}, {Kueter-Young}, {Lan},
  {Lauer}, {Le Guillou}, {Le Van Suu}, {Lee}, {Lesser}, {Perreault Levasseur},
  {Li}, {Mann}, {Marshall}, {Mart{\'\i}nez-V{\'a}zquez}, {Martini}, {du Mas des
  Bourboux}, {McManus}, {Meier}, {M{\'e}nard}, {Metcalfe},
  {Mu{\~n}oz-Guti{\'e}rrez}, {Najita}, {Napier}, {Narayan}, {Newman}, {Nie},
  {Nord}, {Norman}, {Olsen}, {Paat}, {Palanque-Delabrouille}, {Peng},
  {Poppett}, {Poremba}, {Prakash}, {Rabinowitz}, {Raichoor}, {Rezaie},
  {Robertson}, {Roe}, {Ross}, {Ross}, {Rudnick}, {Safonova}, {Saha},
  {S{\'a}nchez}, {Savary}, {Schweiker}, {Scott}, {Seo}, {Shan}, {Silva},
  {Slepian}, {Soto}, {Sprayberry}, {Staten}, {Stillman}, {Stupak}, {Summers},
  {Sien Tie}, {Tirado}, {Vargas-Maga{\~n}a}, {Vivas}, {Wechsler}, {Williams},
  {Yang}, {Yang}, {Yapici}, {Zaritsky}, {Zenteno}, {Zhang}, {Zhang}, {Zhou}, \&
  {Zhou}}]{dey19}
{Dey}, A., {Schlegel}, D.~J., {Lang}, D., {et~al.} 2019, \aj, 157, 168

\bibitem[{{Diemer}(2018)}]{diemer18}
{Diemer}, B. 2018, \apjs, 239, 35

\bibitem[{{Diemer} \& {Kravtsov}(2015)}]{diemer15}
{Diemer}, B. \& {Kravtsov}, A.~V. 2015, \apj, 799, 108

\bibitem[{{Dietrich} {et~al.}(2019){Dietrich}, {Bocquet}, {Schrabback},
  {Applegate}, {Hoekstra}, {Grandis}, {Mohr}, {Allen}, {Bayliss}, {Benson},
  {Bleem}, {Brodwin}, {Bulbul}, {Capasso}, {Chiu}, {Crawford}, {Gonzalez}, {de
  Haan}, {Klein}, {von der Linden}, {Mantz}, {Marrone}, {McDonald},
  {Raghunathan}, {Rapetti}, {Reichardt}, {Saro}, {Stalder}, {Stark}, {Stern},
  \& {Stubbs}}]{dietrich19}
{Dietrich}, J.~P., {Bocquet}, S., {Schrabback}, T., {et~al.} 2019, \mnras, 483,
  2871

\bibitem[{{Dietrich} {et~al.}(2012){Dietrich}, {B{\"o}hnert}, {Lombardi},
  {Hilbert}, \& {Hartlap}}]{dietrich12}
{Dietrich}, J.~P., {B{\"o}hnert}, A., {Lombardi}, M., {Hilbert}, S., \&
  {Hartlap}, J. 2012, \mnras, 419, 3547

\bibitem[{{Dolag} {et~al.}(in preparation){Dolag}, {Name}, \& et~al.}]{dolag21}
{Dolag}, N., {Name}, N., \& et~al. in preparation

\bibitem[{Eckert {et~al.}(2016)Eckert, Ettori, Coupon, Gastaldello, Pierre,
  Melin, {Le Brun}, McCarthy, Adami, Chiappetti, Faccioli, Giles, Lavoie,
  Lef{\`{e}}vre, Lieu, Mantz, Maughan, McGee, Pacaud, Paltani, Sadibekova,
  Smith, \& Ziparo}]{eckert16}
Eckert, D., Ettori, S., Coupon, J., {et~al.} 2016, \aap, 592, A12

\bibitem[{Eddington(1913)}]{eddington13}
Eddington, A. 1913, \mnras, 73, 359

\bibitem[{{Evrard} {et~al.}(2014){Evrard}, {Arnault}, {Huterer}, \&
  {Farahi}}]{evrard14}
{Evrard}, A.~E., {Arnault}, P., {Huterer}, D., \& {Farahi}, A. 2014, \mnras,
  441, 3562

\bibitem[{Finoguenov {et~al.}(2001)Finoguenov, Reiprich, \&
  B{\"{o}}hringer}]{finoguenov01}
Finoguenov, A., Reiprich, T.~H., \& B{\"{o}}hringer, H. 2001, \aap, 368, 749

\bibitem[{{Foreman-Mackey} {et~al.}(2019){Foreman-Mackey}, {Farr}, {Sinha},
  {Archibald}, {Hogg}, {Sanders}, {Zuntz}, {Williams}, {Nelson}, {de
  Val-Borro}, {Erhardt}, {Pashchenko}, \& {Pla}}]{foreman19}
{Foreman-Mackey}, D., {Farr}, W., {Sinha}, M., {et~al.} 2019, The Journal of
  Open Source Software, 4, 1864

\bibitem[{Foreman-Mackey {et~al.}(2013)Foreman-Mackey, Hogg, Lang, \&
  Goodman}]{foreman13}
Foreman-Mackey, D., Hogg, D., Lang, D., \& Goodman, J. 2013, \pasp, 125, 306

\bibitem[{{Furusawa} {et~al.}(2018){Furusawa}, {Koike}, {Takata}, {Okura},
  {Miyatake}, {Lupton}, {Bickerton}, {Price}, {Bosch}, {Yasuda}, {Mineo},
  {Yamada}, {Miyazaki}, {Nakata}, {Koshida}, {Komiyama}, {Utsumi},
  {Kawanomoto}, {Jeschke}, {Noumaru}, {Schubert}, {Iwata}, {Finet},
  {Fujiyoshi}, {Tajitsu}, {Terai}, \& {Lee}}]{furusawa18}
{Furusawa}, H., {Koike}, M., {Takata}, T., {et~al.} 2018, \pasj, 70, S3

\bibitem[{{Gatti} {et~al.}(2020){Gatti}, {Giannini}, {Bernstein}, {Alarcon},
  {Myles}, {Amon}, {Cawthon}, {Troxel}, {DeRose}, {Everett}, {Ross}, {Rykoff},
  {Elvin-Poole}, {Cordero}, {Harrison}, {Sanchez}, {Prat}, {Gruen}, {Lin},
  {Crocce}, {Rozo}, {Abbott}, {Aguena}, {Allam}, {Annis}, {Avila}, {Bacon},
  {Bertin}, {Brooks}, {Burke}, {Carnero Rosell}, {Carrasco Kind}, {Carretero},
  {Castander}, {Choi}, {Conselice}, {Costanzi}, {Crocce}, {da Costa},
  {Pereira}, {Dawson}, {Desai}, {Diehl}, {Eckert}, {Eifler}, {Evrard},
  {Ferrero}, {Flaugher}, {Fosalba}, {Frieman}, {Garcia-Bellido}, {Gaztanaga},
  {Giannantonio}, {Gruendl}, {Gschwend}, {Hinton}, {Hollowood}, {Honscheid},
  {Hoyle}, {Huterer}, {James}, {Kuehn}, {Kuropatkin}, {Lahav}, {Lima},
  {MacCrann}, {Maia}, {March}, {Marshall}, {Melchior}, {Menanteau}, {Miquel},
  {Mohr}, {Morgan}, {Ogando}, {Palmese}, {Paz-Chinchon}, {Percival}, {Plazas},
  {Rodriguez-Monroy}, {Roodman}, {Rossi}, {Samuroff}, {Sanchez}, {Scarpine},
  {Secco}, {Serrano}, {Sevilla-Noarbe}, {Smith}, {Soares-Santos}, {Suchyta},
  {Swanson}, {Tarle}, {Thomas}, {To}, {Varga}, {Weller}, \&
  {Wilkinson}}]{gatti20}
{Gatti}, M., {Giannini}, G., {Bernstein}, G.~M., {et~al.} 2020, arXiv e-prints,
  arXiv:2012.08569

\bibitem[{{Ghirardini} {et~al.}(2021){Ghirardini}, {Bulbul}, {Hoang}, {Klein},
  {Okabe}, {Biffi}, {Br{\"u}ggen}, {Ramos-Ceja}, {Comparat}, {Oguri},
  {Shimwell}, {Basu}, {Bonafede}, {Botteon}, {Brunetti}, {Cassano}, {de
  Gasperin}, {Dennerl}, {Gatuzz}, {Gastaldello}, {Intema}, {Merloni}, {Nandra},
  {Pacaud}, {Predehl}, {Reiprich}, {Robrade}, {R{\"o}ttgering}, {Sanders}, {van
  Weeren}, \& {Williams}}]{ghirardini21}
{Ghirardini}, V., {Bulbul}, E., {Hoang}, D.~N., {et~al.} 2021, \aap, 647, A4

\bibitem[{{Giles} {et~al.}(2016){Giles}, {Maughan}, {Pacaud}, {Lieu}, {Clerc},
  {Pierre}, {Adami}, {Chiappetti}, {D{\'e}mocl{\'e}s}, {Ettori}, {Le
  F{\'e}vre}, {Ponman}, {Sadibekova}, {Smith}, {Willis}, \& {Ziparo}}]{giles16}
{Giles}, P.~A., {Maughan}, B.~J., {Pacaud}, F., {et~al.} 2016, \aap, 592, A3

\bibitem[{{Giodini} {et~al.}(2013){Giodini}, {Lovisari}, {Pointecouteau},
  {Ettori}, {Reiprich}, \& {Hoekstra}}]{giodini13}
{Giodini}, S., {Lovisari}, L., {Pointecouteau}, E., {et~al.} 2013, \ssr, 177,
  247

\bibitem[{{Grandis} {et~al.}(2021){Grandis}, {Bocquet}, {Mohr}, {Klein}, \&
  {Dolag}}]{grandis21}
{Grandis}, S., {Bocquet}, S., {Mohr}, J.~J., {Klein}, M., \& {Dolag}, K. 2021,
  \mnras [\eprint[arXiv]{2103.16212}]

\bibitem[{{Grandis} {et~al.}(2020){Grandis}, {Klein}, {Mohr}, {Bocquet},
  {Paulus}, {Abbott}, {Aguena}, {Allam}, {Annis}, {Benson}, {Bertin},
  {Bhargava}, {Brooks}, {Burke}, {Carnero Rosell}, {Carrasco Kind},
  {Carretero}, {Capasso}, {Costanzi}, {da Costa}, {De Vicente}, {Desai},
  {Dietrich}, {Doel}, {Eifler}, {Evrard}, {Flaugher}, {Fosalba}, {Frieman},
  {Garc{\'\i}a-Bellido}, {Gaztanaga}, {Gerdes}, {Gruen}, {Gruendl}, {Gschwend},
  {Gutierrez}, {Hartley}, {Hinton}, {Hollowood}, {Honscheid}, {James},
  {Jeltema}, {Kuehn}, {Kuropatkin}, {Lima}, {Maia}, {Marshall}, {Melchior},
  {Menanteau}, {Miquel}, {Ogando}, {Palmese}, {Paz-Chinch{\'o}n}, {Plazas},
  {Romer}, {Roodman}, {Sanchez}, {Saro}, {Scarpine}, {Schubnell}, {Serrano},
  {Sheldon}, {Smith}, {Stark}, {Suchyta}, {Swanson}, {Tarle}, {Thomas},
  {Tucker}, {Varga}, {Weller}, \& {Wilkinson}}]{grandis20}
{Grandis}, S., {Klein}, M., {Mohr}, J.~J., {et~al.} 2020, \mnras, 498, 771

\bibitem[{{Grandis} {et~al.}(2019){Grandis}, {Mohr}, {Dietrich}, {Bocquet},
  {Saro}, {Klein}, {Paulus}, \& {Capasso}}]{grandis19}
{Grandis}, S., {Mohr}, J.~J., {Dietrich}, J.~P., {et~al.} 2019, \mnras, 488,
  2041

\bibitem[{{Gruen} \& {Brimioulle}(2017)}]{gruen17}
{Gruen}, D. \& {Brimioulle}, F. 2017, \mnras, 468, 769

\bibitem[{{Gruen} {et~al.}(2014){Gruen}, {Seitz}, {Brimioulle}, {Kosyra},
  {Koppenhoefer}, {Lee}, {Bender}, {Riffeser}, {Eichner}, {Weidinger}, \&
  {Bierschenk}}]{gruen14}
{Gruen}, D., {Seitz}, S., {Brimioulle}, F., {et~al.} 2014, \mnras, 442, 1507

\bibitem[{{Hamana} {et~al.}(2020){Hamana}, {Shirasaki}, {Miyazaki}, {Hikage},
  {Oguri}, {More}, {Armstrong}, {Leauthaud}, {Mandelbaum}, {Miyatake},
  {Nishizawa}, {Simet}, {Takada}, {Aihara}, {Bosch}, {Komiyama}, {Lupton},
  {Murayama}, {Strauss}, \& {Tanaka}}]{hamana20}
{Hamana}, T., {Shirasaki}, M., {Miyazaki}, S., {et~al.} 2020, \pasj, 72, 16

\bibitem[{{Hasinger} {et~al.}(2018){Hasinger}, {Capak}, {Salvato}, {Barger},
  {Cowie}, {Faisst}, {Hemmati}, {Kakazu}, {Kartaltepe}, {Masters}, {Mobasher},
  {Nayyeri}, {Sanders}, {Scoville}, {Suh}, {Steinhardt}, \&
  {Yang}}]{hasinger18}
{Hasinger}, G., {Capak}, P., {Salvato}, M., {et~al.} 2018, \apj, 858, 77

\bibitem[{{Henden} {et~al.}(2019){Henden}, {Puchwein}, \& {Sijacki}}]{henden19}
{Henden}, N.~A., {Puchwein}, E., \& {Sijacki}, D. 2019, \mnras, 489, 2439

\bibitem[{{Hikage} {et~al.}(2019){Hikage}, {Oguri}, {Hamana}, {More},
  {Mandelbaum}, {Takada}, {K{\"o}hlinger}, {Miyatake}, {Nishizawa}, {Aihara},
  {Armstrong}, {Bosch}, {Coupon}, {Ducout}, {Ho}, {Hsieh}, {Komiyama},
  {Lanusse}, {Leauthaud}, {Lupton}, {Medezinski}, {Mineo}, {Miyama},
  {Miyazaki}, {Murata}, {Murayama}, {Shirasaki}, {Sif{\'o}n}, {Simet},
  {Speagle}, {Spergel}, {Strauss}, {Sugiyama}, {Tanaka}, {Utsumi}, {Wang}, \&
  {Yamada}}]{hikage19}
{Hikage}, C., {Oguri}, M., {Hamana}, T., {et~al.} 2019, \pasj, 71, 43

\bibitem[{{Hildebrandt} {et~al.}(2020){Hildebrandt}, {K{\"o}hlinger}, {van den
  Busch}, {Joachimi}, {Heymans}, {Kannawadi}, {Wright}, {Asgari}, {Blake},
  {Hoekstra}, {Joudaki}, {Kuijken}, {Miller}, {Morrison}, {Tr{\"o}ster},
  {Amon}, {Archidiacono}, {Brieden}, {Choi}, {de Jong}, {Erben}, {Giblin},
  {Mead}, {Peacock}, {Radovich}, {Schneider}, {Sif{\'o}n}, \&
  {Tewes}}]{hildebrandt20}
{Hildebrandt}, H., {K{\"o}hlinger}, F., {van den Busch}, J.~L., {et~al.} 2020,
  \aap, 633, A69

\bibitem[{{Hilton} {et~al.}(2021){Hilton}, {Sif{\'o}n}, {Naess},
  {Madhavacheril}, {Oguri}, {Rozo}, {Rykoff}, {Abbott}, {Adhikari}, {Aguena},
  {Aiola}, {Allam}, {Amodeo}, {Amon}, {Annis}, {Ansarinejad}, {Aros-Bunster},
  {Austermann}, {Avila}, {Bacon}, {Battaglia}, {Beall}, {Becker}, {Bernstein},
  {Bertin}, {Bhandarkar}, {Bhargava}, {Bond}, {Brooks}, {Burke}, {Calabrese},
  {Carrasco Kind}, {Carretero}, {Choi}, {Choi}, {Conselice}, {da Costa},
  {Costanzi}, {Crichton}, {Crowley}, {D{\"u}nner}, {Denison}, {Devlin},
  {Dicker}, {Diehl}, {Dietrich}, {Doel}, {Duff}, {Duivenvoorden}, {Dunkley},
  {Everett}, {Ferraro}, {Ferrero}, {Fert{\'e}}, {Flaugher}, {Frieman},
  {Gallardo}, {Garc{\'\i}a-Bellido}, {Gaztanaga}, {Gerdes}, {Giles}, {Golec},
  {Gralla}, {Grandis}, {Gruen}, {Gruendl}, {Gschwend}, {Gutierrez}, {Han},
  {Hartley}, {Hasselfield}, {Hill}, {Hilton}, {Hincks}, {Hinton}, {Ho},
  {Honscheid}, {Hoyle}, {Hubmayr}, {Huffenberger}, {Hughes}, {Jaelani}, {Jain},
  {James}, {Jeltema}, {Kent}, {Knowles}, {Koopman}, {Kuehn}, {Lahav}, {Lima},
  {Lin}, {Lokken}, {Loubser}, {MacCrann}, {Maia}, {Marriage}, {Martin},
  {McMahon}, {Melchior}, {Menanteau}, {Miquel}, {Miyatake}, {Moodley},
  {Morgan}, {Mroczkowski}, {Nati}, {Newburgh}, {Niemack}, {Nishizawa},
  {Ogando}, {Orlowski-Scherer}, {Page}, {Palmese}, {Partridge},
  {Paz-Chinch{\'o}n}, {Phakathi}, {Plazas}, {Robertson}, {Romer}, {Carnero
  Rosell}, {Salatino}, {Sanchez}, {Schaan}, {Schillaci}, {Sehgal}, {Serrano},
  {Shin}, {Simon}, {Smith}, {Soares-Santos}, {Spergel}, {Staggs}, {Storer},
  {Suchyta}, {Swanson}, {Tarle}, {Thomas}, {To}, {Trac}, {Ullom}, {Vale}, {Van
  Lanen}, {Vavagiakis}, {De Vicente}, {Wilkinson}, {Wollack}, {Xu}, \&
  {Zhang}}]{hilton21}
{Hilton}, M., {Sif{\'o}n}, C., {Naess}, S., {et~al.} 2021, \apjs, 253, 3

\bibitem[{{Hinton}(2016)}]{hinton2016}
{Hinton}, S.~R. 2016, The Journal of Open Source Software, 1, 00045

\bibitem[{{Hoekstra}(2003)}]{hoekstra03}
{Hoekstra}, H. 2003, \mnras, 339, 1155

\bibitem[{Hoekstra {et~al.}(2013)Hoekstra, Bartelmann, Dahle, Israel, Limousin,
  \& Meneghetti}]{hoekstra13}
Hoekstra, H., Bartelmann, M., Dahle, H., {et~al.} 2013, \ssr, 177, 75

\bibitem[{Hoekstra {et~al.}(2015)Hoekstra, Herbonnet, Muzzin, Babul, Mahdavi,
  Viola, \& Cacciato}]{hoekstra15}
Hoekstra, H., Herbonnet, R., Muzzin, A., {et~al.} 2015, \mnras, 449, 685

\bibitem[{{Hoshino} {et~al.}(2015){Hoshino}, {Leauthaud}, {Lackner}, {Hikage},
  {Rozo}, {Rykoff}, {Mand elbaum}, {More}, {More}, {Saito}, \&
  {Vulcani}}]{hoshino15}
{Hoshino}, H., {Leauthaud}, A., {Lackner}, C., {et~al.} 2015, \mnras, 452, 998

\bibitem[{{Hsieh} \& {Yee}(2014)}]{hsieh14}
{Hsieh}, B.~C. \& {Yee}, H.~K.~C. 2014, \apj, 792, 102

\bibitem[{{Huang} {et~al.}(2018){Huang}, {Leauthaud}, {Murata}, {Bosch},
  {Price}, {Lupton}, {Mandelbaum}, {Lackner}, {Bickerton}, {Miyazaki},
  {Coupon}, \& {Tanaka}}]{huang18}
{Huang}, S., {Leauthaud}, A., {Murata}, R., {et~al.} 2018, \pasj, 70, S6

\bibitem[{Hunter(2007)}]{Hunter:2007}
Hunter, J.~D. 2007, Computing In Science \& Engineering, 9, 90

\bibitem[{{Ikebe} {et~al.}(2002){Ikebe}, {Reiprich}, {B{\"o}hringer}, {Tanaka},
  \& {Kitayama}}]{ikebe02}
{Ikebe}, Y., {Reiprich}, T.~H., {B{\"o}hringer}, H., {Tanaka}, Y., \&
  {Kitayama}, T. 2002, \aap, 383, 773

\bibitem[{{Ilbert} {et~al.}(2009){Ilbert}, {Capak}, {Salvato}, {Aussel},
  {McCracken}, {Sanders}, {Scoville}, {Kartaltepe}, {Arnouts}, {Le Floc'h},
  {Mobasher}, {Taniguchi}, {Lamareille}, {Leauthaud}, {Sasaki}, {Thompson},
  {Zamojski}, {Zamorani}, {Bardelli}, {Bolzonella}, {Bongiorno}, {Brusa},
  {Caputi}, {Carollo}, {Contini}, {Cook}, {Coppa}, {Cucciati}, {de la Torre},
  {de Ravel}, {Franzetti}, {Garilli}, {Hasinger}, {Iovino}, {Kampczyk},
  {Kneib}, {Knobel}, {Kovac}, {Le Borgne}, {Le Brun}, {Le F{\`e}vre}, {Lilly},
  {Looper}, {Maier}, {Mainieri}, {Mellier}, {Mignoli}, {Murayama}, {Pell{\`o}},
  {Peng}, {P{\'e}rez-Montero}, {Renzini}, {Ricciardelli}, {Schiminovich},
  {Scodeggio}, {Shioya}, {Silverman}, {Surace}, {Tanaka}, {Tasca}, {Tresse},
  {Vergani}, \& {Zucca}}]{ilbert09}
{Ilbert}, O., {Capak}, P., {Salvato}, M., {et~al.} 2009, \apj, 690, 1236

\bibitem[{{Ivezic} {et~al.}(2008){Ivezic}, {Axelrod}, {Brandt}, {Burke},
  {Claver}, {Connolly}, {Cook}, {Gee}, {Gilmore}, {Jacoby}, {Jones}, {Kahn},
  {Kantor}, {Krabbendam}, {Lupton}, {Monet}, {Pinto}, {Saha}, {Schalk},
  {Schneider}, {Strauss}, {Stubbs}, {Sweeney}, {Szalay}, {Thaler}, {Tyson}, \&
  {LSST Collaboration}}]{ivezic08}
{Ivezic}, Z., {Axelrod}, T., {Brandt}, W.~N., {et~al.} 2008, Serbian
  Astronomical Journal, 176, 1

\bibitem[{{Ivezic} {et~al.}(2019){Ivezic}, {Kahn}, {Tyson}, {Abel}, {Acosta},
  {Allsman}, {Alonso}, {AlSayyad}, {Anderson}, {Andrew}, {Angel}, {Angeli},
  {Ansari}, {Antilogus}, {Araujo}, {Armstrong}, {Arndt}, {Astier}, {Aubourg},
  {Auza}, {Axelrod}, {Bard}, {Barr}, {Barrau}, {Bartlett}, {Bauer}, {Bauman},
  {Baumont}, {Bechtol}, {Bechtol}, {Becker}, {Becla}, {Beldica}, {Bellavia},
  {Bianco}, {Biswas}, {Blanc}, {Blazek}, {Bland ford}, {Bloom}, {Bogart},
  {Bond}, {Booth}, {Borgland}, {Borne}, {Bosch}, {Boutigny}, {Brackett},
  {Bradshaw}, {Brand t}, {Brown}, {Bullock}, {Burchat}, {Burke}, {Cagnoli},
  {Calabrese}, {Callahan}, {Callen}, {Carlin}, {Carlson}, {Chand rasekharan},
  {Charles-Emerson}, {Chesley}, {Cheu}, {Chiang}, {Chiang}, {Chirino}, {Chow},
  {Ciardi}, {Claver}, {Cohen-Tanugi}, {Cockrum}, {Coles}, {Connolly}, {Cook},
  {Cooray}, {Covey}, {Cribbs}, {Cui}, {Cutri}, {Daly}, {Daniel}, {Daruich},
  {Daubard}, {Daues}, {Dawson}, {Delgado}, {Dellapenna}, {de Peyster}, {de
  Val-Borro}, {Digel}, {Doherty}, {Dubois}, {Dubois-Felsmann}, {Durech},
  {Economou}, {Eifler}, {Eracleous}, {Emmons}, {Fausti Neto}, {Ferguson},
  {Figueroa}, {Fisher-Levine}, {Focke}, {Foss}, {Frank}, {Freemon}, {Gangler},
  {Gawiser}, {Geary}, {Gee}, {Geha}, {Gessner}, {Gibson}, {Gilmore},
  {Glanzman}, {Glick}, {Goldina}, {Goldstein}, {Goodenow}, {Graham},
  {Gressler}, {Gris}, {Guy}, {Guyonnet}, {Haller}, {Harris}, {Hascall},
  {Haupt}, {Hernand ez}, {Herrmann}, {Hileman}, {Hoblitt}, {Hodgson}, {Hogan},
  {Howard}, {Huang}, {Huffer}, {Ingraham}, {Innes}, {Jacoby}, {Jain}, {Jammes},
  {Jee}, {Jenness}, {Jernigan}, {Jevremovi{\'c}}, {Johns}, {Johnson},
  {Johnson}, {Jones}, {Juramy-Gilles}, {Juri{\'c}}, {Kalirai}, {Kallivayalil},
  {Kalmbach}, {Kantor}, {Karst}, {Kasliwal}, {Kelly}, {Kessler}, {Kinnison},
  {Kirkby}, {Knox}, {Kotov}, {Krabbendam}, {Krughoff}, {Kub{\'a}nek},
  {Kuczewski}, {Kulkarni}, {Ku}, {Kurita}, {Lage}, {Lambert}, {Lange},
  {Langton}, {Le Guillou}, {Levine}, {Liang}, {Lim}, {Lintott}, {Long},
  {Lopez}, {Lotz}, {Lupton}, {Lust}, {MacArthur}, {Mahabal}, {Mand elbaum},
  {Markiewicz}, {Marsh}, {Marshall}, {Marshall}, {May}, {McKercher}, {McQueen},
  {Meyers}, {Migliore}, {Miller}, {Mills}, {Miraval}, {Moeyens}, {Moolekamp},
  {Monet}, {Moniez}, {Monkewitz}, {Montgomery}, {Morrison}, {Mueller},
  {Muller}, {Mu{\~n}oz Arancibia}, {Neill}, {Newbry}, {Nief}, {Nomerotski},
  {Nordby}, {O'Connor}, {Oliver}, {Olivier}, {Olsen}, {O'Mullane}, {Ortiz},
  {Osier}, {Owen}, {Pain}, {Palecek}, {Parejko}, {Parsons}, {Pease},
  {Peterson}, {Peterson}, {Petravick}, {Libby Petrick}, {Petry},
  {Pierfederici}, {Pietrowicz}, {Pike}, {Pinto}, {Plante}, {Plate}, {Plutchak},
  {Price}, {Prouza}, {Radeka}, {Rajagopal}, {Rasmussen}, {Regnault}, {Reil},
  {Reiss}, {Reuter}, {Ridgway}, {Riot}, {Ritz}, {Robinson}, {Roby}, {Roodman},
  {Rosing}, {Roucelle}, {Rumore}, {Russo}, {Saha}, {Sassolas}, {Schalk},
  {Schellart}, {Schindler}, {Schmidt}, {Schneider}, {Schneider}, {Schoening},
  {Schumacher}, {Schwamb}, {Sebag}, {Selvy}, {Sembroski}, {Seppala}, {Serio},
  {Serrano}, {Shaw}, {Shipsey}, {Sick}, {Silvestri}, {Slater}, {Smith},
  {Smith}, {Sobhani}, {Soldahl}, {Storrie-Lombardi}, {Stover}, {Strauss},
  {Street}, {Stubbs}, {Sullivan}, {Sweeney}, {Swinbank}, {Szalay}, {Takacs},
  {Tether}, {Thaler}, {Thayer}, {Thomas}, {Thornton}, {Thukral}, {Tice},
  {Trilling}, {Turri}, {Van Berg}, {Vanden Berk}, {Vetter}, {Virieux},
  {Vucina}, {Wahl}, {Walkowicz}, {Walsh}, {Walter}, {Wang}, {Wang}, {Warner},
  {Wiecha}, {Willman}, {Winters}, {Wittman}, {Wolff}, {Wood-Vasey}, {Wu},
  {Xin}, {Yoachim}, \& {Zhan}}]{ivezic19}
{Ivezic}, {\v{Z}}., {Kahn}, S.~M., {Tyson}, J.~A., {et~al.} 2019, \apj, 873,
  111

\bibitem[{{Jee} {et~al.}(2011){Jee}, {Dawson}, {Hoekstra}, {Perlmutter},
  {Rosati}, {Brodwin}, {Suzuki}, {Koester}, {Postman}, {Lubin}, {Meyers},
  {Stanford}, {Barbary}, {Barrientos}, {Eisenhardt}, {Ford}, {Gilbank},
  {Gladders}, {Gonzalez}, {Harris}, {Huang}, {Lidman}, {Rykoff}, {Rubin}, \&
  {Spadafora}}]{jee11}
{Jee}, M.~J., {Dawson}, K.~S., {Hoekstra}, H., {et~al.} 2011, \apj, 737, 59

\bibitem[{{Johnston} {et~al.}(2007{\natexlab{a}}){Johnston}, {Sheldon},
  {Tasitsiomi}, {Frieman}, {Wechsler}, \& {McKay}}]{johnston07a}
{Johnston}, D.~E., {Sheldon}, E.~S., {Tasitsiomi}, A., {et~al.}
  2007{\natexlab{a}}, \apj, 656, 27

\bibitem[{{Johnston} {et~al.}(2007{\natexlab{b}}){Johnston}, {Sheldon},
  {Wechsler}, {Rozo}, {Koester}, {Frieman}, {McKay}, {Evrard}, {Becker}, \&
  {Annis}}]{johnston07b}
{Johnston}, D.~E., {Sheldon}, E.~S., {Wechsler}, R.~H., {et~al.}
  2007{\natexlab{b}}, arXiv e-prints, arXiv:0709.1159

\bibitem[{{Juric} {et~al.}(2017){Juric}, {Kantor}, {Lim}, {Lupton},
  {Dubois-Felsmann}, {Jenness}, {Axelrod}, {Aleksi{\'c}}, {Allsman},
  {AlSayyad}, {Alt}, {Armstrong}, {Basney}, {Becker}, {Becla}, {Biswas},
  {Bosch}, {Boutigny}, {Kind}, {Ciardi}, {Connolly}, {Daniel}, {Daues},
  {Economou}, {Chiang}, {Fausti}, {Fisher-Levine}, {Freemon}, {Gris},
  {Hernandez}, {Hoblitt}, {Ivezi{\'c}}, {Jammes}, {Jevremovi{\'c}}, {Jones},
  {Kalmbach}, {Kasliwal}, {Krughoff}, {Lurie}, {Lust}, {MacArthur}, {Melchior},
  {Moeyens}, {Nidever}, {Owen}, {Parejko}, {Peterson}, {Petravick},
  {Pietrowicz}, {Price}, {Reiss}, {Shaw}, {Sick}, {Slater}, {Strauss},
  {Sullivan}, {Swinbank}, {Van Dyk}, {Vuj{\v{c}}i{\'c}}, {Withers}, \&
  {Yoachim}}]{juric17}
{Juric}, M., {Kantor}, J., {Lim}, K.~T., {et~al.} 2017, Astronomical Society of
  the Pacific Conference Series, Vol. 512, {The LSST Data Management System},
  ed. N.~P.~F. {Lorente}, K.~{Shortridge}, \& R.~{Wayth}, 279

\bibitem[{Kaiser(1986)}]{kaiser1986}
Kaiser, N. 1986, \mnras, 222, 323

\bibitem[{{Kawanomoto} {et~al.}(2018){Kawanomoto}, {Uraguchi}, {Komiyama},
  {Miyazaki}, {Furusawa}, {Finet}, {Hattori}, {Wang}, {Yasuda}, \&
  {Suzuki}}]{kawanomoto18}
{Kawanomoto}, S., {Uraguchi}, F., {Komiyama}, Y., {et~al.} 2018, \pasj, 70, 66

\bibitem[{{Kelly} {et~al.}(2014){Kelly}, {von der Linden}, {Applegate},
  {Allen}, {Allen}, {Burchat}, {Burke}, {Ebeling}, {Capak}, {Czoske},
  {Donovan}, {Mantz}, \& {Morris}}]{kelly14}
{Kelly}, P.~L., {von der Linden}, A., {Applegate}, D.~E., {et~al.} 2014,
  \mnras, 439, 28

\bibitem[{{Kettula} {et~al.}(2013){Kettula}, {Finoguenov}, {Massey}, {Rhodes},
  {Hoekstra}, {Taylor}, {Spinelli}, {Tanaka}, {Ilbert}, {Capak}, {McCracken},
  \& {Koekemoer}}]{kettula13}
{Kettula}, K., {Finoguenov}, A., {Massey}, R., {et~al.} 2013, \apj, 778, 74

\bibitem[{{King} \& {Corless}(2007)}]{king07}
{King}, L. \& {Corless}, V. 2007, \mnras, 374, L37

\bibitem[{{Klein} {et~al.}(2019){Klein}, {Grandis}, {Mohr}, {Paulus}, {Abbott},
  {Annis}, {Avila}, {Bertin}, {Brooks}, {Buckley-Geer}, {Rosell}, {Kind},
  {Carretero}, {Castander}, {Cunha}, {D'Andrea}, {da Costa}, {De Vicente},
  {Desai}, {Diehl}, {Dietrich}, {Doel}, {Evrard}, {Flaugher}, {Fosalba},
  {Frieman}, {Garc{\'\i}a-Bellido}, {Gaztanaga}, {Giles}, {Gruen}, {Gruendl},
  {Gschwend}, {Gutierrez}, {Hartley}, {Hollowood}, {Honscheid}, {Hoyle},
  {James}, {Jeltema}, {Kuehn}, {Kuropatkin}, {Lima}, {Maia}, {March},
  {Marshall}, {Menanteau}, {Miquel}, {Ogando}, {Plazas}, {Romer}, {Roodman},
  {Sanchez}, {Scarpine}, {Schindler}, {Serrano}, {Sevilla-Noarbe}, {Smith},
  {Smith}, {Soares-Santos}, {Sobreira}, {Suchyta}, {Swanson}, {Tarle},
  {Thomas}, {Vikram}, \& {DES Collaboration}}]{klein19}
{Klein}, M., {Grandis}, S., {Mohr}, J.~J., {et~al.} 2019, \mnras, 488, 739

\bibitem[{{Klein} {et~al.}(2018){Klein}, {Mohr}, {Desai}, {Israel}, {Allam},
  {Benoit-L{\'e}vy}, {Brooks}, {Buckley-Geer}, {Carnero Rosell}, {Carrasco
  Kind}, {Cunha}, {da Costa}, {Dietrich}, {Eifler}, {Evrard}, {Frieman},
  {Gruen}, {Gruendl}, {Gutierrez}, {Honscheid}, {James}, {Kuehn}, {Lima},
  {Maia}, {March}, {Melchior}, {Menanteau}, {Miquel}, {Plazas}, {Reil},
  {Romer}, {Sanchez}, {Santiago}, {Scarpine}, {Schubnell}, {Sevilla-Noarbe},
  {Smith}, {Soares-Santos}, {Sobreira}, {Suchyta}, {Swanson}, {Tarle}, \&
  {Collaboration}}]{klein18}
{Klein}, M., {Mohr}, J.~J., {Desai}, S., {et~al.} 2018, \mnras, 474, 3324

\bibitem[{{Klein} {et~al.}(2021){Klein}, {Oguri}, {Mohr}, {Grandis},
  {Ghirardini}, {Liu}, {Liu}, {Bulbul}, {Wolf}, {Comparat}, {Ramos-Ceja},
  {Buchner}, {Chiu}, {Clerc}, {Merloni}, {Miyatake}, {Miyazaki}, {Okabe},
  {Ota}, {Pacaud}, {Salvato}, \& {Driver}}]{klein21}
{Klein}, M., {Oguri}, M., {Mohr}, J.~J., {et~al.} 2021, arXiv e-prints,
  arXiv:2106.14519

\bibitem[{{Kodwani} {et~al.}(2019){Kodwani}, {Alonso}, \&
  {Ferreira}}]{kodwani19}
{Kodwani}, D., {Alonso}, D., \& {Ferreira}, P. 2019, The Open Journal of
  Astrophysics, 2, 3

\bibitem[{{Komiyama} {et~al.}(2018){Komiyama}, {Obuchi}, {Nakaya}, {Kamata},
  {Kawanomoto}, {Utsumi}, {Miyazaki}, {Uraguchi}, {Furusawa}, {Morokuma},
  {Uchida}, {Miyatake}, {Mineo}, {Fujimori}, {Aihara}, {Karoji}, {Gunn}, \&
  {Wang}}]{komiyama18}
{Komiyama}, Y., {Obuchi}, Y., {Nakaya}, H., {et~al.} 2018, \pasj, 70, S2

\bibitem[{Kravtsov {et~al.}(2006)Kravtsov, Vikhlinin, \& Nagai}]{kravtsov06a}
Kravtsov, A., Vikhlinin, A., \& Nagai, D. 2006, \apj, 650, 128

\bibitem[{{Kravtsov} {et~al.}(2018){Kravtsov}, {Vikhlinin}, \&
  {Meshcheryakov}}]{kravtsov18}
{Kravtsov}, A.~V., {Vikhlinin}, A.~A., \& {Meshcheryakov}, A.~V. 2018,
  Astronomy Letters, 44, 8

\bibitem[{{Laigle} {et~al.}(2016){Laigle}, {McCracken}, {Ilbert}, {Hsieh},
  {Davidzon}, {Capak}, {Hasinger}, {Silverman}, {Pichon}, {Coupon}, {Aussel},
  {Le Borgne}, {Caputi}, {Cassata}, {Chang}, {Civano}, {Dunlop}, {Fynbo},
  {Kartaltepe}, {Koekemoer}, {Le Fevre}, {Le Floc'h}, {Leauthaud}, {Lilly},
  {Lin}, {Marchesi}, {Milvang-Jensen}, {Salvato}, {Sanders}, {Scoville},
  {Smolcic}, {Stockmann}, {Taniguchi}, {Tasca}, {Toft}, {Vaccari}, \&
  {Zabl}}]{laigle16}
{Laigle}, C., {McCracken}, H.~J., {Ilbert}, O., {et~al.} 2016, \apjs, 224, 24

\bibitem[{{Lauer} {et~al.}(2014){Lauer}, {Postman}, {Strauss}, {Graves}, \&
  {Chisari}}]{lauer14}
{Lauer}, T.~R., {Postman}, M., {Strauss}, M.~A., {Graves}, G.~J., \& {Chisari},
  N.~E. 2014, \apj, 797, 82

\bibitem[{Laureijs {et~al.}(2010)Laureijs, Duvet, {Escudero Sanz}, Gondoin,
  Lumb, Oosterbroek, \& {Saavedra Criado}}]{laureijs10}
Laureijs, R., Duvet, L., {Escudero Sanz}, I., {et~al.} 2010, in Society of
  Photo-Optical Instrumentation Engineers (SPIE) Conference Series, Vol. 7731,
  Society of Photo-Optical Instrumentation Engineers (SPIE) Conference Series,
  1

\bibitem[{{Le Brun} {et~al.}(2017){Le Brun}, {McCarthy}, {Schaye}, \&
  {Ponman}}]{lebrun17}
{Le Brun}, A. M.~C., {McCarthy}, I.~G., {Schaye}, J., \& {Ponman}, T.~J. 2017,
  \mnras, 466, 4442

\bibitem[{Lewis {et~al.}(2000)Lewis, Challinor, \& Lasenby}]{lewis99}
Lewis, A., Challinor, A., \& Lasenby, A. 2000, Astrophys. J., 538, 473

\bibitem[{{Li} {et~al.}(2021){Li}, {Miyatake}, {Luo}, {More}, {Oguri},
  {Hamana}, {Mandelbaum}, {Shirasaki}, {Takada}, {Armstrong}, {Kannawadi},
  {Takita}, {Miyazaki}, {Nishizawa}, {Plazas Malag{\'o}n}, {Strauss}, {Tanaka},
  \& {Yoshida}}]{li21}
{Li}, X., {Miyatake}, H., {Luo}, W., {et~al.} 2021, arXiv e-prints,
  arXiv:2107.00136

\bibitem[{{Lieu} {et~al.}(2016){Lieu}, {Smith}, {Giles}, {Ziparo}, {Maughan},
  {D{\'e}mocl{\`e}s}, {Pacaud}, {Pierre}, {Adami}, {Bah{\'e}}, {Clerc},
  {Chiappetti}, {Eckert}, {Ettori}, {Lavoie}, {Le Fevre}, {McCarthy},
  {Kilbinger}, {Ponman}, {Sadibekova}, \& {Willis}}]{lieu16}
{Lieu}, M., {Smith}, G.~P., {Giles}, P.~A., {et~al.} 2016, \aap, 592, A4

\bibitem[{{Lilly} {et~al.}(2009){Lilly}, {Le Brun}, {Maier}, {Mainieri},
  {Mignoli}, {Scodeggio}, {Zamorani}, {Carollo}, {Contini}, {Kneib}, {Le
  F{\`e}vre}, {Renzini}, {Bardelli}, {Bolzonella}, {Bongiorno}, {Caputi},
  {Coppa}, {Cucciati}, {de la Torre}, {de Ravel}, {Franzetti}, {Garilli},
  {Iovino}, {Kampczyk}, {Kovac}, {Knobel}, {Lamareille}, {Le Borgne}, {Pello},
  {Peng}, {P{\'e}rez-Montero}, {Ricciardelli}, {Silverman}, {Tanaka}, {Tasca},
  {Tresse}, {Vergani}, {Zucca}, {Ilbert}, {Salvato}, {Oesch}, {Abbas},
  {Bottini}, {Capak}, {Cappi}, {Cassata}, {Cimatti}, {Elvis}, {Fumana},
  {Guzzo}, {Hasinger}, {Koekemoer}, {Leauthaud}, {Maccagni}, {Marinoni},
  {McCracken}, {Memeo}, {Meneux}, {Porciani}, {Pozzetti}, {Sanders},
  {Scaramella}, {Scarlata}, {Scoville}, {Shopbell}, \& {Taniguchi}}]{lilly09}
{Lilly}, S.~J., {Le Brun}, V., {Maier}, C., {et~al.} 2009, \apjs, 184, 218

\bibitem[{Lin {et~al.}(2003)Lin, Mohr, \& Stanford}]{lin03b}
Lin, Y., Mohr, J., \& Stanford, S. 2003, \apj, 591, 749

\bibitem[{{Lin} {et~al.}(2017){Lin}, {Hsieh}, {Lin}, {Oguri}, {Chen}, {Tanaka},
  {Chiu}, {Huang}, {Kodama}, {Leauthaud}, {More}, {Nishizawa}, {Bundy}, {Lin},
  \& {Miyazaki}}]{lin17}
{Lin}, Y.-T., {Hsieh}, B.-C., {Lin}, S.-C., {et~al.} 2017, \apj, 851, 139

\bibitem[{Lin \& Mohr(2004)}]{lin04b}
Lin, Y.-T. \& Mohr, J. 2004, \apj, 617, 879

\bibitem[{{Liu} {et~al.}(2021{\natexlab{a}}){Liu}, {Bulbul}, {Ghirardini},
  {Liu}, {Klein}, {Clerc}, {Oezsoy}, {Ramos-Ceja}, {Pacaud}, {Comparat},
  {Okabe}, {Bahar}, {Biffi}, {Brunner}, {Brueggen}, {Buchner}, {Ider Chitham},
  {Chiu}, {Dolag}, {Gatuzz}, {Gonzalez}, {Hoang}, {Lamer}, {Merloni}, {Nandra},
  {Oguri}, {Ota}, {Predehl}, {Reiprich}, {Salvato}, {Schrabback}, {Sanders},
  {Seppi}, \& {Thibaud}}]{liu21}
{Liu}, A., {Bulbul}, E., {Ghirardini}, V., {et~al.} 2021{\natexlab{a}}, arXiv
  e-prints, arXiv:2106.14518

\bibitem[{{Liu} {et~al.}(2012){Liu}, {Mao}, \& {Meng}}]{liu12}
{Liu}, F.~S., {Mao}, S., \& {Meng}, X.~M. 2012, \mnras, 423, 422

\bibitem[{{Liu} {et~al.}(2015){Liu}, {Mohr}, {Saro}, {Aird}, {Ashby}, {Bautz},
  {Bayliss}, {Benson}, {Bleem}, {Bocquet}, {Brodwin}, {Carlstrom}, {Chang},
  {Chiu}, {Cho}, {Clocchiatti}, {Crawford}, {Crites}, {de Haan}, {Desai},
  {Dietrich}, {Dobbs}, {Foley}, {Gangkofner}, {George}, {Gladders}, {Gonzalez},
  {Halverson}, {Hennig}, {Hlavacek-Larrondo}, {Holder}, {Holzapfel}, {Hrubes},
  {Jones}, {Keisler}, {Lee}, {Leitch}, {Lueker}, {Luong-Van}, {McDonald},
  {McMahon}, {Meyer}, {Mocanu}, {Murray}, {Padin}, {Pryke}, {Reichardt},
  {Rest}, {Ruel}, {Ruhl}, {Saliwanchik}, {Sayre}, {Schaffer}, {Shirokoff},
  {Spieler}, {Stalder}, {Staniszewski}, {Stark}, {Story}, {{\v{S}}uhada},
  {Vanderlinde}, {Vieira}, {Vikhlinin}, {Williamson}, {Zahn}, \&
  {Zenteno}}]{liu15a}
{Liu}, J., {Mohr}, J., {Saro}, A., {et~al.} 2015, \mnras, 448, 2085

\bibitem[{{Liu} {et~al.}(2021{\natexlab{b}}){Liu}, {Merloni}, {Comparat},
  {Nandra}, {Sanders}, {Lamer}, {Buchner}, {Dwelly}, {Freyberg}, {Malyali},
  {Georgakakis}, {Salvato}, {Brunner}, {Brusa}, {Klein}, {Ghirardini}, {Clerc},
  {Pacaud}, {Bulbul}, {Liu}, {Schwope}, {Robrade}, {Wilms}, {Dauser},
  {Ramos-Ceja}, {Reiprich}, {Boller}, \& {Wolf}}]{liuteng21}
{Liu}, T., {Merloni}, A., {Comparat}, J., {et~al.} 2021{\natexlab{b}}, arXiv
  e-prints, arXiv:2106.14528

\bibitem[{{Lovisari} {et~al.}(2015){Lovisari}, {Reiprich}, \&
  {Schellenberger}}]{lovisari15}
{Lovisari}, L., {Reiprich}, T.~H., \& {Schellenberger}, G. 2015, \aap, 573,
  A118

\bibitem[{{Magnier} {et~al.}(2013){Magnier}, {Schlafly}, {Finkbeiner}, {Juric},
  {Tonry}, {Burgett}, {Chambers}, {Flewelling}, {Kaiser}, {Kudritzki},
  {Morgan}, {Price}, {Sweeney}, \& {Stubbs}}]{magnier13}
{Magnier}, E.~A., {Schlafly}, E., {Finkbeiner}, D., {et~al.} 2013, \apjs, 205,
  20

\bibitem[{{Mahdavi} {et~al.}(2013){Mahdavi}, {Hoekstra}, {Babul}, {Bildfell},
  {Jeltema}, \& {Henry}}]{mahdavi13}
{Mahdavi}, A., {Hoekstra}, H., {Babul}, A., {et~al.} 2013, \apj, 767, 116

\bibitem[{{Mandelbaum} {et~al.}(2005){Mandelbaum}, {Hirata}, {Seljak}, {Guzik},
  {Padmanabhan}, {Blake}, {Blanton}, {Lupton}, \& {Brinkmann}}]{mandalbaum05b}
{Mandelbaum}, R., {Hirata}, C.~M., {Seljak}, U., {et~al.} 2005, \mnras, 361,
  1287

\bibitem[{{Mandelbaum} {et~al.}(2018{\natexlab{a}}){Mandelbaum}, {Lanusse},
  {Leauthaud}, {Armstrong}, {Simet}, {Miyatake}, {Meyers}, {Bosch}, {Murata},
  {Miyazaki}, \& {Tanaka}}]{mandelbaum18b}
{Mandelbaum}, R., {Lanusse}, F., {Leauthaud}, A., {et~al.} 2018{\natexlab{a}},
  \mnras, 481, 3170

\bibitem[{{Mandelbaum} {et~al.}(2018{\natexlab{b}}){Mandelbaum}, {Lanusse},
  {Leauthaud}, {Armstrong}, {Simet}, {Miyatake}, {Meyers}, {Bosch}, {Murata},
  {Miyazaki}, \& {Tanaka}}]{mandalbaum18b}
{Mandelbaum}, R., {Lanusse}, F., {Leauthaud}, A., {et~al.} 2018{\natexlab{b}},
  \mnras, 481, 3170

\bibitem[{{Mandelbaum} {et~al.}(2018{\natexlab{c}}){Mandelbaum}, {Miyatake},
  {Hamana}, {Oguri}, {Simet}, {Armstrong}, {Bosch}, {Murata}, {Lanusse},
  {Leauthaud}, {Coupon}, {More}, {Takada}, {Miyazaki}, {Speagle}, {Shirasaki},
  {Sif{\'o}n}, {Huang}, {Nishizawa}, {Medezinski}, {Okura}, {Okabe}, {Czakon},
  {Takahashi}, {Coulton}, {Hikage}, {Komiyama}, {Lupton}, {Strauss}, {Tanaka},
  \& {Utsumi}}]{mandelbaum18}
{Mandelbaum}, R., {Miyatake}, H., {Hamana}, T., {et~al.} 2018{\natexlab{c}},
  \pasj, 70, S25

\bibitem[{Mantz {et~al.}(2016)Mantz, Allen, Morris, von~der Linden, Applegate,
  Kelly, Burke, Donovan, \& Ebeling}]{mantz16b}
Mantz, A., Allen, S., Morris, R., {et~al.} 2016, \mnras, 463, 3582

\bibitem[{{Mantz} {et~al.}(2015){Mantz}, {von der Linden}, {Allen},
  {Applegate}, {Kelly}, {Morris}, {Rapetti}, {Schmidt}, {Adhikari}, {Allen},
  {Burchat}, {Burke}, {Cataneo}, {Donovan}, {Ebeling}, {Shandera}, \&
  {Wright}}]{mantz15}
{Mantz}, A.~B., {von der Linden}, A., {Allen}, S.~W., {et~al.} 2015, \mnras,
  446, 2205

\bibitem[{{Masters} {et~al.}(2015){Masters}, {Capak}, {Stern}, {Ilbert},
  {Salvato}, {Schmidt}, {Longo}, {Rhodes}, {Paltani}, {Mobasher}, {Hoekstra},
  {Hildebrandt}, {Coupon}, {Steinhardt}, {Speagle}, {Faisst}, {Kalinich},
  {Brodwin}, {Brescia}, \& {Cavuoti}}]{masters15}
{Masters}, D., {Capak}, P., {Stern}, D., {et~al.} 2015, \apj, 813, 53

\bibitem[{{Masters} {et~al.}(2019){Masters}, {Stern}, {Cohen}, {Capak},
  {Stanford}, {Hernitschek}, {Galametz}, {Davidzon}, {Rhodes}, {Sanders},
  {Mobasher}, {Castander}, {Pruett}, \& {Fotopoulou}}]{masters19}
{Masters}, D.~C., {Stern}, D.~K., {Cohen}, J.~G., {et~al.} 2019, \apj, 877, 81

\bibitem[{Maughan(2007)}]{maughan07}
Maughan, B. 2007, \apj, 668, 772

\bibitem[{{McClintock} {et~al.}(2019){McClintock}, {Varga}, {Gruen}, {Rozo},
  {Rykoff}, {Shin}, {Melchior}, {DeRose}, {Seitz}, {Dietrich}, {Sheldon},
  {Zhang}, {von der Linden}, {Jeltema}, {Mantz}, {Romer}, {Allen}, {Becker},
  {Bermeo}, {Bhargava}, {Costanzi}, {Everett}, {Farahi}, {Hamaus}, {Hartley},
  {Hollowood}, {Hoyle}, {Israel}, {Li}, {MacCrann}, {Morris}, {Palmese},
  {Plazas}, {Pollina}, {Rau}, {Simet}, {Soares-Santos}, {Troxel}, {Vergara
  Cervantes}, {Wechsler}, {Zuntz}, {Abbott}, {Abdalla}, {Allam}, {Annis},
  {Avila}, {Bridle}, {Brooks}, {Burke}, {Carnero Rosell}, {Carrasco Kind},
  {Carretero}, {Castander}, {Crocce}, {Cunha}, {D'Andrea}, {da Costa}, {Davis},
  {De Vicente}, {Diehl}, {Doel}, {Drlica-Wagner}, {Evrard}, {Flaugher},
  {Fosalba}, {Frieman}, {Garc{\'\i}a-Bellido}, {Gaztanaga}, {Gerdes},
  {Giannantonio}, {Gruendl}, {Gutierrez}, {Honscheid}, {James}, {Kirk},
  {Krause}, {Kuehn}, {Lahav}, {Li}, {Lima}, {March}, {Marshall}, {Menanteau},
  {Miquel}, {Mohr}, {Nord}, {Ogando}, {Roodman}, {Sanchez}, {Scarpine},
  {Schindler}, {Sevilla-Noarbe}, {Smith}, {Smith}, {Sobreira}, {Suchyta},
  {Swanson}, {Tarle}, {Tucker}, {Vikram}, {Walker}, {Weller}, \& {DES
  Collaboration}}]{mcclintock19}
{McClintock}, T., {Varga}, T.~N., {Gruen}, D., {et~al.} 2019, \mnras, 482, 1352

\bibitem[{{McKerns} {et~al.}(2012){McKerns}, {Strand}, {Sullivan}, {Fang}, \&
  {Aivazis}}]{pathos}
{McKerns}, M.~M., {Strand}, L., {Sullivan}, T., {Fang}, A., \& {Aivazis}, M.
  A.~G. 2012, arXiv e-prints, arXiv:1202.1056

\bibitem[{{Medezinski} {et~al.}(2018{\natexlab{a}}){Medezinski}, {Battaglia},
  {Umetsu}, {Oguri}, {Miyatake}, {Nishizawa}, {Sif{\'o}n}, {Spergel}, {Chiu},
  {Lin}, {Bahcall}, \& {Komiyama}}]{medezinski18b}
{Medezinski}, E., {Battaglia}, N., {Umetsu}, K., {et~al.} 2018{\natexlab{a}},
  \pasj, 70, S28

\bibitem[{{Medezinski} {et~al.}(2018{\natexlab{b}}){Medezinski}, {Oguri},
  {Nishizawa}, {Speagle}, {Miyatake}, {Umetsu}, {Leauthaud}, {Murata},
  {Mandelbaum}, {Sif{\'o}n}, {Strauss}, {Huang}, {Simet}, {Okabe}, {Tanaka}, \&
  {Komiyama}}]{medezinski18a}
{Medezinski}, E., {Oguri}, M., {Nishizawa}, A.~J., {et~al.} 2018{\natexlab{b}},
  \pasj, 70, 30

\bibitem[{{Melchior} {et~al.}(2017){Melchior}, {Gruen}, {McClintock}, {Varga},
  {Sheldon}, {Rozo}, {Amara}, {Becker}, {Benson}, {Bermeo}, {Bridle},
  {Clampitt}, {Dietrich}, {Hartley}, {Hollowood}, {Jain}, {Jarvis}, {Jeltema},
  {Kacprzak}, {MacCrann}, {Rykoff}, {Saro}, {Suchyta}, {Troxel}, {Zuntz},
  {Bonnett}, {Plazas}, {Abbott}, {Abdalla}, {Annis}, {Benoit-L{\'e}vy},
  {Bernstein}, {Bertin}, {Brooks}, {Buckley-Geer}, {Carnero Rosell}, {Carrasco
  Kind}, {Carretero}, {Cunha}, {D'Andrea}, {da Costa}, {Desai}, {Eifler},
  {Flaugher}, {Fosalba}, {Garc{\'{\i}}a-Bellido}, {Gaztanaga}, {Gerdes},
  {Gruendl}, {Gschwend}, {Gutierrez}, {Honscheid}, {James}, {Kirk}, {Krause},
  {Kuehn}, {Kuropatkin}, {Lahav}, {Lima}, {Maia}, {March}, {Martini},
  {Menanteau}, {Miller}, {Miquel}, {Mohr}, {Nichol}, {Ogando}, {Romer},
  {Sanchez}, {Scarpine}, {Sevilla-Noarbe}, {Smith}, {Soares-Santos},
  {Sobreira}, {Swanson}, {Tarle}, {Thomas}, {Walker}, {Weller}, \&
  {Zhang}}]{melchior17}
{Melchior}, P., {Gruen}, D., {McClintock}, T., {et~al.} 2017, \mnras, 469, 4899

\bibitem[{{Merloni} {et~al.}(2012){Merloni}, {Predehl}, {Becker},
  {B{\"o}hringer}, {Boller}, {Brunner}, {Brusa}, {Dennerl}, {Freyberg},
  {Friedrich}, {Georgakakis}, {Haberl}, {Hasinger}, {Meidinger}, {Mohr},
  {Nandra}, {Rau}, {Reiprich}, {Robrade}, {Salvato}, {Santangelo}, {Sasaki},
  {Schwope}, {Wilms}, \& {German eROSITA Consortium}}]{merloni12}
{Merloni}, A., {Predehl}, P., {Becker}, W., {et~al.} 2012, ArXiv e-prints
  [\eprint[arXiv]{1209.3114}]

\bibitem[{{Miyatake} {et~al.}(2019){Miyatake}, {Battaglia}, {Hilton},
  {Medezinski}, {Nishizawa}, {More}, {Aiola}, {Bahcall}, {Bond}, {Calabrese},
  {Choi}, {Devlin}, {Dunkley}, {Dunner}, {Fuzia}, {Gallardo}, {Gralla},
  {Hasselfield}, {Halpern}, {Hikage}, {Hill}, {Hincks}, {Hlo{\v{z}}ek},
  {Huffenberger}, {Hughes}, {Koopman}, {Kosowsky}, {Louis}, {Madhavacheril},
  {McMahon}, {Mandelbaum}, {Marriage}, {Maurin}, {Miyazaki}, {Moodley},
  {Murata}, {Naess}, {Newburgh}, {Niemack}, {Nishimichi}, {Okabe}, {Oguri},
  {Osato}, {Page}, {Partridge}, {Robertson}, {Sehgal}, {Sherwin}, {Shirasaki},
  {Sievers}, {Sif{\'o}n}, {Simon}, {Spergel}, {Staggs}, {Stein}, {Takada},
  {Trac}, {Umetsu}, {van Engelen}, \& {Wollack}}]{miyatake19}
{Miyatake}, H., {Battaglia}, N., {Hilton}, M., {et~al.} 2019, \apj, 875, 63

\bibitem[{{Miyazaki}(2015)}]{miyazaki15}
{Miyazaki}, S. 2015, IAU General Assembly, 22, 2255916

\bibitem[{{Miyazaki} {et~al.}(2018){Miyazaki}, {Komiyama}, {Kawanomoto}, {Doi},
  {Furusawa}, {Hamana}, {Hayashi}, {Ikeda}, {Kamata}, {Karoji}, {Koike},
  {Kurakami}, {Miyama}, {Morokuma}, {Nakata}, {Namikawa}, {Nakaya}, {Nariai},
  {Obuchi}, {Oishi}, {Okada}, {Okura}, {Tait}, {Takata}, {Tanaka}, {Tanaka},
  {Terai}, {Tomono}, {Uraguchi}, {Usuda}, {Utsumi}, {Yamada}, {Yamanoi},
  {Aihara}, {Fujimori}, {Mineo}, {Miyatake}, {Oguri}, {Uchida}, {Tanaka},
  {Yasuda}, {Takada}, {Murayama}, {Nishizawa}, {Sugiyama}, {Chiba}, {Futamase},
  {Wang}, {Chen}, {Ho}, {Liaw}, {Chiu}, {Ho}, {Lai}, {Lee}, {Jeng}, {Iwamura},
  {Armstrong}, {Bickerton}, {Bosch}, {Gunn}, {Lupton}, {Loomis}, {Price},
  {Smith}, {Strauss}, {Turner}, {Suzuki}, {Miyazaki}, {Muramatsu}, {Yamamoto},
  {Endo}, {Ezaki}, {Ito}, {Kawaguchi}, {Sofuku}, {Taniike}, {Akutsu}, {Dojo},
  {Kasumi}, {Matsuda}, {Imoto}, {Miwa}, {Suzuki}, {Takeshi}, \&
  {Yokota}}]{miyazaki18}
{Miyazaki}, S., {Komiyama}, Y., {Kawanomoto}, S., {et~al.} 2018, \pasj, 70, S1

\bibitem[{Mohr \& Evrard(1997)}]{mohr97}
Mohr, J.~J. \& Evrard, A.~E. 1997, \apj, 491, 38

\bibitem[{Mohr {et~al.}(1999)Mohr, Mathiesen, \& Evrard}]{mohr99}
Mohr, J.~J., Mathiesen, B., \& Evrard, A.~E. 1999, \apj, 517, 627

\bibitem[{{Momcheva} {et~al.}(2016){Momcheva}, {Brammer}, {van Dokkum},
  {Skelton}, {Whitaker}, {Nelson}, {Fumagalli}, {Maseda}, {Leja}, {Franx},
  {Rix}, {Bezanson}, {Da Cunha}, {Dickey}, {F{\"o}rster Schreiber},
  {Illingworth}, {Kriek}, {Labb{\'e}}, {Ulf Lange}, {Lundgren}, {Magee},
  {Marchesini}, {Oesch}, {Pacifici}, {Patel}, {Price}, {Tal}, {Wake}, {van der
  Wel}, \& {Wuyts}}]{momcheva16}
{Momcheva}, I.~G., {Brammer}, G.~B., {van Dokkum}, P.~G., {et~al.} 2016, \apjs,
  225, 27

\bibitem[{{Murata} {et~al.}(2019){Murata}, {Oguri}, {Nishimichi}, {Takada},
  {Mandelbaum}, {More}, {Shirasaki}, {Nishizawa}, \& {Osato}}]{murata19}
{Murata}, R., {Oguri}, M., {Nishimichi}, T., {et~al.} 2019, \pasj, 71, 107

\bibitem[{Nagai {et~al.}(2007)Nagai, Kravtsov, \& Vikhlinin}]{nagai07}
Nagai, D., Kravtsov, A., \& Vikhlinin, A. 2007, \apj, 668, 1

\bibitem[{Navarro {et~al.}(1997)Navarro, Frenk, \& White}]{navarro1997}
Navarro, J., Frenk, C., \& White, S. 1997, \apj, 490, 493

\bibitem[{{Oguri} {et~al.}(2018{\natexlab{a}}){Oguri}, {Lin}, {Lin},
  {Nishizawa}, {More}, {More}, {Hsieh}, {Medezinski}, {Miyatake}, {Jian},
  {Lin}, {Takada}, {Okabe}, {Speagle}, {Coupon}, {Leauthaud}, {Lupton},
  {Miyazaki}, {Price}, {Tanaka}, {Chiu}, {Komiyama}, {Okura}, {Tanaka}, \&
  {Usuda}}]{oguri18}
{Oguri}, M., {Lin}, Y.-T., {Lin}, S.-C., {et~al.} 2018{\natexlab{a}}, \pasj,
  70, S20

\bibitem[{{Oguri} {et~al.}(2018{\natexlab{b}}){Oguri}, {Miyazaki}, {Hikage},
  {Mand elbaum}, {Utsumi}, {Miyatake}, {Takada}, {Armstrong}, {Bosch},
  {Komiyama}, {Leauthaud}, {More}, {Nishizawa}, {Okabe}, \&
  {Tanaka}}]{oguri18b}
{Oguri}, M., {Miyazaki}, S., {Hikage}, C., {et~al.} 2018{\natexlab{b}}, \pasj,
  70, S26

\bibitem[{{Oguri} {et~al.}(2010){Oguri}, {Takada}, {Okabe}, \&
  {Smith}}]{oguri10}
{Oguri}, M., {Takada}, M., {Okabe}, N., \& {Smith}, G.~P. 2010, \mnras, 405,
  2215

\bibitem[{O'Hara {et~al.}(2006)O'Hara, Mohr, Bialek, \& Evrard}]{ohara06}
O'Hara, T., Mohr, J., Bialek, J., \& Evrard, A. 2006, \apj, 639, 64

\bibitem[{{Okabe} {et~al.}(2019){Okabe}, {Oguri}, {Akamatsu}, {Hamabata},
  {Nishizawa}, {Medezinski}, {Koyama}, {Hayashi}, {Okabe}, {Ueda}, {Mitsuishi},
  \& {Ota}}]{okabe19}
{Okabe}, N., {Oguri}, M., {Akamatsu}, H., {et~al.} 2019, \pasj, 71, 79

\bibitem[{{Okabe} \& {Smith}(2016)}]{okabe16}
{Okabe}, N. \& {Smith}, G.~P. 2016, \mnras, 461, 3794

\bibitem[{{Pacaud} {et~al.}(2018){Pacaud}, {Pierre}, {Melin}, {Adami},
  {Evrard}, {Galli}, {Gastaldello}, {Maughan}, {Sereno}, {Alis}, {Altieri},
  {Birkinshaw}, {Chiappetti}, {Faccioli}, {Giles}, {Horellou}, {Iovino},
  {Koulouridis}, {Le F{\`e}vre}, {Lidman}, {Lieu}, {Maurogordato},
  {Moscardini}, {Plionis}, {Poggianti}, {Pompei}, {Sadibekova}, {Valtchanov},
  \& {Willis}}]{pacaud18}
{Pacaud}, F., {Pierre}, M., {Melin}, J.~B., {et~al.} 2018, \aap, 620, A10

\bibitem[{{P{\^a}ris} {et~al.}(2018){P{\^a}ris}, {Petitjean}, {Aubourg},
  {Myers}, {Streblyanska}, {Lyke}, {Anderson}, {Armengaud}, {Bautista},
  {Blanton}, {Blomqvist}, {Brinkmann}, {Brownstein}, {Brandt}, {Burtin},
  {Dawson}, {de la Torre}, {Georgakakis}, {Gil-Mar{\'\i}n}, {Green}, {Hall},
  {Kneib}, {LaMassa}, {Le Goff}, {MacLeod}, {Mariappan}, {McGreer}, {Merloni},
  {Noterdaeme}, {Palanque-Delabrouille}, {Percival}, {Ross}, {Rossi},
  {Schneider}, {Seo}, {Tojeiro}, {Weaver}, {Weijmans}, {Y{\`e}che}, {Zarrouk},
  \& {Zhao}}]{paris18}
{P{\^a}ris}, I., {Petitjean}, P., {Aubourg}, {\'E}., {et~al.} 2018, \aap, 613,
  A51

\bibitem[{P\'erez \& Granger(2007)}]{PER-GRA:2007}
P\'erez, F. \& Granger, B.~E. 2007, Computing in Science and Engineering, 9, 21

\bibitem[{{Pierre} {et~al.}(2016){Pierre}, {Pacaud}, {Adami}, {Alis},
  {Altieri}, {Baran}, {Benoist}, {Birkinshaw}, {Bongiorno}, {Bremer}, {Brusa},
  {Butler}, {Ciliegi}, {Chiappetti}, {Clerc}, {Corasaniti}, {Coupon}, {De
  Breuck}, {Democles}, {Desai}, {Delhaize}, {Devriendt}, {Dubois}, {Eckert},
  {Elyiv}, {Ettori}, {Evrard}, {Faccioli}, {Farahi}, {Ferrari}, {Finet},
  {Fotopoulou}, {Fourmanoit}, {Gandhi}, {Gastaldello}, {Gastaud},
  {Georgantopoulos}, {Giles}, {Guennou}, {Guglielmo}, {Horellou}, {Husband},
  {Huynh}, {Iovino}, {Kilbinger}, {Koulouridis}, {Lavoie}, {Le Brun}, {Le
  Fevre}, {Lidman}, {Lieu}, {Lin}, {Mantz}, {Maughan}, {Maurogordato},
  {McCarthy}, {McGee}, {Melin}, {Melnyk}, {Menanteau}, {Novak}, {Paltani},
  {Plionis}, {Poggianti}, {Pomarede}, {Pompei}, {Ponman}, {Ramos-Ceja},
  {Ranalli}, {Rapetti}, {Raychaudury}, {Reiprich}, {Rottgering}, {Rozo},
  {Rykoff}, {Sadibekova}, {Santos}, {Sauvageot}, {Schimd}, {Sereno}, {Smith},
  {Smol{\v{c}}i{\'c}}, {Snowden}, {Spergel}, {Stanford}, {Surdej}, {Valageas},
  {Valotti}, {Valtchanov}, {Vignali}, {Willis}, \& {Ziparo}}]{peirre16}
{Pierre}, M., {Pacaud}, F., {Adami}, C., {et~al.} 2016, \aap, 592, A1

\bibitem[{{Piffaretti} {et~al.}(2011){Piffaretti}, {Arnaud}, {Pratt},
  {Pointecouteau}, \& {Melin}}]{piffaretti11}
{Piffaretti}, R., {Arnaud}, M., {Pratt}, G.~W., {Pointecouteau}, E., \&
  {Melin}, J.~B. 2011, \aap, 534, A109

\bibitem[{{Pillepich} {et~al.}(2012){Pillepich}, {Porciani}, \&
  {Reiprich}}]{pillepich12}
{Pillepich}, A., {Porciani}, C., \& {Reiprich}, T.~H. 2012, \mnras, 422, 44

\bibitem[{{Pillepich} {et~al.}(2018){Pillepich}, {Reiprich}, {Porciani},
  {Borm}, \& {Merloni}}]{pillepich18}
{Pillepich}, A., {Reiprich}, T.~H., {Porciani}, C., {Borm}, K., \& {Merloni},
  A. 2018, \mnras, 481, 613

\bibitem[{{Planck Collaboration} {et~al.}(2016){Planck Collaboration}, {Ade},
  {Aghanim}, {Arnaud}, {Ashdown}, {Aumont}, {Baccigalupi}, {Banday},
  {Barreiro}, {Bartlett}, {Bartolo}, {Battaner}, {Battye}, {Benabed},
  {Beno{\^\i}t}, {Benoit-L{\'e}vy}, {Bernard}, {Bersanelli}, {Bielewicz},
  {Bock}, {Bonaldi}, {Bonavera}, {Bond}, {Borrill}, {Bouchet}, {Bucher},
  {Burigana}, {Butler}, {Calabrese}, {Cardoso}, {Catalano}, {Challinor},
  {Chamballu}, {Chary}, {Chiang}, {Christensen}, {Church}, {Clements},
  {Colombi}, {Colombo}, {Combet}, {Comis}, {Couchot}, {Coulais}, {Crill},
  {Curto}, {Cuttaia}, {Danese}, {Davies}, {Davis}, {de Bernardis}, {de Rosa},
  {de Zotti}, {Delabrouille}, {D{\'e}sert}, {Diego}, {Dolag}, {Dole},
  {Donzelli}, {Dor{\'e}}, {Douspis}, {Ducout}, {Dupac}, {Efstathiou}, {Elsner},
  {En{\ss}lin}, {Eriksen}, {Falgarone}, {Fergusson}, {Finelli}, {Forni},
  {Frailis}, {Fraisse}, {Franceschi}, {Frejsel}, {Galeotta}, {Galli}, {Ganga},
  {Giard}, {Giraud-H{\'e}raud}, {Gjerl{\o}w}, {Gonz{\'a}lez-Nuevo},
  {G{\'o}rski}, {Gratton}, {Gregorio}, {Gruppuso}, {Gudmundsson}, {Hansen},
  {Hanson}, {Harrison}, {Henrot-Versill{\'e}}, {Hern{\'a}ndez-Monteagudo},
  {Herranz}, {Hildebrandt}, {Hivon}, {Hobson}, {Holmes}, {Hornstrup}, {Hovest},
  {Huffenberger}, {Hurier}, {Jaffe}, {Jaffe}, {Jones}, {Juvela},
  {Keih{\"a}nen}, {Keskitalo}, {Kisner}, {Kneissl}, {Knoche}, {Kunz},
  {Kurki-Suonio}, {Lagache}, {L{\"a}hteenm{\"a}ki}, {Lamarre}, {Lasenby},
  {Lattanzi}, {Lawrence}, {Leonardi}, {Lesgourgues}, {Levrier}, {Liguori},
  {Lilje}, {Linden-V{\o}rnle}, {L{\'o}pez-Caniego}, {Lubin},
  {Mac{\'\i}as-P{\'e}rez}, {Maggio}, {Maino}, {Mandolesi}, {Mangilli}, {Maris},
  {Martin}, {Mart{\'\i}nez-Gonz{\'a}lez}, {Masi}, {Matarrese}, {McGehee},
  {Meinhold}, {Melchiorri}, {Melin}, {Mendes}, {Mennella}, {Migliaccio},
  {Mitra}, {Miville-Desch{\^e}nes}, {Moneti}, {Montier}, {Morgante},
  {Mortlock}, {Moss}, {Munshi}, {Murphy}, {Naselsky}, {Nati}, {Natoli},
  {Netterfield}, {N{\o}rgaard-Nielsen}, {Noviello}, {Novikov}, {Novikov},
  {Oxborrow}, {Paci}, {Pagano}, {Pajot}, {Paoletti}, {Partridge}, {Pasian},
  {Patanchon}, {Pearson}, {Perdereau}, {Perotto}, {Perrotta}, {Pettorino},
  {Piacentini}, {Piat}, {Pierpaoli}, {Pietrobon}, {Plaszczynski},
  {Pointecouteau}, {Polenta}, {Popa}, {Pratt}, {Pr{\'e}zeau}, {Prunet},
  {Puget}, {Rachen}, {Rebolo}, {Reinecke}, {Remazeilles}, {Renault}, {Renzi},
  {Ristorcelli}, {Rocha}, {Roman}, {Rosset}, {Rossetti}, {Roudier},
  {Rubi{\~n}o-Mart{\'\i}n}, {Rusholme}, {Sandri}, {Santos}, {Savelainen},
  {Savini}, {Scott}, {Seiffert}, {Shellard}, {Spencer}, {Stolyarov}, {Stompor},
  {Sudiwala}, {Sunyaev}, {Sutton}, {Suur-Uski}, {Sygnet}, {Tauber}, {Terenzi},
  {Toffolatti}, {Tomasi}, {Tristram}, {Tucci}, {Tuovinen}, {T{\"u}rler},
  {Umana}, {Valenziano}, {Valiviita}, {Van Tent}, {Vielva}, {Villa}, {Wade},
  {Wandelt}, {Wehus}, {Weller}, {White}, {Yvon}, {Zacchei}, \&
  {Zonca}}]{PlanckCollaboration2015b}
{Planck Collaboration}, {Ade}, P.~A.~R., {Aghanim}, N., {et~al.} 2016, \aap,
  594, A24

\bibitem[{{Poole} {et~al.}(2008){Poole}, {Babul}, {McCarthy}, {Sanderson}, \&
  {Fardal}}]{poole08}
{Poole}, G.~B., {Babul}, A., {McCarthy}, I.~G., {Sanderson}, A.~J.~R., \&
  {Fardal}, M.~A. 2008, \mnras, 391, 1163

\bibitem[{Pratt {et~al.}(2009)Pratt, Croston, Arnaud, \&
  B{\"{o}}hringer}]{pratt09}
Pratt, G., Croston, J., Arnaud, M., \& B{\"{o}}hringer, H. 2009, \aap, 498, 361

\bibitem[{{Pratt} {et~al.}(2019){Pratt}, {Arnaud}, {Biviano}, {Eckert},
  {Ettori}, {Nagai}, {Okabe}, \& {Reiprich}}]{pratt19}
{Pratt}, G.~W., {Arnaud}, M., {Biviano}, A., {et~al.} 2019, \ssr, 215, 25

\bibitem[{{Predehl} {et~al.}(2021){Predehl}, {Andritschke}, {Arefiev},
  {Babyshkin}, {Batanov}, {Becker}, {B{\"o}hringer}, {Bogomolov}, {Boller},
  {Borm}, {Bornemann}, {Br{\"a}uninger}, {Br{\"u}ggen}, {Brunner}, {Brusa},
  {Bulbul}, {Buntov}, {Burwitz}, {Burkert}, {Clerc}, {Churazov}, {Coutinho},
  {Dauser}, {Dennerl}, {Doroshenko}, {Eder}, {Emberger}, {Eraerds},
  {Finoguenov}, {Freyberg}, {Friedrich}, {Friedrich}, {F{\"u}rmetz},
  {Georgakakis}, {Gilfanov}, {Granato}, {Grossberger}, {Gueguen}, {Gureev},
  {Haberl}, {H{\"a}lker}, {Hartner}, {Hasinger}, {Huber}, {Ji}, {Kienlin},
  {Kink}, {Korotkov}, {Kreykenbohm}, {Lamer}, {Lomakin}, {Lapshov}, {Liu},
  {Maitra}, {Meidinger}, {Menz}, {Merloni}, {Mernik}, {Mican}, {Mohr},
  {M{\"u}ller}, {Nandra}, {Nazarov}, {Pacaud}, {Pavlinsky}, {Perinati},
  {Pfeffermann}, {Pietschner}, {Ramos-Ceja}, {Rau}, {Reiffers}, {Reiprich},
  {Robrade}, {Salvato}, {Sanders}, {Santangelo}, {Sasaki}, {Scheuerle},
  {Schmid}, {Schmitt}, {Schwope}, {Shirshakov}, {Steinmetz}, {Stewart},
  {Str{\"u}der}, {Sunyaev}, {Tenzer}, {Tiedemann}, {Tr{\"u}mper}, {Voron},
  {Weber}, {Wilms}, \& {Yaroshenko}}]{predehl21}
{Predehl}, P., {Andritschke}, R., {Arefiev}, V., {et~al.} 2021, \aap, 647, A1

\bibitem[{{Raihan} {et~al.}(2020){Raihan}, {Schrabback}, {Hildebrandt},
  {Applegate}, \& {Mahler}}]{raihan20}
{Raihan}, S.~F., {Schrabback}, T., {Hildebrandt}, H., {Applegate}, D., \&
  {Mahler}, G. 2020, \mnras, 497, 1404

\bibitem[{{Reichert} {et~al.}(2011){Reichert}, {B{\"o}hringer}, {Fassbender},
  \& {M{\"u}hlegger}}]{reichert11}
{Reichert}, A., {B{\"o}hringer}, H., {Fassbender}, R., \& {M{\"u}hlegger}, M.
  2011, \aap, 535, A4

\bibitem[{Reiprich \& B{\"{o}}hringer(2002)}]{reiprich02}
Reiprich, T. \& B{\"{o}}hringer, H. 2002, \apj, 567, 716

\bibitem[{{Ritchie} \& {Thomas}(2002)}]{ritchie02}
{Ritchie}, B.~W. \& {Thomas}, P.~A. 2002, \mnras, 329, 675

\bibitem[{{Rozo} {et~al.}(2011){Rozo}, {Rykoff}, {Koester}, {Nord}, {Wu},
  {Evrard}, \& {Wechsler}}]{rozo11b}
{Rozo}, E., {Rykoff}, E., {Koester}, B., {et~al.} 2011, \apj, 740, 53

\bibitem[{{Rozo} \& {Rykoff}(2014)}]{rozo14a}
{Rozo}, E. \& {Rykoff}, E.~S. 2014, \apj, 783, 80

\bibitem[{{Saro} {et~al.}(2015){Saro}, {Bocquet}, {Rozo}, {Benson}, {Mohr},
  {Rykoff}, {Soares-Santos}, {Bleem}, {Dodelson}, {Melchior}, {Sobreira},
  {Upadhyay}, {Weller}, {Abbott}, {Abdalla}, {Allam}, {Armstrong}, {Banerji},
  {Bauer}, {Bayliss}, {Benoit-L{\'e}vy}, {Bernstein}, {Bertin}, {Brodwin},
  {Brooks}, {Buckley-Geer}, {Burke}, {Carlstrom}, {Capasso}, {Capozzi},
  {Carnero Rosell}, {Carrasco Kind}, {Chiu}, {Covarrubias}, {Crawford},
  {Crocce}, {D'Andrea}, {da Costa}, {DePoy}, {Desai}, {de Haan}, {Diehl},
  {Dietrich}, {Doel}, {Cunha}, {Eifler}, {Evrard}, {Fausti Neto}, {Fernand ez},
  {Flaugher}, {Fosalba}, {Frieman}, {Gangkofner}, {Gaztanaga}, {Gerdes},
  {Gruen}, {Gruendl}, {Gupta}, {Hennig}, {Holzapfel}, {Honscheid}, {Jain},
  {James}, {Kuehn}, {Kuropatkin}, {Lahav}, {Li}, {Lin}, {Maia}, {March},
  {Marshall}, {Martini}, {McDonald}, {Miller}, {Miquel}, {Nord}, {Ogando},
  {Plazas}, {Reichardt}, {Romer}, {Roodman}, {Sako}, {Sanchez}, {Schubnell},
  {Sevilla}, {Smith}, {Stalder}, {Stark}, {Strazzullo}, {Suchyta}, {Swanson},
  {Tarle}, {Thaler}, {Thomas}, {Tucker}, {Vikram}, {von der Linden}, {Walker},
  {Wechsler}, {Wester}, {Zenteno}, \& {Ziegler}}]{saro15}
{Saro}, A., {Bocquet}, S., {Rozo}, E., {et~al.} 2015, \mnras, 454, 2305

\bibitem[{{Schellenberger} \& {Reiprich}(2017)}]{schellenberger17}
{Schellenberger}, G. \& {Reiprich}, T.~H. 2017, \mnras, 471, 1370

\bibitem[{{Schlafly} {et~al.}(2012){Schlafly}, {Finkbeiner}, {Juri{\'c}},
  {Magnier}, {Burgett}, {Chambers}, {Grav}, {Hodapp}, {Kaiser}, {Kudritzki},
  {Martin}, {Morgan}, {Price}, {Rix}, {Stubbs}, {Tonry}, \&
  {Wainscoat}}]{schlafly12}
{Schlafly}, E.~F., {Finkbeiner}, D.~P., {Juri{\'c}}, M., {et~al.} 2012, \apj,
  756, 158

\bibitem[{{Schrabback} {et~al.}(2018{\natexlab{a}}){Schrabback}, {Applegate},
  {Dietrich}, {Hoekstra}, {Bocquet}, {Gonzalez}, {von der Linden}, {McDonald},
  {Morrison}, {Raihan}, {Allen}, {Bayliss}, {Benson}, {Bleem}, {Chiu}, {Desai},
  {Foley}, {de Haan}, {High}, {Hilbert}, {Mantz}, {Massey}, {Mohr},
  {Reichardt}, {Saro}, {Simon}, {Stern}, {Stubbs}, \& {Zenteno}}]{schrabback18}
{Schrabback}, T., {Applegate}, D., {Dietrich}, J.~P., {et~al.}
  2018{\natexlab{a}}, \mnras, 474, 2635

\bibitem[{{Schrabback} {et~al.}(2021){Schrabback}, {Bocquet}, {Sommer},
  {Zohren}, {van den Busch}, {Hern{\'a}ndez-Mart{\'\i}n}, {Hoekstra}, {Raihan},
  {Schirmer}, {Applegate}, {Bayliss}, {Benson}, {Bleem}, {Dietrich}, {Floyd},
  {Hilbert}, {Hlavacek-Larrondo}, {McDonald}, {Saro}, {Stark}, \&
  {Weissgerber}}]{schrabback21}
{Schrabback}, T., {Bocquet}, S., {Sommer}, M., {et~al.} 2021, \mnras, 505, 3923

\bibitem[{{Schrabback} {et~al.}(2010){Schrabback}, {Hartlap}, {Joachimi},
  {Kilbinger}, {Simon}, {Benabed}, {Brada{\v{c}}}, {Eifler}, {Erben},
  {Fassnacht}, {High}, {Hilbert}, {Hildebrandt}, {Hoekstra}, {Kuijken},
  {Marshall}, {Mellier}, {Morganson}, {Schneider}, {Semboloni}, {van Waerbeke},
  \& {Velander}}]{schrabback10}
{Schrabback}, T., {Hartlap}, J., {Joachimi}, B., {et~al.} 2010, \aap, 516, A63

\bibitem[{{Schrabback} {et~al.}(2018{\natexlab{b}}){Schrabback}, {Schirmer},
  {van der Burg}, {Hoekstra}, {Buddendiek}, {Applegate}, {Brada{\v{c}}},
  {Eifler}, {Erben}, {Gladders}, {Hern{\'a}ndez-Mart{\'\i}n}, {Hildebrandt},
  {Hoag}, {Klaes}, {von der Linden}, {Marchesini}, {Muzzin}, {Sharon}, \&
  {Stefanon}}]{schrabback18b}
{Schrabback}, T., {Schirmer}, M., {van der Burg}, R. F.~J., {et~al.}
  2018{\natexlab{b}}, \aap, 610, A85

\bibitem[{{Sehgal} {et~al.}(2013){Sehgal}, {Addison}, {Battaglia},
  {Battistelli}, {Bond}, {Das}, {Devlin}, {Dunkley}, {D{\"u}nner}, {Gralla},
  {Hajian}, {Halpern}, {Hasselfield}, {Hilton}, {Hincks}, {Hlozek}, {Hughes},
  {Kosowsky}, {Lin}, {Louis}, {Marriage}, {Marsden}, {Menanteau}, {Moodley},
  {Niemack}, {Page}, {Partridge}, {Reese}, {Sherwin}, {Sievers}, {Sif{\'o}n},
  {Spergel}, {Staggs}, {Swetz}, {Switzer}, \& {Wollack}}]{sehgal13}
{Sehgal}, N., {Addison}, G., {Battaglia}, N., {et~al.} 2013, \apj, 767, 38

\bibitem[{Seitz \& Schneider(1997)}]{seitz97}
Seitz, C. \& Schneider, P. 1997, \aap, 318, 687

\bibitem[{{Sereno} {et~al.}(2020){Sereno}, {Umetsu}, {Ettori}, {Eckert},
  {Gastaldello}, {Giles}, {Lieu}, {Maughan}, {Okabe}, {Birkinshaw}, {Chiu},
  {Fujita}, {Miyazaki}, {Rapetti}, {Koulouridis}, \& {Pierre}}]{sereno20}
{Sereno}, M., {Umetsu}, K., {Ettori}, S., {et~al.} 2020, \mnras, 492, 4528

\bibitem[{{Silverman} {et~al.}(2015){Silverman}, {Kashino}, {Sanders},
  {Kartaltepe}, {Arimoto}, {Renzini}, {Rodighiero}, {Daddi}, {Zahid}, {Nagao},
  {Kewley}, {Lilly}, {Sugiyama}, {Baronchelli}, {Capak}, {Carollo}, {Chu},
  {Hasinger}, {Ilbert}, {Juneau}, {Kajisawa}, {Koekemoer}, {Kovac}, {Le
  F{\`e}vre}, {Masters}, {McCracken}, {Onodera}, {Schulze}, {Scoville},
  {Strazzullo}, \& {Taniguchi}}]{silverman15}
{Silverman}, J.~D., {Kashino}, D., {Sanders}, D., {et~al.} 2015, \apjs, 220, 12

\bibitem[{{Skelton} {et~al.}(2014){Skelton}, {Whitaker}, {Momcheva}, {Brammer},
  {van Dokkum}, {Labb{\'e}}, {Franx}, {van der Wel}, {Bezanson}, {Da Cunha},
  {Fumagalli}, {F{\"o}rster Schreiber}, {Kriek}, {Leja}, {Lundgren}, {Magee},
  {Marchesini}, {Maseda}, {Nelson}, {Oesch}, {Pacifici}, {Patel}, {Price},
  {Rix}, {Tal}, {Wake}, \& {Wuyts}}]{skelton14}
{Skelton}, R.~E., {Whitaker}, K.~E., {Momcheva}, I.~G., {et~al.} 2014, \apjs,
  214, 24

\bibitem[{{Smith} {et~al.}(1979){Smith}, {Mushotzky}, \&
  {Serlemitsos}}]{smith79}
{Smith}, B.~W., {Mushotzky}, R.~F., \& {Serlemitsos}, P.~J. 1979, \apj, 227, 37

\bibitem[{{Sommer} {et~al.}(2021){Sommer}, {Schrabback}, {Applegate},
  {Hilbert}, {Ansarinejad}, {Floyd}, \& {Grandis}}]{sommer21}
{Sommer}, M.~W., {Schrabback}, T., {Applegate}, D.~E., {et~al.} 2021, arXiv
  e-prints, arXiv:2105.08027

\bibitem[{{Song} {et~al.}(2012){Song}, {Zenteno}, {Stalder}, {Desai}, {Bleem},
  {Aird}, {Armstrong}, {Ashby}, {Bayliss}, {Bazin}, {Benson}, {Bertin},
  {Brodwin}, {Carlstrom}, {Chang}, {Cho}, {Clocchiatti}, {Crawford}, {Crites},
  {de Haan}, {Dobbs}, {Dudley}, {Foley}, {George}, {Gettings}, {Gladders},
  {Gonzalez}, {Halverson}, {Harrington}, {High}, {Holder}, {Holzapfel},
  {Hoover}, {Hrubes}, {Joy}, {Keisler}, {Knox}, {Lee}, {Leitch}, {Liu},
  {Lueker}, {Luong-Van}, {Marrone}, {McDonald}, {McMahon}, {Mehl}, {Meyer},
  {Mocanu}, {Mohr}, {Montroy}, {Natoli}, {Nurgaliev}, {Padin}, {Plagge},
  {Pryke}, {Reichardt}, {Rest}, {Ruel}, {Ruhl}, {Saliwanchik}, {Saro}, {Sayre},
  {Schaffer}, {Shaw}, {Shirokoff}, {{\v{S}}uhada}, {Spieler}, {Stanford},
  {Staniszewski}, {Stark}, {Story}, {Stubbs}, {van Engelen}, {Vanderlinde},
  {Vieira}, {Williamson}, \& {Zahn}}]{song12b}
{Song}, J., {Zenteno}, A., {Stalder}, B., {et~al.} 2012, \apj, 761, 22

\bibitem[{{Straatman} {et~al.}(2018){Straatman}, {van der Wel}, {Bezanson},
  {Pacifici}, {Gallazzi}, {Wu}, {Noeske}, {Bari{\v{s}}i{\'c}}, {Bell},
  {Brammer}, {Calhau}, {Chauke}, {Franx}, {van Houdt}, {Labb{\'e}}, {Maseda},
  {Mu{\~n}oz-Mateos}, {Muzzin}, {van de Sande}, {Sobral}, \&
  {Spilker}}]{straatman18}
{Straatman}, C. M.~S., {van der Wel}, A., {Bezanson}, R., {et~al.} 2018, \apjs,
  239, 27

\bibitem[{{Sunayama} {et~al.}(2020){Sunayama}, {Park}, {Takada}, {Kobayashi},
  {Nishimichi}, {Kurita}, {More}, {Oguri}, \& {Osato}}]{sunayama20}
{Sunayama}, T., {Park}, Y., {Takada}, M., {et~al.} 2020, \mnras, 496, 4468

\bibitem[{{Sunyaev} {et~al.}(2021){Sunyaev}, {Arefiev}, {Babyshkin},
  {Bogomolov}, {Borisov}, {Buntov}, {Brunner}, {Burenin}, {Churazov},
  {Coutinho}, {Eder}, {Eismont}, {Freyberg}, {Gilfanov}, {Gureyev}, {Hasinger},
  {Khabibullin}, {Kolmykov}, {Komovkin}, {Krivonos}, {Lapshov}, {Levin},
  {Lomakin}, {Lutovinov}, {Medvedev}, {Merloni}, {Mernik}, {Mikhailov},
  {Molodzov}, {Mzhelsky}, {Mueller}, {Nandra}, {Nazarov}, {Pavlinsky},
  {Poghodin}, {Predehl}, {Robrade}, {Sazonov}, {Scheuerle}, {Shirshakov},
  {Tkachenko}, \& {Voron}}]{sunyaev21}
{Sunyaev}, R., {Arefiev}, V., {Babyshkin}, V., {et~al.} 2021, arXiv e-prints,
  arXiv:2104.13267

\bibitem[{Sunyaev \& Zel'dovich(1972)}]{sunyaev72}
Sunyaev, R. \& Zel'dovich, Y. 1972, Comments on Astrophysics and Space Physics,
  4, 173

\bibitem[{{Tanaka} {et~al.}(2018){Tanaka}, {Coupon}, {Hsieh}, {Mineo},
  {Nishizawa}, {Speagle}, {Furusawa}, {Miyazaki}, \& {Murayama}}]{tanaka18}
{Tanaka}, M., {Coupon}, J., {Hsieh}, B.-C., {et~al.} 2018, \pasj, 70, S9

\bibitem[{{Tanaka} {et~al.}(2017){Tanaka}, {Hasinger}, {Silverman},
  {Bickerton}, {Furusawa}, {Harikane}, {Hu}, {Ikeda}, {Li}, {McCracken},
  {Price}, {Strauss}, {Koike}, {Komiyama}, {Mineo}, {Miyazaki}, {Nishizawa},
  {Takata}, {Utsumi}, {Yamada}, \& {Yasuda}}]{tanaka17}
{Tanaka}, M., {Hasinger}, G., {Silverman}, J.~D., {et~al.} 2017, arXiv
  e-prints, arXiv:1706.00566

\bibitem[{{Taylor}(2005)}]{topcat1}
{Taylor}, M.~B. 2005, in Astronomical Society of the Pacific Conference Series,
  Vol. 347, Astronomical Data Analysis Software and Systems XIV, ed.
  P.~{Shopbell}, M.~{Britton}, \& R.~{Ebert}, 29

\bibitem[{{Taylor}(2006)}]{topcat2}
{Taylor}, M.~B. 2006, in Astronomical Society of the Pacific Conference Series,
  Vol. 351, Astronomical Data Analysis Software and Systems XV, ed.
  C.~{Gabriel}, C.~{Arviset}, D.~{Ponz}, \& S.~{Enrique}, 666

\bibitem[{{The Dark Energy Survey Collaboration}(2005)}]{des05}
{The Dark Energy Survey Collaboration}. 2005, arXiv e-prints, astro

\bibitem[{{To} {et~al.}(2021){To}, {Krause}, {Rozo}, {Wu}, {Gruen}, {Wechsler},
  {Eifler}, {Rykoff}, {Costanzi}, {Becker}, {Bernstein}, {Blazek}, {Bocquet},
  {Bridle}, {Cawthon}, {Choi}, {Crocce}, {Davis}, {DeRose}, {Drlica-Wagner},
  {Elvin-Poole}, {Fang}, {Farahi}, {Friedrich}, {Gatti}, {Gaztanaga},
  {Giannantonio}, {Hartley}, {Hoyle}, {Jarvis}, {MacCrann}, {McClintock},
  {Miranda}, {Pereira}, {Park}, {Porredon}, {Prat}, {Rau}, {Ross}, {Samuroff},
  {S{\'a}nchez}, {Sevilla-Noarbe}, {Sheldon}, {Troxel}, {Varga}, {Vielzeuf},
  {Zhang}, {Zuntz}, {Abbott}, {Aguena}, {Amon}, {Annis}, {Avila}, {Bertin},
  {Bhargava}, {Brooks}, {Burke}, {Carnero Rosell}, {Carrasco Kind},
  {Carretero}, {Chang}, {Conselice}, {da Costa}, {Davis}, {Desai}, {Diehl},
  {Dietrich}, {Everett}, {Evrard}, {Ferrero}, {Flaugher}, {Fosalba}, {Frieman},
  {Garc{\'\i}a-Bellido}, {Gruendl}, {Gutierrez}, {Hinton}, {Hollowood},
  {Honscheid}, {Huterer}, {James}, {Jeltema}, {Kron}, {Kuehn}, {Kuropatkin},
  {Lima}, {Maia}, {Marshall}, {Menanteau}, {Miquel}, {Morgan}, {Muir}, {Myles},
  {Palmese}, {Paz-Chinch{\'o}n}, {Plazas}, {Romer}, {Roodman}, {Sanchez},
  {Santiago}, {Scarpine}, {Serrano}, {Smith}, {Suchyta}, {Swanson}, {Tarle},
  {Thomas}, {Tucker}, {Weller}, {Wester}, {Wilkinson}, \& {DES
  Collaboration}}]{to21}
{To}, C., {Krause}, E., {Rozo}, E., {et~al.} 2021, \prl, 126, 141301

\bibitem[{{Tonry} {et~al.}(2012){Tonry}, {Stubbs}, {Lykke}, {Doherty},
  {Shivvers}, {Burgett}, {Chambers}, {Hodapp}, {Kaiser}, {Kudritzki},
  {Magnier}, {Morgan}, {Price}, \& {Wainscoat}}]{tonry12}
{Tonry}, J.~L., {Stubbs}, C.~W., {Lykke}, K.~R., {et~al.} 2012, \apj, 750, 99

\bibitem[{{Truemper}(1982)}]{rosat}
{Truemper}, J. 1982, Advances in Space Research, 2, 241

\bibitem[{{Umetsu}(2020)}]{umetsu20b}
{Umetsu}, K. 2020, \aapr, 28, 7

\bibitem[{Umetsu {et~al.}(2014)Umetsu, Medezinski, Nonino, Merten, Postman,
  Meneghetti, Donahue, Czakon, Molino, Seitz, Gruen, Lemze, Balestra, Benitez,
  Biviano, Broadhurst, Ford, Grillo, Koekemoer, Melchior, Mercurio, Moustakas,
  Rosati, \& Zitrin}]{umetsu14}
Umetsu, K., Medezinski, E., Nonino, M., {et~al.} 2014, \apj, 795, 163

\bibitem[{{Umetsu} {et~al.}(2020){Umetsu}, {Sereno}, {Lieu}, {Miyatake},
  {Medezinski}, {Nishizawa}, {Giles}, {Gastaldello}, {McCarthy}, {Kilbinger},
  {Birkinshaw}, {Ettori}, {Okabe}, {Chiu}, {Coupon}, {Eckert}, {Fujita},
  {Higuchi}, {Koulouridis}, {Maughan}, {Miyazaki}, {Oguri}, {Pacaud}, {Pierre},
  {Rapetti}, \& {Smith}}]{umetsu20}
{Umetsu}, K., {Sereno}, M., {Lieu}, M., {et~al.} 2020, \apj, 890, 148

\bibitem[{van~der Burg {et~al.}(2014)van~der Burg, Muzzin, Hoekstra, Wilson,
  Lidman, \& Yee}]{burg14}
van~der Burg, R., Muzzin, A., Hoekstra, H., {et~al.} 2014, \aap, 561, A79

\bibitem[{Van Der~Walt {et~al.}(2011)Van Der~Walt, Colbert, \&
  Varoquaux}]{van2011numpy}
Van Der~Walt, S., Colbert, S.~C., \& Varoquaux, G. 2011, Computing in Science
  \& Engineering, 13, 22

\bibitem[{{Varga} {et~al.}(2019){Varga}, {DeRose}, {Gruen}, {McClintock},
  {Seitz}, {Rozo}, {Costanzi}, {Hoyle}, {MacCrann}, {Plazas}, {Rykoff},
  {Simet}, {von der Linden}, {Wechsler}, {Annis}, {Avila}, {Bertin}, {Brooks},
  {Buckley-Geer}, {Burke}, {Carnero Rosell}, {Carrasco Kind}, {Carretero},
  {Cunha}, {D'Andrea}, {da Costa}, {De Vicente}, {Desai}, {Diehl}, {Dietrich},
  {Doel}, {Evrard}, {Flaugher}, {Fosalba}, {Frieman}, {Garc{\'\i}a-Bellido},
  {Gaztanaga}, {Gerdes}, {Gruendl}, {Gschwend}, {Gutierrez}, {Hartley},
  {Hollowood}, {Honscheid}, {James}, {Jeltema}, {Kuehn}, {Kuropatkin}, {Lima},
  {Maia}, {March}, {Marshall}, {Melchior}, {Menanteau}, {Miller}, {Miquel},
  {Ogando}, {Romer}, {Sanchez}, {Scarpine}, {Schubnell}, {Serrano},
  {Sevilla-Noarbe}, {Smith}, {Sobreira}, {Suchyta}, {Swanson}, {Tarle},
  {Thomas}, {Tucker}, {Zhang}, \& {DES Collaboration}}]{varga19}
{Varga}, T.~N., {DeRose}, J., {Gruen}, D., {et~al.} 2019, \mnras, 489, 2511

\bibitem[{Vikhlinin {et~al.}(2009)Vikhlinin, Burenin, Ebeling, Forman,
  Hornstrup, Jones, Kravtsov, Murray, Nagai, Quintana, \&
  Voevodkin}]{vikhlinin09a}
Vikhlinin, A., Burenin, R., Ebeling, H., {et~al.} 2009, \apj, 692, 1033

\bibitem[{Vikhlinin {et~al.}(2006)Vikhlinin, Kravtsov, Forman, Jones,
  Markevitch, Murray, \& {Van Speybroeck}}]{vikhlinin06}
Vikhlinin, A., Kravtsov, A., Forman, W., {et~al.} 2006, \apj, 640, 691

\bibitem[{{Virtanen} {et~al.}(2020){Virtanen}, {Gommers}, {Oliphant},
  {Haberland}, {Reddy}, {Cournapeau}, {Burovski}, {Peterson}, {Weckesser},
  {Bright}, {van der Walt}, {Brett}, {Wilson}, {Jarrod Millman}, {Mayorov},
  {Nelson}, {Jones}, {Kern}, {Larson}, {Carey}, {Polat}, {Feng}, {Moore}, {Vand
  erPlas}, {Laxalde}, {Perktold}, {Cimrman}, {Henriksen}, {Quintero}, {Harris},
  {Archibald}, {Ribeiro}, {Pedregosa}, {van Mulbregt}, \&
  {Contributors}}]{virtanen_scipy}
{Virtanen}, P., {Gommers}, R., {Oliphant}, T.~E., {et~al.} 2020, Nature
  Methods, 17, 261

\bibitem[{{von der Linden} {et~al.}(2014){von der Linden}, {Allen},
  {Applegate}, {Kelly}, {Allen}, {Ebeling}, {Burchat}, {Burke}, {Donovan},
  {Morris}, {Blandford}, {Erben}, \& {Mantz}}]{vonderlinden14a}
{von der Linden}, A., {Allen}, M.~T., {Applegate}, D.~E., {et~al.} 2014,
  \mnras, 439, 2

\bibitem[{von~der Linden {et~al.}(2014)von~der Linden, Mantz, Allen, Applegate,
  Kelly, Morris, Wright, Allen, Burchat, Burke, Donovan, \&
  Ebeling}]{vonderlinden14b}
von~der Linden, A., Mantz, A., Allen, S., {et~al.} 2014, \mnras, 443, 1973

\bibitem[{{Yang} {et~al.}(2006){Yang}, {Mo}, {van den Bosch}, {Jing},
  {Weinmann}, \& {Meneghetti}}]{yang06}
{Yang}, X., {Mo}, H.~J., {van den Bosch}, F.~C., {et~al.} 2006, \mnras, 373,
  1159

\bibitem[{{Zhang} {et~al.}(2019){Zhang}, {Jeltema}, {Hollowood}, {Everett},
  {Rozo}, {Farahi}, {Bermeo}, {Bhargava}, {Giles}, {Romer}, {Wilkinson},
  {Rykoff}, {Mantz}, {Diehl}, {Evrard}, {Stern}, {Gruen}, {von der Linden},
  {Splettstoesser}, {Chen}, {Costanzi}, {Allen}, {Collins}, {Hilton}, {Klein},
  {Mann}, {Manolopoulou}, {Morris}, {Mayers}, {Sahlen}, {Stott}, {Vergara
  Cervantes}, {Viana}, {Wechsler}, {Allam}, {Avila}, {Bechtol}, {Bertin},
  {Brooks}, {Burke}, {Carnero Rosell}, {Carrasco Kind}, {Carretero},
  {Castander}, {da Costa}, {De Vicente}, {Desai}, {Dietrich}, {Doel},
  {Flaugher}, {Fosalba}, {Frieman}, {Garc{\'\i}a-Bellido}, {Gaztanaga},
  {Gruendl}, {Gschwend}, {Gutierrez}, {Hartley}, {Honscheid}, {Hoyle},
  {Krause}, {Kuehn}, {Kuropatkin}, {Lima}, {Maia}, {Marshall}, {Melchior},
  {Menanteau}, {Miller}, {Miquel}, {Ogando}, {Plazas}, {Sanchez}, {Scarpine},
  {Schindler}, {Serrano}, {Sevilla-Noarbe}, {Smith}, {Soares-Santos},
  {Suchyta}, {Swanson}, {Tarle}, {Thomas}, {Tucker}, {Vikram}, {Wester}, \&
  {DES Collaboration}}]{zhang19}
{Zhang}, Y., {Jeltema}, T., {Hollowood}, D.~L., {et~al.} 2019, \mnras, 487,
  2578

\bibitem[{Zhang {et~al.}(2012)Zhang, Lagan{\'{a}}, Pierini, Puchwein,
  Schneider, \& Reiprich}]{zhang12}
Zhang, Y.-Y., Lagan{\'{a}}, T., Pierini, D., {et~al.} 2012, \aap, 544, C3

\bibitem[{{Zitrin} {et~al.}(2012){Zitrin}, {Bartelmann}, {Umetsu}, {Oguri}, \&
  {Broadhurst}}]{zitrin12}
{Zitrin}, A., {Bartelmann}, M., {Umetsu}, K., {Oguri}, M., \& {Broadhurst}, T.
  2012, \mnras, 426, 2944

\bibitem[{{Zu} {et~al.}(2017){Zu}, {Mandelbaum}, {Simet}, {Rozo}, \&
  {Rykoff}}]{zu17}
{Zu}, Y., {Mandelbaum}, R., {Simet}, M., {Rozo}, E., \& {Rykoff}, E.~S. 2017,
  \mnras, 470, 551

\end{thebibliography}

%
%

\appendix

\section{The richness-to-mass-and-redshift relation of the eFEDS sample}
\label{sec:richness}

\begin{figure*}
\centering
\resizebox{0.48\textwidth}{!}{
\includegraphics[scale=1]{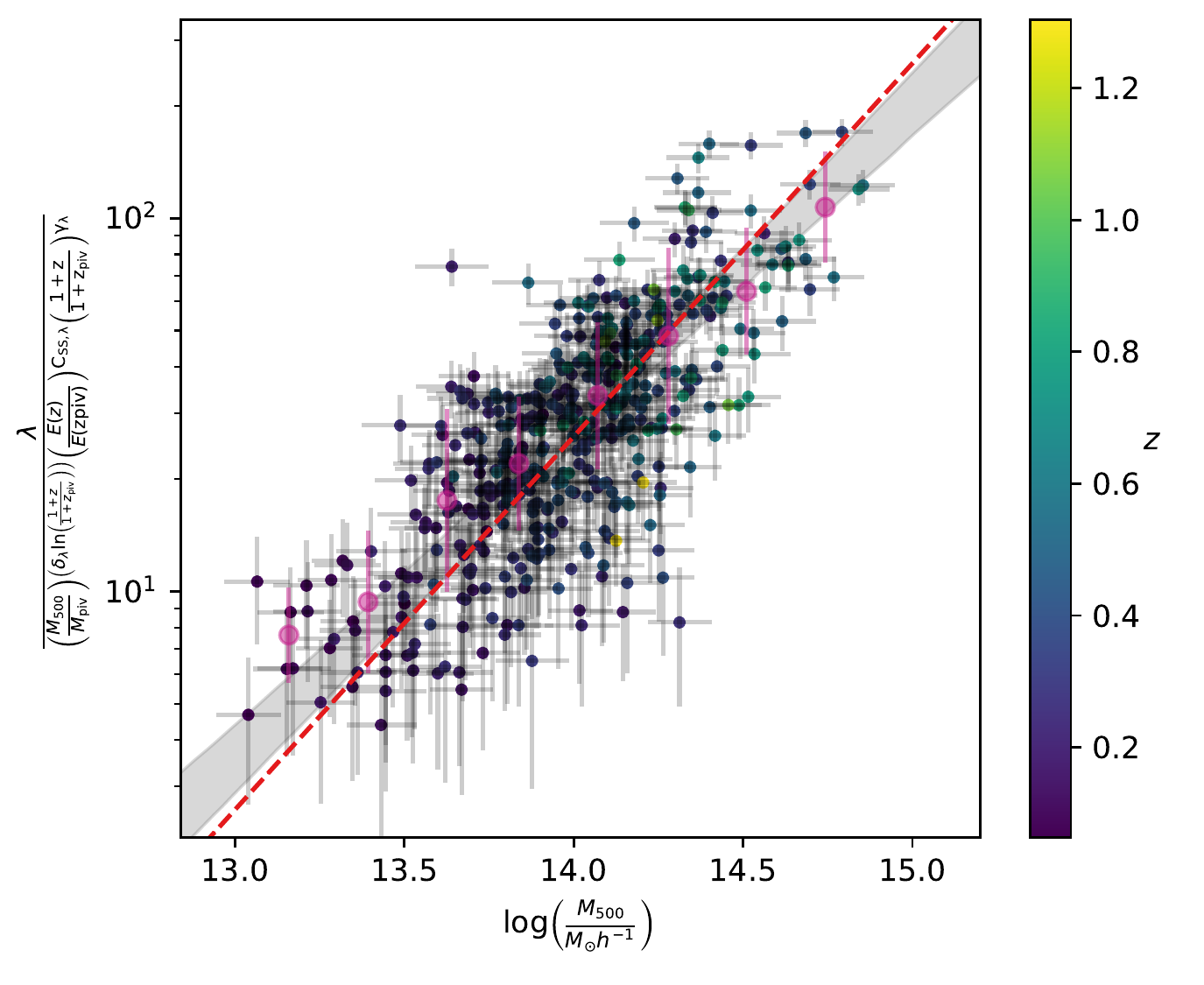}
}
\resizebox{0.48\textwidth}{!}{
\includegraphics[scale=1]{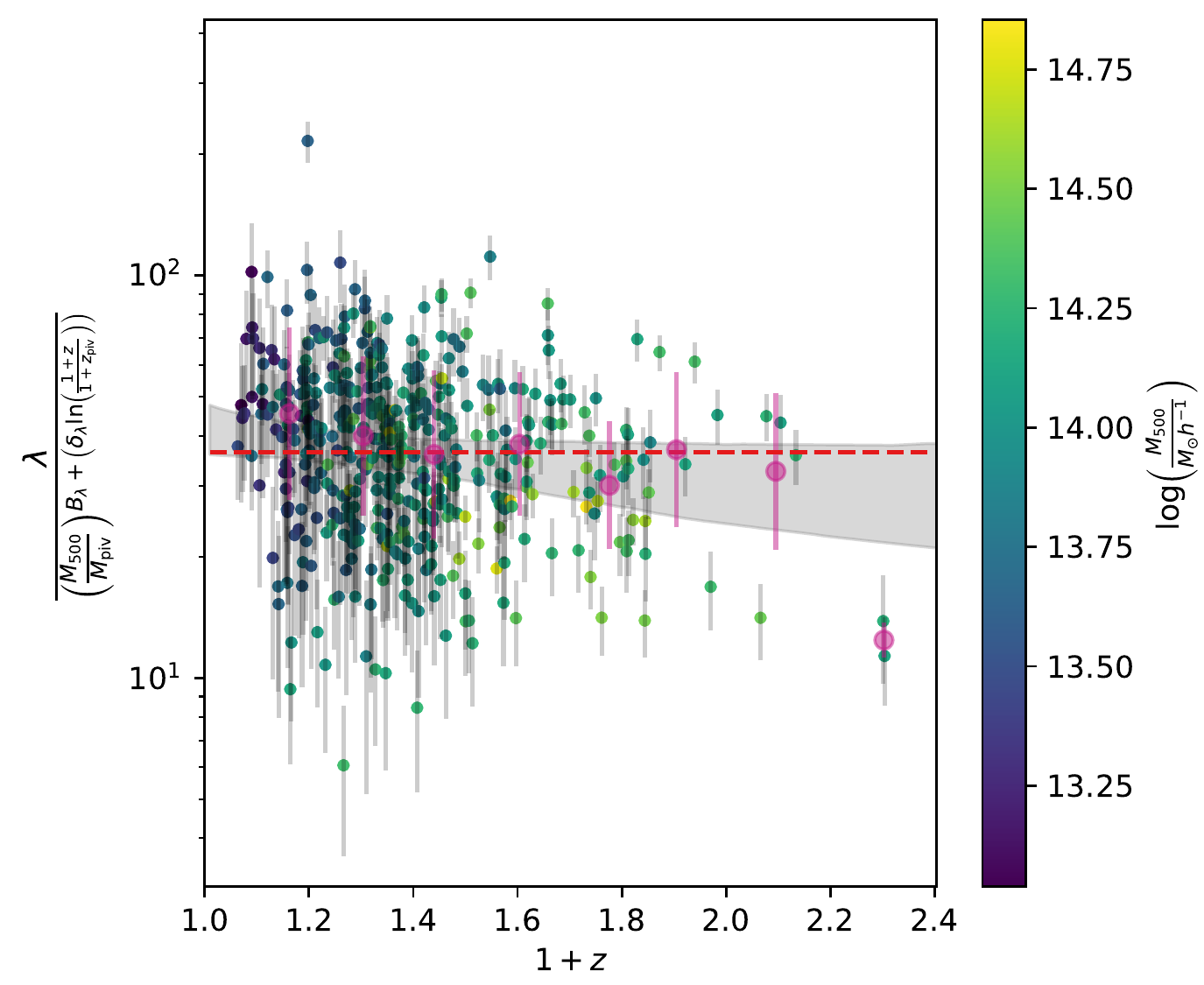}
}
\caption{
Observed richness \rich\ as a function of the cluster mass \Mfiveoo\ (left panel) and redshift \redshift\ (right panel).
This plot is produced and shown in the same way as in Fig.~\ref{fig:lmz}.
}
\label{fig:rchmz}
\end{figure*}

In this section, we show the richness-to-mass-and-redshift (\rich--\Mfiveoo--\redshift) relation of the eFEDS sample, where the richness \rich\ is estimated by the MCMF algorithm (see Sect.~\ref{sec:clustersample}).
We use the same fitting framework as described in Sect.~\ref{sec:wlsrmcalib} with $\Xlabel = \rich$ in Eq.~(\ref{eq:like_sr_single}).
The resulting relation is 
\begin{multline}
\label{eq:rchm}
\left\langle\ln\rich|\Mfiveoo\right\rangle 
= \ln \left( \ansArch \right) + \\
\left[ \left( \ansBrch \right) + \left( \ansdeltarch \right) \ln\left(\frac{1 + \redshift}{1 + \ZPIV}\right) \right] \times
\ln\left(\frac{\Mfiveoo}{\MPIV}\right) \\
+ \left( \ansgammarch \right)
 \times \ln \left(\frac{1 + \redshift}{1 + \ZPIV}\right)
 \, ,
\end{multline}
with log-normal intrinsic scatter of \anssigmarch\ at fixed mass.
At the pivotal mass ($\MPIV = 1.4\times10^{14}\Msunh$) and redshift ($\ZPIV=0.35$), the richness is constrained to be \ansArch\ scaled as ${\Mfiveoo}^{\ansBrch}$ without a significant redshift trend ($\rich\propto (1 + \redshift)^{\ansgammarch}$).
We show the richness of the eFEDS sample as a function of mass (redshift) in the left (right) panel of Fig.~\ref{fig:rchmz}, where we re-normalize the richness to the pivotal redshift (mass) following the prescription in producing Fig.~\ref{fig:lmz}.
The red line in Fig.~\ref{fig:rchmz} represents a self-similar prediction in terms of richness, which is $\rich\propto\Mfiveoo(1+\redshift)^{0}$.
Our result shows a shallower mass trend deviating from one at a level of 
$\approx1.5\sigma$
and no significant redshift-dependent mass trend ($\left\langle\ln\rich|\Mfiveoo\right\rangle  \propto \ansdeltarch \times \ln\left( \Mfiveoo \right) \ln \left( 1 + \redshift \right)$).

We note one caveat as follows. 
The result of the \rich--\Mfiveoo--\redshift\ relation is obtained using the secure sample with a cut on \fcont, for which this is effectively equivalent to a redshift-dependent cut on the observed richness \citep{klein18,klein19}.
The fitting does not account for this richness-dependent selection, for which a more detailed modeling of the selection function is needed \citep[as in][]{grandis20}.
This effect is likely small compared to the current uncertainty, given the sample size used here.
We therefore defer a more complete modeling to a future work.

\section{Post-unblinding analysis}
\label{sec:postblinding}

After the submission of the paper and during the process of the referee review, we discovered two errors in the X-ray analysis that need to be corrected after the unblinding.
The first error is that incorrect redshifts were assigned to a small fraction of clusters ($11$ systems), such that the X-ray follow-up observables of those clusters were incorrect.
The second error is due to a bug in the code used to derive the bolometric luminosity, resulting in a systematic bias in \Lb\ at the level of $\approx-13\percent$.
To correct these errors, we reran the modeling of \Xlabel--\Mfiveoo--\redshift, where $\Xlabel \in \left\lbrace\Lx, \Lb, \Tx, \Mg, \Yx\right\rbrace$, without any further changes except the uniform prior applied on the normalization of the \Lb--\Mfiveoo--\redshift\ relation\footnote{We took this opportunity of the rerun to modify the uniform prior applied on  $A_{\Lb}$ from $\mathcal{U}(10^{41}\mathrm{erg/s},10^{44}\mathrm{erg/s})$ to $\mathcal{U}(10^{41}\mathrm{erg/s},10^{46}\mathrm{erg/s})$, since we found that the former, which was used, is only slightly above the upper limit of the $68\percent$ confidence level of the resulting $A_{\Lb}$.}, $A_{\Lb}$.
We find the results are not significantly affected by these errors, except for the normalization of the \Lb--\Mfiveoo--\redshift\ relation, which becomes higher due to the systematic bias.
We note that only the modeling of the follow-up X-ray scaling relations were affected by these errors, while the mass calibration results are intact in the post-unblinding analysis.

\section{The cluster mass}
\label{sec:clustermasstable}

We present the estimates of the cluster true mass (see Sect.~\ref{sec:mass_calibration_results}) for the eFEDS clusters in Table~\ref{tab:mass}.
In addition to the secure sample of $434$ clusters with $\fcont<0.2$, we also show the mass of clusters with $0.2\leq\fcont<0.3$.
This leads to a total number of $457$ clusters in Table~\ref{tab:mass}.

\begin{table*}
\caption{
The estimates of the cluster true mass \Mfiveoo\ of the eFEDS clusters.
The first column records the cluster name in the format of Jhhmmss.s$\pm$ddmmss, and the complete source name is ``eFEDS~Jhhmmss.s$\pm$ddmmss''.
The resulting mass estimates with (without) the core in the weak-lensing mass calibration are shown in the second and third (fourth and fifth) columns, which are the ensemble mass and the median of the mass posterior, respectively.
The quoted errors are the standard deviation of the mass posterior.
If available, the median and the standard deviation of the mass posteriors estimated with spectroscopic redshifts and without the core in the weak-lensing calibration are shown in the sixth column.
These estimates are expressed in terms of $\log\left(\frac{\Mfiveoo}{\Msunh}\right)$.
We additionally list the cluster with $0.2\leq\fcont<0.3$, marked by $\star$ in the cluster name.
}
\label{tab:mass}
\centering
\resizebox{1\textwidth}{!}{
\begin{tabular}{cccccc|cccccc|cccccc|cccccc|}
\hline\hline
 \multirow{2}{*}{Name} & \multicolumn{5}{c|}{ $\log\left(\frac{\Mfiveoo}{\Msunh}\right)$ } &
 \multirow{2}{*}{Name} & \multicolumn{5}{c|}{ $\log\left(\frac{\Mfiveoo}{\Msunh}\right)$ } &
 \multirow{2}{*}{Name} & \multicolumn{5}{c|}{ $\log\left(\frac{\Mfiveoo}{\Msunh}\right)$ } &
 \multirow{2}{*}{Name} & \multicolumn{5}{c|}{ $\log\left(\frac{\Mfiveoo}{\Msunh}\right)$ } \\
                                    & \multicolumn{2}{c}{  $R > 0.2\Mpch$ } & \multicolumn{2}{c}{  $R > 0.5\Mpch$ } & \multicolumn{1}{c|}{  with spec-\redshift }&
                                    & \multicolumn{2}{c}{  $R > 0.2\Mpch$ } & \multicolumn{2}{c}{  $R > 0.5\Mpch$ } & \multicolumn{1}{c|}{  with spec-\redshift }&
                                    & \multicolumn{2}{c}{  $R > 0.2\Mpch$ } & \multicolumn{2}{c}{  $R > 0.5\Mpch$ } & \multicolumn{1}{c|}{  with spec-\redshift }&
                                    & \multicolumn{2}{c}{  $R > 0.2\Mpch$ } & \multicolumn{2}{c}{  $R > 0.5\Mpch$ } & \multicolumn{1}{c|}{  with spec-\redshift }\\                                    
(1) & (2) & (3) & (4) & (5) & (6) &
(1) & (2) & (3) & (4) & (5) & (6) &
(1) & (2) & (3) & (4) & (5) & (6) &
(1) & (2) & (3) & (4) & (5) & (6) \\
\hline 
  J093712.8+031651    & $ 13.508 $    & $ 13.680 \pm 0.111 $    & $ 13.516 $    & $ 13.686 \pm 0.111 $   &                       &
  J083811.8-015934    & $ 14.737 $    & $ 14.725 \pm 0.083 $    & $ 14.742 $    & $ 14.730 \pm 0.085 $   &                       &
  J093520.9+023234    & $ 14.406 $    & $ 14.671 \pm 0.083 $    & $ 14.398 $    & $ 14.671 \pm 0.085 $   & $ 14.675 \pm 0.085 $    &
  J092121.2+031726    & $ 14.732 $    & $ 14.692 \pm 0.090 $    & $ 14.715 $    & $ 14.676 \pm 0.091 $   & $ 14.650 \pm 0.092 $    \\
\hline\hline
\end{tabular}
}
\end{table*}

\section{Corrections for the cluster mass in the literature}
\label{sec:correctionM500}

Different methods were used to estimate the cluster mass in the literature, therefore causing systematic discrepancy in \Mfiveoo.
For example, the mass estimated from X-ray assuming the hydrostatic equilibrium in ICM is suggested to be biased low by $\approx20\percent$ compared to that inferred from weak lensing \citep[see e.g.,][]{bocquet15,dietrich19}.
Therefore, we need to correct the different mass estimates from the literature to make a fair comparison in Fig.~\ref{fig:comparisons}.

We follow the scheme introduced in \cite{chiu18a} to correct for the systematic difference in \Mfiveoo\  from the literature.
For \cite{vikhlinin09a}, \cite{pratt09}, \cite{mahdavi13}, and \cite{lovisari15}, the cluster mass \Mfiveoo\ is estimated in X-rays assuming the hydrostatic equilibrium.
We therefore multiply their mass by a constant factor of $1.12$, which is quantified by comparing the masses inferred from X-ray and the weak-lensing technique \citep[see more in][]{dietrich19}.
For \cite{mahdavi13}, \cite{mantz16b}, and \cite{bulbul19}, the mass estimates are obtained from either a direct weak-lensing technique or a mass proxy with the weak-lensing calibration; we therefore do not apply any correction to them, assuming that their weak-lensing inferred mass is the true underlying mass.
For the simulated clusters presented in \cite{barnes17}, the true cluster mass is used, and hence no correction is required.

\end{CJK*}
\end{document}